\PassOptionsToPackage{numbers,square,sort&compress}{natbib}
\documentclass[examplefnt]{nowfnt}

\usepackage[utf8]{inputenc}
\usepackage{tikz}
\usetikzlibrary{arrows.meta, positioning, shapes.multipart, fit, calc}
\usepackage{rotating} 

\usepackage{amssymb,amsmath,amsfonts,eurosym,geometry,ulem,graphicx,caption,color,setspace,sectsty,comment,footmisc,pdflscape,array}
\usepackage[graphicx]{realboxes}
\usepackage{booktabs}
\usepackage{subcaption}
\usepackage{hyperref}
\usepackage{csquotes}

\newcolumntype{L}[1]{>{\raggedright\let\newline\\arraybackslash\hspace{0pt}}m{#1}}
\newcolumntype{C}[1]{>{\centering\let\newline\\arraybackslash\hspace{0pt}}m{#1}}
\newcolumntype{R}[1]{>{\raggedleft\let\newline\\arraybackslash\hspace{0pt}}m{#1}}

\usetikzlibrary{positioning,calc}

\title{Cyber Resilience in Next-Generation Networks: Threat Landscape, Theoretical Foundations, and Design Paradigms}

\maintitleauthorlist{
Junaid Farooq \\
University of Michigan-Dearborn \\
mjfarooq@umich.edu
\and
Quanyan Zhu \\
New York University \\
qz494@nyu.edu
}

\issuesetup
{%
 copyrightowner={J.~Farooq and Q.~Zhu},
 volume        = 17,
 issue         = 1-2,
 pubyear       = 2026,
 isbn          = xxx-x-xxxxx-xxx-x,
 eisbn         = xxx-x-xxxxx-xxx-x,
 doi           = 10.1108/FTNET-12-2025-0081, 
 firstpage     = 1, 
 lastpage      = 145
 }

\usepackage{mwe}

\author[1]{Farooq, Junaid}
\author[2]{Zhu, Quanyan}

\affil[1]{University of Michigan-Dearborn; mjfarooq@umich.edu}
\affil[2]{New York University; qz494@nyu.edu}

\articledatabox{\nowfntstandardcitation}
\journalbibstyle{numeric}

\begin{document}

\makeabstracttitle

\begin{abstract}

The evolution of networked systems, driven by innovations in software-defined networking (SDN), network function virtualization (NFV), open radio access networks (O-RAN), and cloud-native architectures, is redefining both the operational landscape and the threat surface of critical infrastructures. This article offers an in-depth, interdisciplinary examination of how resilience must be reconceptualized and re-engineered to address the multifaceted challenges posed by these transformations.


The article is structured as a cohesive framework that progresses from characterization of the threat landscape to formal definitions and metrics of resilience, followed by theoretical foundations rooted in control theory, game theory, learning, and network science. Building on these foundations, the work develops methodologies for quantitative risk assessment, including probabilistic modeling, network analytics, and emerging agentic AI approaches leveraging large language models (LLMs).

The final part of the article focuses on resilient system design paradigms in NextG networks, with emphasis on risk-aware resource orchestration, multi-agent reinforcement learning for slice management, and LLM-driven adaptive control in O-RAN environments. The overarching objective is to establish a principled connection between threat modeling, risk quantification, and control design, enabling the development of networked systems that can anticipate, withstand, and adapt to evolving adversarial conditions.


\end{abstract}

\chapter{Risk Landscape in Next Generation Networks}

This chapter frames the threat landscape of next-generation (NextG) networks from a system-level perspective, where risk emerges from interactions across tightly coupled domains rather than from isolated vulnerabilities. Architectural features such as virtualization, disaggregation, and programmability not only expand the attack surface but also introduce new pathways through which disruptions propagate and amplify across the network. By characterizing threats in terms of cross-domain interactions and systemic impact, this chapter establishes the foundation for the quantitative risk modeling and resilience design developed in subsequent chapters.

\section{Understanding Next-Generation Networks}

NextG networks, particularly those underpinned by 5G and its successors, represent a transformative shift from traditional telecommunications paradigms~\cite{lin20215g, tataria20216g, giordani2020toward}. They are not merely incremental upgrades in bandwidth or speed, but comprehensive re-architectures of the communication stack. NextG networks introduce architectural innovations such as software-defined networking (SDN)~\cite{benzekki2016software, zaidi2018will}, network function virtualization (NFV)~\cite{hawilo2014nfv, matias2015toward}, and open interfaces (e.g., O\mbox{-}RAN)~\cite{garcia2021ran, polese2023understanding}, coupled with unprecedented service flexibility enabled by network slicing, edge computing, and integration with cloud-native principles. These capabilities allow the network to evolve from a rigid infrastructure into a programmable platform that can dynamically host diverse services: from enhanced mobile broadband (eMBB) to ultra-reliable low-latency communication (URLLC)~\cite{8636206} and massive machine-type communication (mMTC)~\cite{popovski20185g}.

 NextG networks play a central role as the enabling fabric of the digital economy. Beyond supporting consumer connectivity, it supports critical infrastructures such as smart grids~\cite{dragivcevic2019future, leligou2018smart}, intelligent transportation~\cite{gohar2021role, guevara2020role}, remote healthcare~\cite{latif20175g, devi20235g}, and industrial automation~\cite{brown2018ultra, rao2018impact}. In many cases, mission-critical and safety-critical operations, such as autonomous vehicles coordinating in real time, tele-surgery surgeons or factories running with digital twins, depend directly on the deterministic performance and reliability of NextG services. Unlike previous generations, where mobile networks were largely confined to consumer use cases, NextG is tightly interwoven with societal functions, making its dependability a matter of public safety, national security, and economic competitiveness.

The same innovations that make NextG powerful also expand its attack surface~\cite{ahmad2019security}. Virtualization and softwarization increase the number of components and interfaces that must be secured~\cite{PZhang2018,Cui2021}. Disaggregated radio access network (RAN) architectures introduce new trust boundaries between radio units (RU), distributed units (DU), and centralized units (CU), each with their own control and data flows~\cite{groen2024securing, groen2024implementing}. Cloud-native deployments inherit the vulnerabilities of container platforms, orchestration systems, and third-party software supply chains~\cite{lyu2025supplychain}. Multi-tenant network slicing, while enabling efficiency and customization, raises the risk of cross-slice leakage and resource contention attacks~\cite{de2023survey}. Moreover, reliance on third-party vendors and open interfaces means that security is no longer centrally enforced but must be coordinated across heterogeneous ecosystems~\cite{abdalla2024end}.

Adversaries can target not only consumer data, but also industrial processes, public safety systems, and national infrastructure~\cite{konstantopoulou2023securing, suomalainen2021securing}. State-sponsored actors may view NextG as a strategic battleground for espionage and disruption. Criminal organizations may exploit vulnerabilities for ransomware or DDoS campaigns. Even unintentional misconfigurations in such complex systems can cause widespread outages, as recent incidents in cloud services have demonstrated~\cite{barona2017survey}. Thus, the security of NextG networks is not an auxiliary concern, but a prerequisite for their safe adoption.

\begin{figure}[t!]
    \centering
    \includegraphics[width=0.9\linewidth]{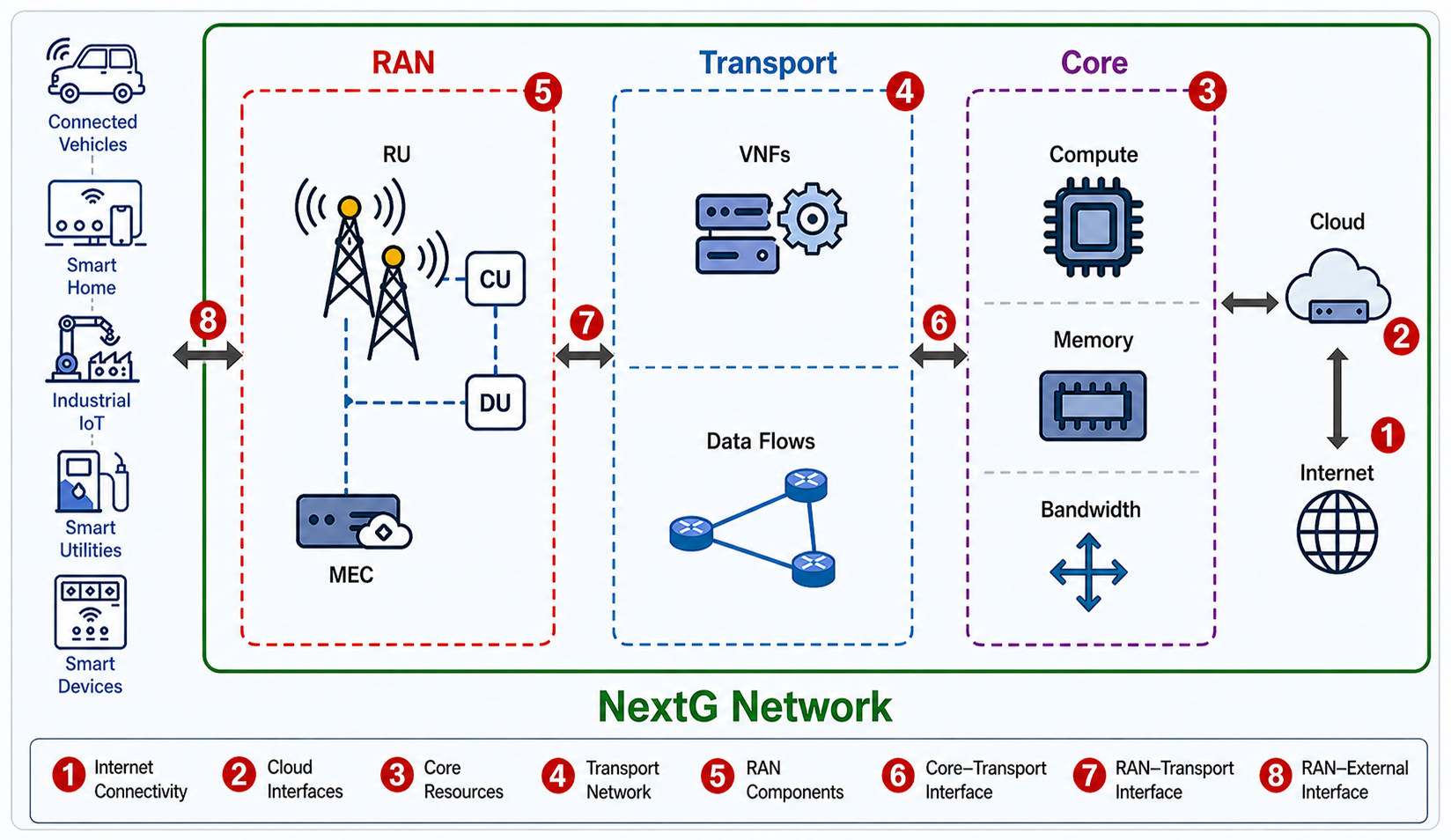}
    \caption{\small 
    Architecture of a NextG network across Core, Transport, and RAN domains, with external interfaces and annotated threat points highlighting cross-domain exposure.
    }
    \label{fig:nextg_network}
\end{figure}

\paragraph{Architectural Overview.}
Fig.~\ref{fig:nextg_network} provides a high-level view of a NextG network, organized into three main domains, core, transport, and RAN, with external interfaces to the Internet, cloud platforms, and end-user devices~\cite{rostami2017orchestration, parvez2018survey}. This layered view reflects both the functional separation of modern mobile networks and the trust boundaries through which data and control traffic must pass. Unlike legacy monolithic architectures, NextG systems are deeply disaggregated and virtualized, meaning that control and user planes are distributed across software-defined functions running on commodity hardware, often orchestrated in multi-vendor, cloud-native environments. This shift amplifies flexibility, but simultaneously creates a much larger and more complex attack surface.

The red numbering in Fig.~\ref{fig:nextg_network} highlights potential threat points, many of which cluster at inter-domain interfaces where assumptions of trust are weakest~\cite{mahyoub2024security}. Specifically, points (1)--(2) capture risks introduced by external connectivity to the Internet and cloud services. These connections expose the NextG infrastructure to distributed denial-of-service (DDoS) attacks, cloud-side supply chain risks, and data exfiltration attempts that originate outside the operator’s administrative control~\cite{agrawal2019defense}. Point (3) highlights vulnerabilities in Core network compute, memory, and bandwidth resources. Here, threats include denial-of-service against virtual network functions of the control-plane (VNFs), exploitation of software bugs in user-plane functions, and abuse of APIs in the service-based architecture (SBA)~\cite{wehbe2022security}.  

Point (4) identifies risks in the Transport domain, which is increasingly virtualized and programmable. Software-defined backhaul and fronthaul links enable flexible routing, but also introduce susceptibility to mis-configuration, routing manipulation, and man-in-the-middle attacks~\cite{pearce2013virtualization}. Points (5)--(7) focus on the disaggregated RAN, where RUs, DUs, CUs, and multi-access edge computing (MEC) nodes interact over open interfaces. While this disaggregation allows multi-vendor deployments and fine-grained scalability, it also exposes additional attack vectors such as rogue RUs injecting false signals, tampering with DU-CU fronthaul protocols, or exploiting MEC workloads that lack strong isolation~\cite{de2025radio}.  

Finally, point (8) highlights the risks at the end-user and critical-infrastructure edges, where heterogeneous devices connect to the network. These include consumer smartphones, IoT sensors, industrial control devices, and safety-critical endpoints in transportation and healthcare~\cite{zhu2021cybersecurity,yang2024game,zhu2018multilayer,oser2022risk}. Many of these devices operate with limited security features, making them attractive entry points for adversaries who can pivot into the core network. Their compromise can lead not only to localized data breaches, but also to systemic effects such as botnet-driven DDoS, signaling storms, or manipulation of mission-critical services~\cite{zhao2024integrated,pawlick2018modeling,mohebbi2015trust,asadi2024botnets}.

In what follows, we provide a detailed description of the background and function of each architectural component, followed by an analysis of its associated threat vectors. This approach highlights how the same architectural features that empower NextG networks with unprecedented flexibility and programmability also introduce novel security challenges that require proactive, responsive, and retrospective resilience mechanisms.

\subsubsection*{External Interfaces: Internet and Cloud}

NextG networks interface with the global Internet and cloud infrastructures to deliver agile, scalable services. The cloud provides elastic compute and storage resources to host network functions, analytics platforms, and AI/ML workloads, while Internet connectivity ensures global reachability and interconnection with third-party services~\cite{Duan2012}. These interfaces form the bridge between the telecom infrastructure and the broader digital ecosystem, allowing for cloud gaming, edge AI inference, and global roaming~\cite{taleb2016service}. In practice, major operators already deploy critical control-plane functions in public or hybrid clouds to take advantage of scalability, but this also shifts the trust boundary beyond traditional telecom domains.

At threat points~1--2, these same interfaces create the most exposed attack surfaces. Malicious or unverified cloud updates can compromise workloads, as seen in recent supply-chain attacks in which adversaries inserted backdoors into software libraries~\cite{yang2024stealthy}. Misconfigured APIs in the SBA may allow unauthorized requests or privilege escalation~\cite{da20245grecon}. Cloud-side compromises, such as a breach of an orchestration platform, can propagate directly into the control and user planes, undermining both confidentiality and availability~\cite{shringarputale2020co}. Peering and interconnection points with the Internet are also targets: attackers may inject spoofed BGP updates to reroute traffic, launch volumetric DDoS attacks that overwhelm ingress points, or exploit DNS misconfigurations to hijack sessions~\cite{suchara2011bgp}. 

Defenses must therefore include strong API authentication with mutual TLS, fine-grained access policies, encrypted traffic inspection to detect anomalies without violating privacy constraints, and hardware-based attestation of workloads to ensure that functions running in the cloud have not been tampered with. Current best practices also involve continuous posture assessment of cloud services~\cite{garg2025cloud}, supply-chain verification pipelines~\cite{williams2025research}, and DDoS scrubbing at peering edges~\cite{yazdani2024glossy}. However, despite these defenses, the speed and scale of cloud-native updates mean that even small lapses can expose entire operator networks to cascading risks.

\subsubsection*{Core Network: SDN and NFV Foundations}

The Core (threat point~3) is the programmable heart of NextG networks. It embodies the shift from hardware appliances to flexible, software-defined infrastructures. SDN centralizes policy and routing control, while NFV allows functions such as user plane functions (UPFs), session management functions (SMFs), and firewalls to run as virtualized workloads on commodity servers~\cite{farris2018survey}. Together, SDN and NFV enable elasticity, service agility, and cost efficiency: operators can spin up new slices for enterprises in minutes rather than months, dynamically reconfigure routing, or scale security functions with changing load~\cite{bannour2017distributed}.

However, the same virtualization and programmability that enable flexibility also concentrate risk. The SDN controller is a single point of control; if compromised, it can alter the forwarding rules to reroute or drop traffic, perform traffic intercept, or disable entire network slices~\cite{xi2022ruleout}. Similarly, a compromised NFV orchestrator could stop service instantiations, inject malicious VNFs, or disable defensive functions~\cite{zoure2022network}. Vulnerabilities in hypervisors or container runtimes further expand the attack surface, allowing adversaries to escape virtualized environments and access host infrastructure~\cite{ngoc2021mitigating}. 

The interfaces between the Core and Transport (threat point~4) are particularly critical, as they connect logical control functions with heterogeneous physical backhaul resources. Attackers who pivot across this boundary can perform lateral movement: e.g., leveraging a compromised control-plane API in the core to disrupt multipath routing in the transport. In the status quo, operators mitigate these risks with role-based access controls~\cite{al2019sdn}, segmented management planes~\cite{alabbad2022dynamic}, and continuous vulnerability scanning of VNFs~\cite{alnaim2024securing}. Yet, many deployments still rely on legacy VNFs with limited patch cycles, leaving core functions exposed to evolving zero-day exploits.

\subsubsection*{Transport Network: Programmable Data Flows}

The Transport domain (threat point~4) provides the backbone that connects the Core and RAN. It carries high-capacity data flows over backhaul and fronthaul links, supporting programmable traffic handling for bandwidth allocation, quality-of-service (QoS) enforcement, and tunneling. The shift to programmable transport allows dynamic reconfiguration of flows, enabling better load balancing and resilience against failures~\cite{alabbad2022dynamic}. In this domain manage traffic engineering, policy enforcement, and encryption services, providing agility that was not possible with static hardware routers.

At the same time, programmable transport introduces new vulnerabilities. Attacks against VNFs, such as traffic shapers or firewalls, can degrade data-plane integrity or bypass enforcement rules. Control-plane signaling in transport networks is also a target; for example, compromised SDN controllers managing transport switches could alter routing policies, enabling blackholing or interception of traffic~\cite{mas2020reliable}. Physical sabotage remains a real threat: cutting fiber links or tampering with roadside transport nodes can cause wide-area outages, while natural disasters (e.g., earthquakes, floods) highlight the fragility of centralized transport hubs~\cite{hayford2017causes}. 

Defensive practices today include encrypted overlays (e.g., IPSec or TLS tunnels) to protect data-plane confidentiality~\cite{tian2023network}, multipath routing to ensure that traffic can be re-routed around compromised or failed links~\cite{raposo2016machete}, and automated recovery protocols that detect faults and restore flows rapidly~\cite{lakshmi2023self}. Some operators are experimenting with AI-based transport anomaly detection, analyzing flow-level measurement to identify rerouting attacks or sudden congestion indicative of malicious interference. However, these systems are often reactive, and ensuring provable guarantees of transport resilience under adversarial conditions remains an open challenge.

\subsubsection*{Radio Access Networks}

The RAN (threat points~5--7) is undergoing a fundamental transformation with the advent of Open RAN (O-RAN)~\cite{polese2023understanding}, which supports disaggregation and openness. In this new architecture, monolithic base stations have been decomposed into Radio Units (RUs), Distributed Units (DUs), and Central Units (CUs), interconnected by standardized fronthaul and midhaul protocols~\cite{garcia2021ran}. This split architecture allows multi-vendor interoperability, flexible deployment, and optimized resource utilization. Multi-access edge computing extends compute and storage to the RAN domain, enabling ultra-low-latency services such as augmented/virtual reality, autonomous driving, and industrial process control~\cite{filali2020multi}. The RAN Intelligent Controller (RIC), both near-real-time and non-real-time, introduces programmability into the RAN, supporting AI-driven optimization of spectrum allocation, handover management, and interference mitigation~\cite{balasubramanian2021ric}.

However, this disaggregation significantly enlarges the attack surface. The interfaces between RU, DU, and CU are susceptible to man-in-the-middle and replay attacks, where adversaries manipulate control messages to degrade performance or exfiltrate data. Rogue base stations (so-called “IMSI catchers” or “stingrays”) can spoof legitimate signals, harvest subscriber data, or force devices into weaker encryption modes~\cite{huang2023developing}. Jamming and interference remain perennial threats, but in NextG they pose increased risks to URLLC applications where even millisecond interruptions can lead to catastrophic outcomes~\cite{abou2024federated}. Vulnerabilities in RIC xApps or rApps could propagate faulty policies throughout the RAN, creating systemic instability~\cite{hung2024security,fernando2025securing}. The MEC nodes, located on the edge, inherit risks from the cloud and RAN domains: a compromised MEC platform could host malicious workloads, manipulate latency-sensitive data flows, or serve as a pivot for lateral movement to core functions~\cite{cheng2022attack,li2024decision, pietrantuono2023testing}.  

In the status quo, operators deploy measures such as mutual authentication between fronthaul links, encryption of control-plane signaling, and geofencing of base stations. RIC security frameworks are still nascent, with ongoing work in the O-RAN Alliance to define trust anchors and policy controls for xApp/rApp onboarding. MEC security largely depends on cloud-native defenses (container isolation, trusted execution environments), but uniform standards are lacking, leaving implementations uneven between vendors.

\subsubsection*{Heterogeneity, Scalability, and Edge Integration}

NextG networks must simultaneously support eMBB, URLLC, and mMTC. This heterogeneity drives the deployment of dense small-cell infrastructures, mobility-aware scheduling, dynamic spectrum sharing, and real-time orchestration at the edge~\cite{popovski20185g}. It also requires the integration of heterogeneous devices, from smartphones and IoT sensors to industrial robots and connected vehicles, each with distinct performance and security profiles.

At the threat point~8, the diversity of endpoints multiplies the attack opportunities. Millions of IoT devices, often designed with minimal security features, represent a vast distributed attack surface~\cite{yu2015handling}. Weak authentication, default credentials, or unpatched firmware allow adversaries to conscript devices into botnets, launch DDoS attacks, or act as footholds for lateral movement into higher-value network domains~\cite{wu2024your}. Smart vehicles and industrial robots, if compromised, can be manipulated to cause physical damage or safety incidents \cite{zhu2021cybersecurity}. Edge integration further complicates matters: workloads that migrate between devices, MEC nodes, and cloud data centers create opportunities for adversarial manipulation of data-in-transit or manipulation of scheduling decisions~\cite{puthal2017seen}.  

In practice, today’s defenses include device attestation protocols~\cite{atalay2024openran}, slice-aware access control in RAN~\cite{alam2025comprehensive}, and anomaly detection for IoT traffic patterns~\cite{aversano2021effective}. However, the scale of device diversity makes centralized approaches difficult to maintain. Emerging research emphasizes AI-driven interference management~\cite{ngo2024ai}, federated learning for anomaly detection at the edge~\cite{attanayaka2023peer}, and trust frameworks that can dynamically assign risk scores to endpoints based on behavioral evidence~\cite{groen2024implementing}. Despite these developments, the heterogeneity of endpoints remains one of the weakest links in NextG security.

\subsubsection*{Interfaces with Critical Infrastructures and Supply Chains}

Beyond consumer connectivity, NextG networks are envisioned as enablers of critical infrastructures such as healthcare, transportation, logistics, and energy systems~\cite{ishii2022security}. These verticals rely on NextG’s low-latency, high-reliability features to deliver real-time monitoring, automation, and control. Examples include telesurgery relying on URLLC slices, autonomous vehicles coordinated through V2X communications, and energy grid management via mMTC-enabled sensors. This deep integration means that NextG is not simply a communications platform, instead, it becomes a foundational pillar for future smart societies.

Therefore, risks at the threat point~8 extend beyond technical failures to systemic consequences. A vulnerability in a baseband unit sourced from an untrusted supplier could be exploited to surveil or disrupt entire regions of critical infrastructure~\cite{lyu2024mapping}. Firmware backdoors in MEC equipment could allow attackers to trigger cascading failures in hospitals or transportation hubs~\cite{kieras2022iot, kieras2021scram}. Supply chain dependencies are especially concerning: as disaggregation encourages multi-vendor ecosystems, ensuring that every component (hardware, firmware, orchestration software) is free from hidden vulnerabilities becomes a daunting task. Nation-state actors may exploit these dependencies to insert long-term backdoors, while criminal organizations may leverage counterfeit or tampered equipment to launch persistent attacks.

Mitigation requires secure development lifecycles between all vendors, rigorous vendor validation and certification, and the application of Zero Trust principles not just within the telecom network but across sectoral boundaries~\cite{lyu2024zero}. Current efforts include initiatives such as the GSMA’s Network Equipment Security Assurance Scheme (NESAS)~\cite{lachmund2015standards} and US / EU proposals for trusted vendor certification~\cite{benzaid2021trust, paskauskas2025preliminary}. However, global supply chains remain opaque and geopolitical tensions complicate the consensus on trust anchors. As a result, securing the integration of critical infrastructure with NextG remains one of the most pressing challenges for both operators and policy makers.

\section{Integrated Risks Across Next-Generation Networks}

Although individual vulnerabilities in NextG networks are serious on their own, the most disruptive scenarios often arise when adversaries combine multiple threat vectors across domains. These integrated risks highlight the systemic nature of next-generation networks and demonstrate why piecemeal defenses are inadequate. Unlike isolated incidents, integrated attacks exploit interdependencies between cloud, Core, Transport, and RAN domains, amplifying their impact and complicating detection and mitigation. Table~\ref{fig:layered-threat-taxonomy} organizes the heterogeneous threat space of NextG networks into a layered taxonomy that links vectors, targets, and representative real world incidents.

\subsubsection*{Advanced Persistent Threats (APTs)}

APTs represent long-term stealthy campaigns typically orchestrated by nation-states or highly resourced adversaries~\cite{alshamrani2019survey}. Their hallmark is patience: adversaries gradually infiltrate, establish persistence mechanisms, and move laterally to expand control~\cite{wang2024combating,zhu2018multi,huang2020dynamic}. In NextG networks, the programmability of SDN and the elasticity of NFV create attractive avenues for APT operators. Similarly, the disaggregated and open interfaces of O-RAN provide additional footholds for exploitation.

A hypothetical but plausible chain of events begins with adversaries gaining initial access through a misconfigured cloud API (threat point~2), as seen in the 2019 Capital One breach, where a misconfigured cloud firewall exposed sensitive data~\cite{khan2022systematic}. From there, attackers could compromise an NFV orchestrator in the Core (threat point~3), similar in spirit to the SolarWinds supply-chain compromise of 2020, where trusted software updates were weaponized to deploy malicious code across thousands of organizations~\cite{tan2025advanced}. Once persistence is established, adversaries could pivot into RAN controllers (threat points~5--7). Rogue base stations such as IMSI catchers have already been widely documented in law enforcement and criminal contexts, showing that manipulation of signaling channels is feasible~\cite{palama2021imsi}. In an integrated APT scenario, this could escalate to manipulating radio scheduling or selectively disrupting connectivity in politically sensitive regions \cite{yang2025toward,yang2025multi}.  

Real-world parallels exist: the U.S. Cybersecurity and Infrastructure Security Agency (CISA) has repeatedly warned that advanced state-backed actors are probing telecom networks for long-term access. Incidents such as the compromise of Ukraine’s telecom providers during the early stages of the 2022 war illustrate how adversaries use cross-domain access, covering network control and radio functions, to monitor and disrupt communications on a scale~\cite{axon2024private}. Such integrated footholds allow covert traffic monitoring, long-term data exfiltration, or synchronized disruption campaigns designed to evade conventional detection, ultimately eroding trust in the entire NextG ecosystem.

\subsubsection*{Ransomware in Telecom Infrastructure}

Ransomware, once restricted to laptops and enterprise IT, has become a strategic weapon against critical infrastructure~\cite{oz2022survey}. In NextG networks, where continuous availability underpins essential services, from emergency dispatch to industrial automation, the stakes are exponentially higher. Telecom operators also manage vast orchestration platforms and virtualized services, making them prime targets for ransomware~\cite{bhardwaj2024cybercrime,zhao2021combating}.

The 2021 Colonial Pipeline attack~\cite{beerman2023review} demonstrated how ransomware against operational technology (OT) can cascade into widespread economic disruption, while the 2021 Kaseya VSA incident~\cite{bhunia2025analyzing} showed how attackers can simultaneously compromise hundreds of organizations by targeting a single orchestration platform. Similar logic applies in NextG contexts. A sophisticated ransomware campaign might launch simultaneously in multiple domains: disrupting Transport VNFs managing QoS and tunneling (threat point~4) ~ \cite{zoure2022network}, degrading connectivity across metropolitan areas, while simultaneously locking MEC nodes (threat point~7) to paralyze latency-sensitive workloads such as telemedicine or AR/VR services~\cite{ranaweera2021mec}.  

Real-world telecom incidents show the plausibility of such campaigns. In 2022, Costa Rica’s government was crippled by Conti ransomware, which disrupted tax collection and customs systems~\cite{alzahrani2022analysis}. In 2021, Irish health services were paralyzed by a ransomware attack, leading to canceled medical appointments and delayed emergency care~\cite{moore2023resilient}. While not limited to NextG infrastructure, these incidents illustrate the societal stakes when connectivity and cloud orchestration are held hostage. In a NextG context, ransomware could encrypt orchestration databases or slice management functions, preventing operators from controlling traffic flows or restoring workloads. The leverage comes not just from data encryption but from paralyzing service continuity across critical sectors. Even short-lived outages could ripple into public safety failures or financial market disruptions, making operators highly vulnerable to extortion.

\begin{sidewaystable}[htbp]
\centering
\includegraphics[width=0.90\textheight]{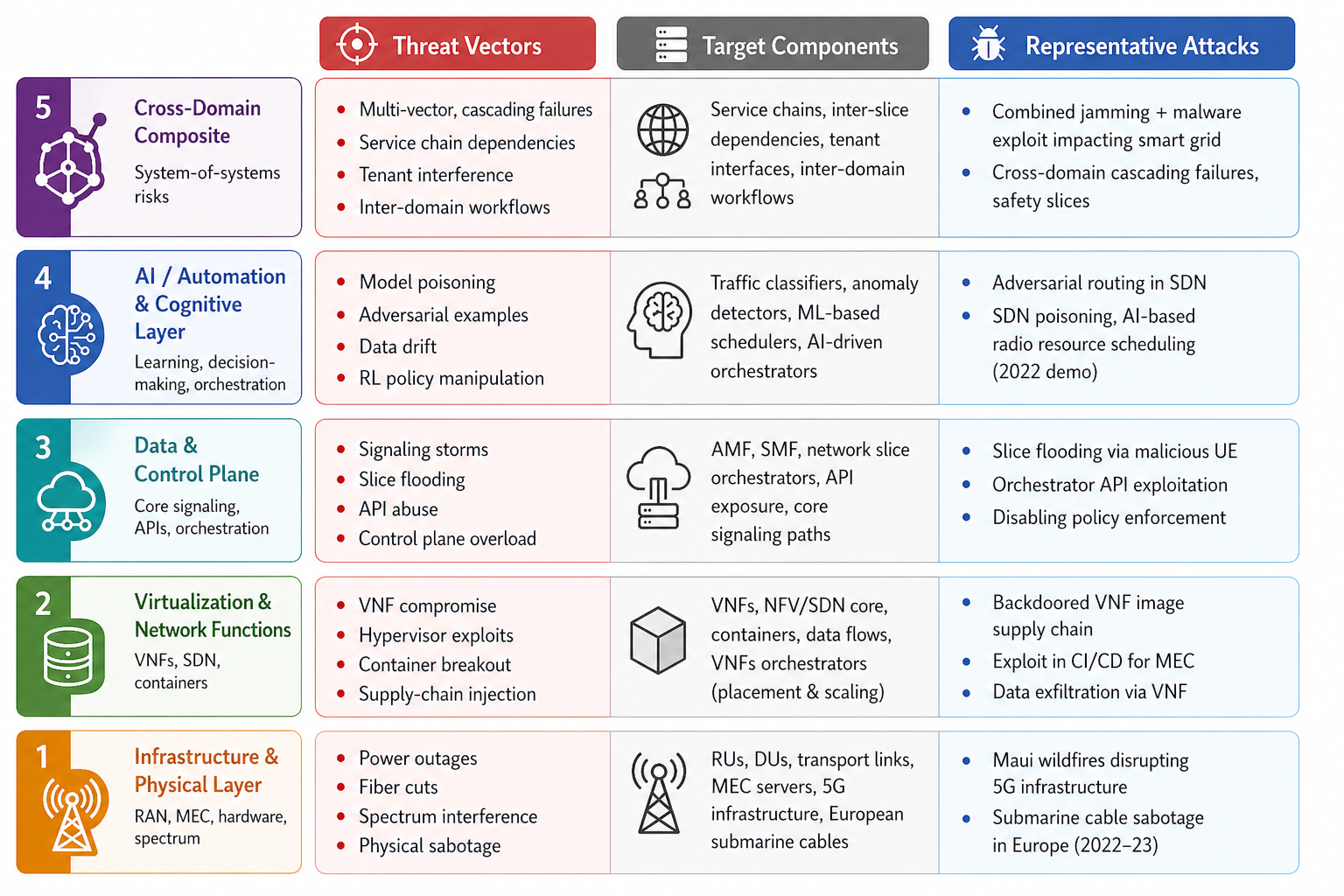}
\caption{\small Layered taxonomy of emerging threats in NextG networks. 
The figure organizes threats across five layers from infrastructure to cross-domain composite systems and presents threat vectors, target components, and representative attacks. 
This visualization emphasizes the structural relationships across layers and improves readability compared to tabular representation.}
\label{fig:layered-threat-taxonomy}
\end{sidewaystable}

\section{Organization of This Article}

This article is organized around a system-level view of cyber resilience in NextG networks. It moves step by step from motivation to measurement, analysis, and design. The main idea is that \emph{risk} connects threats, models, and decisions, while \emph{theoretical foundations} provide the tools to measure, understand, and reduce that risk to achieve resilience.  
The chapters form a connected pipeline, where metrics guide how systems are modeled, models support risk assessment, and risk assessment informs how systems are designed.

\begin{figure}[t]
\centering
\includegraphics[trim={0 0 0 1.3in}, clip, width=\linewidth]{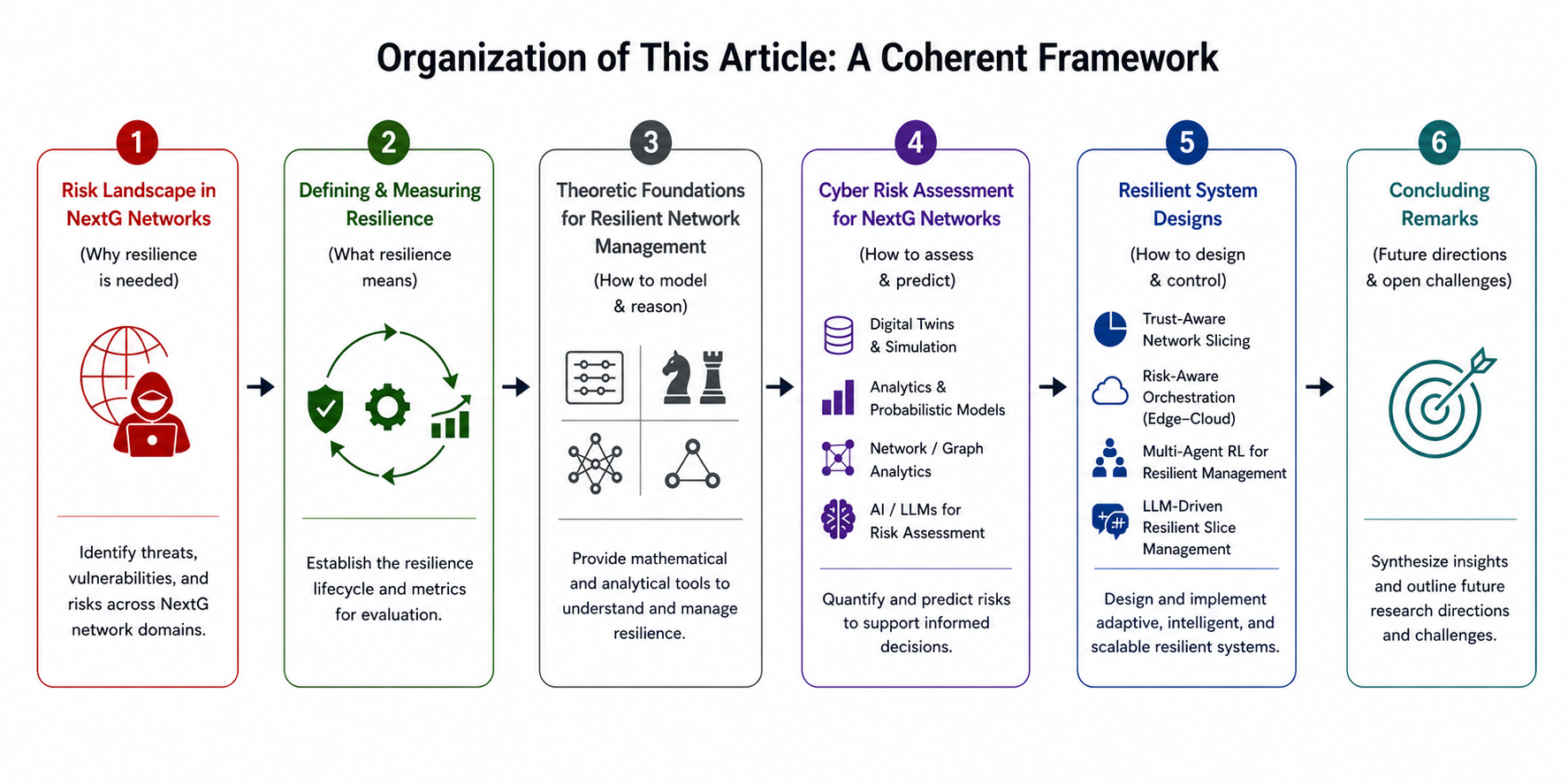}
\caption{\small 
Organization of the article from threat landscape to resilience definition, theoretical foundations, risk assessment, and resilient system design.
}
\label{fig:organization_framework}
\end{figure}

Chapter~1 introduces the \emph{threat landscape}, which motivates the need for resilience. It characterizes the evolving attack surface in NextG architectures, including SDN, NFV, and O-RAN, and highlights how vulnerabilities emerge from cross-layer interactions across core, transport, RAN, and edge/cloud domains. The chapter emphasizes that modern threats are adaptive, interconnected, and system-wide, thereby necessitating a holistic and quantitative treatment of risk.

Chapter~2 defines \emph{resilience} as a measurable and operational concept through proactive, responsive, and retrospective mechanisms. Beyond conceptual definitions, it introduces evaluation metrics and performance criteria that quantify system behavior under disruption. These metrics establish the foundation for subsequent modeling and provide the lens through which risk is interpreted and compared.

Chapter~3 develops the \emph{theoretical foundations} that support both risk assessment and system design. It presents a spectrum of analytical tools, including control theory for feedback and stability, game theory for adversarial and strategic interactions, learning theory for adaptation under uncertainty, and network science for modeling interdependencies and cascading effects. These foundations are not used in isolation; rather, they are combined and leveraged in later chapters to construct quantitative risk models and to design resilient control mechanisms.

Chapter~4 focuses on \emph{risk assessment}, serving as the central analytical layer of the article. It spans a range of methodologies, from classical approaches such as probabilistic risk assessment (PRA) and reliability analysis, to modern techniques including digital twins, network and graph-based analytics, and agentic AI frameworks leveraging large language models. By integrating these approaches, the chapter provides a comprehensive toolkit for quantifying, predicting, and interpreting risk in complex and dynamic environments. Importantly, this chapter operationalizes the metrics and theoretical tools developed earlier, translating them into decision-relevant insights.

Chapter~5 presents \emph{design paradigms} for resilient systems, building directly on the insights from risk assessment and the foundations established in Chapter~3. Rather than prescribing a single solution, it adopts a case-study-driven approach across multiple application domains, such as network slicing, edge-cloud orchestration, and multi-agent system management. These paradigms illustrate how different system contexts require tailored design strategies, all grounded in a common framework of risk-aware decision-making. The chapter highlights that resilience is achieved through the integration of modeling, prediction, and control.

Chapter~6 concludes by synthesizing the framework and outlining future directions, emphasizing emerging challenges such as AI-enabled resilience, domain-specific constraints in cyber-physical systems, and the need for scalable, adaptive, and interpretable solutions. It reinforces the perspective that resilience is not static, but evolves through continuous advances in modeling, data, and system design.

\chapter{Defining and Measuring Resilience}

Chapter 1 established that NextG networks operate as tightly coupled, multi-domain systems in which disruptions propagate across cloud, core, transport, RAN, and edge layers. In such environments, ensuring dependable operation requires a framework that accounts for uncertainty, interdependence, and continuous system evolution.

This chapter provides a formal definition of resilience in NextG networks as a system-level property. It characterizes resilience through proactive, responsive, and retrospective mechanisms operating across different stages of disruption. The chapter further introduces quantitative metrics for evaluating system performance under adverse conditions, establishing the foundation for the modeling and design methodologies developed in subsequent chapters.

\section{Understanding Network Resilience in 5G Networks}

Network resilience refers to the ability of NextG communication systems to {prepare for, withstand, recover from, and adapt to disruptions}, particularly those induced by cyber threats, software misconfigurations, or cascading hardware failures. Unlike previous generations of networks, 5G introduces new complexities through \textit{ultra-dense deployments, virtualization, network slicing, and multi-access edge computing (MEC)}, all of which necessitate a fundamental shift from classical robustness-focused designs to resilience-centered architectures. The dynamic, programmable, and distributed nature of 5G infrastructure demands mechanisms that can autonomously respond to failures, maintain service continuity across heterogeneous domains, and evolve in response to emerging threats.

\subsection*{Resilience vs. Robustness}

{Robustness} in 5G networks typically involves the design of base stations (gNodeBs), core components, and control functions to \textit{withstand predefined disruptions}. For example, a robust 5G base station might incorporate dual power sources, redundant transmission links, and hardened hardware to tolerate power failures or backhaul outages~\cite{tang2020shiftguard,chen2016game,zhu2019multilayer}. Robust control planes can include failover routes and standby network functions to handle predefined load spikes or equipment failures~\cite{fonseca2017survey,zhu_basar_resilience2024}. However, robustness assumes \textit{static failure models} and is primarily aimed at preserving normal operations under known or statistically likely stressors~\cite{baker2008assessment}. It does not fully account for intelligent, evolving, or stealthy threats, such as a novel zero-day exploit or an adversarial AI agent manipulating control messages.

In contrast, {resilience} in 5G embraces the reality that disruptions, especially those of \textit{adaptive cyber attackers, supply-chain exploits, or unforeseen cross-slice dependencies}, cannot always be predicted~\cite{benslimen2021attacks, de2024performance}. Resilience emphasizes the ability of the network to detect anomalies, contain damage, restore functionality, and adapt over time~\cite{windle2011resilience}. If a rogue application in a network slice causes signaling storms that congest shared core functions (such as the Access and Mobility Management Function, AMF), a resilient architecture would leverage real-time measurement and AI-driven diagnostics to detect irregularities, automatically quarantine or rate-limit the faulty slice, and re-balance affected virtual network functions (VNFs) using orchestration platforms like O-RAN Service Management and Orchestration (SMO) or Kubernetes-based network function virtualization (NFV) managers~\cite{taleb2016service, vittal2023revamping}.

 Consider a 5G-enabled autonomous vehicle that relies on an ultra-reliable low-latency (URLLC) slice to communicate with roadside units and edge-hosted control services~\cite{zhou20245g}. Suppose that an adversary initiates a volumetric DDoS attack against the edge server hosting the URLLC slice's control logic, thereby overwhelming the MEC node~\cite{ma2024survey}. In a merely robust system, protective firewalls or ingress filtering might block known traffic sources or hardcoded IP addresses, but this would be insufficient if the attack employs distributed botnets with adaptive spoofing or uses legitimate service ports to evade detection.

In a resilient 5G network, the MEC controller would first detect abnormal traffic patterns through real-time flow analytics and alert the central orchestrator~\cite{dealmeida2021abnormal}. The affected URLLC slice would then be migrated to a neighboring edge site using live VNF migration~\cite{wu2025poster}, preserving critical communication for the autonomous vehicle. Non-essential services such as vehicle infotainment or measurement might be de-prioritized or throttled to ensure that bandwidth and compute resources remain available for safety-critical operations~\cite{moore2023ssxapp}. In addition, the orchestration system could initiate a forensic analysis to trace the origin of the attack~\cite{sharevski2018towards}, update threat intelligence repositories~\cite{brown2015cyber}, and generate new traffic rules to prevent recurrence~\cite{pan2013proactive}. As a result, the vehicle would maintain its control plane connection with negligible latency variation, preserving navigational safety and mission continuity.

This sequence of detection, containment, service continuity, and adaptive learning demonstrates how resilience in 5G systems exceeds classical robustness by enabling autonomous, context-aware, and mission-prioritized responses in the face of unpredictable disruptions.

\subsection{Resilience vs. Fault-Tolerance and Reliability}

5G networks employ \emph{fault-tolerant components}, such as redundant core servers, geographically distributed gNodeBs, and standby VNFs, to ensure that localized faults, such as hardware failures, software crashes, or power disruptions, do not result in a complete service outage~\cite{de2024performance}. These fault-tolerance mechanisms are typically designed to absorb failures within narrowly defined bounds and are often evaluated using \emph{reliability metrics} such as ``five nines'' (99.999\%) availability~\cite{hauer2020meaningful}. These metrics quantify the expected uptime of system components and provide assurance that under nominal operating conditions, critical functions will remain continuously available. This level of availability is particularly important for mission-critical services, including emergency communications~\cite{deepak2019overview}, industrial automation~\cite{brown2018ultra}, and autonomous vehicle coordination~\cite{shah20185g}.

However, fault-tolerance in 5G systems is generally \textit{pre-configured}, meaning that it depends on static redundancy models and assumptions about the nature of failures. For example, a centralized 5G core deployment might rely on active-passive replication, where a primary node manages operations and a standby replica takes over in the event of failure~\cite{ito2025multipath}. While effective for isolated hardware issues or predictable load spikes, this model becomes inadequate in scenarios involving \textit{multi-site failures}, such as those triggered by cascading software misconfigurations introduced through continuous integration/continuous deployment (CI/CD) pipelines~\cite{zhong2020network}. Additionally, coordinated cyber-attacks targeting shared management APIs or exploiting overlooked inter-slice dependencies can propagate rapidly across control and user planes, rendering static redundancy models insufficient.

In contrast to static fault-tolerance, \emph{resilience} in 5G networks encompasses a dynamic and adaptive approach to failure management that spans detection, containment, recovery, and adaptation. Resilience begins with real-time detection, in which measurement data, such as slice-level latency, packet loss, and signaling rate, are continuously monitored by AI-powered analytics engines to uncover abnormal patterns that might signal imminent failures or security breaches~\cite{sevgican2020intelligent}. Upon detection of an anomaly, containment mechanisms are activated. These may involve automated micro-segmentation of affected network functions~\cite{benzaid2025multi}, isolation of suspicious user equipment (UE), or even temporary quarantine of compromised network slices to prevent lateral threat propagation~\cite{wu2025poster}.

Once the immediate impact is contained, resilient 5G systems initiate recovery by reallocating services and workloads to unaffected parts of the infrastructure. This may include dynamic migration of VNFs to alternate MEC sites or centralized clouds using Kubernetes-based orchestration platforms, ensuring service continuity with minimal human intervention~\cite{tsourdinis2023drl}. Finally, adaptation involves post-incident analysis, during which logs, alarms, and incident traces are analyzed to refine detection thresholds, update security policies, and optimize traffic engineering rules, thereby improving the network's ability to cope with similar incidents in the future.

Consider, for example, a 5G-enabled smart manufacturing facility that relies on private network slicing to support distinct workflows, such as robotic control, quality assurance, and remote monitoring~\cite{fowler20235g}. Suppose a routine firmware update introduces unintended latency in a critical robotic controller operating on an URLLC slice. A fault-tolerant system might attempt to roll back the firmware if a snapshot is available, but this assumes that the error was detected quickly and that the backup mechanism is reliable. However, a truly resilient system would respond more intelligently. It could detect the latency anomaly through slice-level measurement~\cite{mekki2021scalable}, deprioritize or throttle non-essential background processes such as periodic sensor uploads, and reallocate compute resources to latency-sensitive control functions~\cite{borsatti2022mission}. Furthermore, the system could autonomously reschedule a full maintenance window during a low-activity shift and use predictive analytics to validate that other slices or services would not be impacted by the update.

\subsection{Multi-Dimensional and Dynamic Nature of 5G Resilience}

Resilience in 5G networks is inherently \emph{multi-dimensional}, operating across cyber, physical, organizational, and temporal layers. In the \emph{cyber domain}, 5G networks rely heavily on software-defined components such as SDN controllers, VNFs, and MEC platforms. These components are vulnerable to zero-day exploits, misconfigurations, and privilege escalations, particularly in containerized or virtualized environments. A resilient system incorporates safeguards such as secure boot protocols, runtime attestation for trust verification, and continuous intrusion detection mechanisms that can flag anomalous behaviors across the service-based architecture (SBA) interfaces~\cite{ortiz2020inspire, amachaghi2024survey}.

In the \emph{physical dimension}, threats such as localized power failures, electromagnetic interference, or RF jamming can compromise the availability of radio resources, especially in dense urban small cell deployments. Resilience strategies at this layer may involve power-aware load balancing, fallback to macro cells with acceptable latency penalties, or even the use of hybrid satellite-terrestrial links for emergency failover~\cite{han2017network}.

On the \emph{human and organizational front}, operator errors, poor policy management, and inadequate role-based access controls continue to pose major risks to network reliability. To mitigate these, 5G management frameworks must integrate tools that enforce policy compliance, audit configuration changes, and provide automated rollback capabilities to revert to known good states when anomalies are detected~\cite{cherrared2019survey,xing2023enabling,huang2021duplicity,casey2016compliance,casey2015compliance}.

Finally, resilience unfolds along the \emph{temporal dimension}. It must account not only for immediate responses to incidents but also for long-term adaptations. Short-term responses may involve sub-second failover to alternate nodes or spectrum slices, while longer-term adaptation might include retraining AI models used for threat detection, refining orchestration workflows based on observed inefficiencies, or reallocating spectrum in anticipation of future usage patterns. These dimensions define the evolving and context-aware character of resilience in 5G systems, distinguishing it fundamentally from static fault-tolerance and traditional reliability engineering.

\begin{figure}[h!]
    \centering
    \includegraphics[width=0.95\linewidth]{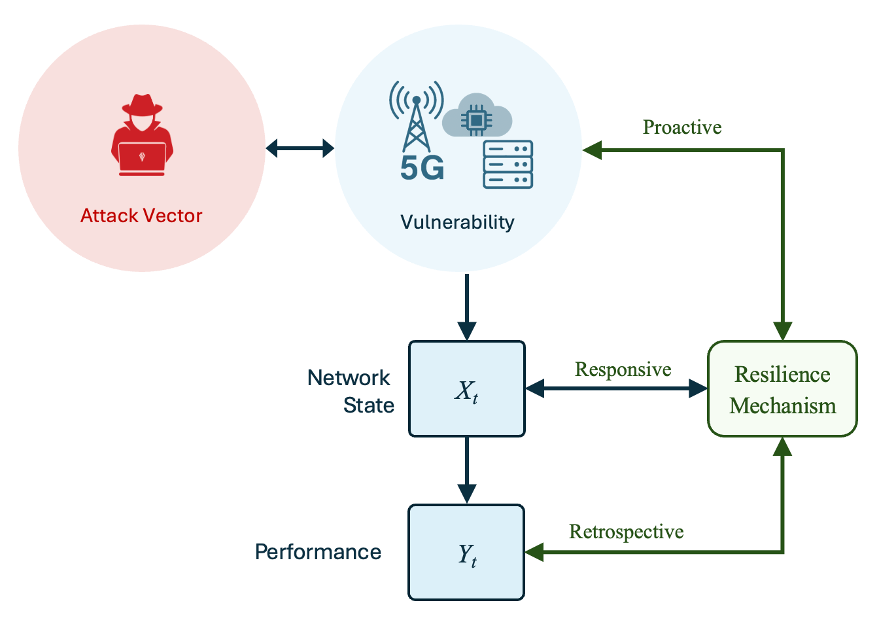}
    \caption{\small Interaction between attack vectors, vulnerabilities, network state, and resilience mechanisms. The diagram highlights three classes of defense: proactive (anticipating and hardening against vulnerabilities), responsive (adapting to the current network state $X_t$), and retrospective (learning from performance outcomes $Y_t$ to improve future resilience).}
    \label{fig:resilience_mechanism}
\end{figure}

\section{Types of Resilience Mechanisms in 5G Networks}

Fig.~\ref{fig:resilience_mechanism} illustrates the dynamic interaction between adversarial attack vectors, system vulnerabilities, network state, and resilience mechanisms. An attacker exploits a vulnerability, driving the system into a perturbed state $X_t$, which in turn degrades the observable performance $Y_t$. Resilience mechanisms intervene at three stages: proactively by anticipating and mitigating vulnerabilities before they are exploited~\cite{vidal2019framework, alnfiai2025ai}, responsively by adapting to changes in the state of the network as attacks unfold~\cite{fazrina2024securing,carvalho2020agile}, and retrospectively by analyzing results to recover performance and strengthen defenses~\cite{fan2025novel, liu2024post}. This tripartite structure highlights that resilience in 5G is not a static attribute, but a dynamic process that unfolds over time.

\subsection{Proactive Resilience}

Proactive resilience corresponds to the upper branch of Fig.~\ref{fig:resilience_mechanism}, where defenses act upstream of disruption, before an attack has an opportunity to take effect. The essence of this stage is to design solutions ahead of time that eliminate or neutralize vulnerabilities that are already known and prevent them from being exploited~\cite{zhu2024disentangling,zhu2020cross,chen2019interdependent,zhu2015game}. In 5G, the importance of proactive resilience is amplified by the vast attack surface created through virtualization, multi-tenant network slicing, and reliance on heterogeneous third-party software and hardware components~\cite{lyu2024mapping}. Because many of these risks stem from well-characterized classes of threats, such as unpatched software vulnerabilities, misconfigured APIs, or insecure interfaces, they can, in principle, be mitigated or prevented through deliberate architectural choices and rigorous security engineering.

In today’s deployments, proactive resilience is realized primarily through layered security hardening and standards-driven compliance. NFV on cloud-native platforms are hardened with continuous patching pipelines~\cite{bell2017agile}, container scanning~\cite{javed2021evaluation}, and automated vulnerability assessments~\cite{pan2024towards}. However, patching cycles often lag behind zero-day exploit timelines, leaving a window of exposure. Similarly, API protection mechanisms such as TLS encryption and access tokens are widely deployed, but misconfiguration or insufficient key management still accounts for a significant portion of breaches. The SBA, while enabling agility and programmability, expands the number of interfaces to be secured, multiplying opportunities for attackers. 

Operators currently mitigate these risks by enforcing security baselines mandated by 3GPP and ETSI standards, but adoption is uneven between vendors and regions. Multi-vendor interoperability, especially in O\mbox{-}RAN ecosystems, complicates proactive resilience: a gNodeB provided by one vendor may not seamlessly integrate with the security hardening protocols of a RAN Intelligent Controller (RIC) from another~\cite{groen2024implementing}. As a result, despite the availability of proactive mechanisms in principle, the operational reality is one of patchwork integration, where the weakest link may dictate the effective resilience of the system~\cite{soltani2025intelligent}.

\paragraph{Examples of Proactive Measures.}
Examples of proactive measures span architectural design, access control, redundancy, and automated assurance. One fundamental mechanism is logical segmentation through network slicing, which confines faults and attacks to the affected slice and prevents lateral propagation across services. For example, a DDoS attack targeting an enhanced mobile broadband (eMBB) slice should not affect URLLC slices that support critical functions such as telemedicine or industrial automation. Slice-aware resource allocation at the RAN and core further reinforces this isolation, ensuring that bandwidth, processing, and control signaling are insulated between tenants.

Another pillar of proactive resilience is \emph{zero-trust authentication}~\cite{he2022survey,ge2022mufaza}, implemented across the SBA. In this model, every API call and microservice interaction must be explicitly authenticated and authorized, eliminating implicit trust assumptions between network functions. Techniques include mutual transport layer security (TLS) for service-to-service authentication, token-based access enforcement through OAuth 2.0 or JSON Web Tokens, and fine-grained policy enforcement points deployed within the Service Mesh. By enforcing least-privilege access across hundreds of microservices, zero-trust frameworks eliminate entire categories of lateral movement attacks that have historically compromised monolithic core networks~\cite{niakanlahiji2020shadowmove}.

Hardware and service redundancy provide a complementary layer of defense. Redundant gNodeBs, distributed UPFs (User Plane Functions), and geo-redundant data centers ensure that failures or targeted attacks against single nodes do not escalate into systemic outages~\cite{ito2025multipath}. High-availability orchestration frameworks (e.g., Kubernetes with self-healing deployments) can restart failed network functions automatically, while distributed consensus mechanisms in control-plane clusters maintain service continuity even under node compromise. This approach shifts the system away from fragile single points of failure and toward fault-tolerant cloud-native operations.

In practice, major operators are also beginning to implement \emph{continuous compliance pipelines}. These pipelines automatically test slice configurations, access policies, and orchestration templates against predefined security baselines before deployment~\cite{xu2012model}. For example, Infrastructure-as-Code templates are validated for adherence to 3GPP, ETSI, and operator-specific security policies, reducing the risk of misconfiguration-induced vulnerabilities. Emerging industry roadmaps go further by integrating AI-driven configuration management. Machine learning models trained on historical misconfigurations and threat intelligence can proactively detect anomalous parameter settings or deviations from expected baselines, flagging them before adversaries can exploit them. Such approaches are already being prototyped within the O\mbox{-}RAN Alliance’s security workgroups, where AI-enabled RICs are envisioned to continuously monitor and enforce compliance~\cite{tang2023ai}.

Collectively, these measures illustrate how proactive resilience transforms knowledge of known vulnerabilities into preventive safeguards embedded at the infrastructure, orchestration, and policy layers. By confining failures, eliminating implicit trust, building redundancy, and automating compliance, proactive strategies reduce the effective attack surface of 5G/O\mbox{-}RAN systems long before disruptions materialize.

\paragraph{Limitations and the Need for Quantitative Modeling.}
Yet, anticipation alone cannot guarantee comprehensive protection. Novel attack vectors inevitably emerge, and vulnerabilities cannot be exhaustively enumerated in such a dynamic and evolving ecosystem. For example, increasing reliance on supply chain components introduces risks that may bypass even rigorously designed controls~\cite{lyu2024mapping}. Moreover, proactive measures may introduce their own trade-offs: extensive redundancy raises cost and energy consumption, while zero-trust architectures increase signaling overhead and complexity~\cite{lyu2024zero}. 

This limitation highlights the importance of quantitative modeling. Proactive resilience must be assessed not only in terms of whether safeguards exist but also in terms of how effectively they reduce risk relative to cost, complexity, and performance constraints. Developing models that capture preparedness as a measurable state, linking vulnerability coverage, attack likelihood, and resilience investment, provides a principled way to evaluate trade-offs. Such models can represent resilience readiness through metrics such as coverage ratios (fraction of known vulnerabilities addressed), protection depth (layers of defense applied), or risk-weighted benefit-cost indices. These foundations allow designers to move beyond heuristic defenses and toward systematically engineered architectures that balance efficiency with security, ensuring that 5G networks enter the operational environment with robust, verifiable resilience against known threats.

\subsection{Responsive Resilience}

Responsive resilience lies at the core of real-time adaptation. It is invoked when proactive defenses prove insufficient, either because adversaries exploit zero-day vulnerabilities that could not be anticipated in advance, or because known threats bypass preventive mechanisms due to configuration gaps or operational oversights. Once an attack perturbs the system state $X_t$, the resilience layer must react rapidly to prevent localized disruptions from escalating into systemic failures. In 5G networks, this requirement is amplified by URLLC applications, where even millisecond-scale delays can compromise safety-critical services such as autonomous vehicle coordination, industrial robot control, or remote telesurgery~\cite{de2024performance}.

Today, responsive resilience is achieved largely through a combination of monitoring platforms, orchestration systems, and security appliances integrated into the 5G/O\mbox{-}RAN stack. Most operators deploy Security Operations Centers (SOCs) with real-time analytics pipelines that ingest measurement from RAN, transport, and core domains~\cite{basta2024open}. Near-real-time RAN Intelligent Controllers (near-RT RICs) use xApps for anomaly detection and resource optimization, while non-real-time RICs (non-RT RICs) provide policy guidance. Cloud-native orchestration frameworks such as Kubernetes add auto-healing capabilities, restarting failed pods and redistributing workloads when anomalies are detected. However, these responses are often best-effort and heuristic rather than quantitatively guaranteed. For example, anomaly detection systems may raise alarms too late to prevent service-level agreement (SLA) violations, and current orchestration policies may not differentiate sufficiently between mission-critical and best-effort slices. This leaves a gap between the vision of responsive resilience and the operational practices that dominate current deployments.

\paragraph{Examples of Responsive Measures.}
Responsive resilience mechanisms act directly on the evolving network state through dynamic orchestration, anomaly detection, fault isolation, and controlled degradation. Their purpose is not necessarily to fully neutralize the attack, but to contain its impact, preserve continuity of essential services, and maintain graceful degradation rather than abrupt collapse.

\emph{Dynamic orchestration} allows workloads to be reallocated in response to localized disruptions~\cite{lian2023dynamic,farooq2026dynamic}. For instance, if an MEC node is compromised, overloaded, or disconnected, latency-tolerant workloads (e.g., content caching, video streaming) can be migrated to centralized cloud resources. This preserves edge capacity for URLLC applications that cannot tolerate disruption. Cloud-native platforms support such failovers through container migration, traffic redirection, and dynamic slice reconfiguration. However, in the status quo, many operators still rely on static orchestration rules, limiting the speed and efficiency of such migrations.

\emph{Anomaly detection} serves as the first line of defense against stealthy or fast-spreading attacks~\cite{aversano2021effective, ranaweera2021mec}. Responsive anomaly detection modules analyze signaling patterns, traffic volumes, and device behaviors in real time to detect events such as signaling storms, botnet-driven DDoS, or misbehaving IoT devices. For example, if a surge of attach requests destabilizes control-plane functions, anomaly detection can trigger throttling or quarantining of suspicious traffic sources. In practice, anomaly detection is typically implemented via rule-based intrusion detection systems (IDS) and threshold monitors~\cite{amachaghi2024survey,fang2022fundamental,balta2023digital,fung2011smurfen}. While these systems are widely deployed, they often suffer from high false positive rates or limited ability to adapt to evolving adversarial strategies. Research prototypes are increasingly exploring machine-learning-based anomaly detectors, but deployment at scale remains limited due to explainability and trust concerns.

\emph{Fault isolation} mechanisms explicitly contain failures or attacks by segmenting affected subsystems~\cite{zidan2016fault}. For example, if a slice orchestration component exhibits abnormal behavior, responsive mechanisms can isolate that slice to prevent control-plane instability across other tenants. Similarly, service mesh firewalls can block compromised microservices from interacting with neighboring functions. Today, such isolation is often manual or policy-driven, with limited automation, which delays the response window and risks cascading failures before isolation takes effect.

\emph{Controlled degradation} ensures continuity of essential services even when full performance cannot be maintained~\cite{zhang2023dependent}. For instance, in the event of resource exhaustion, responsive policies may prioritize URLLC slices by shedding non-critical eMBB traffic, preserving life-critical applications while gracefully degrading best-effort services. Current operator practices include basic traffic shaping and prioritization~\cite{prados2021optimization}, but fine-grained and adaptive degradation strategies, such as dynamically adjusting quality-of-experience targets or selectively offloading workloads, are still in their early stages of adoption.

\paragraph{Need for Quantitative Guarantees.}
The effectiveness of responsive resilience hinges on the ability to provide quantitative guarantees~\cite{de2024performance}. Critical questions include: How quickly must a control action be executed to satisfy stringent service-level agreements? What level of performance loss can be tolerated in a degraded mode before cascading failures occur across interdependent services? How can false positives in anomaly detection be balanced against the risk of delayed containment? Addressing these challenges requires rigorous mathematical formulations that link adversarial actions, state transitions, and control responses. By embedding responsiveness into models of stochastic dynamics, hybrid control, and adversarial interaction, we can move from heuristic responses to provably effective real-time resilience mechanisms. This is particularly vital in 5G/O\mbox{-}RAN, where the margin for error is narrow and disruptions in milliseconds can ripple across safety-critical ecosystems.

\subsection{Retrospective Resilience}

Retrospective resilience closes the loop by addressing the residual consequences of disruptions that could not be fully prevented or neutralized in earlier stages. It begins once performance outcomes $Y_t$ have already been degraded, and its primary function is to transform these outcomes into actionable knowledge for future preparedness~\cite{zhu2025foundations}. Unlike proactive and responsive measures, which focus on anticipation and real-time adaptation, retrospective resilience is inherently reflective. It recognizes that not all failures can be foreseen or contained and therefore emphasizes recovery, learning, and long-term adaptation.  

In today’s networks, retrospective resilience is implemented through post-incident response frameworks that combine operational, organizational, and financial tools. Telecom operators typically maintain Security Operations Centers (SOCs) and Network Operations Centers (NOCs) that conduct forensic analysis after a breach or outage~\cite{muniz2015security}. For example, when a slice outage occurs due to a misconfigured VNF or API abuse, logs and measurements are gathered to reconstruct the sequence of events. Recovery is often achieved through restoring from snapshots or redeploying cloud-native functions using orchestration tools such as Kubernetes~\cite{koziolek2021dynamic}. Vulnerabilities discovered in the aftermath are patched in subsequent software updates, though patch latency remains a critical weakness, as operators must test updates across heterogeneous vendor equipment before deployment. At the organizational level, retrospective resilience often relies on financial instruments such as cyber insurance and compensation schemes for SLA violations. Although these mechanisms are widespread, they remain fragmented, and lessons from one incident are not always systematically incorporated into future architectures.

\paragraph{Examples of Retrospective Measures.}
Retrospective mechanisms span several layers of response and adaptation:  

\emph{Forensic investigations} reconstruct the timeline of a disruption, identifying the initial vector, propagation path, and defensive breakdowns. For example, in the wake of a signaling storm that crashed a control-plane VNF, forensic analysis might reveal abnormal registration requests from compromised IoT devices, alongside misconfigured throttling policies~\cite{sharevski2018towards}. These insights not only resolve attribution but also provide feedback to refine anomaly detection thresholds and slice isolation rules.  

\emph{Recovery and restoration procedures} aim to bring services back online as quickly as possible, even if at reduced capacity~\cite{awoyemi2018network}. In cloud-native 5G, this typically involves redeploying containerized VNFs, restoring MEC workloads from backups, or shifting services to unaffected geographic regions. In practice, operators often employ automated “runbooks” integrated with orchestration frameworks, though their coverage is incomplete. Mission-critical URLLC applications, such as telesurgery or autonomous vehicle platooning, highlight the limits of retrospective recovery: workloads may be restored, but degraded latency or jitter performance may render the restored service unsafe for real-time use.  

\emph{Post-incident patch development} addresses the vulnerabilities exploited in the disruption~\cite{ahmad2019security}. In the status quo, patches are released by vendors on varying timelines and must be tested across diverse hardware and virtual environments before rollout. This lag creates recurring exposure windows. Some operators have begun integrating retrospective vulnerability discovery directly into CI/CD pipelines, where measurement and forensic artifacts automatically trigger patch generation, regression testing, and staged rollout across slices.  

\emph{Financial and organizational mechanisms} mitigate the long-term impact of disruptions. SLA violation penalties provide immediate compensation to customers when performance falls below contractual thresholds. Cyber insurance policies cover broader losses, such as reputational damage or large-scale service outages \cite{zhang2021optimal,liu2023cyber,liu2022mitigating,zhang2017bi,zhang2019mathtt}. Although these mechanisms are widely adopted, they are often reactive and do not inherently improve technical resilience unless coupled with systematic feedback loops that invest the proceeds of such penalties into resilience upgrades.  

\paragraph{Limits of Recovery.}
A defining feature of retrospective resilience is that recovery may not always be complete. Some disruptions result in irrecoverable data loss, long-lasting service degradation, or systemic architectural weaknesses. For example, a ransomware attack on a MEC cluster may allow operators to restore workloads from backups, but applications requiring strict latency guarantees may still be rendered unusable due to degraded performance. Similarly, a compromise of core orchestration software may undermine trust in future updates, necessitating costly revalidation of the entire stack. Retrospective resilience therefore acknowledges these limits and emphasizes not only minimizing immediate damage but also capturing lessons to prevent recurrence.  

\paragraph{Toward Quantitative Retrospective Resilience.}
If left purely qualitative, retrospective resilience risks becoming anecdotal, i.e., dependent on operator intuition rather than embedded in formal design processes. What is required is a quantitative framework that measures how much knowledge is gained from each incident, how effectively this knowledge reduces the probability or severity of future risks, and how quickly resilience mechanisms adapt across successive disruptions. Candidate metrics include:
\begin{itemize}
    \item \emph{Knowledge gain ratio:} fraction of incident root causes for which new defensive rules, patches, or detection signatures are generated.
    \item \emph{Patch latency:} average delay between vulnerability discovery and deployment of effective remediation across all affected nodes.
    \item \emph{Learning curve slope:} rate at which incident recurrence frequency declines across successive disruptions.
\end{itemize}
By formalizing retrospective mechanisms such as forensics, patching, recovery, and insurance, networks can convert disruption into an opportunity for improvement. Every incident becomes a driver for more robust architectures, tighter integration with CI/CD pipelines, and adaptive security policies. In this way, retrospective resilience ensures that 5G/O\mbox{-}RAN and cloud-native infrastructures evolve dynamically, continuously strengthening their preparedness against an ever-changing adversarial landscape.

\subsection{Integration of Proactive, Responsive, and Retrospective Mechanisms}

The resilience of 5G networks cannot be secured by any single class of mechanism in isolation. Instead, it depends on the integration of proactive, responsive, and retrospective approaches into a coherent lifecycle~\cite{jin2025fundamental}. This integration is essential because the threat landscape is characterized not only by well-known vulnerabilities that can be prevented, but also by unknown risks that cannot be anticipated in advance, and by adversarial actions that are difficult to mitigate fully in real time.

Proactive mechanisms form the first line of defense by hardening the system against known threats. In 5G, measures such as network slicing, zero-trust authentication, redundancy in RAN deployments, and rigorous patch management reduce the probability that an attacker can gain a foothold. If proactive mechanisms were perfect, eliminating every vulnerability and anticipating every possible exploit, then responsive and retrospective layers would be unnecessary. In practice, however, this ideal is unattainable. The sheer complexity of cloud-native architectures, the continual discovery of zero-day vulnerabilities, and the sophistication of adaptive adversaries ensure that some threats will bypass even the most carefully engineered proactive safeguards.

Responsive mechanisms serve as the safety net once proactive measures are breached. They detect anomalies, isolate faults, and contain unfolding attacks in real time. For example, when a misconfigured API in the 5G Core is exploited to escalate privileges, responsive defenses, such as anomaly detection and adaptive access control, can contain the intruder before lateral movement compromises multiple slices. Similarly, in the event of jamming attacks on a RAN fronthaul, responsive orchestration can reroute traffic to alternative links or degrade non-essential services to preserve mission-critical URLLC. However, relying too heavily on responsiveness is costly: it requires continuous monitoring, high-speed analytics, and dynamic reallocation of resources under strict latency constraints, which consume substantial computational and operational overhead. Moreover, some attacks, such as stealthy data exfiltration or firmware-level compromises, may not be easily neutralized online.

Retrospective mechanisms address residual consequences that proactive and responsive measures cannot fully prevent. They enable recovery, forensic investigation, patch development, and post-incident learning. For instance, if ransomware encrypts MEC workloads despite anomaly detection, retrospective recovery procedures may restore services from backups while forensic analysis identifies the exploit vector. Post-incident patching then prevents recurrence in future releases. Retrospective mechanisms also include financial and organizational tools such as cyber insurance to manage long-term impacts. However, relying primarily on retrospective resilience is also costly and inefficient, as it accepts disruption as inevitable and shifts the burden to recovery, sometimes partial and incomplete, rather than prevention or containment.

The interdependencies among these mechanisms reveal the trade-offs in designing resilient 5G networks. Strong proactive defenses reduce the frequency of incidents, lowering the reliance on costly responsive and retrospective actions. Robust responsive mechanisms limit the severity of incidents that bypass proactive safeguards, reducing the scale of retrospective recovery needed. Retrospective resilience ensures that every disruption generates knowledge that improves both proactive preparation and responsive agility in the future. The balance among these mechanisms depends on the threat environment: for well-characterized risks such as DDoS attacks, proactive measures like traffic filtering and redundant paths are cost-effective. For unpredictable or rapidly evolving threats such as AI-driven adversarial manipulation of MEC scheduling, responsive and retrospective layers become indispensable. 

In practice, no single mechanism suffices. Proactive defenses can never be perfect, and responsiveness and retrospection are too costly to serve as the sole basis of resilience. It is only through their integration, including anticipating, adapting, and learning, that 5G networks can minimize the overall impact of disruptions while optimizing the trade-offs between cost, complexity, and risk tolerance \cite{li2024conjectural,li2024automated,huang2019adaptive,kamhoua2021game,qian2020receding}.

\subsection{Quantitative Approach to Resilience Design}

The three stages depicted in Fig.~\ref{fig:resilience_mechanism} form a resilience lifecycle that is dynamic and recursive. Proactive mechanisms operate on the time scale of anticipation, reducing the probability that known threats will materialize by hardening the system in advance. Responsive mechanisms operate on the real-time adaptation time scale, intervening in system dynamics as adversarial activity perturbs the state of the network $X_t$. Retrospective mechanisms extend the horizon further by converting performance outcomes $Y_t$ into actionable insights that strengthen future preparedness. Although conceptually distinct, these stages are deeply interdependent. Proactive defenses must be guided by knowledge distilled from retrospective analysis, and responsive strategies must be designed within the constraints created by proactive design choices while also benefiting from retrospective feedback. Resilience is therefore not a static collection of safeguards, but a co-designed adaptive process that evolves with the system it protects~\cite{jin2025fundamental,jin2025cloud}.

This recursive and interdependent structure underscores the urgent need for quantitative modeling and analysis. Qualitative descriptions alone cannot capture the complexity of 5G networks, where adversaries adapt strategically, interdependencies span multiple domains, and service requirements are safety-critical. Quantitative frameworks are needed to characterize how threats propagate through vulnerabilities, how mechanisms interact across timescales, and how resource trade-offs shape achievable resilience~\cite{kott2021improve}. For instance, perfecting proactive defenses would, in theory, reduce reliance on responsive and retrospective mechanisms, but this comes at escalating costs that may be impractical. In contrast, relying too heavily on response or recovery places a high burden on monitoring, orchestration, and post-incident remediation. The optimal balance depends on the type of threat, its likelihood and the cost-effectiveness of each layer of defense. A rigorous analytical foundation is required to formalize and solve these trade-offs.

We will present foundations based on control theory, game theory, learning theory, and network theory to address this challenge \cite{zhu2025foundations}. Control theory provides models of the temporal dynamics of network states and feedback mechanisms that stabilize performance under perturbations \cite{zhu_basar_resilience2024,zhu2022introduction,chen2019control,ishii2022security}. Game theory captures the strategic interplay between attackers and defenders, enabling the study of equilibria, incentives, and adversarial adaptation \cite{huang2020dynamic,xu2016cross,zhu2015game,zhu2013game}. Learning theory supports adaptive mechanisms that improve with experience, extracting patterns from retrospective data to refine proactive planning and responsive detection \cite{zhu2013game,tao23sce,li2024conjectural,zhao2022multi,tao23cola,pan-tao22noneq}. Network theory offers structural insights into interdependencies, cascading effects, and systemic vulnerabilities in interconnected large-scale infrastructures \cite{huang2018distributed,chen2019optimal,chen2021dynamic,nugraha2020dynamic,nugraha2019subgame}. 

To build truly resilient 5G networks, these foundations must be synthesized into principled models that represent resilience as a lifecycle process rather than a static property. Such models must capture adversarial interactions as stochastic or dynamic games, quantify performance trade-offs among security, latency, and availability, and embed learning modules that evolve with the threat landscape. By moving beyond ad hoc defenses and embracing mathematically grounded optimization frameworks, resilience can be designed, analyzed, and assured as a first-class property of next-generation communication systems.

\section{Metrics and Evaluation}

\begin{figure}[h]
    \centering
    \includegraphics[width=0.9\linewidth]{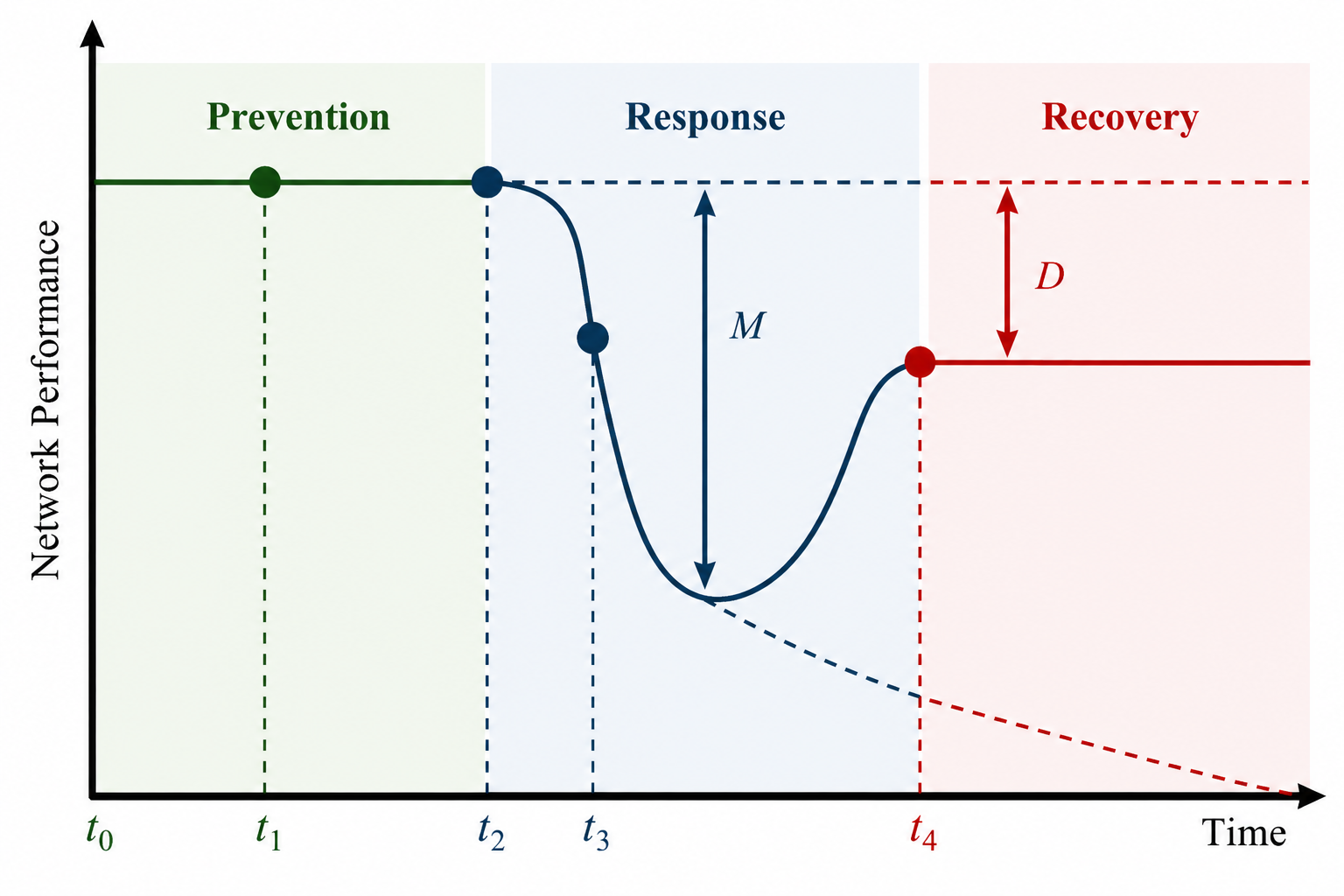}
    \caption{\small Illustration of network resilience across three phases: prevention, response, and recovery. 
    The vertical axis represents network performance, and the horizontal axis represents time. 
    Preventive mechanisms delay or avoid the onset of disruption ($t_0$--$t_1$). 
    Responsive mechanisms act during the disruption window ($t_2$--$t_3$), limiting the maximum performance degradation $M$. 
    Recovery mechanisms operate after $t_4$, aiming to restore performance, though often with a residual deficit $D$. 
    The shaded regions emphasize the temporal domains of each class of resilience mechanism.}
    \label{fig:resilience_metrics}
\end{figure}

Resilience in NextG networks is a multifaceted attribute that captures a system’s ability to withstand, recover from, and adapt to adverse conditions. Unlike classical notions of reliability or robustness, which focus on steady-state operation or tolerance to bounded perturbations, resilience emphasizes the \emph{entire lifecycle} of a disruptive event: anticipation, real-time response, and long-term recovery. Rigorous quantification of resilience is crucial for ensuring operational continuity, guiding architectural decisions, benchmarking infrastructure deployments, and informing investment strategies~\cite{weisman2025experimental, weisman2025quantitative}.  

Fig.~\ref{fig:resilience_metrics} illustrates this lifecycle. During prevention ($t_0$--$t_1$), the system sustains nominal performance despite emerging stressors. At $t_2$, a disruption triggers performance degradation, reaching a maximum shortfall $M$. Responsive mechanisms act during $t_2$--$t_3$ to limit both the depth and duration of this decline. Recovery begins at $t_4$, where performance stabilizes, often with a residual deficit $D$ if the system cannot fully return to baseline. These temporal markers anchor the formal definitions and metrics developed below~\cite{cholda2007survey}.

\subsection{Discrete-Time Formalism for Resilience}

To formalize resilience, we adopt a discrete-time representation that is well suited for cyber-physical networks, where events are logged, monitored, and acted upon at discrete epochs. Let the system evolve at times $t = 0, 1, \dots, T$, where each step represents a measurable operational interval (e.g., seconds, control cycles, or monitoring windows). At each $t$, the system is characterized by a performance level $Q_t$, which is bounded between zero and a nominal maximum $Q_{\max}$. The baseline $Q_{\max}$ corresponds to ideal operating conditions such as target throughput, latency compliance, or service-level guarantees.

A disruptive event is defined by two critical instants. The failure onset occurs at $t_f$ (denoted $t_2$ in Fig.~\ref{fig:resilience_metrics}), marking the departure from nominal behavior. The recovery time is $t_r$ ($t_4$ in the figure), when the system stabilizes in a recovered or steady state, which may or may not fully achieve $Q_{\max}$. Between $t_f$ and $t_r$, the trajectory $\{Q_t\}$ reflects both the severity of the disruptive impact and the efficacy of the detection, mitigation, and recovery mechanisms of the system.

The cumulative resilience loss is then defined as
\begin{equation}
L = \sum_{t = t_f}^{t_r} \left(1 - \frac{Q_t}{Q_{\max}} \right),
\label{eq:resilience_loss}
\end{equation}
which integrates the depth and duration of degradation into a single measure, geometrically corresponding to the shaded area between the degraded trajectory and the baseline in Fig.~\ref{fig:resilience_metrics}. Systems with rapid detection and efficient recovery achieve smaller $L$, whereas systems with sluggish or incomplete responses accumulate a greater loss of resilience.

\subsection{General Temporal Metrics}

While $L$ in~\eqref{eq:resilience_loss} captures the overall severity of a disruption, more granular temporal metrics provide insight into specific phases of the disruption-recovery cycle shown in Fig.~\ref{fig:resilience_metrics}. \emph{Downtime} $D_k$ quantifies the period during which performance falls below an acceptable service threshold $\delta Q_{\max}$ for the $k$-th disruptive event:
\begin{equation}
D_k = \sum_{t = t_f}^{t_r} \mathbb{I}(Q_t < \delta Q_{\max}),
\label{eq:downtime}
\end{equation}
where $\mathbb{I}(\cdot)$ is the indicator function. This corresponds to the interval $t_2$--$t_4$ in Fig.~\ref{fig:resilience_metrics}, where user-perceived quality of service requirements are not met.

\emph{Mean Time to Failure (MTTF)} measures the average length of normal operation between disruptive events, corresponding to the prevention or steady phases ($t_0$--$t_1$). Large MTTF indicates rare disruptions and effective preventive controls, whereas small MTTF indicates recurrent or persistent threats.

\emph{Mean Time to Detect (MTTD)} captures the responsiveness of monitoring mechanisms by measuring the delay between the true failure onset $t_f$ ($t_2$) and the detection instant $t_d$:
\begin{equation}
\text{MTTD}_k = t_d - t_f,
\label{eq:mttd}
\end{equation}
which measures the critical “blind window” during which the system is already degraded but not yet alerted.

\emph{Mean Time to Recovery (MTTR)} reflects the speed of corrective actions by measuring the interval between detection and recovery:
\begin{equation}
\text{MTTR}_k = t_r - t_d,
\label{eq:mttr}
\end{equation}
corresponding in Fig.~\ref{fig:resilience_metrics} to the trajectory from detection at $t_d$ to stabilization at $t_4$.

Two additional structural metrics capture the magnitude of degradation. The \emph{maximum performance drop},
\begin{equation}
M = \max_{t \in [t_f,t_r]} \big(Q_{\max} - Q_t\big),
\label{eq:maxdrop}
\end{equation}
represents the most severe instantaneous degradation (vertical arrow $M$). The \emph{residual deficit},
\begin{equation}
D = Q_{\max} - Q_{t_r},
\label{eq:residual}
\end{equation}
represents the long-term shortfall that persists after recovery (vertical arrow $D$). Together, $M$ and $D$ distinguish between transient acute shocks (large $M$ but small $D$) and lasting structural damage (nonzero $D$).


\subsection{Domain-Specific Metrics}

While general temporal metrics characterize resilience broadly, domain-specific indicators capture properties critical for 5G/O-RAN and cloud-native contexts. These metrics extend the logic of Fig.~\ref{fig:resilience_metrics} by linking performance deviations $Q_t$ to service-level and architectural requirements such as slice isolation, latency guarantees, elastic scaling, and availability. By tailoring resilience assessment to these dimensions, operators can align quantitative measurements with concrete service-level objectives (SLOs) and SLAs.

\paragraph{Slice Isolation Index.}
In network slicing, multiple tenants and services coexist on the same physical infrastructure, relying on logical isolation to ensure that one slice’s disruption does not cascade into others. This makes isolation a cornerstone of resilience in multi-tenant environments. The \emph{Slice Isolation Index} (SII) measures the extent to which a disruption in slice $i$ affects other slices $j \neq i$:
\begin{equation}
\text{SII}_i = 1 - \frac{\sum_{j \neq i} \max_{t \in [t_f,t_r]} \left(1 - \frac{Q_t^{(j)}}{Q_{\max}^{(j)}}\right)}{(N-1)}.
\label{eq:sii}
\end{equation}
High values of $\text{SII}_i$ (close to 1) indicate strong isolation, where slice $i$’s failure is contained. Low values suggest cross-slice propagation, which undermines multi-tenancy guarantees and may lead to cascading failures across RAN, core, or MEC domains.

\paragraph{Latency Compliance.}
URLLC applications, such as autonomous driving or industrial control, impose stringent delay requirements. Even brief violations can cause safety or operational hazards. Let $L_t$ denote the observed latency at time $t$, and $L_{\max}$ the contractual bound. The \emph{Latency Compliance} is defined as
\begin{equation}
\Lambda = \frac{1}{T} \sum_{t=0}^{T} \mathbb{I}(L_t \leq L_{\max}),
\label{eq:latency}
\end{equation}
indicating the fraction of epochs in which latency remains acceptable. In Fig.~\ref{fig:resilience_metrics}, this corresponds to epochs where $Q_t$ is above the latency threshold. A high $\Lambda$ value reflects robust scheduling, congestion control, and RIC policies that maintain end-to-end delay guarantees even during disruptions.

\paragraph{SLA Violation Rate.}
The SLAs in 5G cover multiple KPIs, including throughput, jitter, and reliability, and not meeting them directly impacts operator revenue through penalties or reputational loss. Let $Q_{\text{SLA}}$ denote the minimum contractual guarantee. The \emph{SLA Violation Rate} is given by
\begin{equation}
V_{\text{SLA}} = \frac{1}{T} \sum_{t=0}^{T} \mathbb{I}(Q_t < Q_{\text{SLA}}),
\label{eq:sla_violation}
\end{equation}
which measures the proportion of time contractual guarantees are not satisfied. This metric is closely related to the shaded degradation region in Fig.~\ref{fig:resilience_metrics}. High violation rates signal systemic vulnerabilities in orchestration, scaling, or RAN scheduling that prevent the system from meeting promised service guarantees.

\paragraph{Auto-Scaling Efficiency.}
Cloud-native platforms rely heavily on elasticity: scaling microservices, MEC resources, or RAN functions up or down in response to load variations or disruptive events. Poor scaling may either over-provision (wasting resources) or under-provision (failing to recover quickly). Let $R_t$ denote the actual resources allocated at time $t$, and $R_t^*$ the optimal allocation that minimizes degradation. The \emph{Auto-Scaling Efficiency} is defined as
\begin{equation}
\eta_{\text{scale}} = 1 - \frac{\sum_{t=t_f}^{t_r} \left| R_t - R_t^* \right|}{\sum_{t=t_f}^{t_r} R_t^*}.
\label{eq:scaling}
\end{equation}
High $\eta_{\text{scale}}$ indicates close monitoring of resource demand and rapid adaptation to failures or adversarial load changes. Low values indicate sluggish scaling, which prolongs recovery and increases both resilience loss $L$ and MTTR.

\paragraph{Availability.}
Traditional availability metrics measure uptime, but in resilient systems, availability must account for degraded yet still functional states. Let $Q_{\text{avail}}$ be the minimum performance threshold required for safe or acceptable operation (e.g., minimum data rate for emergency services). The \emph{Availability} metric is
\begin{equation}
A = \frac{1}{T} \sum_{t=0}^{T} \mathbb{I}(Q_t \geq Q_{\text{avail}}).
\label{eq:availability}
\end{equation}
This metric synthesizes both the prevention phase ($t_0$--$t_1$) and the post-recovery plateau ($t > t_4$), providing a global view of how often the system remains usable. In O\mbox{-}RAN settings, high availability reflects effective coordination between RIC policies, SMO functions, and cloud-native orchestration even under compound disruptions.

\medskip

These domain-specific metrics, including slice isolation, latency compliance, SLA violation rate, auto-scaling efficiency, and availability, provide a multidimensional lens on resilience. They not only quantify recovery speed and magnitude of degradation, but also address whether the system continues to uphold the economic and service commitments that define 5G/O\mbox{-}RAN deployments. In practice, operators must jointly monitor these indicators to balance technical robustness with business viability.

\subsection{Composite and Cost-Aware Metrics}

No single metric can fully capture the multifaceted nature of resilience. For example, a system may recover quickly (low MTTR) but experience frequent failures (low MTTF), or maintain high availability while maintaining a large residual deficit $D$. Similarly, minimizing downtime $D_k$ may come at the expense of larger resource expenditures or aggressive fallback mechanisms. This tradeoff motivates the design of composite indices that aggregate multiple resilience indicators into a single figure of merit.

A general formulation is a weighted combination:
\begin{equation}
R_{\text{composite}} = \sum_{i=1}^n w_i \cdot M_i,
\label{eq:composite}
\end{equation}
where $M_i$ are normalized resilience metrics (e.g., $L$, MTTR, MTTD, availability, latency compliance) and $w_i$ are application- or domain-specific weights that satisfy $\sum_i w_i = 1$. The weights allow the composite measure to be tailored to heterogeneous service requirements. For URLLC slices, larger weights may be assigned to latency compliance and SLA violation rate. For eMBB, throughput and availability may dominate. For mMTC, long-term availability and energy efficiency could be emphasized. 

Composite indices are particularly useful for comparing candidate architectures or resiliency mechanisms (e.g., static redundancy vs. dynamic orchestration). By compressing multiple metrics into a single index, $R_{\text{composite}}$ enables benchmarking and ranking of system designs under consistent weighting schemes. However, the choice of weights $w_i$ must be transparent, as they encode implicit priorities and tradeoffs that may differ between operators, regulators, and end-users.

\vspace{1em}
\noindent\textbf{Cost-Aware Resilience.}  
While $L$ in~\eqref{eq:resilience_loss} measures total performance degradation, it does not account for heterogeneous costs associated with different degradation scenarios. A more realistic measure attaches monetary, operational, or societal costs to each unit of performance loss:
\begin{equation}
C_{\text{resilience}} = \sum_{t = t_f}^{t_r} C_t \cdot \left(1 - \frac{Q_t}{Q_{\max}} \right),
\label{eq:costres}
\end{equation}
where $C_t$ is a time-varying cost coefficient. In Fig.~\ref{fig:resilience_metrics}, this corresponds to weighting the shaded degradation area by its impact severity. For example:
\begin{itemize}
    \item In URLLC telemedicine, $C_t$ may represent risk-adjusted costs of violating latency constraints, reflecting potential harm to patients.
    \item In cloud services, $C_t$ can encode SLA penalty charges or reputational costs for downtime.
    \item In industrial automation, $C_t$ may represent opportunity costs due to halted production lines.
\end{itemize}
Thus, $C_{\text{resilience}}$ generalizes resilience loss into an economically meaningful quantity, making it possible to compare strategies not only on technical but also financial grounds. This is critical for operators deciding between competing resilience mechanisms that differ in cost structure (e.g., proactive redundancy vs. on-demand recovery).

\subsection{Integration with Control and Learning Frameworks}

The discrete-time formulation of resilience metrics makes them naturally compatible with control-theoretic and learning-based frameworks. In modern networks, system performance evolves over time as a consequence of both endogenous policies and exogenous disturbances. In this setting, $Q_t$ becomes part of the observable system state, while proactive, responsive, and recovery actions influence its trajectory across disruption and recovery phases. By embedding resilience metrics such as $L$, MTTR, or cost-aware loss $C_{\text{resilience}}$ into optimization objectives or operational constraints, one can obtain principled designs of resilience-enhancing mechanisms that go beyond ad hoc heuristics. This approach provides a systematic bridge between theoretical metrics and the concrete policies executed by controllers, orchestrators, and learning agents in 5G/O\mbox{-}RAN and cloud-native infrastructures.

\paragraph{Control-Theoretic Integration.}
Control theory provides a natural language for modeling resilience as a closed-loop property. In model predictive control (MPC), the controller predicts future trajectories of $Q_t$ under possible failure scenarios and selects actions that minimize a cumulative cost function \cite{xu2015secure,xu2018cross,xu2020secure,miao2014moving}. Minimizing resilience loss $L$ in~\eqref{eq:resilience_loss} can be cast directly as such a cost minimization problem, with additional temporal constraints to ensure recovery within a specified MTTR. Robust control formulations extend this approach by explicitly accounting for uncertainty and adversarial disturbances, guaranteeing that $Q_t$ remains above operational thresholds even in worst-case conditions. Hybrid control models \cite{zhu2011robust,zhu2013resilient}, combining continuous-time traffic dynamics (e.g., queue evolution, SINR variability) with discrete failure events (e.g., gNodeB outage, backhaul disconnection), are particularly relevant in O\mbox{-}RAN, where abrupt failures interact with continuous flows of data and control traffic. In such models, resilience metrics serve both as performance objectives and safety constraints, guiding the system toward guaranteed levels of service continuity.

\paragraph{Reinforcement Learning Integration.}
Resilience can also be optimized through reinforcement learning (RL), where metrics play the role of reward signals \cite{huang2022reinforcement,xu2015cyber}. An RL agent, observing network states $\{Q_t\}$ and exogenous inputs (e.g., traffic load, alarms, or anomaly scores), can be trained to select actions such as spectrum reallocation, slice migration, or MEC resource scaling. For example, during the $t_2$--$t_3$ response window in Fig.~\ref{fig:resilience_metrics}, the agent may learn scaling policies that minimize MTTR or reduce cost-aware loss $C_{\text{resilience}}$ in~\eqref{eq:costres}. Multi-agent RL extends this formulation to settings where distributed controllers (e.g., near-RT and non-RT RICs, or federated operators) must coordinate decisions while optimizing resilience jointly. By aligning resilience metrics with the RL reward design, networks can autonomously discover recovery strategies that adapt to evolving adversarial tactics and traffic dynamics.

\paragraph{Game-Theoretic Integration.}
In adversarial environments, resilience is inherently strategic: attackers attempt to maximize disruption, while defenders attempt to minimize it. Game-theoretic models provide a principled way to capture these interactions. Metrics such as maximum performance drop $M$ in~\eqref{eq:maxdrop} and residual deficit $D$ in~\eqref{eq:residual} can serve directly as attacker and defender payoffs. Zero-sum games formalize the tradeoff: an attacker’s gain (larger $M$ or $D$) is precisely the defender’s loss. More nuanced formulations (e.g., Stackelberg or Bayesian games) capture asymmetric information, where the attacker hides its type or timing \cite{manshaei2013game,zhu2018game,kamhoua2021game}. Equilibrium analysis then identifies achievable resilience levels under rational strategies. Such models motivate the design of adaptive defenses, such as moving-target configurations \cite{zhu2013game}, deception strategies \cite{zhu2012deceptive,jajodia2016cyber,al2019autonomous}, and slice isolation policies \cite{slicing_3,slicing_2,slicing_1}, all of which can be evaluated through the lens of resilience metrics.

\paragraph{Adaptive Resilience.}
The ultimate goal is to move beyond passive measurement toward \emph{adaptive resilience}. By embedding resilience metrics into control and learning objectives, networks gain the ability not only to quantify their performance after a disruption but also to evolve dynamically in response to repeated adversarial encounters. Adaptive resilience entails learning from past disruptions, refining recovery strategies, and reconfiguring policies to preempt future attacks or failures. In 5G/O\mbox{-}RAN and cloud-native architectures, this manifests as self-optimizing loops where RIC controllers, orchestration layers, and learning agents use resilience metrics to continuously update their policies. Thus, resilience becomes both a measurable property and an explicit control objective, shaping the architecture of next-generation communication systems toward self-healing, self-adaptive, and economically sustainable operation.

\chapter{Theoretical Foundations for Resilient Network Management}

Chapter 3 develops the theoretical foundations required to model, analyze, and control resilience in next-generation (NextG) networks. Building on the definitions and metrics introduced in Chapter 2, this chapter formalizes the system dynamics, uncertainty, and adversarial interactions that govern network behavior under disruption.

The discussion integrates control theory, game theory, learning, and network science as complementary frameworks for characterizing system evolution, decision-making, and interdependencies. These foundations provide the analytical basis for the risk assessment and resilience design methodologies developed in the subsequent chapters.

\section{Control-Theoretic Foundation}

Control theory provides a foundational lens for understanding and managing resilience because it formalizes the dynamic, multi-stage evolution of network states under both adversarial attacks and natural disruptions. Unlike static reliability models, control theory explicitly accounts for feedback, adaptation, and time-varying uncertainties, making it ideally suited to capture the continuous interplay between system dynamics and external perturbations~\cite{yuksel2013stochastic,zhang2015survey}.
In networked cyber-physical systems, disruptions emerge from two principal sources. On the one hand, adversarial threats come from intelligent and adaptive opponents that look for vulnerabilities and adjust their strategies in response to defenses. However, natural failures such as hardware crashes, fiber cuts, and power outages also degrade performance and can cascade across interdependent infrastructures. A resilient network must therefore regulate its performance in the presence of both strategic and stochastic disturbances~\cite{long2005denial,dolk2016event}
.
Next-generation networks (5G and beyond) further challenge this challenge. The adoption of SDN, NFV, O-RAN, and MEC introduces programmability and flexibility, but also enlarges the attack surface and creates tighter coupling between components~\cite{chahbar2020comprehensive,lyu2024mapping}. These architectures transform resilience from a static design problem into a real-time control problem, where monitoring, decision-making, and actuation must occur continuously to preserve service continuity.
As a foundation, control theory contributes three essential principles to resilience engineering. First, it provides stability analysis, ensuring that, despite shocks, malicious or accidental, the system returns to an acceptable operating region. Second, it offers feedback mechanisms, allowing the network to sense deviations, adapt policies, and prevent small perturbations from escalating into systemic failures. Third, it enables the design of robust and adaptive controllers, which anticipate uncertainties and adjust to evolving environments without requiring complete foresight of every possible disruption.

\subsection{System Model}
Resilience requires reasoning about how a network evolves under normal operation and disruption. 
A control-theoretic abstraction provides this capability by capturing the 
\emph{system state}, the \emph{defender’s actions}, the \emph{uncertainties} introduced by adversaries and the environment, 
and the \emph{disruption state} that signals mode switches in the network. 
We model the hybrid dynamics as a stochastic process:
\begin{align}
x_{t+1} &= f\!\left(x_t, u_t, w_t, \xi_t, q_t\right), \label{eq:hybrid-dynamics} \\
q_{t+1} &= \phi\!\left(q_t, x_t, u_t, w_t, \xi_t\right), \label{eq:disruption}
\end{align}
where $x_t$ is the continuous-valued \emph{network state}, $u_t$ the defender’s control input, 
$w_t$ an \emph{adversarial disturbance}, $\xi_t$ a \emph{natural disturbance}, and 
$q_t \in \mathcal{Q}$ the \emph{disruption state}. 
The operators $f$ and $\phi$ may be deterministic maps or stochastic kernels, so that both continuous dynamics and mode transitions can evolve probabilistically depending on the current state, inputs, and disturbances~\cite{zhu2013networked,nugraha2022rolling}.  

\begin{figure}[t]
    \centering
    \includegraphics[width=0.65\linewidth]{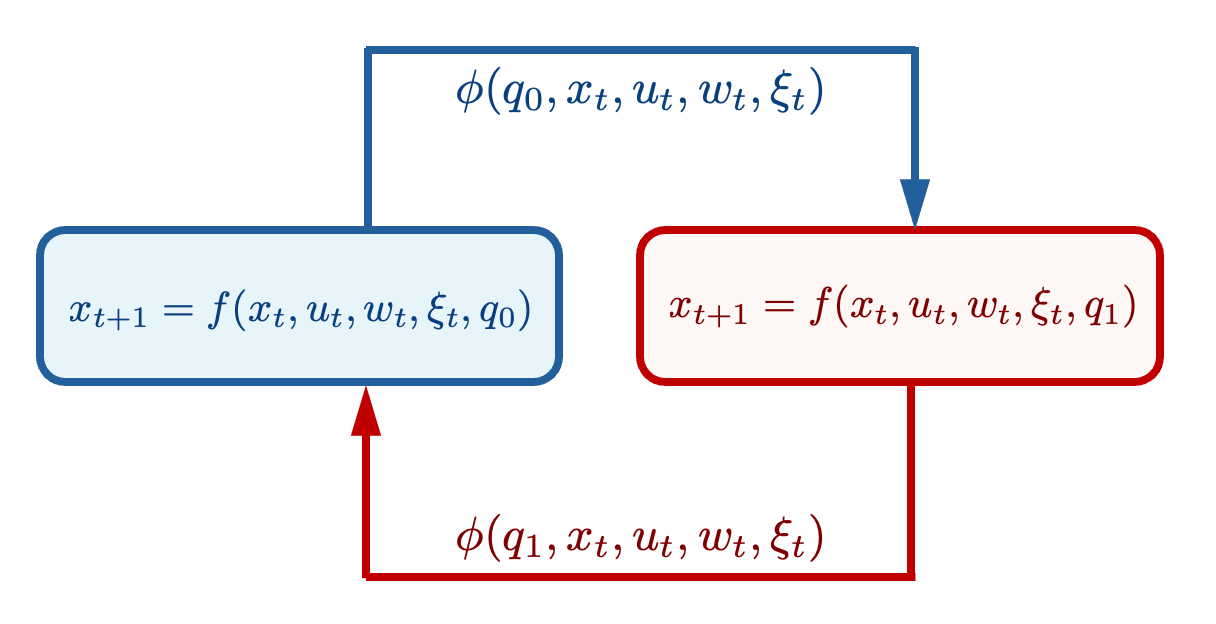}
    \caption{\small Hybrid dynamics with two disruption modes: 
    $q_0$ (normal) and $q_1$ (abnormal). 
    Within each mode, the network state $x_t$ evolves according to the map 
    $f(x_t,u_t,w_t,\xi_t,q_i)$, while transitions between modes are 
    governed by $\phi(q_t,x_t,u_t,w_t,\xi_t)$. 
    This captures how adversarial disturbances, natural hazards, or defensive actions 
    can trigger shifts between normal and abnormal regimes.}
    \label{fig:two-mode-hybrid}
\end{figure}

The set $\mathcal{Q}$ may include labels such as $\{$\text{normal}, \text{degraded}, \text{attack}, \text{failure}$\}$. 
Equation~\eqref{eq:hybrid-dynamics} describes the (possibly random) evolution of performance variables within each regime, 
while Equation~\eqref{eq:disruption} characterizes random mode transitions triggered by adversarial activity, 
natural hazards, or defensive reconfiguration. 
Fig.~\ref{fig:two-mode-hybrid} illustrates this process for the simple case with two disruption modes: 
$q_0$ representing normal operation and $q_1$ representing an abnormal regime. 
Within each mode, the network state $x_t$ evolves according to $f$, while the transition map $\phi$ determines when the system switches between modes.  

By adapting both $x_t$ and $q_t$ to the domain, this hybrid model provides a unifying foundation for designing \emph{tailored resilience strategies} while preserving a coherent framework for analysis. In what follows, we illustrate how different network components induce distinct choices of the continuous state $x_t$, the disruption state $q_t$, adversarial disturbances $w_t$, natural disturbances $\xi_t$, and defender inputs $u_t$. Each example highlights how the abstract hybrid model specializes to concrete resilience objectives in 5G and beyond~\cite{chahbar2020comprehensive,farooq2018secure}.

\paragraph{Core Network (Service-Based Architecture).}
The 5G Core relies on a Service-Based Architecture (SBA), where network functions interact through APIs. 
In this setting, signaling storms or API abuse can overwhelm the control plane. 
The state $x_t$ is described by variables such as the signaling queue length $x_t^{\mathrm{queue}}$, 
the control-plane processing rate $\mu_t$, and the slice-level latency $\ell_t$. 
The disruption mode $m_t$ switches among $\{\text{normal}, \text{overload}, \text{attack}\}$ depending on signaling arrival rates. 
Adversarial disturbances $w_t$ manifest as malicious signaling arrivals $\lambda_t^{q}$, 
while legitimate arrivals are denoted by $\lambda_t^{0}$. 
A representative queueing dynamic is
\[
    x^{\mathrm{queue}}_{t+1} 
    = x^{\mathrm{queue}}_t + \lambda^0_t + \lambda^q_t - \mu(u_t),
\]
where the service rate $\mu(u_t)$ depends on elastic scaling of control-plane VNFs. 
Resilience objectives in this context are twofold: ensuring queue stability, i.e., maintaining bounded 
$x_t^{\mathrm{queue}}$, and enforcing latency guarantees $\ell_t \leq \ell^{\max}$ even under abnormal operating modes.

\paragraph{Backhaul Transport.}
The transport domain carries data between the core and the RAN. Natural disturbances $\xi_t$, such as fiber cuts, reduce capacity, while $q_t \in \{\text{normal}, \text{degraded},$ $\text{partitioned}\}$ indicates disruption status. The state $x_t$ includes available link capacity $B_t$, coverage set $\mathcal{C}_t$, and active connections $N_t$. A fault induces a drop
\[
    B_{t+1} = B_t - \Delta B(\xi_t, q_t),
\]
where $\Delta B$ reflects disruption severity. The defender input $u_t$ consists of rerouting or activating backup links. Resilience requires maintaining coverage $\mathcal{C}_t$ and throughput above thresholds across disruption states.  

\paragraph{Edge Computing (MEC).}
Multi-access edge computing nodes host latency-critical services. The state $x_t$ includes compute capacity $C_t$, workload utilization $\rho_t$, and slice deadlines $\ell^{\mathrm{URLLC}}_t$, while $q_t \in \{\text{normal}, \text{congested}, \text{failure}\}$ captures disruptions. Adversarial inputs $w_t$ may inject excess load $\omega_t$, while natural failures $\xi_t$ reduce available capacity. The utilization evolves as
\[
    \rho_{t+1} = \rho_t + \frac{\lambda^u_t + \omega_t}{C_t - \Delta C(\xi_t,q_t)}.
\]
Resilience requires ensuring that URLLC slice deadlines are met even when the disruption state indicates capacity collapse.  

\paragraph{Radio Access Network (RAN / O-RAN).}
In the RAN, resilience is tied to access reliability. The state $x_t$ includes SINR, spectrum assignment, and coverage areas, while $q_t \in \{\text{normal}, \text{jammed}, \text{outage}\}$ encodes disruption. Natural disturbances $\xi_t$ may disable distributed units (DUs), while adversarial inputs $w_t$ may jam signals or corrupt RIC commands. The SINR dynamics can be written
\[
    \mathrm{SINR}_{t+1} = h\!\left(\mathrm{SINR}_t, -\Delta R(\xi_t,q_t), w_t, u_t\right),
\]
where $\Delta R(\xi_t,q_t)$ captures resource losses in different disruption modes. Defender inputs $u_t$ include adaptive power control and spectrum reassignment. Resilience requires sustaining spectral efficiency and handover stability across disruption states.  

\paragraph{Feedback and OODA Loop for Responsive Resilience.}

\begin{figure}[t]
\centering
\includegraphics[width=\linewidth]{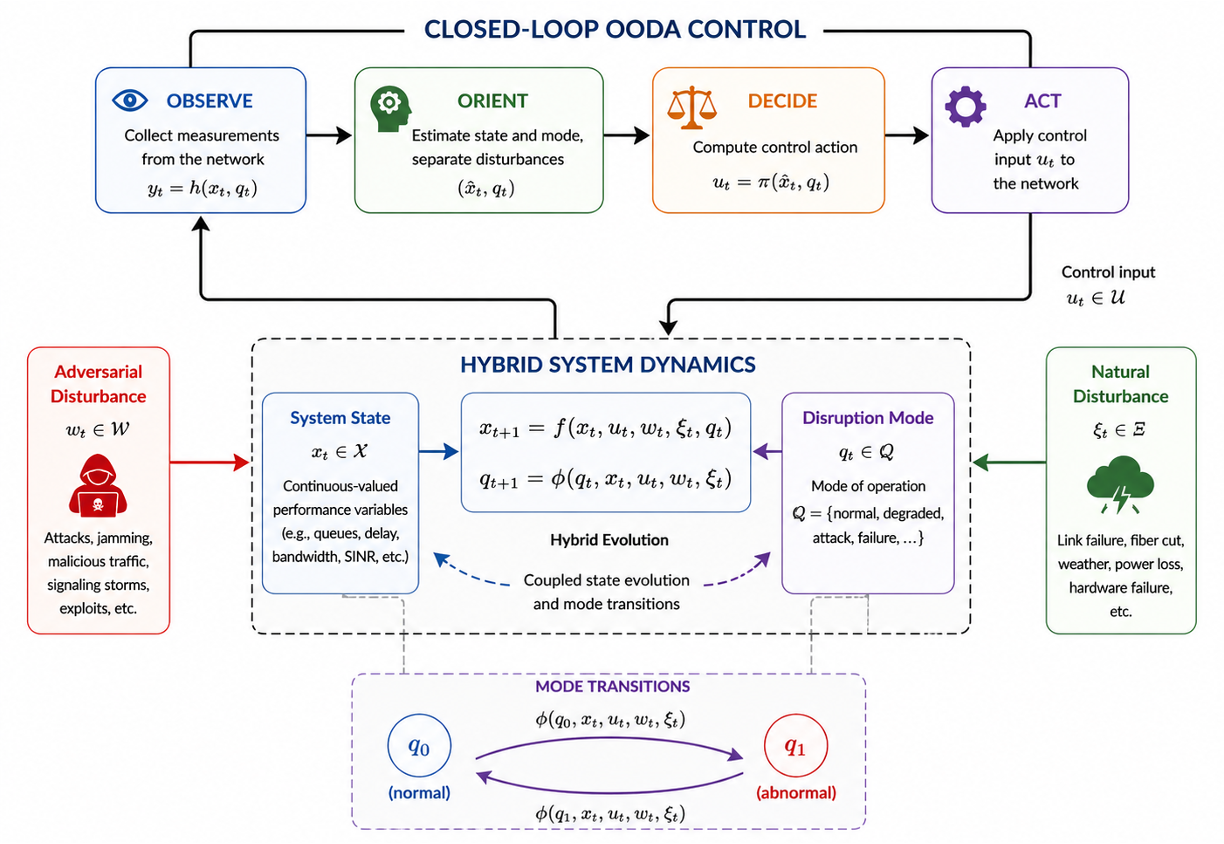}
\caption{\small Closed-loop OODA control for responsive resilience. The loop integrates measurement, estimation, decision, and actuation with hybrid system dynamics. Adversarial and natural disturbances influence both state evolution and mode transitions, while control actions continuously adapt system behavior.}
\label{fig:ooda_closed_loop}
\end{figure}

Responsive resilience in cyber-physical systems is fundamentally a closed-loop process in which sensing, reasoning, and actuation continuously interact with evolving system dynamics. This process is naturally captured by the OODA loop, i.e., \emph{Observe, Orient, Decide, Act}, as illustrated in Fig.~\ref{fig:ooda_closed_loop}. 
In the \emph{Observe} phase, the system collects measurements $y_t = h(x_t,q_t)$ that reflect both continuous performance variables and discrete disruption modes. The \emph{Orient} phase transforms these observations into situational awareness by estimating the system state and identifying the underlying disturbance structure, including both adversarial inputs $w_t$ and natural disturbances $\xi_t$. 

Based on this interpreted state, the \emph{Decide} phase computes control actions $u_t = \pi(\hat{x}_t,q_t)$ that aim to mitigate risk, stabilize the system, or reconfigure resources. The \emph{Act} phase then implements these decisions, directly influencing the hybrid system dynamics and closing the feedback loop.

As shown in Fig.~\ref{fig:ooda_closed_loop}, this loop is embedded within a hybrid system where continuous state evolution and discrete mode transitions are jointly driven by control inputs and disturbances. Adversarial and natural disturbances enter the system through different channels but are unified within the same feedback structure.

At a high level, the OODA loop provides a principled framework for resilience: it enables continual adaptation, supports real-time decision making, and integrates perception, inference, and control into a unified architecture. Unlike static fault-tolerance mechanisms, this closed-loop structure allows the system to respond dynamically to evolving threats, uncertainty, and regime shifts.

\paragraph{Multi-Horizon Optimization for Optimal Responsive Resilience.}
This section focuses on the notion of \emph{responsive resilience}, which emphasizes online adaptation to unfolding disturbances and disruptions. 
In contrast, the \emph{proactive} and \emph{retrospective} resilience approaches are often carried out offline, involving long-term design, planning, and post-event analysis. 
Here, control-theoretic methods are particularly suitable, as they enable online decision-making and closed-loop adjustment in the face of adversarial uncertainty. In the \emph{responsive} stage, closed-loop feedback policies adapt rapidly to disturbances $w_t$ and $\xi_t$, containing degradation in $Q_t$ and steering $(x_t,q_t)$ back toward admissible regions.

When disturbances and disruption modes evolve stochastically, resilience is naturally formulated as a \emph{stochastic} receding-horizon control problem over the hybrid state $(x_t,q_t)$. 
At decision time $t$, given the information set $\mathcal{I}_t$ (measurements, state estimates, and mode), the defender must predict the system’s evolution $H$ steps into the future using models, statistical forecasting, or digital twins. The resulting optimization problem is
\begin{align}
\min_{\,u_{t:t+H-1} \in \mathcal{U}^{H}} \;
&\mathbb{E} \Bigg[
\sum_{\tau=t}^{t+H-1} \big(\ell(x_\tau,u_\tau,q_\tau) + \notag \\
&  c(u_\tau,q_\tau)\big)
\;+\; V(x_{t+H},q_{t+H})
\; \Big \vert \; \mathcal{I}_t
\Bigg], \label{eq:stoch-mpc}
\end{align}
subject to the stochastic hybrid dynamics in 
Equations~\eqref{eq:hybrid-dynamics}--\eqref{eq:disruption}, the initial condition $(x_t,q_t)$, and additional operational constraints
\(
g(\hat{x}_\tau,u_\tau,\hat{q}_\tau) \le 0, \  \forall \tau \in [t,t+H-1],
\)
where $\hat{x}_\tau,\hat{q}_\tau$ denote state and mode estimates obtained from $\mathcal{I}_t$.

Here $H$ is the prediction horizon, and the expectation in \eqref{eq:stoch-mpc} is taken with respect to the distribution of future disturbances and mode transitions $(w_{t:t+H-1},\xi_{t:t+H-1},q_{t+1:t+H})$ conditional on $\mathcal{I}_t$. 
The terminal cost $V(\cdot)$ encodes resilience goals beyond the horizon, such as backlog clearance, state recovery, or returning to a nominal operating mode.

\emph{Receding-horizon implementation.}
Model Predictive Control (MPC) computes an optimal sequence $u_{t:t+H-1}^\star$ conditioned on $\mathcal{I}_t$, applies only the first action $u_t^\star$, and then re-solves the optimization at $t{+}1$ with updated state $(x_{t+1},q_{t+1})$ and information $\mathcal{I}_{t+1}$. 
By combining \eqref{eq:hybrid-dynamics}--\eqref{eq:disruption} with rolling-horizon design, MPC provides a mode-aware and stochastic framework for predictive planning that adapts continually to new observations and regime switches.

\emph{Risk-aware variants.}
When resilience must account for rare but severe events, the expectation in \eqref{eq:stoch-mpc} can be replaced by a risk measure~\cite{yuksel2013stochastic,nugraha2022rolling}. Examples include minimizing Conditional Value-at-Risk (CVaR),
\[
\min \; \mathrm{CVaR}_\alpha\!\left(
\sum_{\tau=t}^{t+H-1} \ell(x_\tau,u_\tau,q_\tau)+c(u_\tau,q_\tau) + V(x_{t+H},q_{t+H})
\;\middle|\; \mathcal{I}_t
\right),
\]
imposing chance constraints such as
\[
\mathbb{P} \Big[g(x_\tau,u_\tau,q_\tau)\le 0 \ \text{for all } \tau \,\Big|\, \mathcal{I}_t\big] \ge 1-\varepsilon,
\]
or adopting distributionally robust formulations,
\[
\sup_{\mathbb{P}\in\mathcal{P}} \; \mathbb{E}_{\mathbb{P}}\!\left[
\sum_{\tau=t}^{t+H-1} \ell(x_\tau,u_\tau,q_\tau)+c(u_\tau,q_\tau) + V(x_{t+H},q_{t+H})
\right],
\]
where $\mathcal{P}$ is an ambiguity set around the nominal disturbance distribution. 
Such variants provide additional protection against tail risks, model uncertainty, and adversarial manipulation of disturbance distributions~\cite{zhu2015game,nugraha2020dynamic}.

\paragraph{Optimal Fallback Switching for Deterministic Systems.}

We consider a hybrid control setting with two modes: an abnormal mode and a safe fallback mode. The defender can either remain in the current mode or incur a switching cost to transition to a safer configuration with improved stability but reduced performance. 
The problem can be formulated as a finite-horizon optimal control problem with mode-dependent dynamics and costs. By optimizing continuous controls and comparing the resulting value functions in each mode, the switching decision is reduced to a comparison between two quadratic cost-to-go functions.

At a high level, the optimal decision admits a simple structure: switching to the safe mode is beneficial when the expected reduction in future state-dependent costs outweighs the immediate switching penalty and service degradation. This leads to a threshold-type policy characterized by a quadratic form in the current state. 
Specifically, there exists a matrix difference that captures the relative advantage of the safe mode such that the fallback switching is optimal whenever
\(
x_t^\top M x_t \;>\; \lambda + (s_1 - s_0),
\) 
where $M$ encodes the difference between the effective cost-to-go under the two modes. 

This condition reveals a clear tradeoff: larger disturbances or system states favor switching, while higher switching penalties or service losses discourage it. In particular, if the safe mode induces more contractive dynamics or lower long-term costs, then a finite state threshold emerges beyond which switching is optimal. 
This result provides a tractable and interpretable rule for resilience: fallback mechanisms should be activated only when the state of the system crosses a risk-dependent limit, balancing immediate costs against long-term recovery benefits.

\subsection{Moving Target Defense.}

Moving Target Defense (MTD) is a key mechanism for achieving \emph{responsive resilience}, in which the system actively reshapes its configuration to deny adversaries a stable attack surface~\cite{zhu2013game}. 
Rather than relying solely on redundancy, MTD introduces controlled nonstationarity, forcing attackers to continuously adapt and thereby increasing their cost and reducing the effectiveness of reconnaissance and exploitation.

\begin{figure}[t]
\centering
\includegraphics[width=0.85\linewidth]{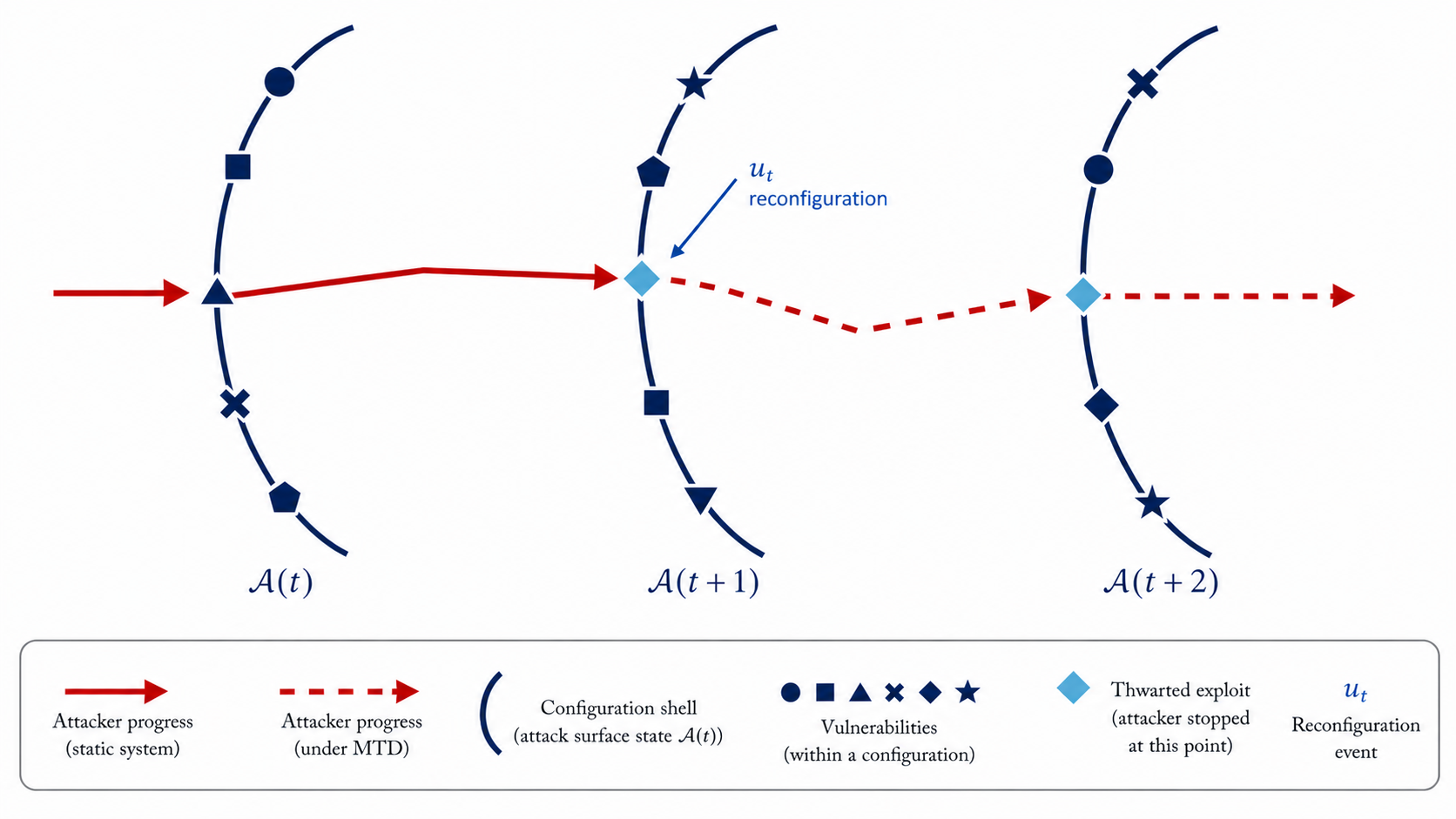}
\caption{\small Moving Target Defense (MTD) as controlled evolution of the attack surface. 
Blue arcs represent configuration shells $\mathcal{A}(t)$, and markers on each shell denote configuration-specific vulnerabilities. 
Red trajectories depict attacker progress: solid for a static system and dashed under MTD. 
Reconfiguration actions $u_t$ shift the system across shells, altering the vulnerability landscape and disrupting attacker trajectories. 
Light-blue markers indicate thwarted exploits, where attacker progress is halted and must restart, thereby reducing effective attack success.}
\label{fig:mtd_concept_clean}
\end{figure}

This principle is illustrated in Fig.~\ref{fig:mtd_concept_clean}. The blue arcs represent successive \emph{configuration shells}, each corresponding to a distinct realization of the system’s exposure. 
When the system remains static, the attacker follows a steady trajectory toward compromise (solid red path). 
In contrast, MTD introduces reconfiguration actions that shift the system across shells, breaking the attacker’s progress and forcing repeated adaptation (dashed red path). 
Each reconfiguration effectively resets the attacker’s information, reducing dwell time and limiting the accumulation of exploit knowledge.

From a control perspective, MTD can be viewed as an input that governs the evolution of the attack surface,
\[
\mathcal{A}(t+1) = g(\mathcal{A}(t), u_t),
\]
where $u_t$ denotes the reconfiguration action. 
This formulation highlights two key effects: it injects uncertainty into the attacker’s decision process and provides the defender with a mechanism to shape system exposure over time.

At a high level, the design of MTD policies involves balancing resilience gains against operational costs. 
Frequent reconfiguration improves security but may degrade service quality or consume resources, while infrequent adaptation leaves the system vulnerable. 
This tradeoff naturally leads to a receding-horizon decision framework, in which the defender selects configurations to minimize risk while accounting for future system evolution and constraints.

Within the broader OODA loop, MTD actions arise in the \emph{Decide} phase and are implemented in the \emph{Act} phase, feeding back into subsequent observations as adversaries react to the changing environment. 
This tight integration of perception, decision, and control makes MTD a canonical example of closed-loop, adaptive defense.

\section{Dynamic Game Foundation}

Cyber resilience in 5G networks is not merely about robustness against random faults; it fundamentally involves \emph{adversarial behavior}. Attackers are intelligent, adaptive agents who observe the system, learn its vulnerabilities, and strategically decide how to maximize disruption. Defenders, in turn, must anticipate these maneuvers while balancing trade-offs between performance, security, and operational cost.  

Game theory provides a natural framework for such interactions. It models the network as a strategic environment in which multiple players (attackers and defenders) repeatedly interact, shaping each other’s payoffs. This perspective captures how resilience mechanisms, such as MTD, slice migration, SDN rerouting, and spectrum hopping, perform not only against average case failures, but under \emph{equilibrium-driven adversarial conditions}. In this sense, game theory embeds resilience into a predictive framework, enabling the design of defense strategies that remain robust against both stochastic disturbances and intelligent adversaries~\cite{zhu2015game,huang2020dynamic}.  

\begin{figure*}[t]
\centering
\includegraphics[width=0.95\linewidth]{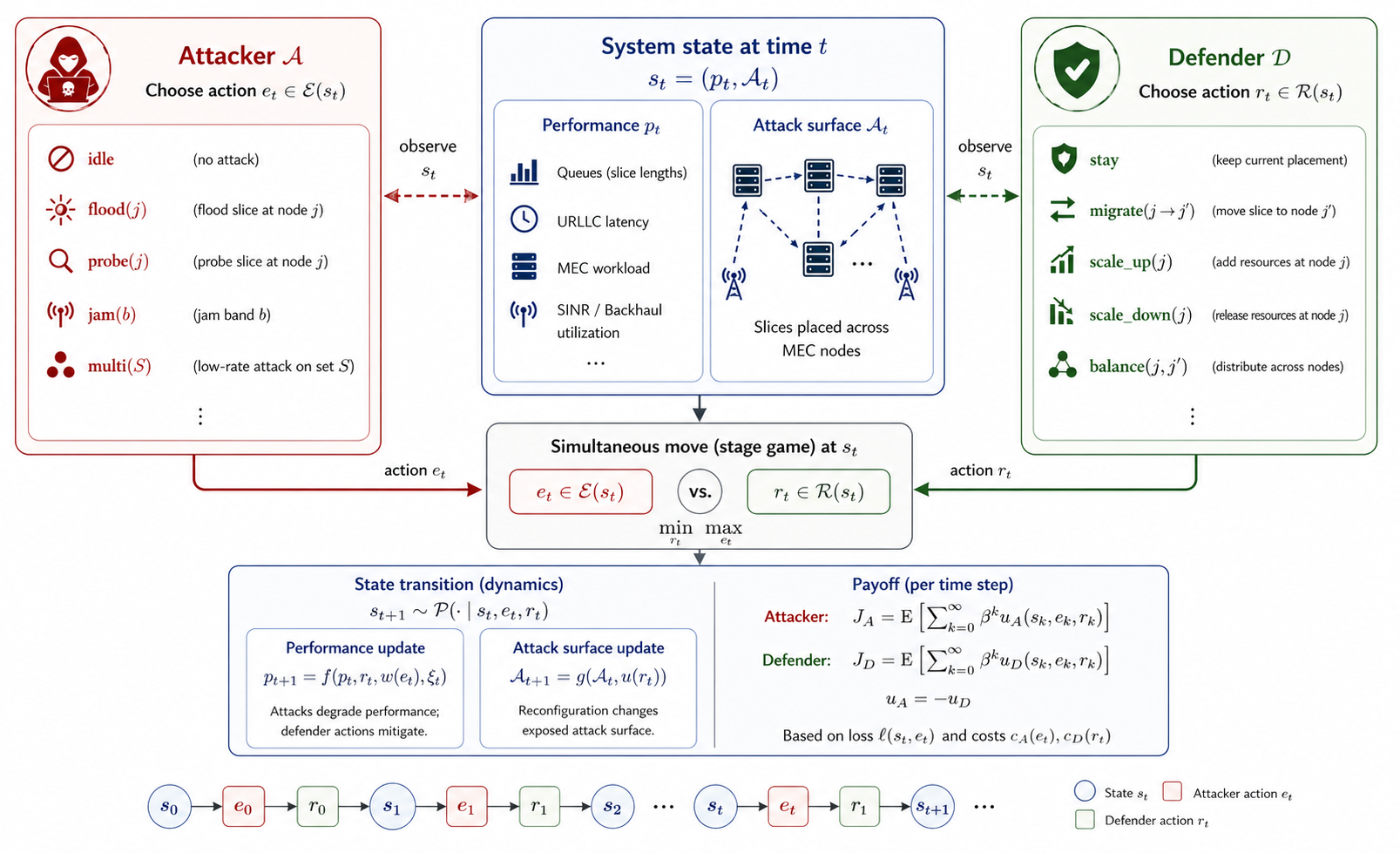}
\caption{\small Dynamic attacker--defender game for 5G network resilience. 
At each time $t$, the system state $s_t=(p_t,\mathcal{A}_t)$ captures both network performance (e.g., queue lengths, latency, workload) and the exposed attack surface (e.g., slice placement across MEC nodes). 
The attacker selects an action $e_t \in \mathcal{E}(s_t)$ (e.g., flooding, probing, jamming), while the defender selects a control $r_t \in \mathcal{R}(s_t)$ (e.g., migration, scaling, load balancing). 
These actions interact through a simultaneous stage game and drive the stochastic state transition $s_{t+1} \sim P(\cdot \mid s_t,e_t,r_t)$, where attacks degrade performance and defensive reconfiguration reshapes the attack surface. 
The resulting repeated interaction defines a discounted stochastic game, in which equilibrium strategies characterize the long-run resilience of the network.}
\label{fig:dynamic_game_5g}
\end{figure*}

\subsection{Model Definition}

We model attacker–defender interactions in 5G as a discounted stochastic game
\(
\Gamma \;=\; \big(\mathcal{S}, \mathcal{E}, \mathcal{R}, P, \{u_A,u_D\}, \beta, \mathcal{I}\big),
\)
where time evolves as $t=0,1,2,\dots$, and $\beta \in (0,1)$ is the discount factor. 
The players are the attacker $A$ and the defender $D$.  
The overall structure of this interaction, including the state, action spaces, and feedback loop, is illustrated in Fig.~\ref{fig:dynamic_game_5g}.

In general, the state at time $t$ can include both the continuous network state $x_t$ and the discrete disruption mode $q_t$. 
The variable $x_t$ can represent the operational condition of the network, such as the lengths of the slicing queue, the latency of URLLC, the workload of the MEC, the SINR or the utilization of the backhaul. 
Alternatively, $x_t$ may capture the configuration of the exposed attack surface, such as the placement of slices across edge clouds or the allocation of spectrum resources. 
The disruption mode $q_t$ encodes the regime in which the system is operating (e.g., normal, degraded, under attack, or in failure), allowing the game to capture both performance evolution and adversarial regime switching.
As depicted in Fig.~\ref{fig:dynamic_game_5g}, these two components are summarized by the state $s_t=(p_t,\mathcal{A}_t)$, which jointly captures performance and exposure.

At each time step $t$, the attacker and defender select actions simultaneously from their respective sets, conditioned on the current state $s_t = (p_t,\mathcal{A}_t)$. 
Here, $p_t$ denotes the network performance state (e.g., slice queue lengths, URLLC latency, MEC workload, SINR, or backhaul utilization), while $\mathcal{A}_t$ describes the exposed attack surface, such as the current placement of slices across MEC nodes.  
This simultaneous-move interaction corresponds to the stage game highlighted in Fig.~\ref{fig:dynamic_game_5g}, where both players act based on their observation of the current state.

The admissible set of the attacker $\mathcal{E}(s_t)$ depends on the placement exposed $\mathcal{A}_t$ and typically includes
\[
\mathcal{E}(s_t) = \{ \texttt{idle},\ \texttt{flood}(j),\ \texttt{probe}(j),\ \texttt{jam}(b),\ \texttt{multi}(S) \}.
\]
Here, $\texttt{idle}$ represents no attack, $\texttt{flood}(j)$ injects adversarial workload into the service-based interface of the slice hosted on MEC node $j$, $\texttt{probe}(j)$ performs low-rate reconnaissance on a slice at node $j$, $\texttt{jam}(b)$ targets a frequency band $b$ relied upon by a slice, and $\texttt{multi}(S)$ distributes simultaneous low-rate attacks across a subset of nodes $S$. 
These actions correspond to the attack primitives illustrated on the left-hand side of Fig.~\ref{fig:dynamic_game_5g}. 
For example, if the slice is deployed at MEC node $j^\star$, then choosing $\texttt{flood}(j^\star)$ directly creates congestion and queue buildup at that node, raising latency and packet loss for URLLC flows. 
Probing actions may cause little immediate damage, but yield information about slice placement or migration patterns.  

The admissible set of the defender $\mathcal{R}(s_t)$ represents feasible resilience controls given the current resource and latency constraints.
$
\mathcal{R}(s_t) = \{ \texttt{stay},\ \texttt{migrate}(j \to j'),\ \texttt{scale\_up}(j), $ $\texttt{scale\_down}(j),$ $ \texttt{balance}(j,j') \}.
$
The action $\texttt{stay}$ keeps the slice at its current location. 
The action $\texttt{migrate}(j \to j')$ moves a slice from node $j$ to a new node $j'$, reducing exposure to targeted floods at the cost of signaling overhead and temporary disruption. 
The action $\texttt{scale\_up}(j)$ allocates additional processing resources at node $j$ to absorb increased load, while $\texttt{scale\_down}(j)$ releases excess resources when demand or attack intensity diminishes. 
Finally, $\texttt{balance}(j,j')$ splits the workload across multiple MEC nodes, lowering the risk of localized congestion at the price of synchronization and coordination costs.  
These control actions correspond to the defensive mechanisms shown on the right-hand side of Fig.~\ref{fig:dynamic_game_5g}, which actively reshape the attack surface.

To illustrate, suppose the slice is currently hosted at node $j^\star$. 
If the attacker selects $\texttt{flood}(j^\star)$, the defender may respond with $\texttt{stay}$, which leaves the slice vulnerable to growing congestion, or with $\texttt{migrate}(j^\star \to j')$, which restores service continuity after a temporary migration delay but incurs overhead. 
Alternatively, the defender may opt for $\texttt{scale\_up}(j^\star)$, which increases local service rate and mitigates flooding at the expense of additional resources. 
Over time, this interplay generates an adaptive cycle: the attacker redirects load toward the current slice placement, while the defender mitigates damage through migration, scaling, or balancing actions across the edge cloud. 
This feedback loop between attack and defense, visualized in Fig.~\ref{fig:dynamic_game_5g}, drives the evolution of both performance and exposure.
The resulting dynamics of $(p_t,\mathcal{A}_t)$ thus embody an arms race between targeted flooding and strategic slice migration.

The system state evolves according to a controlled stochastic kernel,
\(
s_{t+1} \sim P(\,\cdot \mid s_t, e_t, r_t),
\) 
which decomposes into two components:
\[
\begin{cases}
p_{t+1} = f\!\big(p_t, r_t, w(e_t), \xi_t\big), \\[6pt]
\mathcal{A}_{t+1} = g\!\big(\mathcal{A}_t, u(r_t)\big).
\end{cases}
\]
These dynamics correspond directly to the transition block in Fig.~\ref{fig:dynamic_game_5g}, where attacks degrade performance, while defensive reconfiguration alters the future attack surface.
The first component describes the evolution of network performance $p_{t+1}$ under the influence of the defender’s control $r_t$, the disturbance generated by the attacker’s exploit $w(e_t)$, and natural exogenous shocks $\xi_t$ such as fiber cuts or sudden power outages. 
The second component governs the evolution of the exposed attack surface $\mathcal{A}_{t+1}$, which depends directly on the migration or reconfiguration choices $u(r_t)$ embedded in the defender’s resilience action. 

To make this concrete, consider the slice queueing dynamics illustrated in Fig.~\ref{fig:dynamic_game_5g}, where attacker actions inject load and defender controls regulate service. 
If $x_t^{\mathrm{q}}$ denotes the queue length of a slice, then its evolution can be written as
\[
x_{t+1}^{\mathrm{q}} 
= x_t^{\mathrm{q}} + \lambda_t^{0} + \lambda_t^{q}(e_t) - \mu(r_t),
\]
where $\lambda_t^{0}$ is the legitimate user arrival rate, 
$\lambda_t^{q}(e_t)$ represents the adversarial injection induced by attack $e_t$, 
and $\mu(r_t)$ is the service rate determined by the defender’s action $r_t$ (e.g., migration or scaling).  

As reflected in the dynamics block of Fig.~\ref{fig:dynamic_game_5g}, attacker actions increase load while defender actions reshape service capacity. 
Migration to a less congested MEC node can increase $\mu(r_t)$, albeit with temporary disruption, whereas stronger attacks increase $\lambda_t^{q}(e_t)$ and force costly defensive responses. 
These coupled dynamics capture the fundamental arms race between adversarial load injection and adaptive resource control.

Each stage of the game generates immediate rewards and costs, summarized in the payoff block of Fig.~\ref{fig:dynamic_game_5g}. 
The loss function $\ell(x_t,e_t,r_t)$ quantifies performance degradation (e.g., latency violations, queue buildup), while the defender incurs operational cost $c_D(r_t)$ and the attacker incurs cost $c_A(e_t)$.  

For clarity, we consider the zero-sum formulation,
\[
u_D(s_t,e_t,r_t) = -\ell(x_t,e_t,r_t) - c_D(r_t), 
\quad
u_A(s_t,e_t,r_t) = -u_D(s_t,e_t,r_t),
\]
which encodes the adversarial tension depicted in Fig.~\ref{fig:dynamic_game_5g}: performance loss and defensive overhead directly benefit the attacker. 
This formulation can be relaxed to general-sum settings when external costs affect both players.

The cumulative discounted payoff of player $i \in \{A,D\}$ under a strategy profile $\pi = (\pi_A,\pi_D)$ is
\[
J_i^\pi(s_0) = \mathbb{E}^\pi \!\left[ \sum_{t=0}^\infty \beta^t \, u_i(s_t,e_t,r_t) \,\bigg|\, s_0 \right],
\]
capturing the repeated interaction shown along the temporal evolution in Fig.~\ref{fig:dynamic_game_5g}. 
The discount factor $\beta \in (0,1)$ balances immediate disruptions (e.g., transient congestion under flooding) with long-term adaptation costs such as repeated migration and resource reallocation.

The information structure $\mathcal{I}$ specifies what each player observes, corresponding to the observation links in Fig.~\ref{fig:dynamic_game_5g}. 
In the simplest case, both players observe the full state $s_t=(p_t,\mathcal{A}_t)$, yielding a complete-information Markov game. 
While analytically convenient, this assumption is rarely realistic in practice.

A more realistic model is a Bayesian game, where each player has a private type $\theta_i \in \Theta_i$ and receives noisy observations $o_t \sim O(\cdot \mid s_t)$. 
Players maintain beliefs $\mu_t$ over the state and types and act accordingly. 
In the slice setting, this reflects uncertainty about whether congestion is adversarial or benign, or whether a slice is real or a honeypot.

The most general formulation is a partially observable stochastic game (POSG)~\cite{hansen2004dynamic,horak2017manipulating,yang2025transparent,yang2025deceive}, where both the state and types are only partially observed. 
As suggested by the observation–action feedback loop in Fig.~\ref{fig:dynamic_game_5g}, both attacker and defender must continuously update beliefs and adapt strategies under uncertainty.

Continuing the slice migration example, a skilled attacker can rapidly track migrations and redirect attacks, while a constrained defender must balance responsiveness with cost. 
Thus, the equilibrium outcome depends not only on physical dynamics, but also on how information is inferred and acted upon, thereby closing the loop between observation, decision, and system evolution depicted in Fig.~\ref{fig:dynamic_game_5g}.

\subsection{Resilience as an equilibrium property}

Within the dynamic game framework, resilience corresponds to the defender’s ability to implement a policy $\pi_D$ that ensures bounded network loss $\ell$ and recovery of the performance state $x_t$ under worst-case attacker strategies $\pi_A$. In other words, resilience is not defined only by the robustness of the system in response to random shocks, but by the equilibrium outcome of the strategic interaction between attacker and defender.  

To make this precise, recall that a \emph{strategy} for player $i\in\{A,D\}$ is a mapping $\pi_i$ from the observed history (or belief state, under incomplete information) to a probability distribution over actions. A pair of strategies $\pi=(\pi_A,\pi_D)$ induces a stochastic process $\{s_t\}$ through the transition kernel $P$. The performance of a strategy profile is measured by the cumulative discounted payoff
\[
J_i^\pi(s_0) \;=\; \mathbb{E}^\pi \!\left[ \sum_{t=0}^\infty \beta^t \, u_i(s_t,e_t,r_t) \,\bigg|\, s_0 \right].
\]

A fundamental equilibrium notion is the \emph{Nash equilibrium}~\cite{holt2004nash}. A strategy profile $\pi^\star = (\pi_A^\star,\pi_D^\star)$ is a Nash equilibrium if neither player can improve their expected payoff by deviating unilaterally, i.e.,
\[
J_A^{(\pi_A^\star,\pi_D^\star)}(s) \;\geq\; J_A^{(\pi_A,\pi_D^\star)}(s), \quad
J_D^{(\pi_A^\star,\pi_D^\star)}(s) \;\geq\; J_D^{(\pi_A^\star,\pi_D)}(s),
\]
for all states $s \in \mathcal{S}$ and all admissible alternative strategies $\pi_A,\pi_D$. In dynamic games, one often considers refinements such as \emph{Markov perfect equilibrium}~\cite{maskin2001markov}, where strategies depend only on the current state, or \emph{Perfect Bayesian equilibrium}~\cite{fudenberg1991perfect}, where strategies depend on beliefs in incomplete-information settings.  

Resilience can then be understood as a property of the equilibrium value function $V^\star(s)$ of the defender, defined by
\[
V^\star(s) \;=\; J_D^{\pi^\star}(s),
\]
where $\pi^\star$ is an equilibrium strategy profile. The system is resilient if
\[
-\,V^\star(s) \;\leq\; \overline{L}, \qquad \forall s \in \mathcal{S},
\]
for some acceptable threshold $\overline{L}$ that encodes performance and service-level guarantees. This inequality states that at equilibrium, where the attacker has no incentive to reduce pressure and the defender has no incentive to over- or under-react—the long-run loss remains bounded within tolerable levels.  

In the slice migration example, equilibrium captures the stable operating point of the repeated attacker–defender interaction. A powerful attacker may repeatedly flood whichever MEC node currently hosts a slice, while the defender weighs whether to stay, migrate, or scale resources. If the defender migrates too frequently, service degrades due to excessive handoff costs; if the defender rarely migrates, the attacker succeeds in maintaining prolonged congestion. At equilibrium, the defender’s policy $\pi_D^\star$ specifies the optimal balance, such as migrating only when queue lengths exceed a critical threshold, while the attacker’s policy $\pi_A^\star$ specifies the intensity and frequency of floods consistent with maximizing disruption under this defense. The resulting equilibrium value function predicts the average long-run loss to the network.  

By interpreting resilience as an equilibrium property, one obtains both a normative and predictive perspective. Normatively, equilibrium analysis provides a benchmark for the best achievable defense under rational adaptation of the attacker. Predictively, if one can compute or approximate the equilibrium strategies and values, one can forecast how resilient the system will remain under sustained adversarial campaigns and repeated natural disturbances. Thus, equilibrium is not only a mathematical solution concept, but also a quantitative measure of the defender’s ability to maintain service quality in the face of strategic and persistent attacks.

\subsection{Equilibrium Solution via Dynamic Programming}

Game-theoretic models of resilience in 5G networks ultimately require solution concepts that
predict the outcome of attacker--defender interactions. Because these interactions unfold over
time and depend on evolving network states, equilibrium notions must capture both the
dynamic nature of play and the information structure available to each participant. 
Dynamic programming plays a central role, as equilibrium computation is typically reduced
to recursive value functions or belief updates. In this subsection, we provide a rigorous account 
of Markov perfect equilibrium, Bayesian and epistemic equilibria, and signaling games, 
highlighting their relevance to resilience analysis.

\paragraph{Markov perfect equilibrium (MPE).}
In dynamic stochastic games, players repeatedly act in response to the current state of the system. 
A \emph{Markov strategy} for player $i \in \{A,D\}$ is defined as a mapping
\(
\pi_i : \mathcal{S} \longrightarrow \Delta(\mathcal{A}_i),
\)
which prescribes a probability distribution over actions based only on the present state $s\in\mathcal{S}$. 
This restriction to memoryless strategies is particularly natural in large-scale 5G systems, where keeping track of full action histories is computationally infeasible and unnecessary for optimality in discounted settings.  

In the zero-sum case with finite state and action spaces, the defender’s equilibrium value function
$V(s)$ satisfies the \emph{Shapley equation}~\cite{winter2002shapley}:
\[
V(s) \;=\; \min_{r \in \mathcal{R}(s)} \; \max_{e \in \mathcal{E}(s)} 
\Big\{ u_D(s,e,r) + \beta \, \mathbb{E}\big[ V(s') \mid s,e,r \big] \Big\},
\]
where $\beta\in(0,1)$ is the discount factor. The minimization reflects the defender’s objective to minimize loss, while the maximization represents the attacker’s goal of maximizing disruption. 
The recursive structure encodes the trade-off between immediate costs and future consequences, allowing resilience strategies to be optimized over time.  

Shapley’s classical result guarantees that for any finite discounted zero-sum stochastic game \cite{filar2012competitive,shapley1953stochastic},
there exists a stationary Markov strategy profile $(\pi_A^\star, \pi_D^\star)$ that forms an equilibrium,
and that the associated value function $V(s)$ is unique.  
In general-sum games, the definition extends to a \emph{Markov perfect equilibrium}, a stationary profile 
$(\pi_A^\star,\pi_D^\star)$ such that each player’s value function $V_i(s)$ satisfies its own Bellman optimality condition given the opponent’s policy:
\[
V_i(s) \;=\; \mathbb{E}\!\left[ u_i(s,e,r) + \beta V_i(s') \,\middle|\, e \sim \pi_A^\star(s),\ r \sim \pi_D^\star(s) \right],
\] $\pi_i^\star(s) \in \arg\max_{\pi_i} V_i(s).$
While existence of MPE is guaranteed in finite discounted games, uniqueness of equilibrium strategies is not,
and multiple equilibria may yield different resilience outcomes in practice.

\paragraph{Bayesian and epistemic equilibria.}
In realistic 5G networks, players rarely have complete knowledge of the environment or of each other’s capabilities. 
Attackers may be uncertain about whether a given slice is genuine or a honeypot, while defenders may lack clarity about the attacker’s type, for instance, whether the adversary is an unsophisticated botnet operator, a financially motivated cybercriminal, or a nation-state actor with advanced resources. 
To capture such uncertainties, one must move beyond complete-information games and employ \emph{Bayesian games}~\cite{harsanyi1967games}, where players hold private information about their own characteristics, called \emph{types}. 

Formally, let $\theta_A \in \Theta_A$ denote the attacker’s type and $\theta_D \in \Theta_D$ the defender’s type. 
The pair $(\theta_A,\theta_D)$ is drawn from a common prior distribution $p \in \Delta(\Theta_A \times \Theta_D)$. 
Types influence payoffs and feasible actions: a high-skill attacker may have access to advanced jamming techniques, while a low-skill attacker may only generate simple floods. 
Similarly, a resource-rich defender may afford dynamic slice migration, whereas a constrained defender may rely only on throttling.

A \emph{strategy} for player $i \in \{A,D\}$ is a mapping $\sigma_i : \mathcal{H}_t \times \Theta_i \longrightarrow \Delta(\mathcal{A}_i), $
where $\mathcal{H}_t$ denotes the public history of play up to stage $t$, and $\theta_i$ is the player’s private type. 
Thus, strategies are type-dependent: players condition their actions on what they know privately about themselves. 

Because types are unobserved by the opponent, players form and update beliefs over the possible types of the other side. 
Let $\mu_t(\theta_j \mid h_t)$ be the defender’s (or attacker’s) belief about the other player’s type $\theta_j$, conditional on the history $h_t \in \mathcal{H}_t$. 
Beliefs evolve through Bayes’ rule whenever the history $h_t$ occurs with positive probability under the players’ strategies. 
If $O(o_t \mid s_t)$ is the observation kernel describing how observable signals $o_t$ arise from the true state $s_t$, then the posterior distribution satisfies
\[
\mu_{t+1}(\theta_j \mid h_t, o_t) \;=\; \frac{\mu_t(\theta_j \mid h_t)\, O(o_t \mid s_t, \theta_j)}{\sum_{\theta_j'} \mu_t(\theta_j' \mid h_t)\, O(o_t \mid s_t, \theta_j')},
\]
whenever the denominator is nonzero. 
This recursive update reflects how defenders in 5G adjust their threat models in real time by combining prior expectations with newly observed measurement (e.g., abnormal signaling rates or jamming patterns).

\medskip

A central solution concept is the \emph{Perfect Bayesian Equilibrium (PBE)}~\cite{fudenberg1991perfect}. 
A PBE is a triple $(\sigma_A^\star, \sigma_D^\star, \mu^\star)$ consisting of attacker and defender strategies and a system of beliefs such that:

\begin{enumerate}
\item[(i)]\emph{Sequential rationality.} At every history $h_t$ that arises with positive probability under $(\sigma_A^\star,\sigma_D^\star)$, each player’s strategy is optimal given their beliefs. Formally, for each player $i$,
\begin{align}
\sigma_i^\star(\cdot \mid h_t, \theta_i) \in \arg\max_{\sigma_i} \, \mathbb{E}\!\left[\sum_{k=0}^\infty \beta^k u_i(s_{t+k},e_{t+k},r_{t+k}) \,\Big|\, h_t,\theta_i,\sigma_j^\star,\mu^\star \right].
\end{align}

\item[(ii)] \emph{Belief consistency.} Beliefs $\mu^\star$ are updated from the prior $p$ using Bayes’ rule whenever possible, given the equilibrium strategies. 
On equilibrium paths, beliefs coincide with Bayesian posteriors; off equilibrium paths, beliefs are assigned in a way consistent with the spirit of rational inference. 
\end{enumerate}

The PBE concept generalizes Nash equilibrium by requiring not only that strategies are mutual best responses, but also that beliefs are internally consistent with observed actions. This refinement is essential in resilience analysis: a defender deploying honeypots must consider how attackers update their beliefs upon observing suspicious slice behaviors, and whether these beliefs lead attackers to avoid, engage, or ignore the decoys.

\medskip

An alternative but equivalent framework is provided by \emph{epistemic game theory} \cite{aumann1995epistemic}. 
Instead of focusing directly on type distributions, epistemic models describe what players know and believe through \emph{information partitions}. 
The information of each player is represented by a partition of the state space: the player knows in which partition cell they are, but not in which exact state. 
Aumann (1976) formalized the notion of \emph{common knowledge} as information that is not only known by each player but also known to be known, and so on recursively. 
When combined with a common prior over states, epistemic models are equivalent to Harsanyi’s Bayesian games.

\subsection{Equilibria and dynamic programming.}  

Equilibrium analysis in dynamic games is inherently multistage, since players repeatedly interact over time and the outcome of each stage influences future opportunities and payoffs. At each stage $t=0,1,2,\dots$, the system is in a state $s_t \in \mathcal{S}$, and players simultaneously select actions $a_t = (a_t^1,\dots,a_t^N) \in \mathcal{A}^1 \times \cdots \times \mathcal{A}^N$. The state evolves according to a controlled transition kernel
\(
s_{t+1} \sim P(\cdot \mid s_t, a_t),
\)
and each player $i$ accrues a stage payoff $u^i(s_t,a_t)$. The infinite-horizon discounted return of the player $i$ is
\(
J^i(s_0, \sigma) = \mathbb{E} \Bigg[ \sum_{t=0}^\infty \gamma^t u^i(s_t,a_t) \,\Big|\, s_0, \sigma \Bigg],
\)
where $\sigma = (\sigma^1,\dots,\sigma^N)$ is a stationary strategy profile and $\gamma \in (0,1)$ is the discount factor.  

The multistage structure implies that equilibrium computations are inherently recursive: each player’s optimal strategy today depends on the expected continuation value of future states, which in turn depend on the strategies of others. This recursive interdependence is the essence of the connection between dynamic programming and equilibrium analysis.  

\medskip

\noindent\textbf{Zero-sum discounted stochastic games.}  
In the two-player zero-sum case ($u^1 = -u^2$), Shapley’s celebrated recursion generalizes the Bellman equation. Let $V(s)$ denote the value function for the maximizing player. Then
\[
V(s) = \max_{\sigma^1(\cdot \mid s)} \min_{\sigma^2(\cdot \mid s)}
\Bigg\{ \mathbb{E}_{a \sim \sigma^1 \times \sigma^2} \big[u^1(s,a)\big]
+ \gamma \sum_{s' \in \mathcal{S}} P(s' \mid s,a) V(s') \Bigg\}.
\]
This operator is a $\gamma$-contraction on the space of bounded functions, so repeated iteration converges to the unique fixed point $V^\star(s)$. Computationally, this suggests a value iteration scheme:
\[
V_{k+1}(s) = \max_{\sigma^1} \min_{\sigma^2}
\left\{ \mathbb{E}[u^1(s,a)] + \gamma \sum_{s'} P(s' \mid s,a) V_k(s') \right\},
\]
which converges at rate $O(\gamma^k)$. The equilibrium strategies $(\sigma^{1\star}, \sigma^{2\star})$ are obtained by solving the stage game defined by the right-hand side at each state.  

\medskip

\noindent\textbf{General-sum stochastic games.}  
When payoffs are not strictly opposed, each player $i$ seeks to maximize their own return. The value function for player $i$ satisfies the Bellman recursion
\[
V^i(s) = \max_{\sigma^i(\cdot \mid s)}
\mathbb{E}_{a \sim \sigma^i \times \sigma^{-i}}
\Big[ u^i(s,a) + \gamma \sum_{s'} P(s' \mid s,a) V^i(s') \Big].
\]
Unlike the zero-sum case, these equations are coupled across players: the optimality condition for $\sigma^i$ depends on $\sigma^{-i}$, and vice versa. A profile of stationary strategies $\sigma^\star = (\sigma^{1\star},\dots,\sigma^{N\star})$ is a \emph{Markov Perfect Equilibrium (MPE)} if the Bellman recursion for each player is simultaneously satisfied.  

The existence of an MPE in finite-state, finite-action discounted stochastic games is guaranteed, but uniqueness is not. Computation typically requires iterative fixed-point algorithms (e.g., policy iteration, Newton–Kantorovich methods, or reinforcement learning approximations). Complexity grows sharply with the number of players, as solving the stage game at each iteration involves computing Nash equilibria of finite normal-form games.  

\medskip

\noindent\textbf{Bayesian dynamic games.}  
When players have private types $\theta_i \in \Theta_i$, the effective state includes not only physical conditions but also beliefs about types. Let $\mu_t \in \Delta(\Theta)$ denote the distribution of public belief in the type profiles $\theta = (\theta_1,\dots,\theta_N)$. Given public history $h_t$, players update their beliefs using the Bayes rule. The dynamic programming recursion is now expressed in the belief space:
\[
V^i(\mu) = \max_{\sigma^i(\cdot \mid \mu)} 
\; \mathbb{E}_{\theta \sim \mu,\,a \sim \sigma^i \times \sigma^{-i}} 
\Big[ u^i(\theta,a) + \gamma \sum_{\mu'} P(\mu' \mid \mu,a) V^i(\mu') \Big],
\]
where $P(\mu' \mid \mu,a)$ is the transition kernel over beliefs induced by Bayes’ rule. Equilibria of such Bayesian dynamic games refine the Perfect Bayesian Equilibrium to the infinite-horizon setting, requiring consistency of strategies and beliefs at every stage.  

Computationally, belief spaces are continuous and high-dimensional, so approximate dynamic programming or point-based value iteration (adapted from POMDP methods) is often necessary. This highlights the intrinsic difficulty of incomplete-information dynamic games compared to fully observed ones.  

\medskip

\section{Learning as the Foundation}

Game-theoretic modeling provides a rigorous language for analyzing adversarial interactions, but its practical effectiveness is constrained by uncertainties in both model structure and parameters. Threat models rarely capture the full spectrum of adversarial behavior, especially in the presence of zero-day exploits or highly adaptive attackers. Even if models are initially accurate, adversaries may evolve faster than the models can be updated. Furthermore, aleatoric uncertainties in payoffs, admissible action sets, common priors, and network dynamics introduce further mismatches between prescriptive models and operational realities.  

Learning emerges as the essential complement to game and control theory. By continuously adapting policies based on feedback, refining beliefs, and recalibrating models, learning enables defenders to sustain resilience in nonstationary and uncertain environments. In this sense, learning is not an afterthought but the foundational layer that operationalizes resilience. It transforms equilibrium analysis from a static prediction into a dynamic process of adaptation, recovery, and stability.

\begin{figure}[t]
\centering
\includegraphics[trim={0 0 0 0.62in}, clip, width=0.99\linewidth]{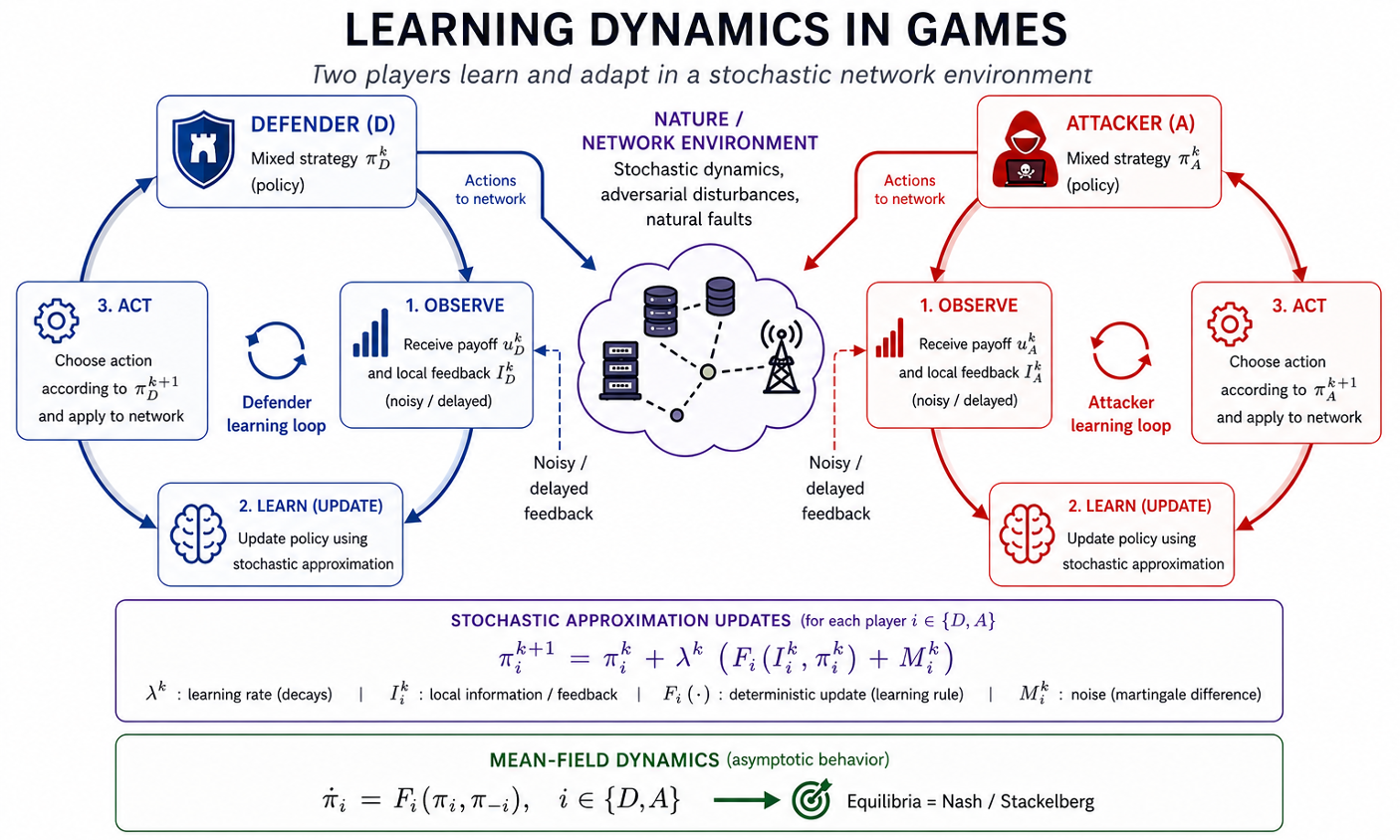}
\caption{\small Learning dynamics in adversarial networked systems. Both defender and attacker operate through closed-loop observe--learn--act cycles under noisy and delayed feedback. Their stochastic approximation updates induce a mean-field dynamical system whose equilibria correspond to Nash or Stackelberg outcomes.}
\label{fig:learning_dynamics_games}
\end{figure}

Figure~\ref{fig:learning_dynamics_games} illustrates this perspective. Rather than assuming that players instantaneously reach equilibrium, both the defender and the attacker evolve their strategies through coupled feedback loops. Each player observes local signals from the network, updates its policy based on noisy information, and acts on the system, thereby influencing future observations. The resulting interaction is inherently dynamical and stochastic, embedding game-theoretic reasoning within a learning-driven feedback architecture.

\subsection{Learning dynamics in games}

Learning in games provides a mathematical framework to describe how players iteratively adjust strategies over repeated interactions \cite{zhu2012hybrid,li2022confluence}. Let $\pi^k_i \in \Delta(\mathcal{A}_i)$ denote the mixed strategy of player $i \in \{A,D\}$ at stage $k$, where $\Delta(\mathcal{A}_i)$ is the simplex over the action set. As depicted in Fig.~\ref{fig:learning_dynamics_games}, each iteration consists of an observation of payoffs and feedback, followed by a policy update and an action applied to the network.

Suppose the defender receives a payoff $u_D^k$ at stage $k$, generated from the stochastic dynamics of the network, and updates its policy according to
\[
\pi^{k+1}_D = \pi^k_D + \lambda^k \big(F_D(I^k_D,\pi^k_D) + M^k_D\big).
\]
Here $\lambda^k$ is the learning rate, $I^k_D$ is the defender’s feedback (such as local queue lengths, delays, or alerts), $F_D$ is the deterministic update term encoding the learning rule, and $M^k_D$ is a martingale-difference noise term capturing randomness in observations. The attacker updates $\pi_A^k$ analogously, forming the coupled learning loops shown in the figure.

This recursion belongs to the class of stochastic approximation algorithms. Under standard conditions on step sizes ($\lambda^k \to 0$ with $\sum_k \lambda^k = \infty$) and boundedness of noise, the asymptotic behavior of $\{\pi^k_A, \pi^k_D\}$ can be studied via the associated mean-field ordinary differential equation (ODE):
\[
\dot{\pi}_i = F_i(\pi_i, \pi_{-i}), \quad i \in \{A,D\}.
\]
As highlighted at the bottom of Fig.~\ref{fig:learning_dynamics_games}, equilibria of this ODE correspond to Nash or Stackelberg equilibria of the underlying game~\cite{holt2004nash}. The key question is whether these equilibria are stable: if deviations caused by shocks or adversarial perturbations are corrected by the learning dynamics, then the system converges back toward equilibrium. Lyapunov analysis provides the mathematical tool to certify this stability.

\subsection{Resilience control as a learning process}

Resilience should not be understood only as the asymptotic convergence to equilibrium, but as the ability of the learning process itself to adapt, recover, and stabilize performance in the presence of disturbances. Formally, let $V^\star(s)$ denote the defender’s equilibrium value function under the limiting strategy profile $\pi^\star = (\pi_A^\star, \pi_D^\star)$. The system is said to be \emph{resilient} if the learning trajectories $(\pi_A^k,\pi_D^k)$ ensure both bounded long-run losses and effective short-run recovery whenever disruptions occur.  

Suppose there exists a Lyapunov function $L(\pi_A,\pi_D) \geq 0$ such that along the trajectories of the learning process,
\[
\dot{L}(\pi_A,\pi_D) \;\leq\; -\alpha \Big(\ell(x_t,e_t,r_t) - \overline{L}\Big),
\]
for some $\alpha > 0$, where $\ell(x_t,e_t,r_t)$ is the instantaneous loss and $\overline{L}$ is a performance threshold. This inequality implies that whenever loss exceeds acceptable levels, the adaptive process acts to reduce it. Resilience here is not a fixed robustness property but a dynamic guarantee: the learning system continuously monitors, adapts, and recovers, ensuring that even under repeated adversarial shocks or random faults, the process is driven back toward safe operation.  
Thus, resilience is inherently a \emph{learning process}. The system does not simply resist attacks; it observes deviations, updates strategies, and recovers performance in real time. This aligns resilience with the asymptotic stability of equilibria while also emphasizing transient adaptation and recovery.

\subsection{Feedback structures and observability}

The convergence and quality of learning dynamics depend critically on the information structure. In complete-information settings, where both players observe the full state and payoffs, learning rules such as fictitious play or regret minimization can converge to equilibrium under broad conditions. In realistic cyber-defense scenarios, however, feedback is local, noisy, and often delayed.  

For example, a defender may only observe local queue lengths and slice performance, not the attacker’s action directly. Conversely, the attacker may infer migration policies indirectly through probing rather than observing them outright. Formally, if $I^k_i$ denotes the local signal available to player $i$, then the update rule depends on conditional expectations $\mathbb{E}[u_i \mid I^k_i]$ rather than exact utilities. Stability in such decentralized settings typically requires monotonicity or a potential structure in the induced ODE, ensuring that distributed learning rules do not diverge or cycle.  

This feedback limitation makes conjectural learning particularly relevant: defenders and attackers both form forecasts about the opponent’s strategy, update them based on local signals, and adapt accordingly. The accuracy and adaptability of these conjectures directly influence resilience.

\subsection{Reinforcement and Meta-Learning}

Reinforcement learning (RL) arises when agents update strategies directly from payoff signals without relying on explicit models of system dynamics or adversary behavior~\cite{8714026}. In the context of resilience, the defender faces a sequential decision problem: at each state $s_t$, actions $a_t$ correspond to mitigation mechanisms such as workload migration, resource scaling, slice balancing, or spectrum reassignment. The environment responds stochastically due to both adversarial disturbances and natural faults, generating a new state $s_{t+1}$ and payoff $u_t$. The defender updates its value estimates through experience, gradually improving policy quality.

A canonical update rule is the $Q$-learning recursion,
\begin{equation}
Q_{t+1}(s,a) \;=\; (1-\alpha_t)\,Q_t(s,a) \;+\; 
\alpha_t \Big(u_t + \beta \max_{a'} Q_t(s',a')\Big),
\end{equation}
where $\alpha_t$ is a decaying learning rate and $\beta$ the discount factor. This update allows the defender to learn effective resilience policies online even when attacker strategies are unknown or non-stationary. For instance, if an adversary launches jamming bursts intermittently, RL enables the defender to discover spectrum-hopping patterns that maximize service continuity while minimizing unnecessary overhead.

\paragraph{Meta-Learning.}
Classical RL optimizes for a single environment, but cyber adversaries evolve, and novel exploits continuously emerge. Meta-learning extends RL by training across a distribution of tasks (attack scenarios), producing a meta-policy that encodes adaptation strategies rather than fixed responses. In practice, the meta-policy is parameterized by $\theta$, which is updated in two loops: an \emph{inner loop} that adapts to a specific task (e.g., denial-of-service against MEC nodes), and an \emph{outer loop} that optimizes $\theta$ across tasks to enable rapid adaptation. Formally, the outer update takes the form
\[
\theta \leftarrow \theta - \gamma \nabla_\theta \mathbb{E}_{\mathcal{T} \sim p(\mathcal{T})}\big[ \mathcal{L}_{\mathcal{T}}(\theta - \eta \nabla_\theta \mathcal{L}_{\mathcal{T}}(\theta)) \big],
\]
where $\mathcal{T}$ indexes attack tasks, $\mathcal{L}_{\mathcal{T}}$ is the loss for task $\mathcal{T}$, and $(\eta,\gamma)$ are step sizes. This Model-Agnostic Meta-Learning (MAML) formulation \cite{pan2025model,pan2024model,pan2023first} is particularly suited to resilience, as it equips defenders to handle zero-day exploits with minimal retraining, critical in environments where reaction time directly impacts service-level compliance.

\paragraph{Conjectural Online Learning.}
In asymmetric-information settings, defenders must anticipate attacker strategies while updating their own. Conjectural Online Learning (COL) generalizes RL by incorporating forecasts of opponent behavior into the learning loop \cite{li2024conjectural,tao23cola}. Architectures such as forecaster–actor–critic models maintain beliefs $\mu_t$ about attacker actions, update them as evidence accumulates, and adapt policies accordingly. Concretely, the defender maintains:
\begin{itemize}
\item a \emph{forecaster} that predicts the attacker’s next action distribution given past observations,
\item an \emph{actor} that selects a mitigation action $a_t$ based on both current state and forecasts,
\item a \emph{critic} that evaluates outcomes to refine value estimates and update beliefs.
\end{itemize}
This creates a closed-loop adaptation scheme where resilience is maintained not only against known attack models but also against adaptive and persistent adversaries who themselves learn over time.

\subsection{Toward Predictive Resilience}

The integration of RL, meta-learning, and conjectural online adaptation highlights that resilience can be both \emph{predicted} and \emph{designed}. Specifically, resilience corresponds to the ability of learning dynamics to converge toward equilibria that sustain bounded loss while satisfying service-level guarantees. If trajectories of $(s_t,a_t)$ remain within a basin of attraction of resilient equilibria, disruptions are contained; if they cycle or diverge, resilience collapses even if equilibria exist in principle.

\paragraph{Equilibria and Learning Dynamics.}
In a slice-mitigation example, the defender learns a threshold-based migration rule: when load exceeds a critical level due to adversarial flooding, workloads are shifted to backup MEC nodes. If the attacker stabilizes at a sustainable level of flooding intensity, the system converges to a stable equilibrium characterized by a steady-state value function $V^\star(s)$ satisfying the Bellman equation
\[
V^\star(s) = \max_{a \in \mathcal{A}} \Big( u(s,a) + \beta \mathbb{E}[V^\star(s') \mid s,a] \Big).
\]
Here, $V^\star(s)$ serves both as a predictive metric of resilience and as a design target.  

\paragraph{Basins of Attraction.}
The resilience of a system depends not only on the equilibria but also on whether learning-driven trajectories converge into them. For example, policies that yield oscillatory migration patterns (frequent switching of slices) may technically satisfy SLA thresholds on average but erode resilience due to instability and wasted resources. Designing reward shaping, step-size schedules, and regularization terms ensures convergence into stable equilibria rather than into pathological cycles~\cite{nugraha2022rolling,huang2020dynamic}.

\paragraph{Predictive Implications.}
Resilience is therefore defined not simply by static metrics (e.g., MTTR, downtime) but by the dynamical properties of learning. Predictive resilience analysis asks: given a learning architecture (RL, meta-learning, COL) and an adversarial strategy space, does the joint adaptation process converge, recover after perturbations, and sustain service guarantees? By embedding resilience objectives into the learning dynamics, networks can be designed to self-stabilize under uncertainty, thereby making resilience a controllable and verifiable property of next-generation infrastructures.

\section{Network-Theoretic Foundation}

\begin{figure}[h!]
    \centering
    \includegraphics[width=0.85\linewidth]{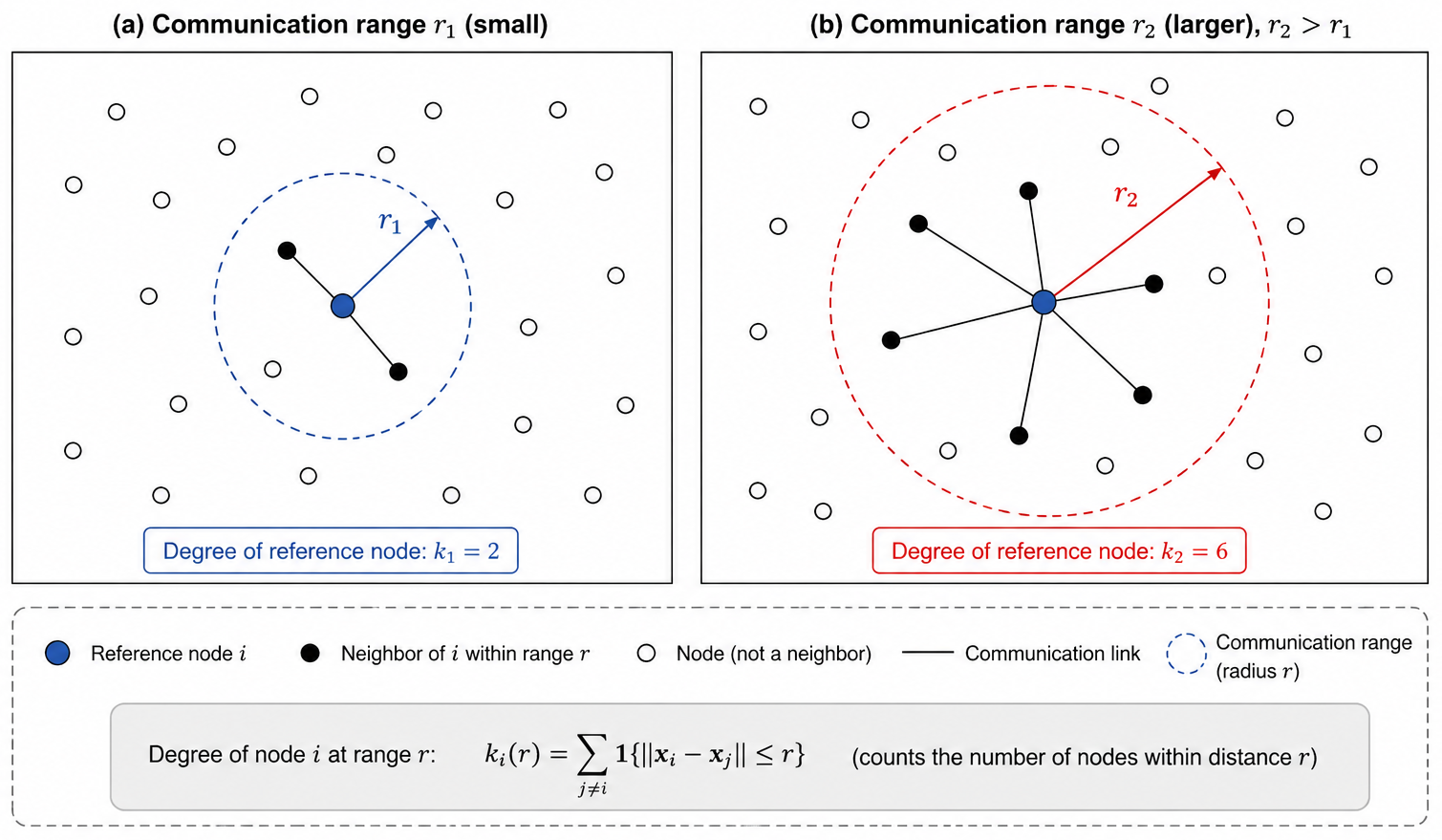}
    \caption{\small Effect of communication range on node connectivity. Increasing the range from $r_1$ to $r_2$ increases the degree of the reference node from $k=1$ to $k=6$, illustrating the dependence of connectivity on spatial density and transmission range.}
    \label{fig:connectivity_fig}
\end{figure}

The resilience of complex communication infrastructures is intrinsically shaped by the structural and spatial organization of their connectivity.  
Network science provides a rigorous mathematical language to describe, analyze, and quantify these interconnections and their influence on the ability of a system to maintain functionality under failures or disturbances~\cite{barabasi2013network}.  
This section presents the foundational concepts underlying network connectivity, spatial geometry, percolation, and propagation, which together form the analytical basis for resilience modeling. The role of spatial embedding and heterogeneous communication ranges in shaping local connectivity is illustrated in Fig.~\ref{fig:connectivity_fig}, where the degree of a reference node varies with transmission range.


\subsubsection{Foundations of Network Connectivity}
\begin{figure}[h!]
    \centering
    \includegraphics[width=0.9\linewidth]{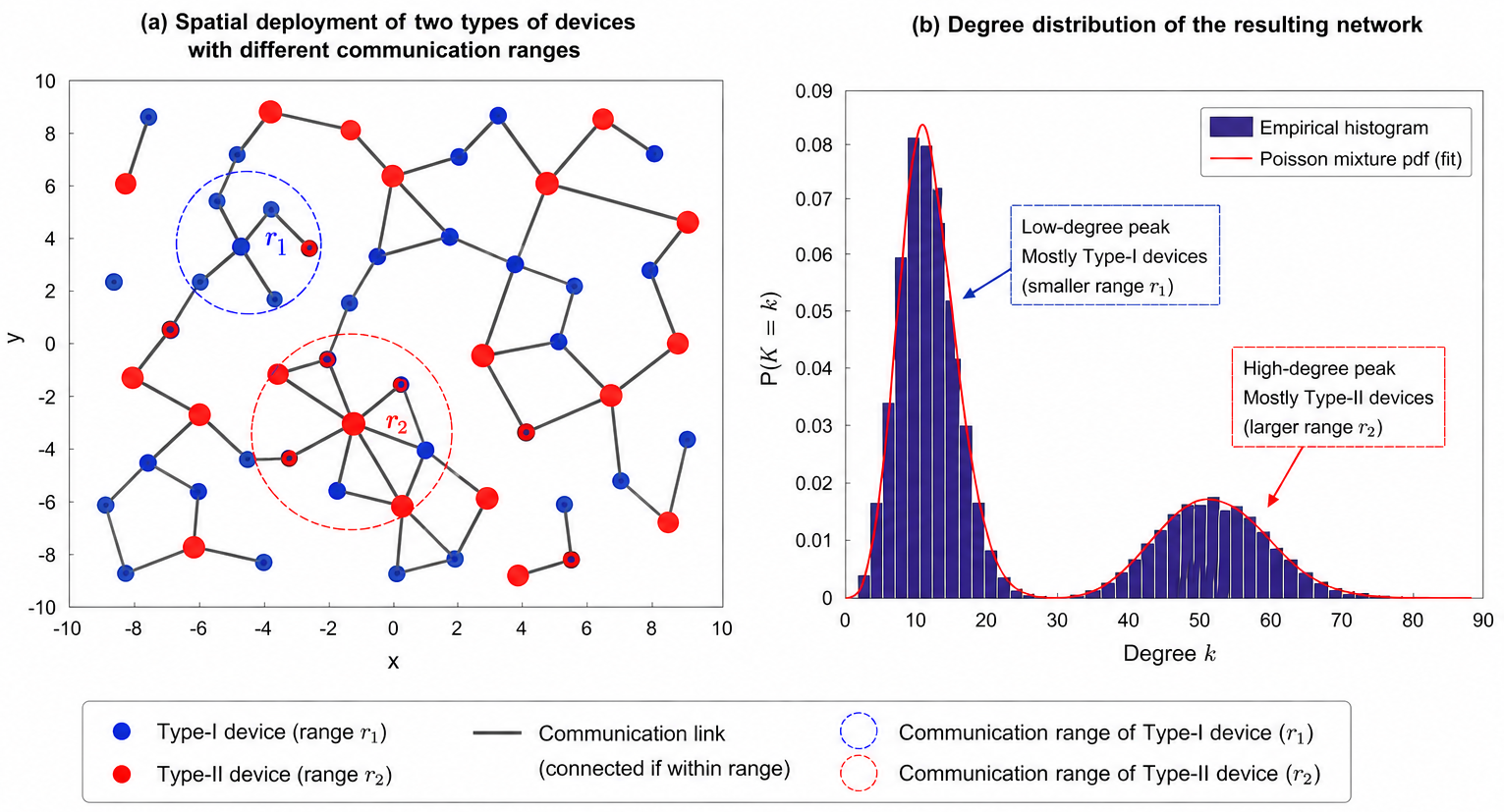}
    \caption{\small Spatial heterogeneity and its impact on network connectivity. (a) Spatial deployment of two device types with communication ranges $r_1$ and $r_2$, forming a heterogeneous random geometric network. (b) Resulting degree distribution, showing a bimodal structure corresponding to low-degree (Type-I) and high-degree (Type-II) devices, with a Poisson mixture fit.}
    \label{fig:degree_distribution}
\end{figure}

Let a network be represented as a graph $\mathcal{G}=(\mathcal{V},\mathcal{E})$, where $\mathcal{V}$ is the set of nodes and $\mathcal{E}$ is the set of links between them.  
The local connectivity of the graph is characterized by the degree distribution $P(k)$, which defines the probability that a randomly selected node has $k$ connections.  
The global structure of $\mathcal{G}$ determines how efficiently information or failures can propagate. The impact of heterogeneous spatial deployment and communication ranges on connectivity is illustrated in Fig.~\ref{fig:degree_distribution}, which exhibits a bimodal degree distribution corresponding to devices with different transmission ranges.

In the classical Erdős–Rényi (ER) random graph~\cite{erdHos1960evolution}, each pair of nodes is connected independently with probability $p$, producing a binomial degree distribution that approaches a Poisson law for large networks.  
Although mathematically tractable, the ER model assumes homogeneous connectivity, which seldom reflects the irregular and heterogeneous structures observed in real communication systems.  

Empirical studies have demonstrated that many natural and engineered networks, including the Internet and social interaction graphs, exhibit heavy-tailed degree distributions that follow a power law,
\[
P(k) \sim k^{-\gamma}, \qquad 2 < \gamma < 3.
\]
Such \emph{scale-free} networks emerge from growth and preferential attachment mechanisms and possess a small number of highly connected hubs~\cite{barabasi2003scale}.  
This topology improves efficiency and fault tolerance against random failures yet increases vulnerability to targeted attacks on hubs.  
Structural robustness can be quantitatively evaluated using spectral metrics such as the algebraic connectivity $\lambda_2(L)$ of the Laplacian matrix, which measures global cohesion, and the spectral radius $\rho(A)$ of the adjacency matrix, which determines the potential for rapid diffusion of perturbations.

\subsubsection*{Spatial Embedding and Random Geometric Networks}

While abstract topological models reveal structural heterogeneity, real communication infrastructures are spatially embedded and constrained by physical proximity, propagation, and interference.  
Spatial randomness is commonly captured using the homogeneous Poisson point process (PPP)~\cite{heath2013modeling},
\[
\Phi = \{x_i\}_{i \ge 1} \subset \mathbb{R}^2, \qquad x_i \sim \text{PPP}(\lambda),
\]
where $\lambda$ denotes the node density in nodes per unit area.  
Two nodes $x_i$ and $x_j$ are connected if $\|x_i - x_j\| \le r$, where $r$ represents the communication range.  
The resulting random geometric graph (RGG)~\cite{dall2002random} exhibits a Poisson degree distribution
\[
\mathbb{P}[K=k] = e^{-\lambda \pi r^2}\frac{(\lambda \pi r^2)^k}{k!}, \qquad \mathbb{E}[K] = \lambda \pi r^2.
\]
This spatial model establishes a direct correspondence between physical deployment parameters $(\lambda, r)$ and network connectivity.  
As the density $\lambda$ or range $r$ increases, the network transitions from a fragmented topology to a fully connected one.  
This phase transition is described by percolation theory.

\subsubsection*{Percolation and Structural Transitions}

Percolation theory provides the mathematical framework to study the emergence of large-scale connectivity as links or nodes are randomly added or removed~\cite{perc1}.  
Let $p$ denote the probability that a link or node is functional.  
A critical value $p_c$ marks the percolation threshold that separates disconnected and connected regimes.  
For $p < p_c$, the network consists of isolated clusters of limited size; for $p > p_c$, a macroscopic connected component spans the system, enabling long-range communication or cascading influence.  
In random geometric networks, $p_c$ depends on the product $\lambda \pi r^2$, which represents the expected number of neighbors per node.  
The percolation transition thus reflects a geometric phase change controlled by deployment density and communication range, providing an interpretable link between physical layout and resilience.

In the context of communication systems, operating well above the percolation threshold ensures redundant paths and fault tolerance.  
Conversely, systems near the critical regime are sensitive to small perturbations and therefore exhibit fragile connectivity.  
Percolation theory thereby formalizes the notion of \emph{structural resilience}, where connectivity and reachability are preserved despite random disruptions.

\subsubsection*{Dynamic Processes and Propagation Models}

Structural connectivity defines the potential for interaction, but resilience also depends on how dynamic processes evolve over the network.  
Information dissemination, congestion, cascading failures, and cyber threats can all be modeled as state transitions among interacting nodes~\cite{fu2013propagation}.  
Let $x_i(t)$ denote the state of node $i$ at time $t$.  
A general state-space representation is given by
\[
\frac{dx_i(t)}{dt} = \sum_{j} a_{ij} \, f\big(x_i(t),x_j(t)\big),
\]
where $a_{ij}$ are elements of the adjacency matrix $A$ and $f(\cdot)$ specifies the interaction dynamics.  
Epidemic-inspired formulations provide an intuitive basis for these interactions.  
In the susceptible–infected–susceptible (SIS) model~\cite{parshani2010epidemic}, the rate of change of the infected fraction in the population is
\[
\frac{dI(t)}{dt} = \beta (1 - I(t)) \Theta(t) - \mu I(t),
\]
where $\beta$ is the transmission rate, $\mu$ the recovery rate, and $\Theta(t)$ the probability that a randomly selected neighbor is infected.  
This abstraction applies not only to malware propagation but also to the diffusion of information or overload within a networked infrastructure.

Linearization around equilibrium yields a stability condition governed by the spectral radius $\rho(A)$.  
The process remains subcritical, and disturbances dissipate if the effective ratio $\beta / \mu < 1/\rho(A)$, whereas supercritical dynamics ($\beta / \mu > 1/\rho(A)$) lead to self-sustaining propagation.  
This threshold provides a direct connection between structural properties and dynamic stability.

\subsubsection*{Equilibrium Analysis and Design Interpretation}

The equilibrium of the propagation process characterizes the steady-state composition of network states, such as the proportion of active, inactive, or compromised nodes, as functions of structural and dynamical parameters.  
At equilibrium, the condition
\[
\mathcal{R}_0 = \frac{\beta \, \mathbb{E}[K]}{\mu} < 1
\]
ensures that the system remains resilient, with perturbations decaying to zero~\cite{farooq2018secure,farooq2019modeling}.  
Here $\mathcal{R}_0$ represents the effective reproduction number, which integrates both topological and temporal effects.  
The threshold $\mathcal{R}_0 = 1$ delineates the boundary between stability and systemic propagation.  
This framework leads to an equilibrium-centric view of network design:  
resilience can be engineered by adjusting controllable parameters such as density $\lambda$, communication range $r$, and recovery rate $\mu$ to ensure that the equilibrium remains within the subcritical regime.

\chapter{Cyber Risk Assessment for NextG Networks}


Chapter 4 develops the framework for cyber risk assessment in next-generation (NextG) networks. Building on the structural and dynamical models introduced in Chapter 3, risk is treated as a quantitative measure that captures the likelihood and impact of disruptions on system performance.

This chapter formulates risk as a function of network state, threat processes, and system response, enabling prediction and comparison of adverse outcomes across different operating conditions. These formulations provide the basis for risk-informed decision-making and directly support the resilient system design methodologies developed in the subsequent chapter.

\section{Probabilistic Risk Assessment with Digital Twins}

Probabilistic Risk Assessment (PRA) provides a principled framework for quantifying the likelihood and consequences of disruptive events in complex infrastructures~\cite{smidts2022cyber,zhao2020finite}. Unlike classical reliability analysis, which assumes deterministic failures or component-level probabilities, PRA explicitly accounts for uncertainty by assigning probabilities to a family of disruption scenarios and evaluating their systemic effects. This probabilistic treatment is particularly critical in 5G/NextG networks, where virtualization, disaggregation, and open programmability create broad and interdependent attack surfaces.

 Let $\mathcal{Q} = \{q^{(1)}, q^{(2)}, \dots, q^{(m)}\}$ denote the set of disruptive scenarios, each of which occurs with probability $p\big(q^{(i)}\big)$. A scenario $q^{(i)}$ can represent, for example, a fronthaul jamming attack, a CU software exploit, or simultaneous slice-level faults. For a given defense policy $\pi$, the loss of resilience in scenario $q^{(i)}$ is denoted by $L^{(q^{(i)})}(\pi)$. PRA then evaluates the expected resilience loss
\begin{equation}
    \mathbb{E}[L(\pi)] = \sum_{i=1}^m p \big(q^{(i)}\big)\, L^{(q^{(i)})}(\pi),
\end{equation}
and may further assess higher-order risk measures such as variance or Conditional Value-at-Risk (CVaR)~\cite{uryasev2001conditional}. These provide operators with insight into which scenarios dominate risk, which investments in resilience yield the largest marginal benefit, and how combinations of hazards can amplify fragility. 

Within the digital-twin (DT) dynamics of Chapter~3, disruptions evolve in time according to \eqref{eq:disruption}, where $q_t$ describes the temporal evolution of disruptions within the system at time $t$, while $q^{(i)}$ is used as a scenario label in PRA. This distinction allows us to integrate the probabilistic scenario analysis of PRA with the dynamic models of Chapter~3.

 DTs complement PRA by simulating the impact of scenarios $q^{(i)}$ on the dynamics of the evolving network over time. A DT maintains a synchronized network state $\widehat{x}_t$ that mirrors the real state $x_t$ and generates synthetic measurements $\widehat{y}_t$ that align with the observed outputs $y_t$. PRA provides probabilities over scenarios, while DTs instantiate these scenarios in realistic 5G dynamics, enabling operators to explore ``what-if'' conditions without endangering live systems.

\subsection{State–Space View and Twin Synchronization}

DTs augment PRA by embedding disruption scenarios into the hybrid dynamics introduced in Chapter~3. While PRA quantifies risks at the level of probabilities, DTs instantiate them within a stochastic state–space model that mirrors live 5G networks. The real network state evolution is modeled as
\begin{equation}
\label{eq:real}
x_{t+1} = f\!\left(x_t, u_t, w_t, \xi_t, q_t\right),
\end{equation}
where $x_t$ is the network state, $u_t$ is the control action generated by the policy $\pi$, $w_t$ denotes adversarial inputs (e.g., jamming, API abuse), $\xi_t$ is the natural disturbance (e.g., traffic bursts, random failures), and $q_t$ is the disruption process describing abnormal operating modes. In this section, we treat $w_t$ as fixed by the scenario, so PRA evaluates how different possible trajectories of $q_t$ affect resilience. More generally, disruptions evolve dynamically according to
\begin{equation}
\label{eq:disruption-general}
q_{t+1} = \phi\!\left(q_t, x_t, u_t, w_t, \xi_t\right),
\end{equation}
so adversarial, environmental, and control factors jointly shape future disruptions.

The DT runs in parallel as a surrogate model,
\begin{equation}
\label{eq:twin}
\widehat{x}_{t+1} = \widehat{f}\!\left(\widehat{x}_t, u_t, \widehat{w}_t, \xi_t, q_t\right),
\end{equation}
driven by the same controls $u_t$ and by replayed or simulated attacks $\widehat{w}_t$. Synchronization is quantified by
\begin{equation}
e_t = \|x_t - \widehat{x}_t\|,
\qquad
\mathcal{F}_T = 1 - \frac{1}{T}\sum_{t=1}^T \frac{e_t}{e_{\max}},
\end{equation}
where $\mathcal{F}_T \in [0,1]$ is a normalized fidelity score. High values of $\mathcal{F}_T$ indicate that the twin closely captures the evolution of the real network, ensuring that the PRA results remain representative of the 5G dynamics.

 Let $\mathcal{Q}=\{q^{(1)},\dots,q^{(m)}\}$ denote the disruption \emph{scenarios} (labels) with probabilities $p\!\left(q^{(i)}\right)$. A scenario $q^{(i)}$ specifies a fixed adversarial input path $w_{0:T-1}^{(i)}$, initial conditions $(x_0^{(i)},q_0^{(i)})$, and, when feedback evolution is enabled, the disruption law $\phi^{(i)}$ used in \eqref{eq:disruption-general}. Given a policy $u_t=\pi(\widehat{x}_t)$, the twin rolls out
\[
\widehat{x}_{t+1}^{(i)}=\widehat{f}\!\left(\widehat{x}_t^{(i)},\,\pi(\widehat{x}_t^{(i)}),\,\widehat{w}_t^{(i)},\,\xi_t,\,q_t\right),
\]
producing a trajectory $\{\widehat{x}_t^{(i)}\}_{t=0}^T$ for the scenario $q^{(i)}$. (When feedback evolution is used, $q_t$ is propagated via \\$q_{t+1}=\phi^{(i)}(q_t,\widehat{x}_t^{(i)},\pi(\widehat{x}_t^{(i)}),\widehat{w}_t^{(i)},\xi_t)$.) 
 Let $h:\mathbb{R}^n\!\to\!\mathbb{R}$ map the state to a scalar \emph{measurement} (e.g., slice latency, URLLC reliability, throughput). Define the normalized shortfall and cumulative loss in a window $[t_f,t_r]$:
\begin{equation}
\widehat{Q}_t^{(i)}=h\!\big(\widehat{x}_t^{(i)}\big),
\ 
s_t^{(i)}(\pi) \;=\; 1-\frac{\widehat{Q}_t^{(i)}}{Q_{\max}},
\ 
L^{(q^{(i)})}(\pi)=\sum_{t=t_f}^{t_r}s_t^{(i)}(\pi).
\label{eq:resilience-from-traj}
\end{equation}
Two complementary summaries are often reported:
\[
M^{(q^{(i)})}(\pi)=\max_{t\in[t_f,t_r]} s_t^{(i)}(\pi)\quad\text{(maximum drop)}, 
\]
\[
D^{(q^{(i)})}(\pi)=s_{t_r}^{(i)}(\pi)\quad\text{(residual deficit)}.
\]
For availability-style evaluation with threshold $Q_{\min}$:
\[
T_{\textrm{d}}^{(q^{(i)})}(\pi)=\sum_{t=t_f}^{t_r} \mathbf{1}\!\left\{\widehat{Q}_t^{(i)}<Q_{\min}\right\}.
\]

 To account for randomness in $\xi_t$, draw $N$ i.i.d.\ samples $\{\xi_{0:T-1}^{(i,k)}\}_{k=1}^N$ and average the losses:
\begin{align}
&\widehat{\mathbb{E}}_{\xi}\!\big[L^{(q^{(i)})}(\pi)\big]
=\frac{1}{N}\sum_{k=1}^N 
L^{(q^{(i)}),\,\xi^{(i,k)}}(\pi),
\\
&\widehat{\mathrm{Var}}_{\xi}\!\big[L^{(q^{(i)})}(\pi)\big]
=\frac{1}{N-1}\sum_{k=1}^N 
\Big(L^{(q^{(i)}),\,\xi^{(i,k)}}(\pi)-\widehat{\mathbb{E}}_{\xi}[L^{(q^{(i)})}]\Big)^2.
\end{align}

 PRA combines per-scenario statistics using the scenario probabilities:
\begin{equation}
\label{eq:aggregate-risk}
\begin{aligned}
\mathbb{E}[L(\pi)]
&= \sum_{i=1}^m p\!\big(q^{(i)}\big)\,
   \mathbb{E}_{\xi}\!\big[L^{(q^{(i)})}(\pi)\big], \\
\mathrm{CVaR}_{\alpha}[L(\pi)]
&= \min_{\eta}\Big\{
    \eta + \tfrac{1}{1-\alpha}\,
    \mathbb{E}\big[(L(\pi)-\eta)_+\big]
   \Big\}.
\end{aligned}
\end{equation}
These quantities assess how resilient the system is, on average and in the tail, across the set of scenarios $\mathcal{Q}$.

In URLLC, $h(x_t)$ might return the reliability of the meeting deadline; shortfalls $s_t^{(i)}$ then capture reliability dips during fronthaul jamming plus CU overload. In eMBB, $h(x_t)$ can be throughput; the same pipeline quantifies drops and recovery after the transport-layer congestion. Because the twin uses the same policy $u_t=\pi(\widehat{x}_t)$ as the real network and maintains high synchronization fidelity, the resulting metrics $L^{(q^{(i)})}, M^{(q^{(i)})}, D^{(q^{(i)})}, T_{\textrm{d}}^{(q^{(i)})}$ provide actionable evidence about how \emph{state} $x_t$ evolves under each scenario and how robust the system is across the mixture defined by $p(q^{(i)})$.

\section{Strategic Risk Assessment}

Strategic risk assessment assesses resilience against \emph{adaptive} adversaries who plan, coordinate, and optimize their actions. It extends the PRA framework by replacing exogenous scenario probabilities with adversarial decision-making. Whereas PRA quantifies expected loss across random disruptions, strategic risk asks: \textit{What is the performance degradation when attacker and defender both act strategically?} In 5G/NextG networks, this formulation is crucial: adversaries can investigate vulnerabilities, adapt to defenses, and optimize their disruption strategies over time. 
Formally, we embed the attacker-defender interaction in the same hybrid dynamics as in Chapter~3, but with $w_t$ (attacker actions) chosen strategically rather than probabilistically.  

\subsection{Red–Blue Teaming as a Game}

The network evolves according to \eqref{eq:hybrid-dynamics} and \eqref{eq:disruption}, where $u_t$ is the defender's action, $w_t$ is the attacker's action. Both sides choose actions through policies $\pi_b$ and $\pi_r$, possibly under partial observations (as in Chapter~3). 
 Let $h:\mathbb{R}^n \to \mathbb{R}$ be a scalar performance measure (e.g., URLLC reliability, eMBB throughput), normalized by $Q_{\max}>0$.  The defender seeks to preserve performance. Define stage loss
\[
\ell_b(x,u) \;=\; 1 - \frac{h(x)}{Q_{\max}} + c_b(u),
\]
where $c_b(u)$ is the operational cost of defense (e.g., scaling or rerouting overhead). The defender minimizes cumulative loss
\begin{equation}
L_b(\pi_b,\pi_r) \;=\; \sum_{t=t_f}^{t_r} \ell_b(x_t,u_t).
\label{eq:defender-loss}
\end{equation}

The adversary may pursue a different objective from simply maximizing $L_b$. For example, it may wish to maximize disruption subject to resource constraints or balance stealth with impact. Define the stage utility
\[
u_r(x,w) \;=\; \alpha \,\Big(1 - \tfrac{h(x)}{Q_{\max}}\Big) - c_r(w),
\]
where $\alpha \ge 0$ weights the attacker's preference for disruption and $c_r(w)$ is the cost of the attack effort (for example, power for jamming, risk of detection). The attacker maximizes cumulative utility
\begin{equation}
L_r(\pi_b,\pi_r) \;=\; \sum_{t=t_f}^{t_r} u_r(x_t,w_t).
\label{eq:attacker-utility}
\end{equation}

This formulation admits both \emph{zero-sum} (when $L_r=-L_b$ and costs align) and \emph{general-sum} (when attacker’s goal differs) structures.

Strategic risk formalizes resilience evaluation when the attacker and the defender act strategically rather than randomly \cite{pawlick2015flip,pawlick2017strategic}.  
Let $\Pi_b$ and $\Pi_r$ denote the sets of admissible defender and attacker policies, respectively.  
Each policy profile $(\pi_b,\pi_r)\in \Pi_b\times\Pi_r$ induces a stochastic trajectory $\{x_t\}$ through the dynamics of Eq.~\eqref{eq:hybrid-dynamics}, and the corresponding performance metrics such as the cumulative loss of defenders $L_b(\pi_b,\pi_r)$ and the utility of attackers $L_r(\pi_b,\pi_r)$.  

\paragraph{Definition (Strategic Risk).}  
The \emph{strategic risk} of a system under the policy profile $(\pi_b,\pi_r)$ is the vector
\[
\mathcal{R}(\pi_b,\pi_r) 
= \Big( \; \mathbb{E}[L_b(\pi_b,\pi_r)], \;\; \mathbb{E}[L_r(\pi_b,\pi_r)] \;\Big),
\]
where expectations are taken over exogenous uncertainties $\xi_t$ (and possibly over randomized strategies).  
The strategic risk profile of the system is the set
\[
\mathcal{R}^\star 
= \big\{ \mathcal{R}(\pi_b^\star,\pi_r^\star) \;\mid\; (\pi_b^\star,\pi_r^\star) \text{ arises from a chosen solution concept} \big\}.
\]
The choice of solution concept, robust min–max or game-theoretic equilibrium, determines how $u_t$ and $w_t$ are generated and thus how resilience is assessed.

\paragraph{1) Robust Worst-Case Risk.}  
The conservative approach assumes that the attacker will always choose the most damaging policy, regardless of its actual incentives.  
The defender’s task is therefore to minimize the maximum possible expected loss:
\begin{equation}
V_{\mathrm{robust}}
= \inf_{\pi_b \in \Pi_b} \; \sup_{\pi_r \in \Pi_r} \; \mathbb{E}\!\big[L_b(\pi_b,\pi_r)\big].
\label{eq:robust-risk}
\end{equation}
This formulation guarantees a performance bound under the most harmful rational adversary.  
The quantity $V_{\mathrm{robust}}$ defines a resilience \emph{floor}: the defender can always ensure that its expected loss does not exceed this value, no matter how the attacker behaves.

\paragraph{2) Equilibrium Strategic Risk.}  
When the attacker and the defender pursue distinct objectives, a more balanced formulation evaluates risk under equilibrium play.  
A policy pair $(\pi_b^\star,\pi_r^\star)$ is a Nash equilibrium if neither side can improve its expected outcome by deviating:
\begin{equation}
\begin{aligned}
\mathbb{E}[L_b(\pi_b^\star,\pi_r^\star)] &\le \mathbb{E}[L_b(\pi_b,\pi_r^\star)] 
&& \forall \pi_b \in \Pi_b, \\
\mathbb{E}[L_r(\pi_b^\star,\pi_r^\star)] &\ge \mathbb{E}[L_r(\pi_b^\star,\pi_r)] 
&& \forall \pi_r \in \Pi_r.
\end{aligned}
\label{eq:equilibrium}
\end{equation}
In sequential settings, a Stackelberg equilibrium may be more suitable: the defender pre-commits to a policy, and the attacker best-responds.  At equilibrium, the outcomes are as follows.
\[
V_b^{\mathrm{eq}} = \mathbb{E}[L_b(\pi_b^\star,\pi_r^\star)], 
\qquad 
V_r^{\mathrm{eq}} = \mathbb{E}[L_r(\pi_b^\star,\pi_r^\star)]
\]
jointly characterize the \emph{strategic risk profile} of the system.  
These equilibrium values are less conservative than the robust bound but more realistic when adversaries and defenders adapt strategically within operational constraints.

 The {robust} approach protects the system against the most harmful adversary, providing guarantees even when the attacker’s incentives are unknown.  
The {equilibrium} approach captures the likely outcome of a rational contest in which both sides adapt and optimize with their own cost–benefit tradeoffs.  

Both perspectives are valuable. Robust risk highlights the guaranteed resilience floor of the system, while equilibrium risk provides insight into performance degradation in realistic adversarial environments. They define the spectrum of \emph{strategic risk} faced by 5G/NextG networks.

\subsection{Integration with Digital Twins}

Once the attacker and defender strategies $(u_t^\star, w_t^\star)$ have been determined, either from the robust solution \eqref{eq:robust-risk} or from the equilibrium solution \eqref{eq:equilibrium}, these actions are treated as fixed inputs to the digital twin. The twin then evolves according to
\begin{equation}
\widehat{x}_{t+1} \;=\; \widehat{f}\!\big(\widehat{x}_t,\, u_t^\star,\, w_t^\star,\, \xi_t,\, q_t\big),
\label{eq:dt-fixed}
\end{equation}
where $\xi_t$ captures random disturbances and $q_t$ evolves according to the disruption dynamics.  

In this way, the game specifies the adversarial strategies, while the digital twin simulates how the system responds under uncertainty. This two–stage procedure, i.e., first solving the game, then simulating the residual uncertainty, ensures that resilience is assessed under adversarially realistic conditions, consistent with the hybrid modeling foundations of Chapter~3.

\subsection{Evaluating Resilience Metrics}

For each scenario $q^{(i)}$ with initialization $(x_0^{(i)},q_0^{(i)})$, the DT generates a trajectory $\{\widehat{x}_t^{(i)}\}_{t=0}^T$ by rolling out Eq.~\eqref{eq:dt-fixed}. Monte Carlo simulations of $\xi_t$ provide multiple realizations for each scenario, allowing statistical evaluation.

From the simulated state, define
\(
\widehat{Q}_t = h(\widehat{x}_t), \ 
s_t = 1-\frac{\widehat{Q}_t}{Q_{\max}}.
\)
These shortfalls produce the following resilience metrics:
\[
L = \sum_{t=t_f}^{t_r} s_t \  \text{(cumulative loss)}, \ 
M=\max_{t\in[t_f,t_r]} s_t \  \text{(maximum drop)},
\]
\[
D=s_{t_r} \quad \text{(residual deficit)}, \ 
T_{\textrm{d}}=\sum_{t=t_f}^{t_r}\mathbf{1}\{\widehat{Q}_t<Q_{\min}\} \quad \text{(downtime)}.
\]
Aggregating across samples gives estimates of $\mathbb{E}[L]$, $\mathrm{Var}[L]$, and tail measures such as $\mathrm{CVaR}_\alpha$.

 To validate fidelity, the DT is run with the same defender policy $u_t^\star$ as in the live system and the fixed attacker action $w_t^\star$. The synchronization error
\[
e_t=\|x_t-\widehat{x}_t\|,
\ 
\mathcal{F}_T = 1-\frac{1}{T}\sum_{t=1}^T \frac{e_t}{e_{\max}},
\]
quantifies how closely the twin tracks reality, with $\mathcal{F}_T \approx 1$ indicating high accuracy.

\subsection{Insights from the Game to DT Pipeline}

The integration of game-theoretic solutions with digital twin simulations provides several important insights into system resilience.   First, it delivers \emph{worst-case guarantees}. By solving the underlying game before simulation, the attacker’s actions $w_t^\star$ are chosen to be adversarially optimal, either under the robust formulation or at equilibrium. The defender’s actions $u_t^\star$ are similarly determined by the solution concept. When these actions are held fixed and rolled out in the digital twin, the resulting resilience metrics represent conservative yet decision-relevant estimates. This ensures that the outcomes reflect realistic adversarial behavior rather than arbitrary scenario design.  

Second, the approach highlights \emph{scenario sensitivity}. With the strategic actions $(u_t^\star,w_t^\star)$ fixed by the game, the digital twin isolates the effects of disruption laws $q^{(i)}$ and natural disturbances $\xi_t$. By varying these elements across simulations, operators can observe how different attack–disturbance combinations shape degradation and recovery patterns. This makes it possible to identify not only worst-case impacts, but also the range of plausible system responses across diverse operating conditions.  

Third, the pipeline supports \emph{policy refinement}. If the simulated outcomes, measured by cumulative loss $L$, maximum drop $M$, residual deficit $D$, or downtime $T_{\textrm{d}}$, exceed acceptable thresholds, the defender can adjust its policy $\pi_b$ and re-solve the game in either the robust or equilibrium framework. The updated strategies are then tested again in the digital twin. This iterative cycle of solving, simulating, and refining closes the loop between theoretical analysis, policy design, and practical validation.

\section{Network Analytics for Risk Analysis}

Next generation communication infrastructures such as 5G and beyond cellular networks are assembled from many interacting hardware and software components that are procured and operated through complex supply and service chains. A quantitative treatment of cyber risk in such systems benefits from the machinery of reliability theory, where a system is represented as an interconnection of components, and a structure function maps component states to system success or failure. In the present context, component states capture the success or failure of security relevant events such as compromise of a network function, exploitation of an interface, or malicious behavior of a supplier, and the top event corresponds to violation of a specified security property of the network.

\subsubsection*{Attack Tree and System Graph Representation}

\begin{figure}
    \centering
    \includegraphics[width=0.3\linewidth]{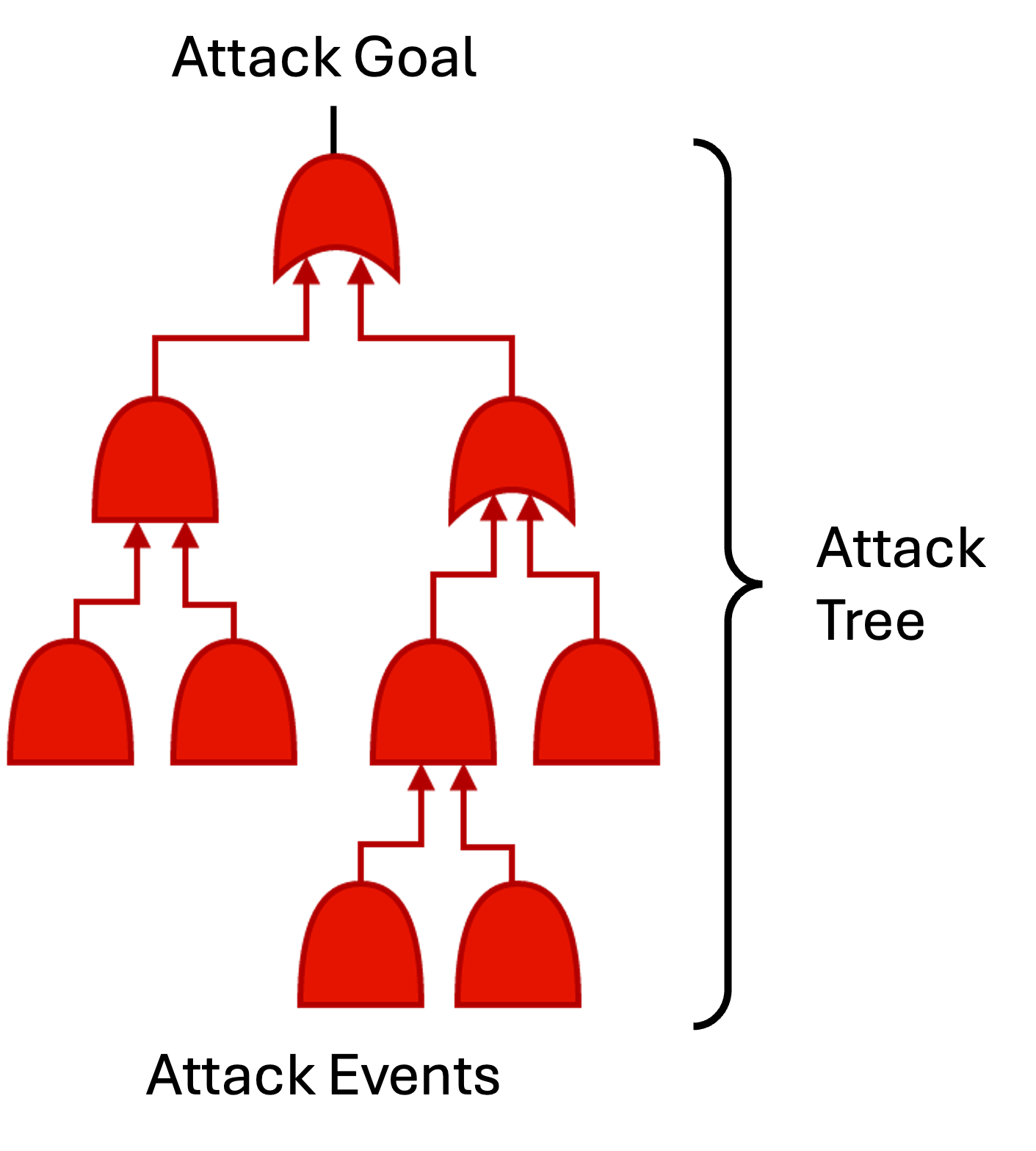}
    \caption{\small Schematic representation of an attack tree capturing the hierarchical decomposition of a high level compromise objective into intermediate goals and atomic attack steps. AND and OR relations encode the logical structure of adversarial progress, providing a foundation for quantitative evaluation of system level security risk in next generation networks.}
    \label{fig:attack_tree}
\end{figure}

A natural entry point is the attack tree representation of a high level compromise scenario. An attack tree is a rooted directed acyclic graph whose root encodes a system level goal such as loss of integrity of the core network or compromise of the slice orchestrator. The internal nodes represent intermediate attack goals and are connected to their children through logical relations of type AND or OR. A child set that is connected to its parent through an AND relation must all be achieved to fulfill the parent goal, whereas an OR relation means that any one of the children suffices. Leaves represent atomic attack steps such as exploitation of a protocol vulnerability in a baseband unit, insertion of malicious code in a virtualized network function, or misuse of a management application programming interface in the service based architecture. An illustrative example of an attack tree is shown in Fig.~\ref{fig:attack_tree}.

In order to connect these logical constructions to concrete assets, the attack tree is coupled with a system graph that represents the components and suppliers of the network. Let $\mathcal{C}$ denote the set of components and $\mathcal{S}$ the set of suppliers. A system graph is a directed graph
\(
\mathsf{SG} = (\mathcal{C} \cup \mathcal{S}, \mathcal{E}),
\)
where edges in $\mathcal{E}$ capture functional or supply dependencies among components and suppliers. For example, an edge from a cloud platform supplier to a user plane function indicates that the correct and trustworthy operation of the function depends on that platform. The decomposition of complex components into subcomponents yields a hierarchical system graph that exposes the internal structure of network elements such as base stations, distributed units, and core functions. The atomic attack steps in the attack tree are mapped to the nodes or edges in this system graph. This mapping creates a bridge from logical attack paths to the concrete elements of a fifth generation or next generation network and its supply chain.

\subsubsection*{Structure Function and Minimal Cut Sets}

Reliability theory represents the behavior of a coherent system through a structure function. In the risk context, each node $v$ in the system graph is associated with a binary state variable $X_v \in \{0,1\}$, where $X_v = 1$ denotes that the node is uncompromised and $X_v = 0$ denotes that it has failed from a security perspective. A structure function
\(
\varphi : \{0,1\}^{|\mathcal{C} \cup \mathcal{S}|} \to \{0,1\}
\)
maps the vector $x = (x_v)_{v \in \mathcal{C} \cup \mathcal{S}}$ to a system level indicator that equals one if the network preserves the specified security property and zero if the adverse event occurs. The function $\varphi$ is induced by the AND and OR relations in the attack tree and by the logical dependencies in the system graph. For example, if the security of the core network depends on the integrity of the access and control planes through an AND relation, then the corresponding component states appear as a logical conjunction in $\varphi$.

A central concept in reliability theory is the minimal cut set~\cite{rausand2003system}. A cut set $w$ is a set of nodes whose simultaneous failure guaranties system failure, that is, $\varphi(x) = 0$ whenever $x_v = 0$ for all $v \in w$. A minimal cut set is a cut set that contains no proper subset that is itself a cut set. The family of all minimal cut sets is denoted by $\mathcal{W}$. In the network security setting, a minimal cut set describes a smallest combination of components and suppliers whose compromise suffices to realize the top level attack goal. For instance, the set containing a particular cloud infrastructure supplier and a specific authentication service may constitute a minimal cut if their joint compromise always implies a loss of trust in the control plane.

Minimal cut sets connect the logical description of attack scenarios to a quantitative measure of risk. Let $r_v \in [0,1]$ denote the probability that node $v$ is successfully attacked or behaves maliciously during the time horizon of interest. The vector of risk values at the node level is $r = (r_v)_{v \in \mathcal{C} \cup \mathcal{S}}$. Under the standard assumption that attacks on different nodes are conditionally independent given $r$, the probability that all nodes in a minimal cut set $w$ fail is
$\prod_{v \in w} r_v$,
and the probability that at least one minimal cut set fails is
\begin{equation}
R(r) = 1 - \prod_{w \in \mathcal{W}}
\left(
1 - \prod_{v \in w} r_v
\right).
\label{eq:system_risk}
\end{equation}
The quantity $R(r)$ is the systemic risk function. It represents the probability that the adverse security event occurs through at least one minimal combination of component and supplier failures. In a 5G network, $R(r)$ can encode events such as joint compromise of the user plane and an orchestrator, or simultaneous manipulation of a radio unit and its management interface, as long as the corresponding minimal cut sets are enumerated.

In practice, direct computation of all minimal cut sets can be challenging for large networks. Algorithms such as \emph{Method of Obtaining CUt Sets (MOCUS)} and other fault tree based procedures are used to obtain $\mathcal{W}$ from the attack tree and system graph while handling general AND and OR logic~\cite{rauzy2003mocus}. Once $\mathcal{W}$ is available, the systemic risk function in \eqref{eq:system_risk} provides a compact summary of how individual node risks combine into a network level metric.

\subsubsection*{Importance Measures from Reliability Theory}

The systemic risk function is the starting point for prioritizing defensive actions. Reliability theory offers a family of importance measures that quantify how sensitive the system risk is to the risk of individual nodes~\cite{johansen2014foundations}. These measures are particularly useful in complex next generation architectures where intuitive criticality rankings may be misleading due to nontrivial logical structure.

Consider first the improvement potential of node $i$. One interprets improvement as a reduction of its compromise probability from its current value $r_i$ to a lower value, conceptually to zero for maximal improvement. Let $s_i^0$ denote the current vector of node risks and $s_i^1$ the hypothetical vector after hardening node $i$, defined as
\begin{align}
s_i^0 = (r_1, \ldots, r_i, \ldots, r_N),
\qquad
s_i^1 = (r_1, \ldots, 0, \ldots, r_N),
\end{align}
where $N$ is the number of nodes and only the $i$th entry is changed. The improvement potential of node $i$ is then
\begin{align}
\mathrm{IP}_i = R(s_i^0) - R(s_i^1).
\end{align}
This is the reduction in systemic risk that would be achieved if node $i$ could be made perfectly secure while leaving all other node risks unchanged. In a 5G network, $\mathrm{IP}_i$ measures the benefit of fully hardening a particular base station vendor, virtualized network function, or cloud platform.

A complementary and widely used measure is the Birnbaum importance~\cite{miziula2019birnbaum}. In reliability theory, the Birnbaum importance of node $i$ is the partial derivative of the system failure probability with respect to the failure probability of that node. Translating to the risk notation above, one defines
\begin{align}
\mathrm{BI}_i = \frac{\partial R(r)}{\partial r_i},
\end{align}
evaluated at the nominal risk vector $r$. This derivative can be interpreted as the probability that node $i$ is critical for system failure, which means that changing its state from secure to compromised toggles the state of the system while all other node states are fixed. For coherent systems, the Birnbaum importance is nonnegative and can be computed without explicit differentiation by evaluating the systemic risk function at two auxiliary vectors that differ only in the value of $r_i$. A common choice is
\begin{align}
s_i^0 = (0, r_{-i}), 
\qquad
s_i^1 = (1, r_{-i}),
\end{align}
where $r_{-i}$ collects the risks of all nodes other than $i$. The Birnbaum importance is then
\begin{align}
\mathrm{BI}_i = R(s_i^1) - R(s_i^0).
\end{align}
Nodes with large Birnbaum importance are those whose risk has the strongest marginal effect on the overall network risk and, therefore, represent structurally critical components or suppliers.

For decision making, it is often helpful to combine structural importance with the inherent risk of a node. A risk importance measure~\cite{kieras2020riots, kieras2020modeling} can be defined as
\begin{align}
\mathrm{RI}_i = r_i \, \mathrm{BI}_i,
\end{align}
which assigns high values to nodes that are both structurally significant and presently risky. In a next generation network this measure highlights, for instance, suppliers that control components belonging to many minimal cut sets and that are currently assessed to have elevated compromise probabilities. Optimization formulations that reduce the sum of risk importance values across the system can then be used to select mitigation actions that approximately minimize the systemic risk without explicitly recomputing the full function \eqref{eq:system_risk} for every candidate configuration~\cite{kieras2021scram}.

\section{Agentic AI and LLMs for Risk Assessment}

\begin{figure}[h!]
\centering
\includegraphics[width=0.9\linewidth]{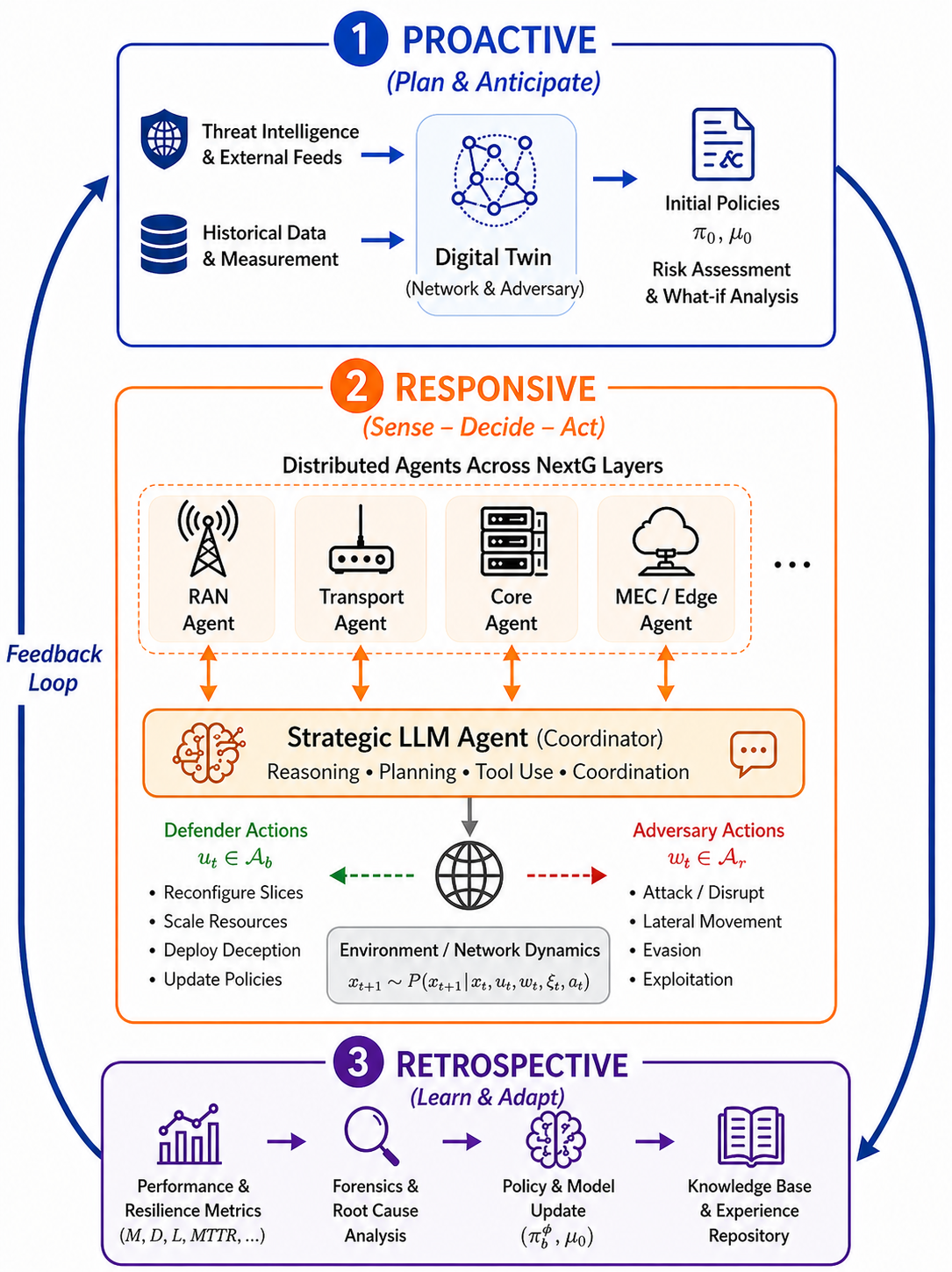}
\caption{Agentic AI workflow for NextG resilience. The system operates in a closed loop across three stages: \textbf{(1) Proactive}: digital twin–driven planning using threat intelligence and measurement; \textbf{(2) Responsive}: distributed agents coordinated by a strategic LLM to sense–decide–act under adversarial interaction; \textbf{(3) Retrospective}: metrics, forensics, and learning update policies and beliefs.}
\label{fig:agentic_workflow}
\end{figure}

NextG networks exhibit unprecedented scale, heterogeneity, and tight coupling across RAN, transport, core, and edge domains. Virtualization, cloud–edge integration, and software-defined programmability create flexibility but also enlarge the attack surface and introduce complex cross-domain vulnerabilities. Traditional static or rule-based defenses are therefore inadequate.   Resilience must be \emph{closed-loop, adaptive, and continually learning}. This motivates the use of \emph{agentic artificial intelligence (AI)} coordinated by \emph{large language models (LLMs)}, enabling distributed, goal-directed agents to perceive, reason, act, and learn across layers \cite{li2025texts}. Fig.~\ref{fig:agentic_workflow} captures the architecture. Resilience emerges from the interaction of three tightly coupled stages: proactive planning, responsive control, and retrospective learning.

\subsection{Background on Agentic AI}

Agentic AI represents a shift from static predictive models to \emph{autonomous, decision-making systems} that operate continuously in dynamic environments. Unlike conventional machine learning models that map inputs to outputs in isolation, agentic systems maintain an internal state, update beliefs, and select actions over time to achieve objectives under uncertainty \cite{li2025texts}.

At a foundational level, agentic AI integrates ideas from control theory, reinforcement learning, and game theory. An agent interacts with an environment characterized by partially observable states $x_t$, stochastic disturbances, and potentially adversarial dynamics. The decision process forms a closed loop:
\[
\text{observe} \;\rightarrow\; \text{infer} \;\rightarrow\; \text{decide} \;\rightarrow\; \text{act} \;\rightarrow\; \text{learn}.
\]

In adversarial settings such as cybersecurity, this interaction becomes a dynamic game between the defender and the attacker. The agent must reason about both the uncertainty of the system and strategic behavior. 
A key enabler of modern agentic AI is the integration of foundation models, particularly LLMs, which provide reasoning, abstraction, and coordination capabilities. Within the architecture of Fig.~\ref{fig:agentic_workflow}, agentic AI is not a single model but a \emph{multi-layer decision system}: digital twins support proactive reasoning, distributed agents execute responsive control, and learning modules drive retrospective adaptation, with LLMs providing a unifying cognitive layer.

\subsection{Agentic AI as Closed-Loop Decision Making}

Agentic AI systems operate through continuous interaction with the environment, aligning naturally with the three-stage workflow in Fig.~\ref{fig:agentic_workflow}. Agents collect heterogeneous observations $o_t$ from measurement, logs and external intelligence, forming beliefs about the state of the system $x_t$ and adversarial modes $q_t$. These beliefs provide a dynamic representation of both the operational environment and the potential threat conditions.

Building on these beliefs, agents evaluate future risks and synthesize policies that balance performance and resilience. In the proactive stage, this reasoning is supported by digital twin simulations that enable a forward-looking risk assessment. In the responsive stage, it is enhanced by LLM-based reasoning, which allows agents to interpret complex, multi-modal information and coordinate across domains. The interaction between defender and adversary is captured by the stochastic dynamics
\[
x_{t+1} \sim P(x_{t+1}\mid x_t, u_t, w_t, \xi_t, q_t),
\quad
u_t \in \mathcal{A}_b,\; w_t \in \mathcal{A}_r,
\]
where defender actions $u_t$ shape system evolution in the presence of adversarial actions $w_t$ and environmental uncertainty.

Learning closes the loop by enabling continual adaptation. Policies are updated through feedback mappings of the form
\(
\pi_b^\phi : (o_t,h_t) \mapsto \Delta(\mathcal{A}_b),
\)
guided by resilience metrics and retrospective analysis. This perception–decision–action–learning cycle corresponds directly to the feedback structure shown in Fig.~\ref{fig:agentic_workflow}, which embeds intelligence in the operational dynamics of the network.

\subsection{Role of LLMs: Cognitive Coordination Layer}

LLMs serve as the strategic coordinator in the center of Fig.~\ref{fig:agentic_workflow}. Formally, they can be viewed as probabilistic mappings
\(
f_\theta : \mathcal{X} \rightarrow \Delta(\mathcal{Y}),
\)
but their role in the system extends far beyond prediction. They provide a mechanism for integrating heterogeneous information, converting measurement data, logs, and threat intelligence into a unified representation that can be acted upon.

Through this unified representation, LLMs enable reasoning and planning across multiple layers of the network. They support hypothesis generation, adversary modeling, and policy synthesis, allowing the system to anticipate and respond to evolving threats. At the same time, they facilitate coordination by translating global objectives into structured actions that can be executed by distributed agents. In this sense, LLMs act as the cognitive layer that links perception, decision-making, and execution into a coherent system.

\subsection{Distributed Agent Roles (Responsive Layer)}

The responsive layer in Fig.~\ref{fig:agentic_workflow} consists of distributed agents embedded across network domains. These include agents operating in the RAN, transport, core, and MEC layers, each responsible for localized monitoring, control, and mitigation. They respond to real-time conditions while interacting with the strategic LLM, which provides guidance and coordination across layers.

The actions of these agents implement control input $u_t$, while adversarial actions $w_t$ capture disruptions such as attacks, lateral movement, and exploitation. The architecture achieves a balance between speed and coherence: local agents provide rapid, domain-specific responses, while centralized reasoning through the LLM ensures that these responses remain aligned with global resilience objectives.

\subsection{Three-Stage Resilience Workflow}

The system operates as a recursive loop structured by Fig.~\ref{fig:agentic_workflow}. In the proactive stage, threat intelligence and measurement are fused through a digital twin to construct prior beliefs $\mu_0$ and initial policies $\pi_b^\phi$. This corresponds to solving
\[
\phi^\star = \arg\min_{\phi} \mathbb{E}_{w \sim \mu_0}[L(x_t)],
\]
where the objective is to minimize the expected resilience loss under anticipated adversarial behavior.

In the responsive stage, agents interact with the environment in real time based on observed signals and historical context. Actions are selected according to
\[
u_t \sim \pi_b^\phi(o_t,h_t), 
\qquad 
x_{t+1}=f(x_t,u_t,w_t,\xi_t,q_t),
\]
with distributed agents executing localized decisions and the LLM coordinating responses across layers to ensure consistency and effectiveness.

In the retrospective stage, performance is evaluated using resilience metrics such as $(M,D,L,\text{MTTR})$, and forensic analysis is used to uncover the causal relationships that underlie the disruptions. These insights drive updates to both belief models and policies,
\(
(\pi_b,\mu) \;\longrightarrow\; (\pi_b',\mu'),
\) 
thereby closing the loop and enabling continual improvement of the system.

In summary, resilience is not a static property but a recursive optimization process that unfolds over time. Proactive planning reduces exposure to potential threats, responsive control mitigates real-time impact, and retrospective learning improves future decision-making. This integration naturally combines control theory, which governs system dynamics, game theory, which captures adversarial interaction, and learning theory, which supports policy adaptation. With LLMs that enable high-level reasoning and coordination, resilience evolves into an anticipatory and self-improving property of next-generation networks rather than a purely reactive defense mechanism.

\chapter{Resilient System Designs}

This chapter develops the system-level design dimension of cyber resilience in NextG networks by building directly on the foundations established in the preceding chapters. Chapter~1 characterizes the evolving threat landscape and expanded attack surface, Chapter~2 formalizes resilience as a multi-stage lifecycle, Chapter~3 provides the theoretical foundations for modeling dynamics and adversarial interactions, and Chapter~4 develops analytical tools for risk assessment under uncertainty. 

This chapter synthesizes these elements by translating them into concrete design and operational mechanisms. It shows how threat models, resilience metrics, theoretical insights, and risk assessments jointly inform the design of control, orchestration, and decision-making strategies in complex networked environments.

From this perspective, resilience is not a static property, but an evolving system capability shaped by continuous monitoring, feedback, and adaptation. Design must therefore account for how systems anticipate disruptions, respond to disturbances, and improve through experience.

The mechanisms developed here reflect the resilience lifecycle. Proactive resilience is achieved through risk-aware planning and resource allocation. Responsive resilience arises from adaptive control and learning under stochastic and adversarial conditions. Retrospective resilience is enabled through reconfiguration and policy updates based on observed system behavior.

To operationalize these dimensions, the chapter integrates three complementary approaches. Optimization-based frameworks provide principled resource allocation under performance and risk constraints. Learning-based methods, particularly multi-agent reinforcement learning, enable adaptive decision-making in dynamic and partially observable environments. LLM-driven agentic control introduces a higher-level semantic layer that translates intent and context into actionable policies.

These approaches operate across multiple timescales, from fast control loops to high-level reasoning, illustrating how resilience emerges from the coupling of optimization, learning, and intelligent decision-making.

The remainder of this chapter develops these paradigms in detail. Section~5.1 presents risk-aware resource orchestration in multi-tier edge-cloud systems. Section~5.2 introduces multi-agent reinforcement learning for resilient slice management. Section~5.3 incorporates LLM-driven agentic intelligence for intent-aware control in O-RAN architectures.

\section{Risk-Aware Resource Orchestration in Multi-Tier 5G Edge-Cloud Systems}

\begin{figure}[h!]
    \centering
    \includegraphics[width = 0.8\columnwidth]{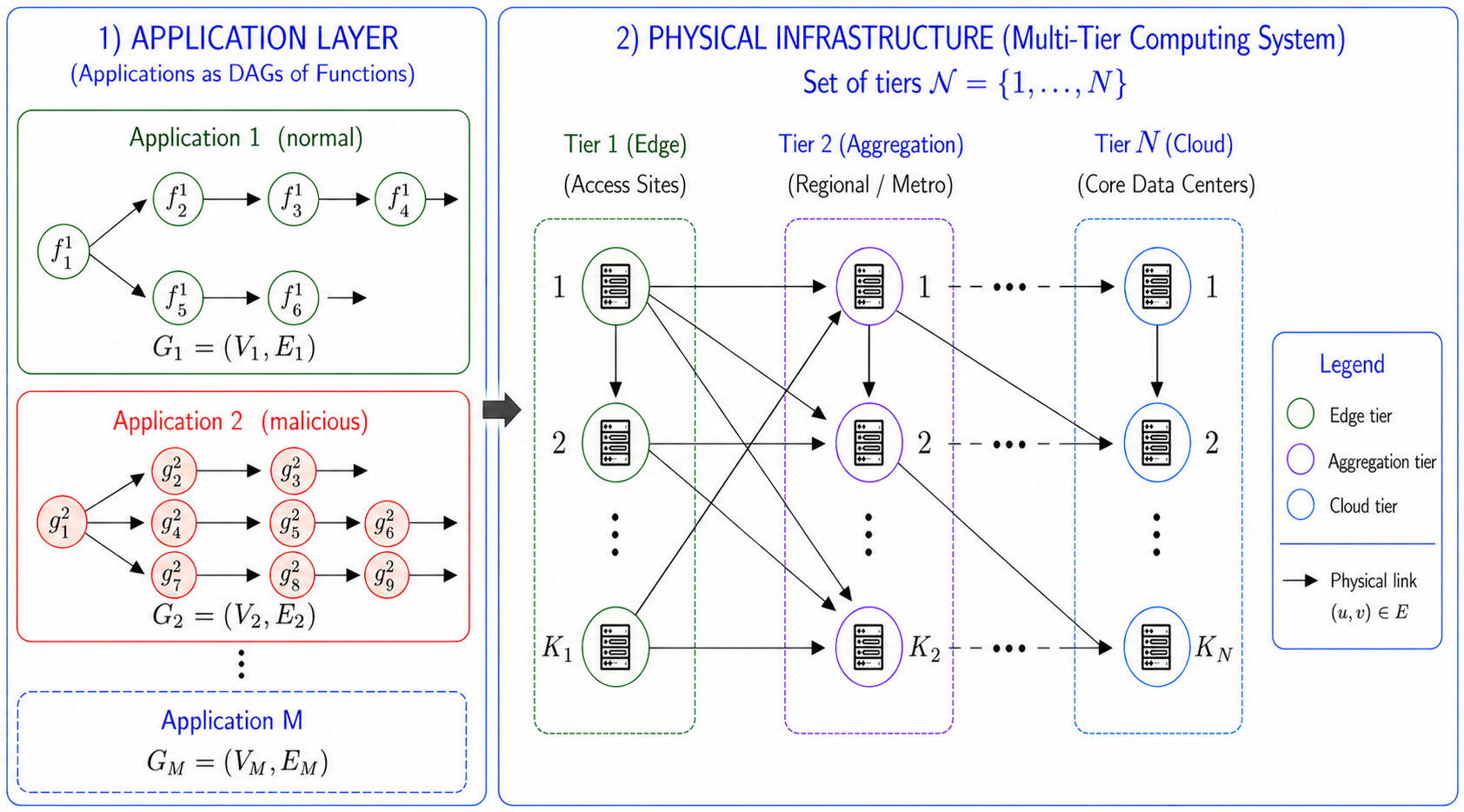}
    \caption{\small Schematic representation of the resource orchestrator managing resources between normal and malicious applications considering the workflows of apps over the set of microservices $f_i$.\vspace{-0.1in}}
    \label{fig:sys_model}
\end{figure}

The evolution of 5G networks toward softwarized and virtualized architectures introduces a multi-tier computing substrate that spans the radio edge, mobile edge computing layers, and centralized cloud infrastructures. Applications deployed in these environments increasingly adopt microservices-based designs, where each service is decomposed into a collection of fine-grained computational functions with explicit data dependencies. This decomposition improves modularity and scalability, yet introduces intricate orchestration challenges because the end-to-end behavior of an application depends strongly on the tier in which each function is placed, the amount of compute and network resources it receives, and the coupling among resources across different layers. In addition, the openness of the 5G ecosystem raises the possibility that adversarial or malicious applications may request resources. It creates the need for a resource allocation strategy that jointly evaluates performance objectives and the risk associated with each application. This formulation directly aligns with the resilience framework developed in Chapter~2, where system robustness is achieved through proactive, responsive, and adaptive mechanisms. In particular, the joint consideration of performance and adversarial risk operationalizes the notion of resilience as maintaining service quality under uncertainty and disruption.
Fig.~\ref{fig:sys_model} illustrates the orchestration setting based on the work in~\cite{wu2024adaptive}, where the resource orchestrator manages the compute and network resources across levels while accounting for normal and malicious applications.

The orchestration process can be interpreted as a closed-loop control system, consistent with the control-theoretic formulation introduced in Chapter~3. The placement and allocation variables act as control inputs, while latency and accuracy constitute system outputs evolving over time under resource constraints.

\subsection*{System Structure and Multi-Tier Execution Model}
We consider a set of computing tiers indexed by $n \in \mathcal{N}=\{1,\ldots,N\}$. Each level $n$ provides finite computing capacity $C_n \in \mathbb{R}_{+}$ and finite network capacity $B_n \in \mathbb{R}_{+}$. The system hosts a set of applications indexed by $m \in \mathcal{M}=\{1,\ldots,M\}$, where each application $m$ is represented as a directed acyclic graph (DAG) $G_m=(V_m,E_m)$, with vertices $V_m$ denoting functions (microservices) and directed edges $E_m$ denoting data dependencies. The placement of a function onto a specific tier determines both its baseline execution characteristics and its contribution to end-to-end latency and accuracy. Lower tiers offer reduced transport delay due to proximity to users, while higher tiers offer greater compute abundance at the expense of longer transport delays. This tradeoff is central to the orchestration problem.

A key characteristic of this setting is the usage of shared functions. A function may appear in multiple application graphs, inducing coupling across workloads. Let $I_v$ denote the aggregate intensity of invocation of the function $v$ induced by all applications. Increased load on shared functions increases contention and affects both latency and accuracy, thereby creating multi-dimensional interactions between compute and network resources that must be modeled explicitly.

\subsection{Risk Modeling and Application Beliefs}
To incorporate security considerations, each application $m$ is associated with a belief value $\pi_m \in [0,1]$ that quantifies the probability that application $m$ is malicious. These belief values serve as soft risk indicators that modulate orchestration decisions. Rather than enforcing hard exclusions, the orchestrator penalizes resource allocations to high-risk applications through the objective function. This enables continuous risk-aware adaptation and provides robustness under uncertainty. The belief-driven representation is consistent with the risk modeling framework developed in Chapter~4, where uncertainty is captured through probabilistic characterizations of adversarial behavior and incorporated into decision-making through weighted objectives.

\subsection*{Latency and Accuracy Representation with Resource Coupling}
For each function $v \in V_m$ placed on tier $n$, let $r^{\mathrm{net}}_{v,m} \ge 0$ and $r^{\mathrm{com}}_{v,m} \ge 0$ denote the allocated network and compute resources. The latency of a function decreases monotonically with the network allocation, while the accuracy increases monotonically with the compute allocation. To reflect coupling due to shared usage, we parameterize the performance of a function by the aggregate intensity of invocation $I_v$.

We adopt the following tier-dependent models for latency and accuracy:
\begin{align}
\ell_{v,n} \bigl(r^{\mathrm{net}}_{v,m}, I_v \bigr)
&= a_n - b \, \frac{r^{\mathrm{net}}_{v,m}}{I_v},
\label{eq:func_latency} \\
\alpha_{v,n} \bigl(r^{\mathrm{com}}_{v,m}, I_v \bigr)
&= a'_n + b' \, \frac{r^{\mathrm{com}}_{v,m}}{I_v},
\label{eq:func_accuracy}
\end{align}
where $a_n$ captures the tier baseline latency, $a'_n$ captures the tier baseline accuracy, and $b>0$, $b'>0$ quantify the marginal effects of network and compute allocations. These models operationalize the principle that contention increases with $I_v$, requiring additional resources to maintain performance.

Let $x^n_{v,m} \in \{0,1\}$ indicate whether function $v \in V_m$ is placed on tier $n$, with $\sum_{n \in \mathcal{N}} x^n_{v,m}~=~1$. The application end-to-end latency and accuracy are modeled as
\begin{align}
L_m &= \sum_{v \in V_m}\sum_{n \in \mathcal{N}} x^n_{v,m}\, \ell_{v,n}\bigl(r^{\mathrm{net}}_{v,m}, I_v \bigr),
\label{eq:app_latency} \\
A_m &= \sum_{v \in V_m}\sum_{n \in \mathcal{N}} x^n_{v,m}\, \alpha_{v,n}\bigl(r^{\mathrm{com}}_{v,m}, I_v \bigr).
\label{eq:app_accuracy}
\end{align}
This composition captures how local function performance aggregates across the application workflow, and why critical-path functions dominate the end-to-end delay.

\subsection*{Risk-Aware Optimization Formulation}
The orchestrator jointly decides function-to-tier placement and resource allocations while accounting for risk beliefs. Define the effective resource consumption of application $m$ as
\begin{equation}
R_m \triangleq \sum_{v \in V_m}\Bigl(r^{\mathrm{com}}_{v,m} + r^{\mathrm{net}}_{v,m}\Bigr).
\label{eq:resource_sum}
\end{equation}
Let $\phi(L_m,A_m)$ denote a monotone performance utility that increases with higher accuracy and improves with lower latency. From a risk-aware decision-making perspective (Chapter~4), the objective function reflects a trade-off between resource efficiency and exposure to adversarial risk. The belief-weighted penalty term internalizes the expected impact of malicious applications, while the performance utility term captures system-level service objectives.

The risk-aware orchestration problem is posed as
\begin{align}
\min_{\substack{x,r,L,A}}
\quad & \sum_{m \in \mathcal{M}} \bigl(\theta + \lambda \pi_m\bigr)\, R_m
\;-\; \kappa \sum_{m \in \mathcal{M}} \phi(L_m,A_m)
\label{eq:obj_risk_orch} \\
\text{s.t.}\quad
& \sum_{n \in \mathcal{N}} x^n_{v,m} = 1,
\quad \forall m \in \mathcal{M},\ \forall v \in V_m,
\label{eq:place_one} \\
& \sum_{m \in \mathcal{M}}\sum_{v \in V_m} r^{\mathrm{com}}_{v,m}\, x^n_{v,m} \le C_n,
\quad \forall n \in \mathcal{N},
\label{eq:cap_comp} \\
& \sum_{m \in \mathcal{M}}\sum_{v \in V_m} r^{\mathrm{net}}_{v,m}\, x^n_{v,m} \le B_n,
\quad \forall n \in \mathcal{N},
\label{eq:cap_net} \\
& L_m \le \bar{L}_m,\quad A_m \ge \bar{A}_m,
\quad \forall m \in \mathcal{M},
\label{eq:qos_constraints} \\
& \text{constraints \eqref{eq:app_latency} and \eqref{eq:app_accuracy} hold for all } m \in \mathcal{M}.
\nonumber
\end{align}
The weight $(\theta + \lambda \pi_m)$ penalizes resource allocations to high-risk applications. Parameter $\lambda \ge 0$ tunes the sensitivity to risk, while $\kappa > 0$ balances performance utility against resource consumption. The presence of binary placement variables and nonlinear coupling in \eqref{eq:app_latency} to \eqref{eq:app_accuracy} yields a mixed-integer nonlinear program that is NP-hard in general.

\subsection*{Iterative Solution via Decomposition}
A practical solution approach adopts a two-step iterative procedure. For fixed performance targets $(\bar{L}_m,\bar{A}_m)$, the algorithm computes a placement and allocation solution by relaxing tier capacity constraints using Lagrangian multipliers. The achieved performance then informs updated targets, and the process repeats until convergence. This structure enables tractable numerical solutions while preserving the coupled nature of compute and network resources across tiers.


\begin{figure*}[h!]
    \centering
    
    \begin{subfigure}[t]{0.48\textwidth}
        \centering
        \includegraphics[width=\linewidth]{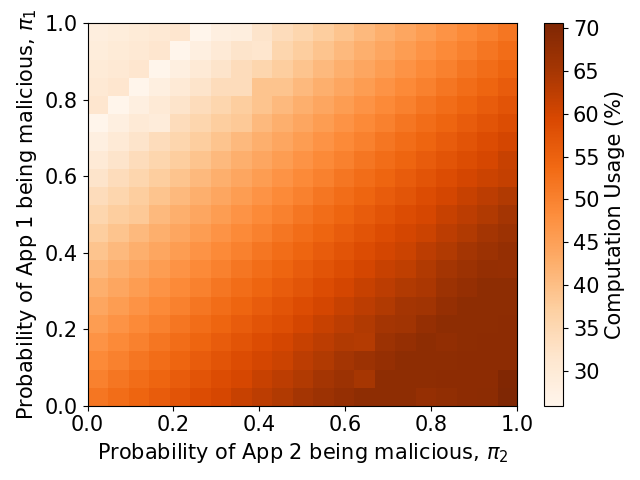}
        \caption{}
    \end{subfigure}
    \hfill
    \begin{subfigure}[t]{0.48\textwidth}
        \centering
        \includegraphics[width=\linewidth]{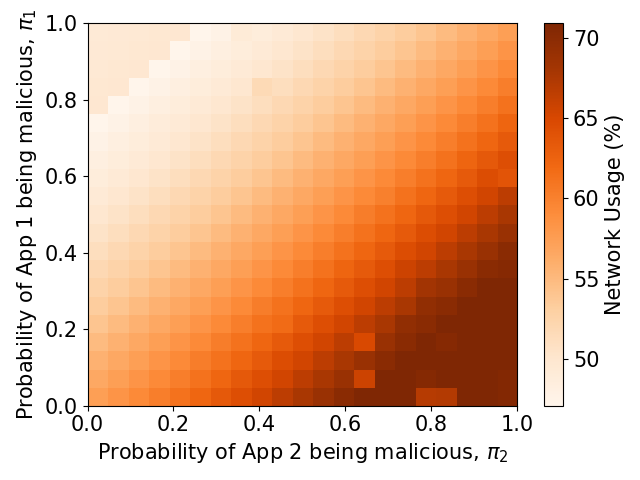}
        \caption{}
    \end{subfigure}
    \hfill
    \begin{subfigure}[t]{0.48\textwidth}
        \centering
        \includegraphics[width=\linewidth]{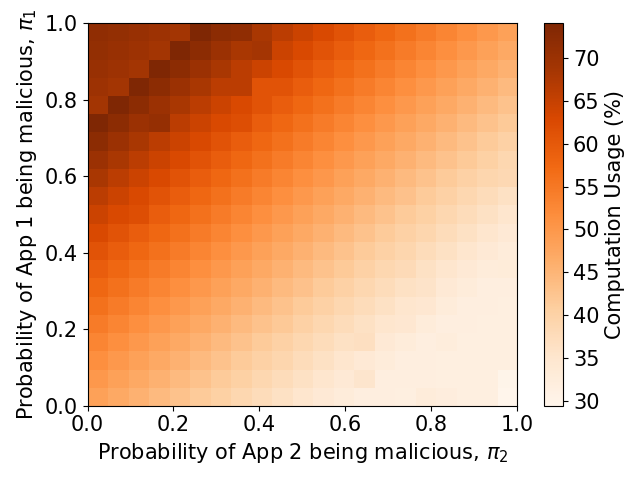}
        \caption{}
    \end{subfigure}
    \hfill
    \begin{subfigure}[t]{0.48\textwidth}
        \centering
        \includegraphics[width=\linewidth]{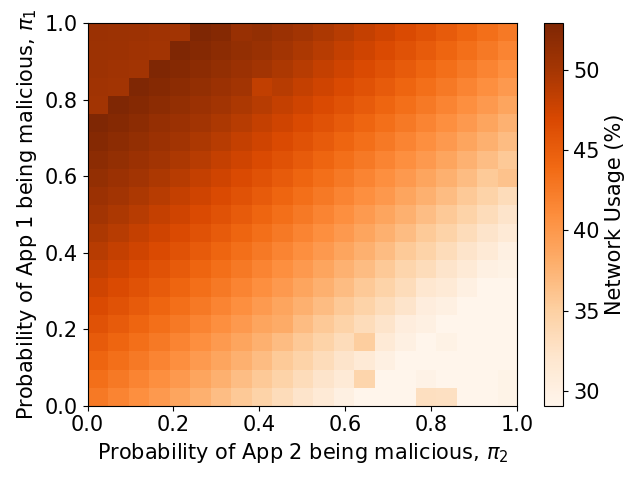}
        \caption{}
    \end{subfigure}

    \caption{\small Comparative analysis of computational and network usage under varying probabilities of App 1 ($\pi_1$) and App 2 ($\pi_2$) being malicious. (a) App 1 computation, (b) App 1 network, (c) App 2 computation, (d) App 2 network.}
    \label{fig:heatmap}
\end{figure*}

Fig.~\ref{fig:heatmap} illustrates how compute and network usage shift as the belief parameters vary across applications. The results indicate that increasing belief for an application systematically reduces its allocated resources, thereby protecting the infrastructure from resource wastage under elevated risk.

\subsection*{Adaptive Re-Orchestration After Malicious Application Removal}
\begin{figure*}[t]
\centering

\begin{subfigure}{0.30\linewidth}
  \centering
  \includegraphics[width=\linewidth]{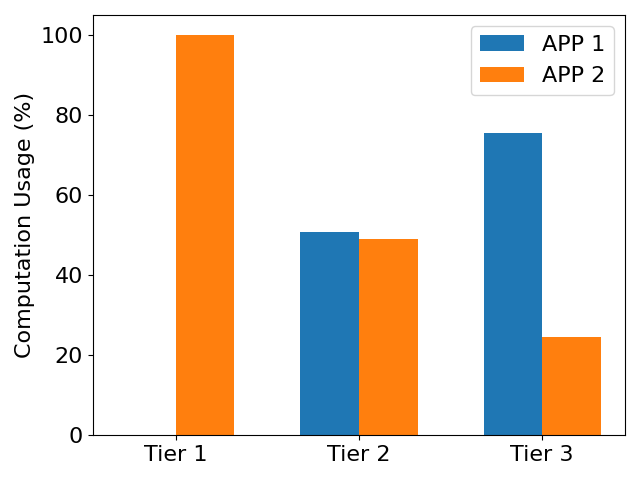}
  \caption{Compute resource usage by tier.}
  \label{fig:tier_usage_comp}
\end{subfigure}
\hfill
\begin{subfigure}{0.30\linewidth}
  \centering
  \includegraphics[width=\linewidth]{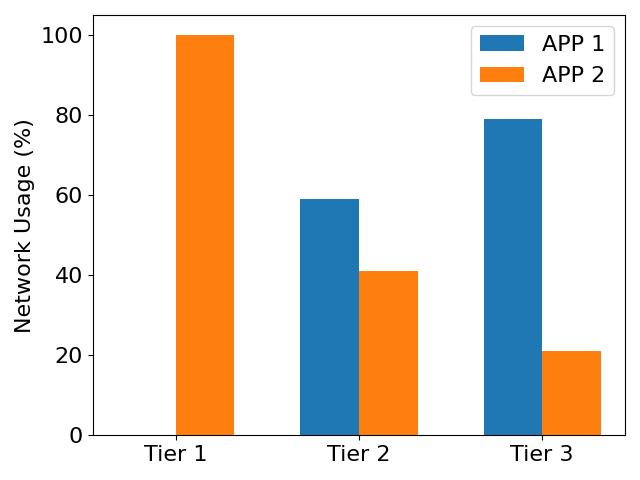}
  \caption{Network resource usage by tier.}
  \label{fig:tier_usage_net}
\end{subfigure}
\hfill
\begin{subfigure}{0.30\linewidth}
  \centering
  \includegraphics[width=\linewidth]{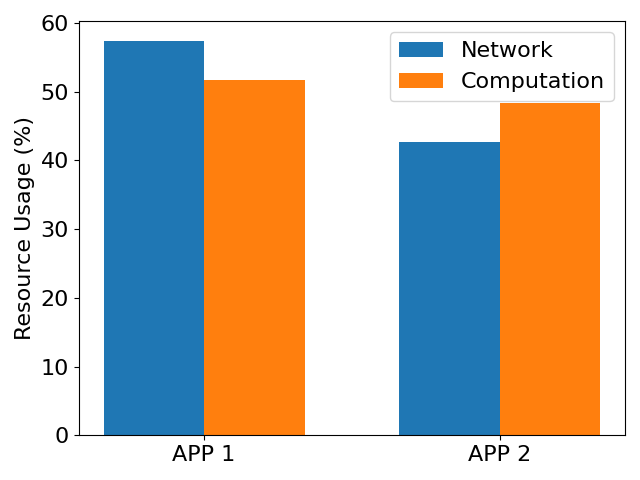}
  \caption{Aggregate resource utilization of App~1 and App~2.}
  \label{fig:app_total_usage}
\end{subfigure}

\caption{Resource utilization analysis across a three-tier architecture for App~1 and App~2. (a) Compute resource usage by tier. (b) Network resource usage by tier. (c) Aggregate resource utilization comparison between applications.}
\label{fig:resource-usage}
\end{figure*}

Once an application is identified as malicious and removed from the system, the remaining resources can be reallocated to improve the performance of benign applications. Since function placement decisions have already been determined, the re-orchestration focuses on reallocating compute and network resources within each tier. Let $\mathcal{D} \subset \mathcal{M}$ denote the set of detected malicious applications, and let $\{x^{n\star}_{v,m}\}$ be the placement solution obtained before removal. The re-orchestration problem can be written as
\begin{align}
\min_{\substack{r,L,A}}
\quad & \sum_{m \in \mathcal{M}\setminus\mathcal{D}} R_m
\;-\; \kappa \sum_{m \in \mathcal{M}\setminus\mathcal{D}} \phi(L_m,A_m)
\label{eq:reorch_obj} \\
\text{s.t.}\quad
& \sum_{m \in \mathcal{M}\setminus\mathcal{D}}\sum_{v \in V_m}
r^{\mathrm{com}}_{v,m}\, x^{n\star}_{v,m} \le C_n,
\quad \forall n \in \mathcal{N},
\label{eq:reorch_cap_comp} \\
& \sum_{m \in \mathcal{M}\setminus\mathcal{D}}\sum_{v \in V_m}
r^{\mathrm{net}}_{v,m}\, x^{n\star}_{v,m} \le B_n,
\quad \forall n \in \mathcal{N},
\label{eq:reorch_cap_net} \\
& L_m \le \bar{L}_m,\quad A_m \ge \bar{A}_m,
\quad \forall m \in \mathcal{M}\setminus\mathcal{D},
\label{eq:reorch_qos}
\end{align}
along with the performance definitions in \eqref{eq:app_latency} and \eqref{eq:app_accuracy} with fixed placements $x^{n\star}_{v,m}$. Under linear or convex performance utilities, \eqref{eq:reorch_obj} to \eqref{eq:reorch_qos} reduces to a convex program, enabling rapid recovery after malicious removal. Fig.~\ref{fig:resource-usage} summarizes tier-wise resource utilization and demonstrates improved efficiency after reallocation.

\subsubsection*{Risk--Performance Tradeoffs and Resource Efficiency}

\begin{figure}[h!]
    \centering
    \includegraphics[width = 3.4in]{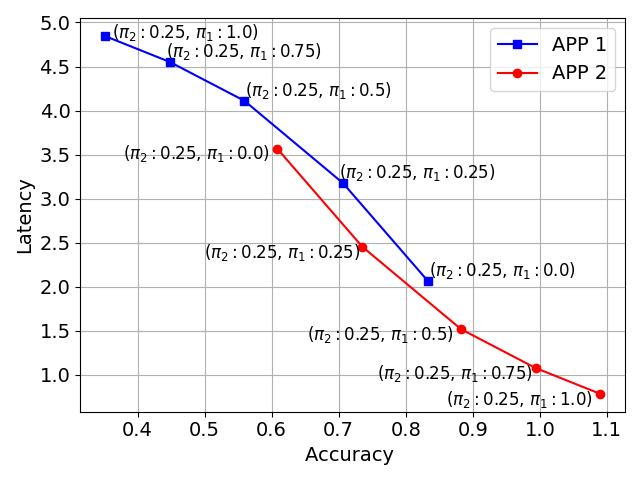}
    \caption{\small Trade-off between latency and accuracy for App 1 and App 2 across varying belief of being malicious ($\pi_2, \pi_1$). Each data point represents a specific configuration of ($\pi_2, \pi_1$) values. \vspace{-0.1in}}
    \label{fig:mapping}
\end{figure}

The impact of belief-driven risk awareness on application-level performance is further illustrated in Fig.~\ref{fig:mapping}. The figure depicts the tradeoff between end-to-end latency and application accuracy as the belief parameters $(\pi_1,\pi_2)$ are varied. As the belief associated with an application increases, the orchestrator systematically deprioritizes its resource allocation, resulting in increased latency and reduced accuracy. Conversely, applications with lower belief values benefit from preferential allocation, achieving improved performance. This behavior confirms that the proposed framework does not impose rigid isolation but instead induces a continuous and controllable degradation of performance for high-risk workloads.

\begin{figure}[h!]
\centering

\begin{subfigure}{0.45\linewidth}
  \centering
  \includegraphics[width=\linewidth]{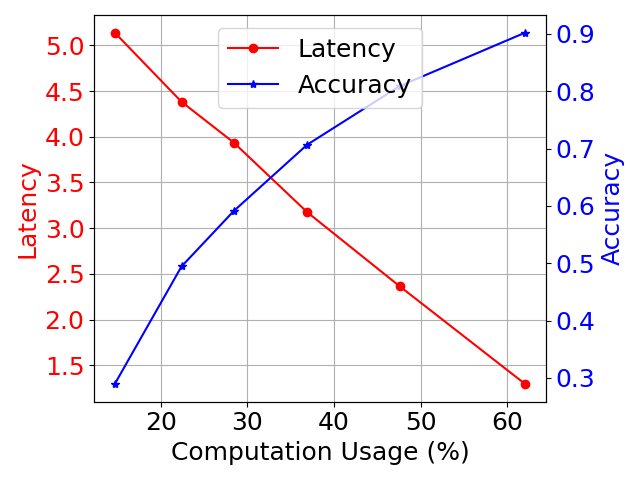}
  \caption{Computation usage (\%).}
  \label{fig:L_A_usage_comp}
\end{subfigure}
\hfill
\begin{subfigure}{0.45\linewidth}
  \centering
  \includegraphics[width=\linewidth]{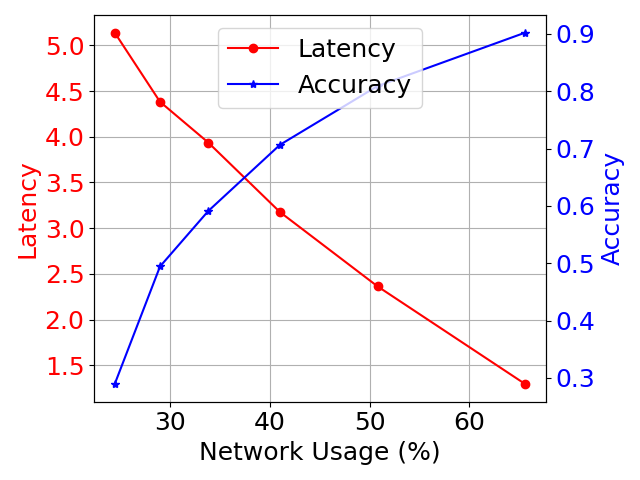}
  \caption{Network usage (\%).}
  \label{fig:L_A_usage_net}
\end{subfigure}

\caption{Trade-off between system latency (seconds) and application accuracy as a function of resource utilization. (a) Computation usage. (b) Network usage.}
\label{fig:L_A_resource}
\end{figure}

Fig.~\ref{fig:L_A_resource} further elucidates the coupling between resource usage and application performance. Specifically, the figure demonstrates how increasing computation or network resource utilization improves accuracy while reducing latency, albeit with diminishing returns. These curves provide empirical justification for the structure of the utility function $\phi(L_m,A_m)$ in \eqref{eq:obj_risk_orch}, and highlight why balanced allocation across compute and network dimensions is essential for efficient orchestration in multi-tier environments.

\begin{figure}[h!]
\centering

\begin{subfigure}{0.45\linewidth}
  \centering
  \includegraphics[width=\linewidth]{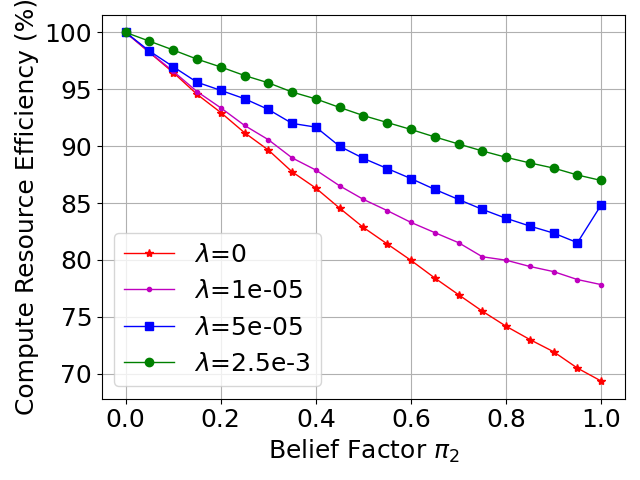}
  \caption{Computation resource efficiency (\%).}
  \label{fig:app2_comp_eff}
\end{subfigure}
\hfill
\begin{subfigure}{0.45\linewidth}
  \centering
  \includegraphics[width=\linewidth]{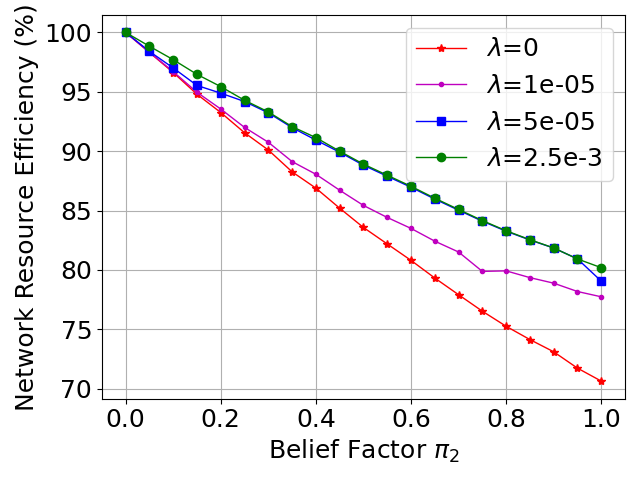}
  \caption{Network resource efficiency (\%).}
  \label{fig:app2_net_eff}
\end{subfigure}

\caption{Variation of resource efficiency with respect to the belief factor for different $\lambda$ values. (a) Computation resource efficiency. (b) Network resource efficiency.}
\label{fig:malicious_attack}
\end{figure}

Finally, Fig.~\ref{fig:malicious_attack} quantifies the effect of increasing belief on overall resource efficiency for different values of the weighting parameter $\lambda$. As the belief of an application increases, both computation and network efficiency decrease, reflecting the intentional withdrawal of resources from potentially malicious workloads. Larger values of $\lambda$ amplify this effect, enabling the system operator to tune the aggressiveness of risk mitigation. These results demonstrate that the proposed formulation offers a principled mechanism to trade resource efficiency for security assurance in a controlled and interpretable manner.

\begin{figure}[h!]
\centering

\begin{subfigure}{0.45\linewidth}
  \centering
  \includegraphics[width=\linewidth]{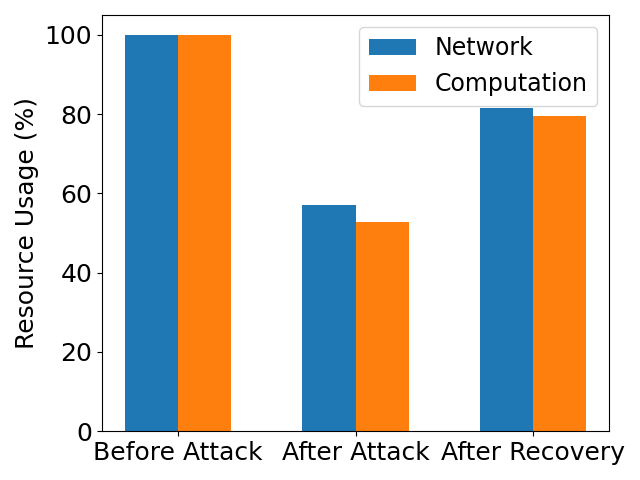}
  \caption{Attack and recovery dynamics over time.}
  \label{fig:attack_recover_dynamics}
\end{subfigure}
\hfill
\begin{subfigure}{0.45\linewidth}
  \centering
  \includegraphics[width=\linewidth]{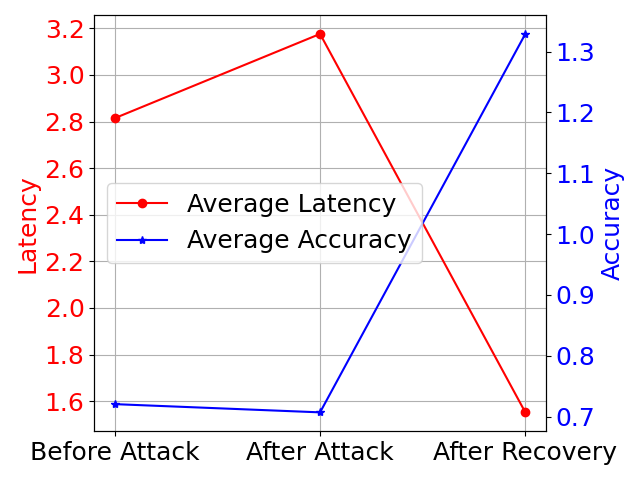}
  \caption{Post-detection performance recovery.}
  \label{fig:attack_recover_performance}
\end{subfigure}

\caption{Resilience of resource orchestration following the detection of malicious applications. (a) Attack and recovery dynamics. (b) System performance recovery after mitigation.}
\label{fig:malicious_attack_recovery}
\end{figure}

Fig.~\ref{fig:malicious_attack_recovery} illustrates the dynamic resilience behavior of the proposed orchestration framework following the detection and removal of malicious applications. Upon attack detection, the system experiences a transient degradation in performance due to resource contention and belief updates. However, adaptive re-orchestration rapidly reallocates reclaimed compute and network resources to benign applications, resulting in a prompt recovery of system performance. The figure highlights both the recovery speed and the stability of the post-attack operating point, demonstrating the ability of the framework to maintain service continuity under adversarial conditions.

\subsection*{Resilience Implications}
This risk-aware orchestration framework illustrates how resilience can be systematically engineered in softwarized networks by integrating performance requirements with explicit modeling of adversarial risk. Proactively, belief-weighted costs reduce expected wastage by steering allocations away from high-risk applications. Responsively, re-orchestration reallocates reclaimed resources to restore and improve benign performance following malicious removal. In retrospect, belief updates and associated penalties discourage persistent misbehavior and improve long-term robustness. Collectively, these mechanisms demonstrate that resilience in multi-tier 5G systems emerges from the coupling of optimization, risk modeling, and adaptive control.

While the preceding formulation provides a structured optimization based approach, it relies on quasi-static assumptions and explicit system models. In contrast, the learning-based frameworks introduced in Chapter~3 motivate adaptive control mechanisms that can operate under stochastic dynamics and delayed feedback, as considered next.

\section{Resilient Slice Management through Multi Agent Reinforcement Learning}

While optimization-based orchestration provides strong guarantees under quasi-static conditions, slice management in O-RAN environments must operate under highly stochastic traffic arrivals, time-varying channel conditions, and delayed feedback. In particular, URLLC slices impose stringent probabilistic delay constraints that cannot be enforced using static admission thresholds or greedy allocation policies. This motivates a sequential decision-making framework capable of adapting admission and resource allocation decisions over time while explicitly respecting reliability and resource constraints.

\begin{figure}[h!]
    \centering
    \includegraphics[width = 3.5in]{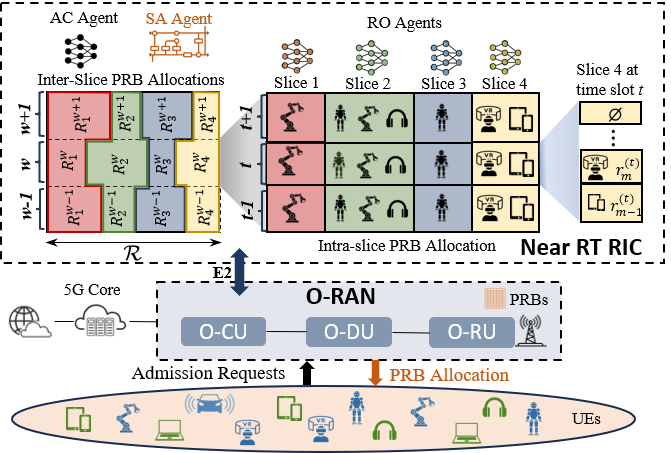}
    \caption{\small Integrated admission control and resource orchestration framework in O-RAN architecture. The figure illustrates a hierarchical control structure implemented within the Near-RT RIC, where admission control (AC), slice-aware (SA), and resource orchestration (RO) agents jointly manage radio resource allocation across multiple network slices.
}
    \label{fig:workflows1}
\end{figure}

To this end, we model resilient slice management as a constrained multi-agent Markov decision process (CMDP), where interacting agents learn coordinated policies that jointly control slice admission and resource orchestration within the near-real-time RAN Intelligent Controller (near-RT RIC)~\cite{xingqi2025joint}. Fig.~\ref{fig:workflows1} illustrates the integrated slice admission control and resource orchestration architecture implemented within the near-RT RIC. The framework operates across both inter-slice and intra-slice resource allocation layers, enabling coordinated control over physical resource blocks (PRBs) in O-RAN. Admission requests originating from heterogeneous URLLC and eMBB users are first processed by the admission control agent, which regulates slice entry based on current system load and reliability conditions. The admitted slices are then managed by a collection of slice-specific resource orchestration agents that dynamically allocate PRBs across time slots and slices. This hierarchical control structure allows the system to jointly manage slice admission and fine-grained radio resource allocation under shared capacity constraints, while preserving the modularity of O-RAN functional splits across the O-RU, O-DU, and O-CU.

\subsection*{System Model and Time-Slotted Operation}

We consider a single gNB serving a set of active user equipments (UEs) over discrete time slots indexed by $t \in \mathbb{N}$. At each time slot, a set of new slice requests arrives according to a stochastic arrival process. Each request belongs to a service class $s \in \mathcal{S}$, where $\mathcal{S}$ includes URLLC and eMBB services.

Let $N_t$ denote the number of active users at time $t$. The system is constrained by total available bandwidth $B_{\max}$ and total computational capacity $C_{\max}$. Each active user $i$ is allocated bandwidth $b_i^t$ and computational resources $c_i^t$, satisfying
\begin{equation}
\sum_{i=1}^{N_t} b_i^t \le B_{\max}, \qquad
\sum_{i=1}^{N_t} c_i^t \le C_{\max}.
\label{eq:resource_constraints}
\end{equation}

The instantaneous latency experienced by user $i$ at time $t$ is denoted by $D_i^t$, which depends on both radio transmission delay and processing delay. For URLLC users, a delay threshold $D_{\text{th}}$ is imposed.

\subsection*{CMDP Formulation}

The resilient slice management problem is modeled as a CMDP defined by the tuple
\[
\mathcal{M} = \bigl( \mathcal{N}, \mathcal{S}, \{\mathcal{O}_i\}_{i\in\mathcal{N}}, \{\mathcal{A}_i\}_{i\in\mathcal{N}}, P, \{r_i\}_{i\in\mathcal{N}}, \{c_i\}_{i\in\mathcal{N}}, \gamma \bigr),
\]
where $\mathcal{N}=\{a,o\}$ denotes the set of agents corresponding to admission control ($a$) and resource orchestration ($o$). This formulation directly instantiates the constrained sequential decision-making framework introduced in Chapter~3.3, where system objectives are optimized under explicit reliability and resource constraints in stochastic environments.

The global system state at time $t$ is
\begin{equation}
s_t = \bigl( N_t, \mathbf{b}^t, \mathbf{c}^t, \mathbf{h}^t, \boldsymbol{\eta}^t \bigr),
\label{eq:global_state}
\end{equation}
where $\mathbf{b}^t$ and $\mathbf{c}^t$ denote current resource allocations, $\mathbf{h}^t$ represents channel quality indicators, and $\boldsymbol{\eta}^t$ captures historical latency fulfillment statistics.

Each agent observes a partial view of the system. The admission agent observes
\begin{equation}
o_a^t = \bigl( N_t, \bar{D}^t, \rho^t \bigr),
\end{equation}
where $\bar{D}^t$ is the average URLLC delay and $\rho^t$ denotes current resource utilization. The orchestration agent observes
\begin{equation}
o_o^t = \bigl( \mathbf{b}^t, \mathbf{c}^t, \mathbf{h}^t \bigr).
\end{equation}

\subsection*{Action Spaces}

The admission control agent selects an action
\begin{equation}
a_a^t \in \mathcal{A}_a = \{0,1,\ldots,K\},
\end{equation}
corresponding to admitting up to $K$ new slice requests at time $t$. The admitted users are selected based on channel quality ordering.

The resource orchestration agent selects continuous-valued actions
\begin{equation}
a_o^t = \bigl\{ \Delta b_i^t, \Delta c_i^t \bigr\}_{i=1}^{N_t},
\end{equation}
which adjust per-user bandwidth and computation allocations subject to \eqref{eq:resource_constraints}.

\begin{figure}[h!]
    \centering
    \includegraphics[width = 3.45in]{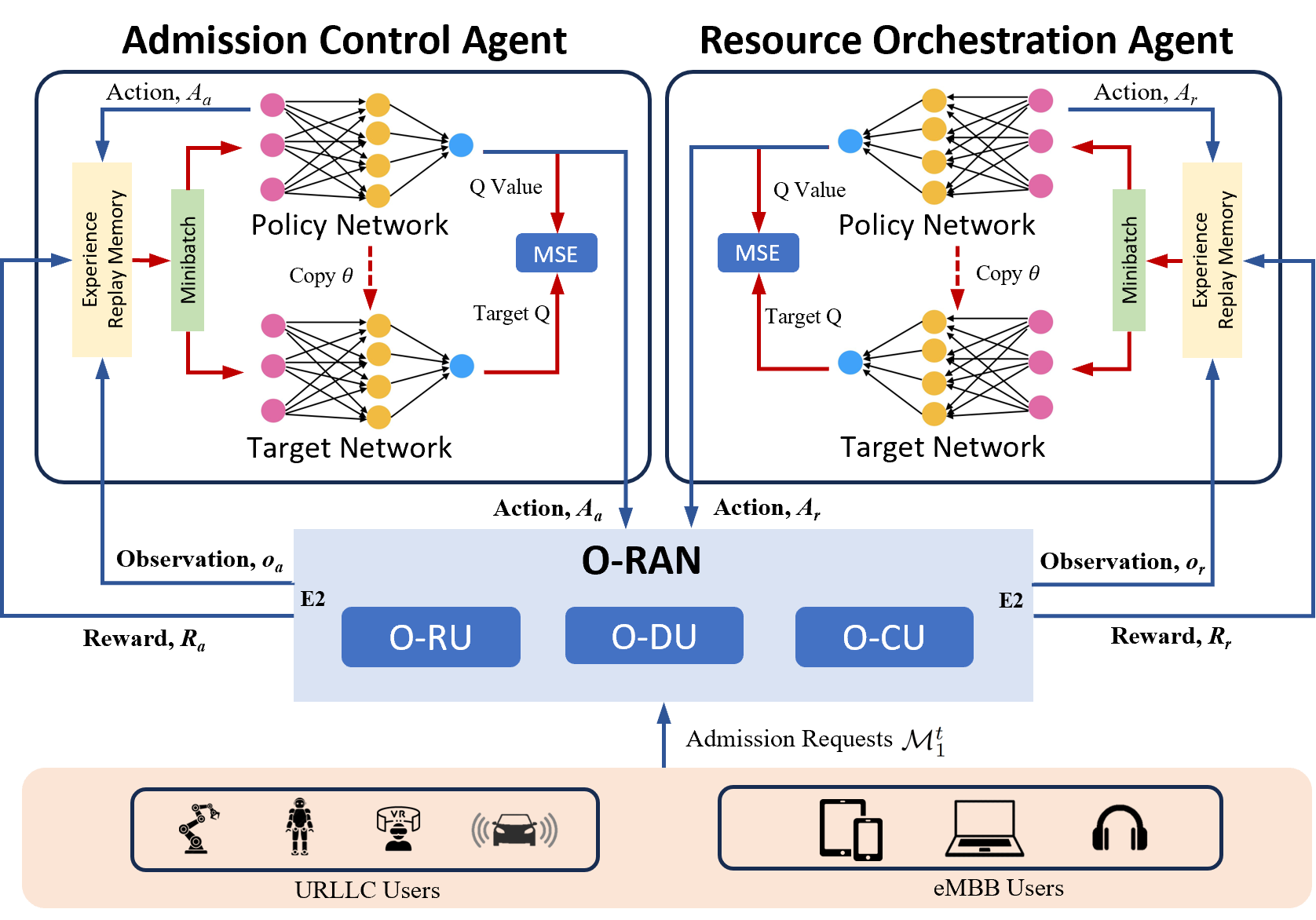}
    \caption{Multi-agent reinforcement learning architecture for admission control and resource orchestration in O-RAN.}
    \label{fig:workflows2}
\end{figure}

The internal learning structure of the proposed multi-agent framework is depicted in Fig.~\ref{fig:workflows2}. The admission control agent and the resource orchestration agent operate as independent yet interacting decision-makers, each equipped with a dueling double deep Q-network architecture. At each time step, both agents receive partial observations from the O-RAN environment via the E2 interface, compute their respective actions, and receive delayed reward feedback reflecting latency reliability and energy efficiency outcomes. The admission control agent focuses on regulating the number of incoming slice requests, while the resource orchestration agent adjusts PRB allocations across admitted users and slices. Although trained independently, the agents are coupled through the shared environment dynamics, resulting in non-stationary observations that necessitate robust learning mechanisms such as target networks and experience replay.

\subsection*{Latency Reliability and Energy Efficiency Metrics}

For URLLC users, reliability is quantified via the latency fulfillment ratio
\begin{equation}
\mathcal{R}^t = \frac{1}{N_t^{\text{URLLC}}} 
\sum_{i \in \mathcal{U}_{\text{URLLC}}}
\mathbb{I}\bigl(D_i^t \le D_{\text{th}}\bigr),
\label{eq:lfr}
\end{equation}
where $\mathbb{I}(\cdot)$ denotes the indicator function.

Energy efficiency at time $t$ is defined as
\begin{equation}
\mathcal{E}^t = \frac{\sum_{i=1}^{N_t} R_i^t}{\sum_{i=1}^{N_t} (b_i^t + c_i^t)},
\end{equation}
where $R_i^t$ denotes the achieved throughput.

\subsection*{Reward and Constraint Functions}

The admission agent reward is defined as
\begin{equation}
r_a^t = \alpha N_t^{\text{admit}} - \beta \bigl(1 - \mathcal{R}^t \bigr),
\label{eq:admission_reward}
\end{equation}
penalizing reliability violations. The orchestration agent reward is
\begin{equation}
r_o^t = \mu \mathcal{R}^t + \nu \mathcal{E}^t - \xi \sum_{i=1}^{N_t} \mathbb{I}(D_i^t > D_{\text{th}}).
\label{eq:orch_reward}
\end{equation}
The structure of the reward functions reflects the resilience objectives defined in Chapter~2, where reliability preservation and efficient resource utilization are treated as primary performance indicators under dynamic operating conditions.

System-level constraints are enforced via expectation bounds
\begin{equation}
\mathbb{E}\bigl[1 - \mathcal{R}^t \bigr] \le \epsilon,
\qquad
\mathbb{E}\bigl[\rho^t\bigr] \le \rho_{\max}.
\label{eq:cmdp_constraints}
\end{equation}

\subsection*{Learning Objective}

Each agent seeks to maximize its expected discounted return
\begin{equation}
\max_{\pi_i} 
\mathbb{E}_{\pi}\!\left[
\sum_{t=0}^{\infty} \gamma^t r_i^t
\right],
\qquad i \in \{a,o\},
\label{eq:marl_objective}
\end{equation}
subject to constraints \eqref{eq:cmdp_constraints}. Policies $\pi_i$ are parameterized using dueling double deep Q-networks, enabling stable learning under partial observability and nonstationary interactions.

\subsection*{Learning Dynamics and Performance Evaluation}

\begin{figure}[h!]
\centering

\begin{subfigure}{0.45\linewidth}
  \centering
  \includegraphics[width=\linewidth]{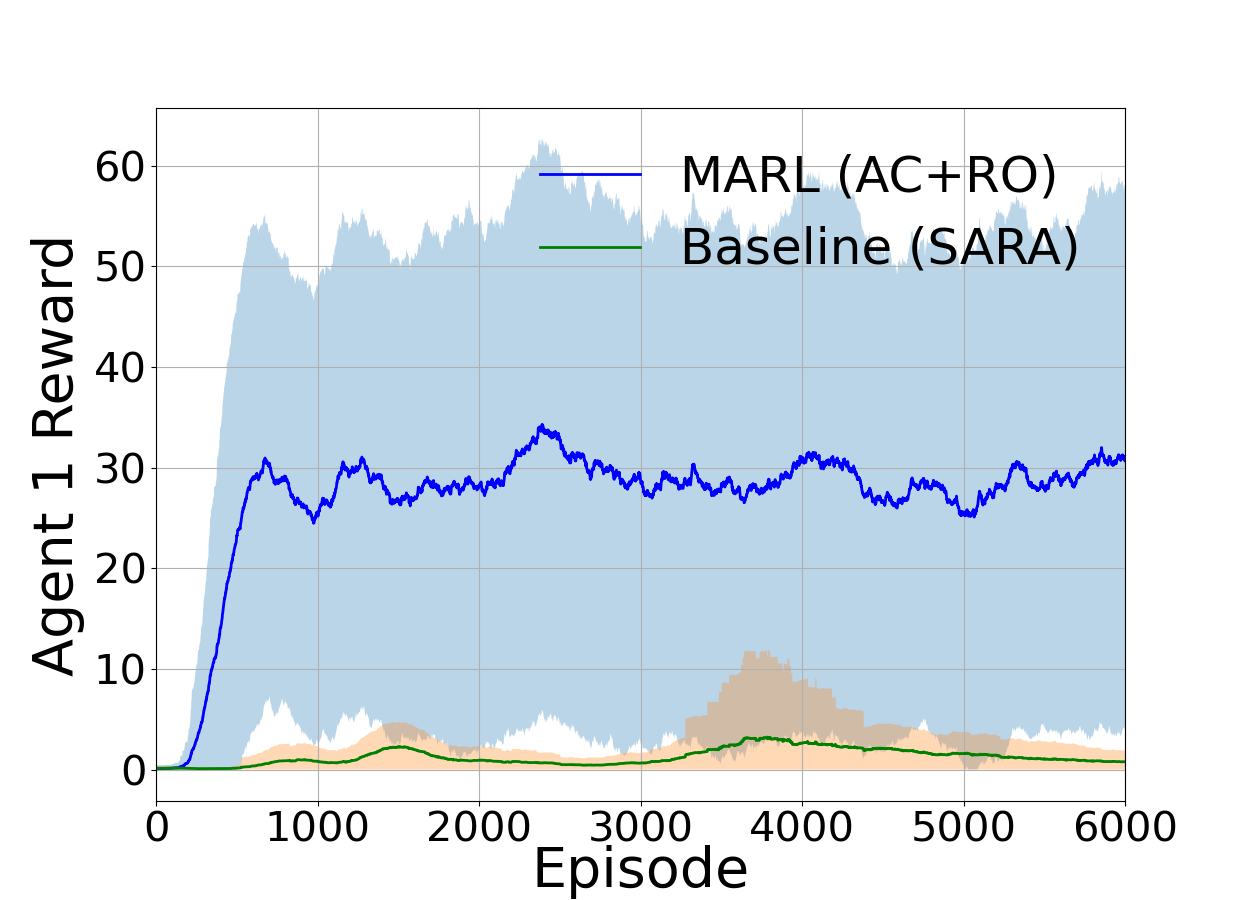}
  \caption{Admission control agent.}
  \label{fig:admission_agent_reward}
\end{subfigure}
\hfill
\begin{subfigure}{0.45\linewidth}
  \centering
  \includegraphics[width=\linewidth]{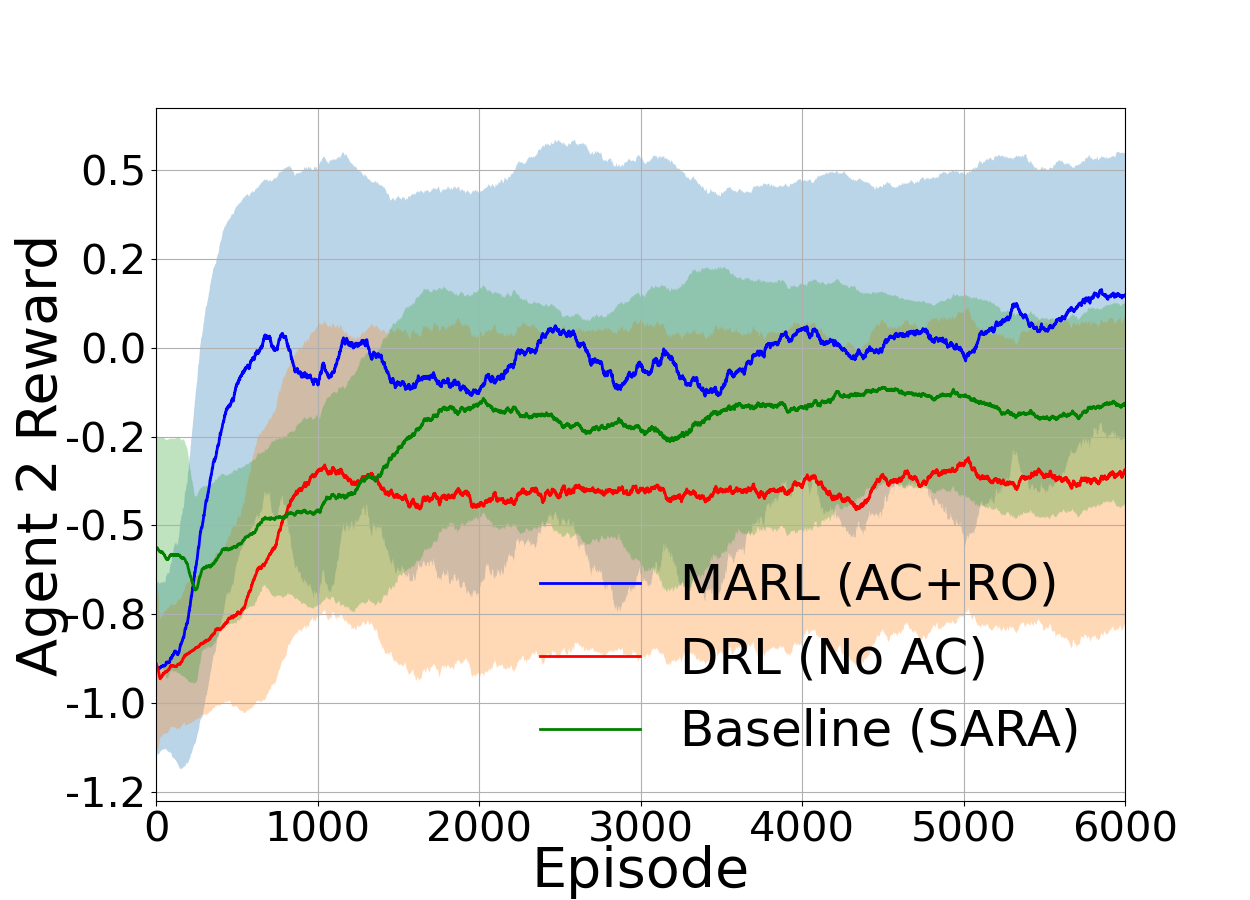}
  \caption{Resource orchestration agent.}
  \label{fig:resource_agent_reward}
\end{subfigure}

\caption{Training reward trajectories of the learning based agents under varying admission strategies. (a) Admission control agent. (b) Resource orchestration agent.}
\label{fig:training_reward}
\end{figure}

Fig.~\ref{fig:training_reward} illustrates the evolution of cumulative training rewards for both the admission control agent and the resource orchestration agent under varying admission strategies. The admission control agent converges toward a stable reward profile as it learns to regulate slice admissions in a manner that balances throughput with reliability preservation. In parallel, the resource orchestration agent exhibits a smoother reward trajectory, reflecting its continuous adaptation of PRB allocations in response to admitted load. The joint convergence behavior confirms that the decentralized learning process stabilizes despite non-stationary interactions between agents, validating the effectiveness of the dueling double deep Q-network architecture in the proposed multi-agent CMDP formulation.

\begin{figure*}[t]
\centering

\begin{subfigure}{0.29\linewidth}
  \centering
  \includegraphics[width=\linewidth]{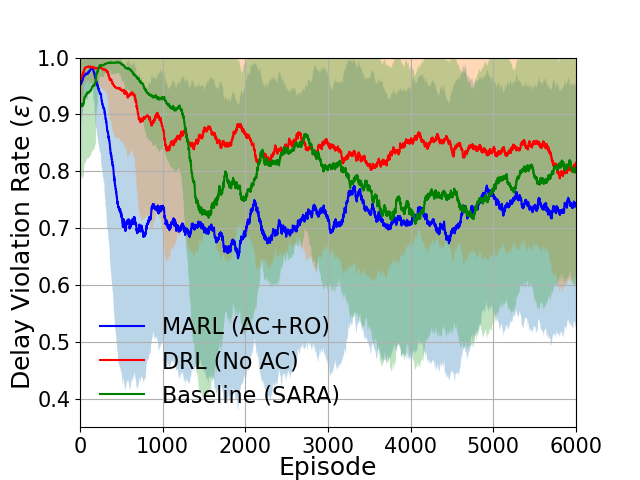}
  \caption{Delay violation rate.}
  \label{fig:delay_violation}
\end{subfigure}
\hfill
\begin{subfigure}{0.29\linewidth}
  \centering
  \includegraphics[width=\linewidth]{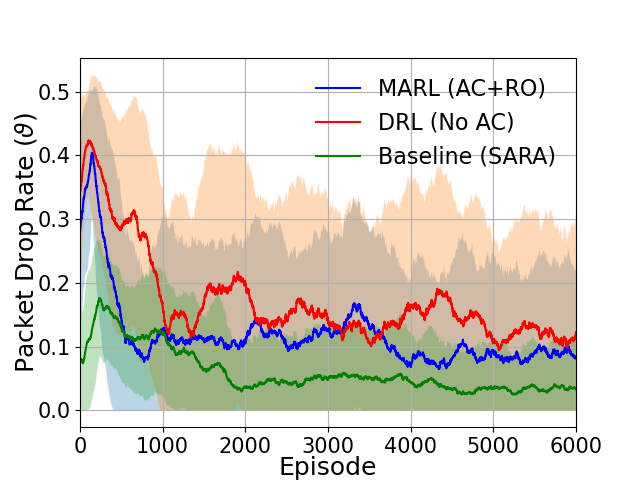}
  \caption{Packet drop rate.}
  \label{fig:pkt_drop_rate}
\end{subfigure}
\hfill
\begin{subfigure}{0.29\linewidth}
  \centering
  \includegraphics[width=\linewidth]{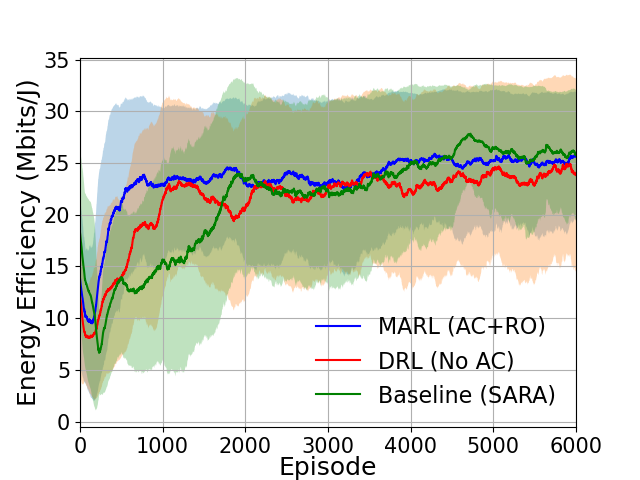}
  \caption{Energy efficiency.}
  \label{fig:energy_efficiency}
\end{subfigure}

\caption{Evolution of training performance metrics. (a) Delay violation rate. (b) Packet drop rate. (c) Energy efficiency.}
\label{fig:training_KPIs}
\end{figure*}

The convergence of key reliability and efficiency indicators during training is further illustrated in Fig.~\ref{fig:training_KPIs}. As learning progresses, the delay violation rate and packet drop rate decrease monotonically, while energy efficiency improves steadily. These trends indicate that the agents progressively internalize the latency reliability constraints encoded in the reward functions and learn to mitigate congestion before violations occur. The simultaneous improvement across all three metrics demonstrates that the learning process does not overfit to a single objective, but instead achieves a balanced operating point consistent with the resilience goals of URLLC slice management.

\begin{figure}[t!]
    \centering
    \begin{subfigure}[b]{0.45\textwidth}
        \includegraphics[width=\textwidth]{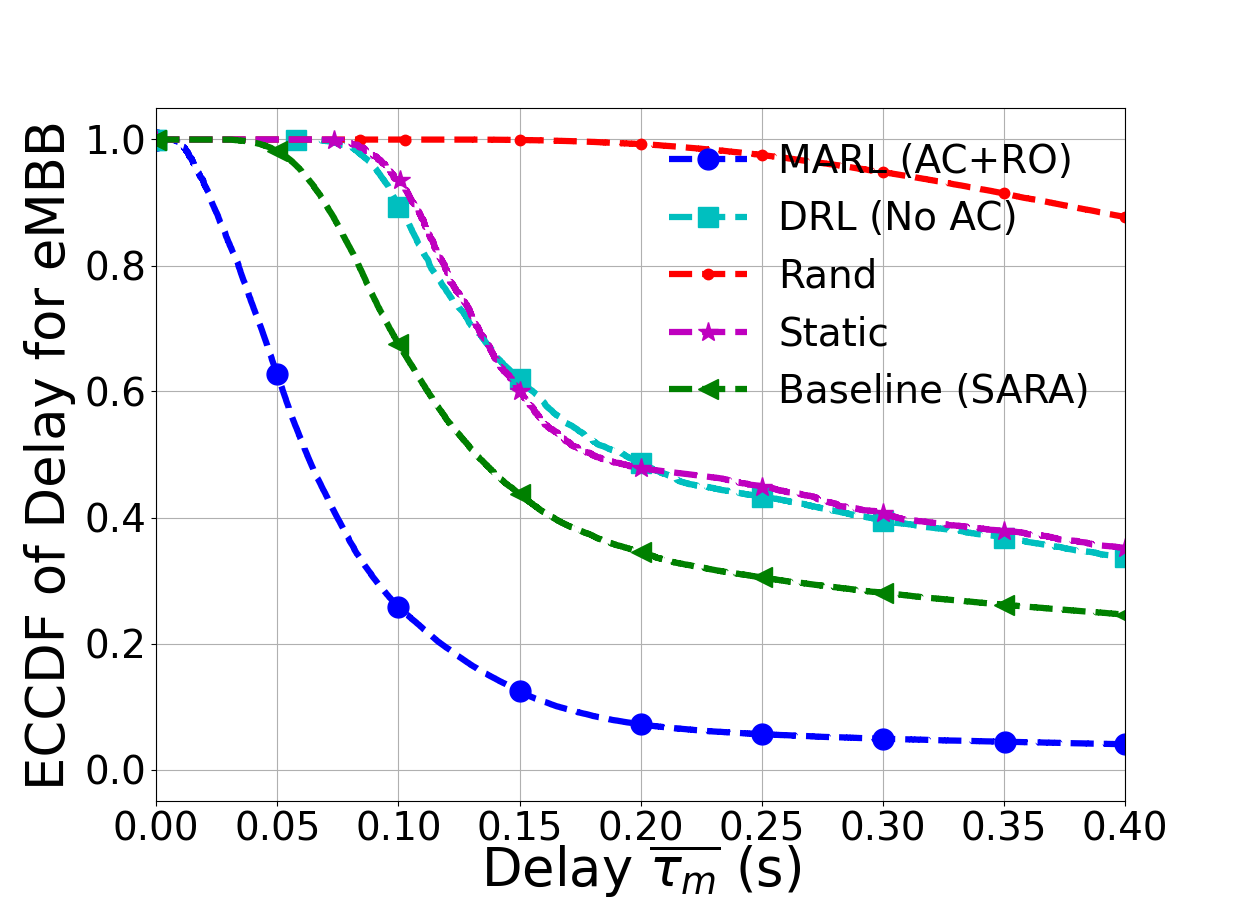}
        \caption{}
    \end{subfigure}
    \hfill
    \begin{subfigure}[b]{0.45\textwidth}
        \includegraphics[width=\textwidth]{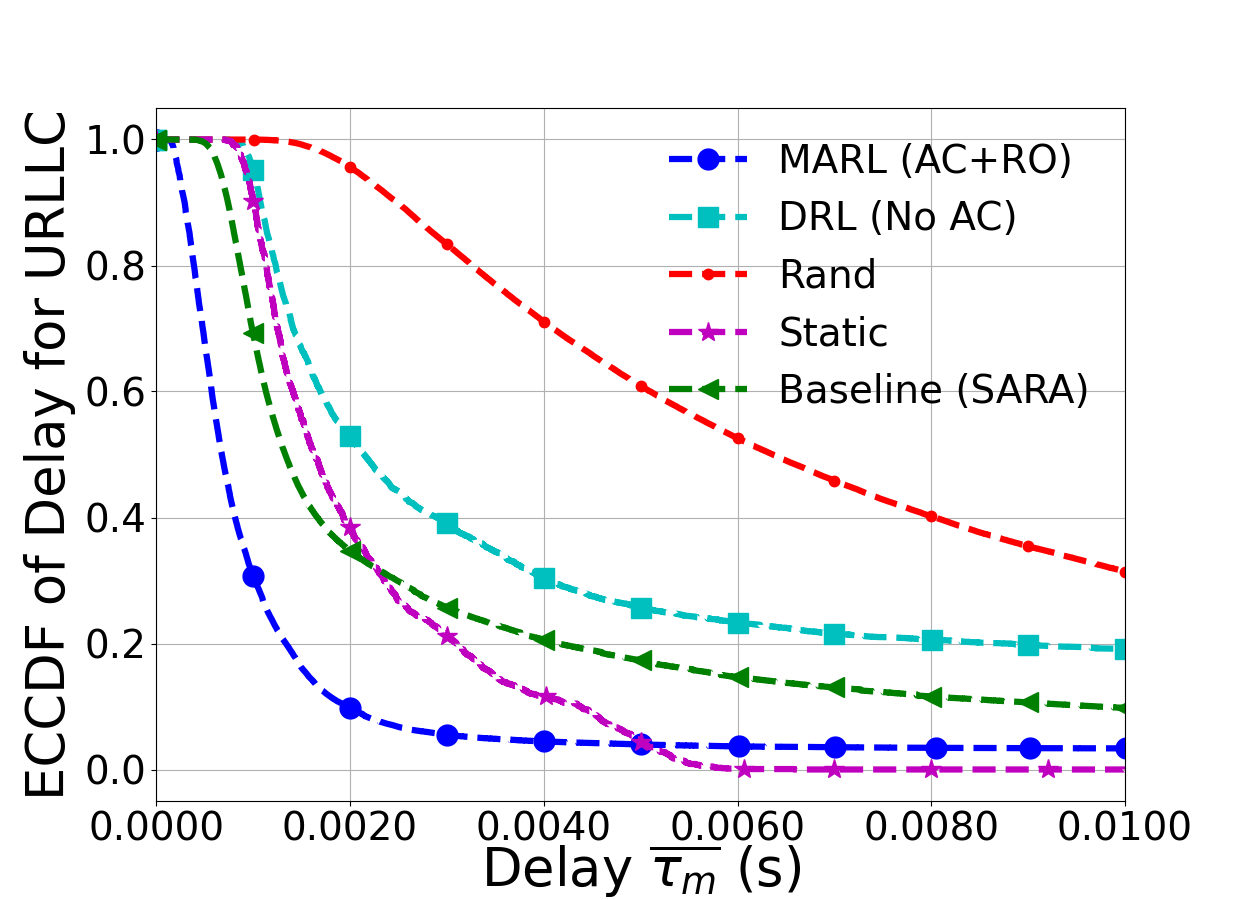}
        \caption{}
    \end{subfigure}
    \caption{ECCDF of delay for admitted slices – (a) eMBB and (b) URLLC.\vspace{-0in}}
    \label{fig:eccdfs}
\end{figure}

The end-to-end delay characteristics of admitted slices are analyzed through the empirical complementary cumulative distribution functions shown in Fig.~\ref{fig:eccdfs}. For both eMBB and URLLC services, the proposed multi-agent framework yields sharply decaying delay tails, with URLLC traffic exhibiting particularly tight concentration below the latency threshold. This behavior confirms that the learned admission and resource orchestration policies successfully suppress extreme delay events, which are the primary drivers of reliability degradation in time-critical services. The separation between eMBB and URLLC delay distributions further demonstrates effective service differentiation under shared radio and compute resources.

\begin{figure}[h!]
\centering

\begin{subfigure}{0.45\linewidth}
  \centering
  \includegraphics[width=\linewidth]{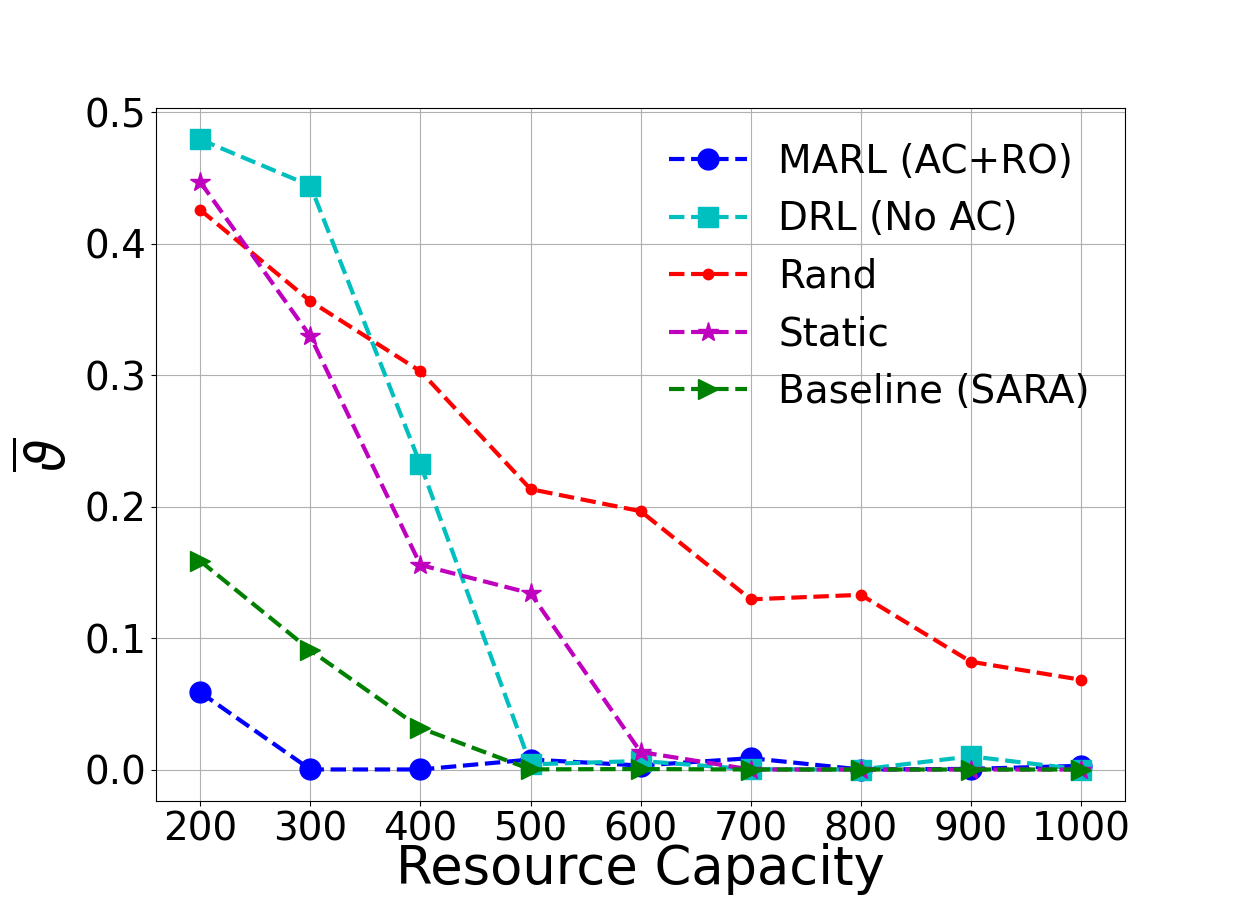}
  \caption{Packet drop rate.}
  \label{fig:kpi_pkt_drop}
\end{subfigure}
\hfill
\begin{subfigure}{0.45\linewidth}
  \centering
  \includegraphics[width=\linewidth]{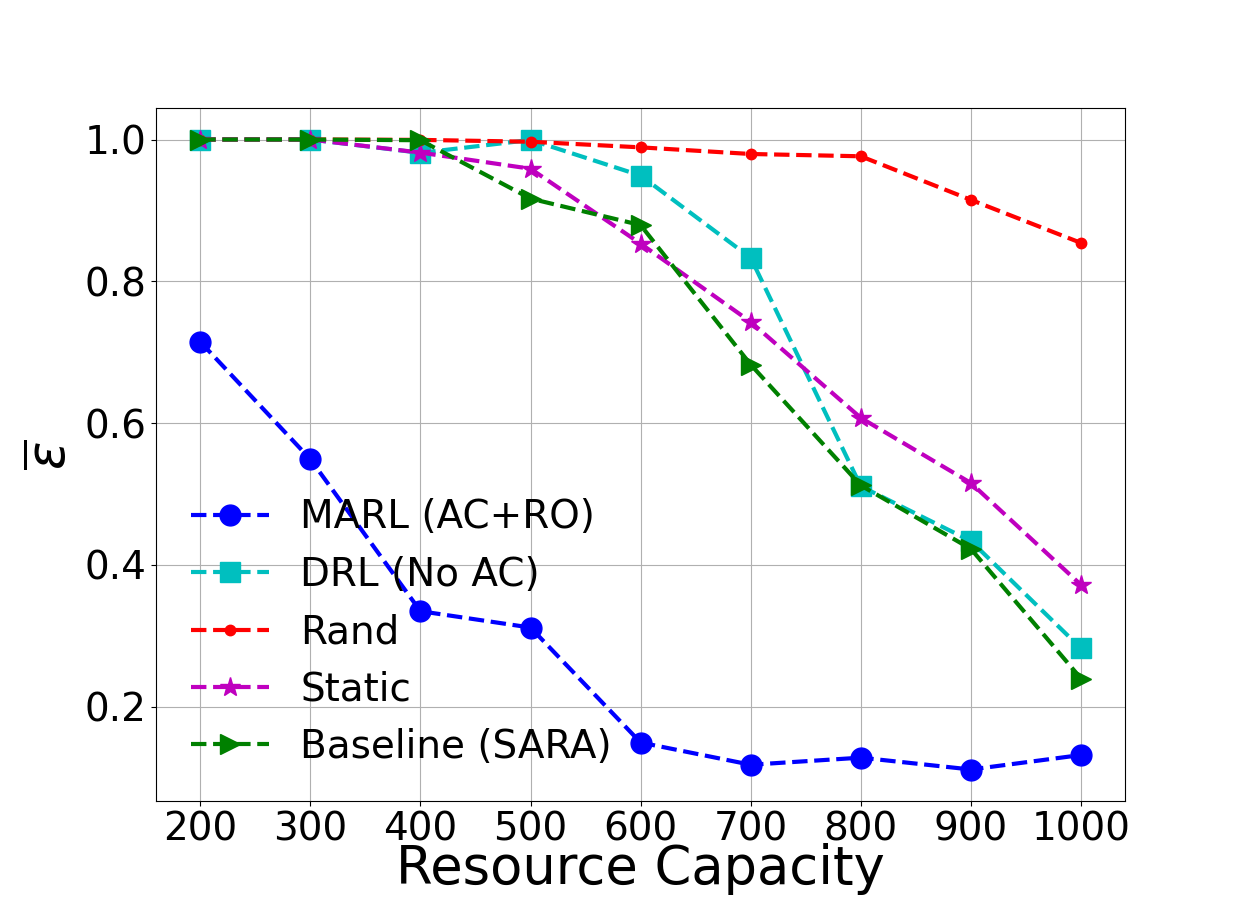}
  \caption{Delay violation rate.}
  \label{fig:kpi_delay_violation}
\end{subfigure}

\vspace{6pt}

\begin{subfigure}{0.45\linewidth}
  \centering
  \includegraphics[width=\linewidth]{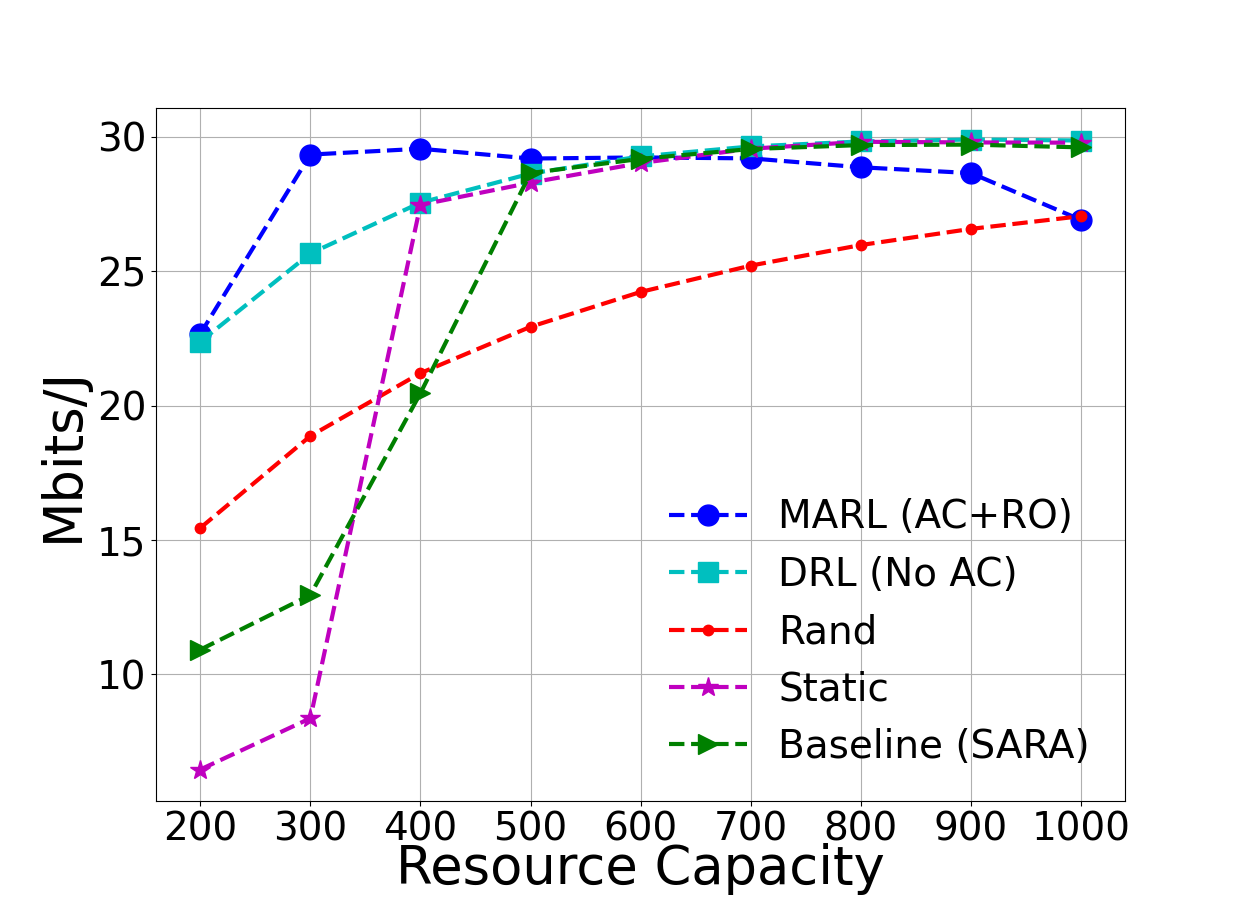}
  \caption{Energy efficiency.}
  \label{fig:kpi_energy_eff}
\end{subfigure}
\hfill
\begin{subfigure}{0.45\linewidth}
  \centering
  \includegraphics[width=\linewidth]{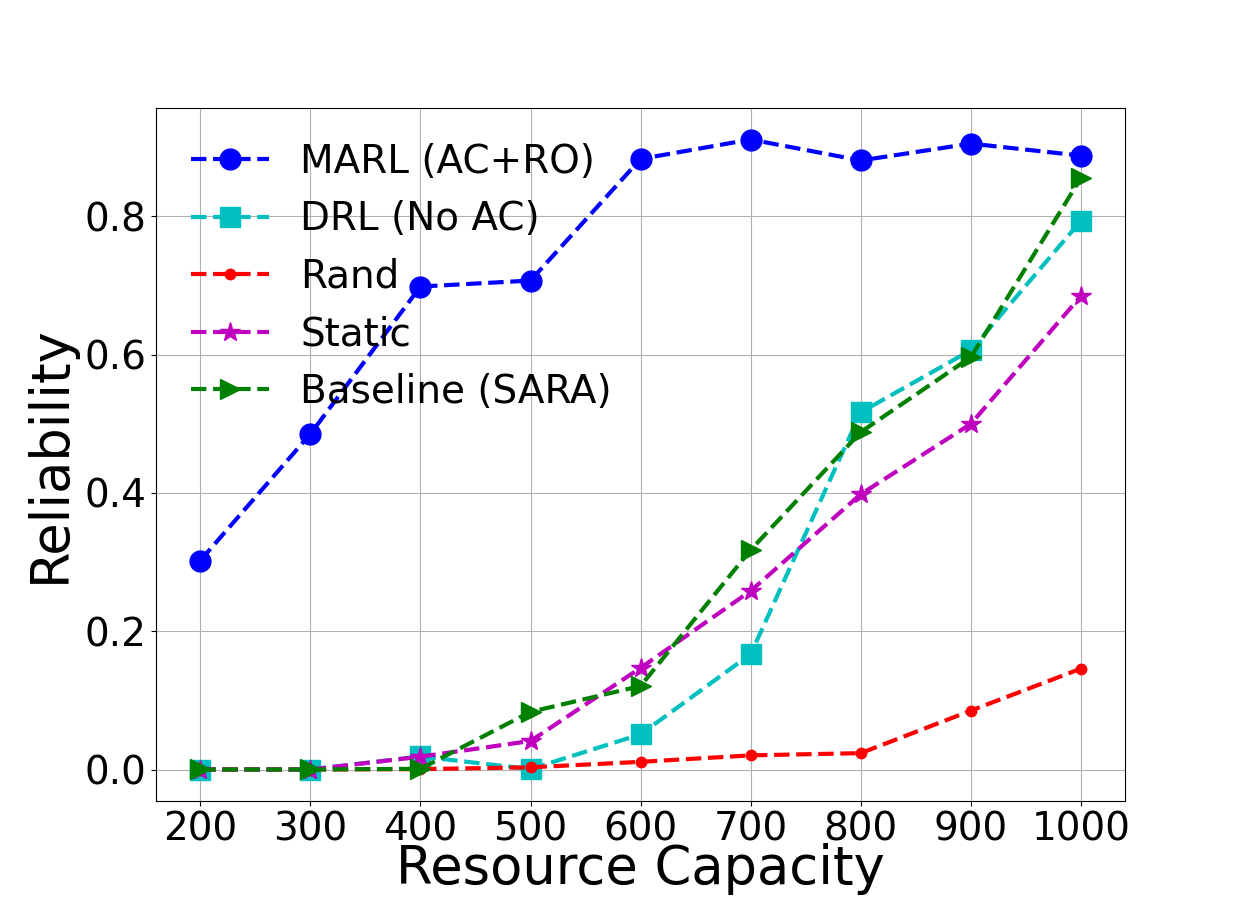}
  \caption{Reliability.}
  \label{fig:kpi_reliability}
\end{subfigure}

\caption{Performance metrics across varying resource capacities. (a) Packet drop rate. (b) Delay violation rate. (c) Energy efficiency. (d) Reliability.}
\label{fig:test_KPIs1}
\end{figure}

Fig.~\ref{fig:test_KPIs1} evaluates system performance under varying resource capacities, highlighting the robustness of the learned policies across different operating regimes. As available resources increase, both packet drop rate and delay violation rate decrease, while energy efficiency and reliability improve correspondingly. Importantly, the framework maintains acceptable reliability even under constrained resource budgets, indicating that the agents learn conservative admission and allocation strategies when capacity is scarce. This adaptability is a key indicator of resilience, as it enables the system to sustain service guarantees under fluctuating infrastructure conditions.

\begin{figure}[h!]
\centering

\begin{subfigure}{0.45\linewidth}
  \centering
  \includegraphics[width=\linewidth]{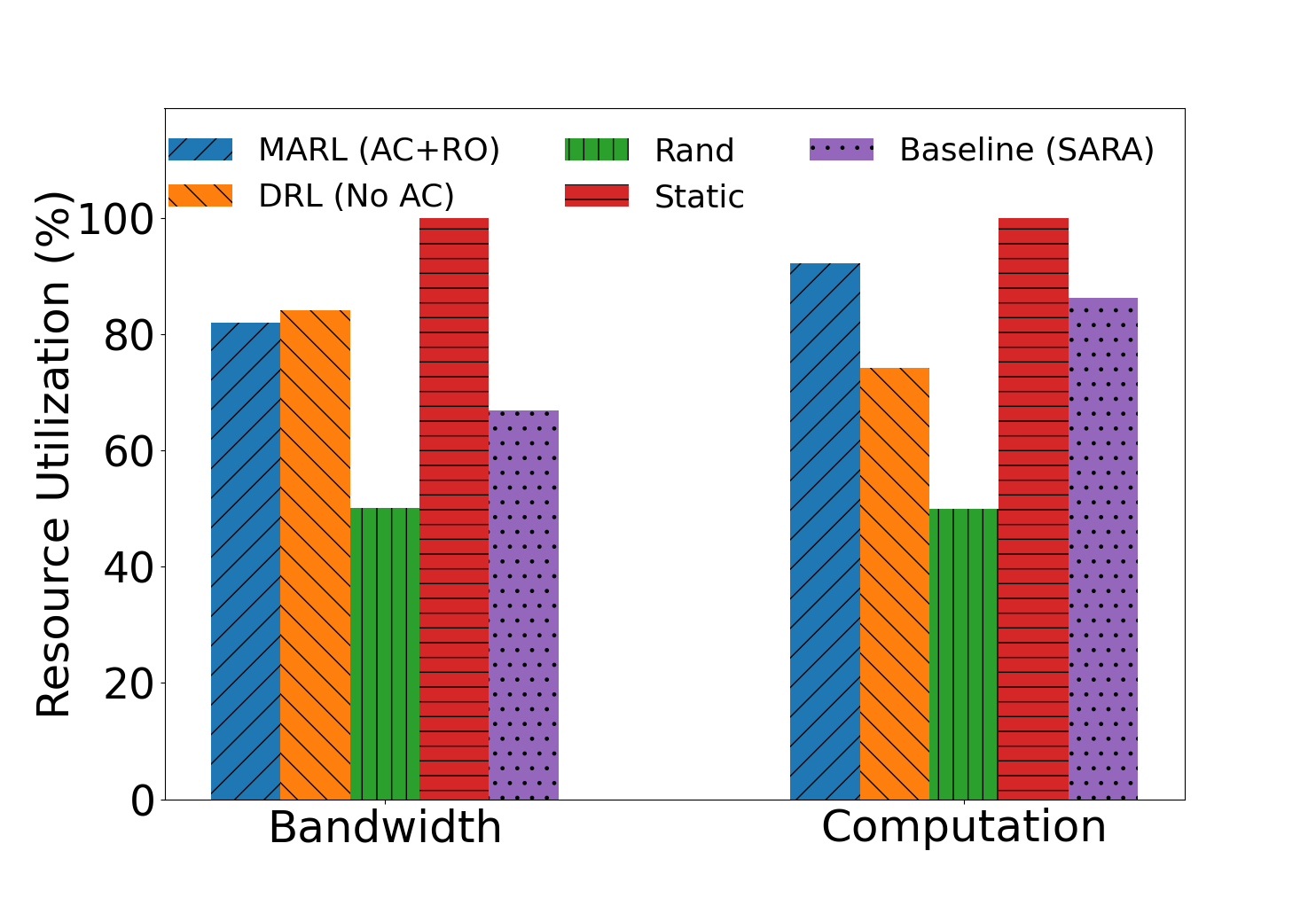}
  \caption{Bandwidth and computation resource utilization.}
  \label{fig:resource_usage}
\end{subfigure}
\hfill
\begin{subfigure}{0.45\linewidth}
  \centering
  \includegraphics[width=\linewidth]{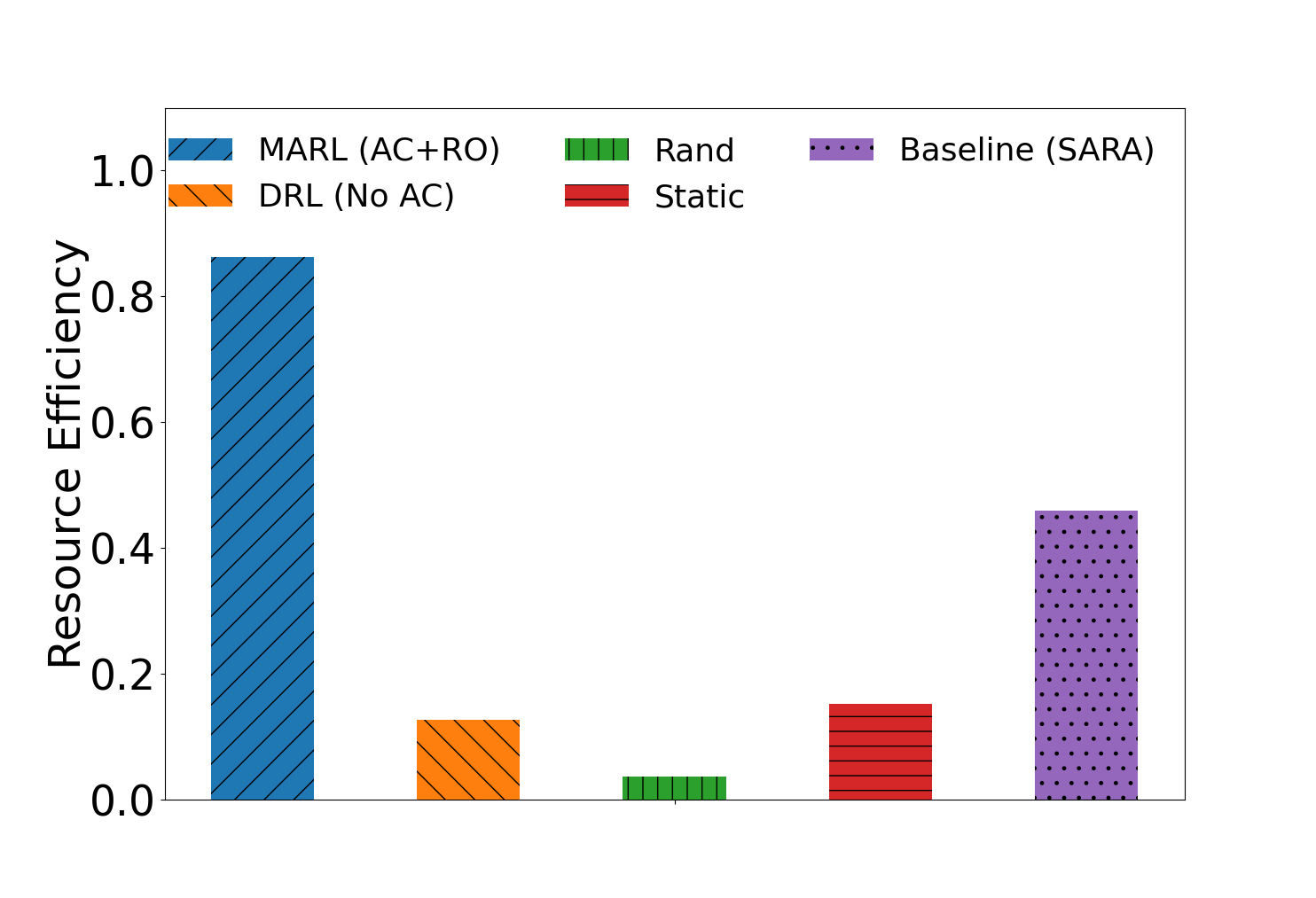}
  \caption{Resource efficiency in terms of reliability.}
  \label{fig:resource_efficiency}
\end{subfigure}

\caption{Comparison of average resource utilization and efficiency. (a) Bandwidth and computation resource utilization. (b) Resource efficiency measured in terms of reliability.}
\label{fig:resource}
\end{figure}

Finally, Fig.~\ref{fig:resource} compares average resource utilization and efficiency achieved by the proposed framework. While resource utilization increases with demand, the corresponding resource efficiency remains high, indicating that additional resources are translated into reliability gains rather than wasted capacity. This behavior reflects the effectiveness of the CMDP constraints in preventing over-allocation and highlights the ability of the multi-agent framework to operate near the Pareto-efficient frontier between utilization and reliability.

\subsection*{Resilience Interpretation}

From a resilience standpoint, the CMDP formulation enables proactive overload avoidance by internalizing reliability constraints into the learning objective. The admission agent learns to limit arrivals before congestion occurs, while the orchestration agent dynamically reallocates resources to absorb traffic surges and channel degradation. Recovery emerges naturally as agents adapt their policies following transient violations, restoring reliability without explicit reconfiguration. This establishes multi-agent reinforcement learning as a complementary resilience mechanism to the optimization-based orchestration strategies developed earlier in this chapter.

\section{LLM Driven Resilient Slice Management}

While the multi-agent reinforcement learning framework in Section~5.3 enables fine-grained, closed-loop adaptation at near real-time timescales, its performance is inherently bounded by the expressiveness of predefined state representations, reward structures, and training regimes. In contrast, large language models (LLMs) introduce a complementary decision-making paradigm that operates at a higher semantic level, enabling intent-aware, context-driven resilience mechanisms that are not easily captured by conventional optimization or reinforcement learning formulations. From a conceptual standpoint, this aligns with the broader shift toward semantic and intent-aware control mechanisms discussed in Chapter~3, where decision-making extends beyond purely numerical optimization to incorporate higher-level system objectives and contextual reasoning.
In this section, we formalize the role of LLMs as cognitive controllers for resilient slice management and analyze their interaction with the O-RAN control stack.

\begin{figure}[h!]
    \centering
    \vspace{-0.0in}
    \includegraphics[width = 0.7\textwidth]{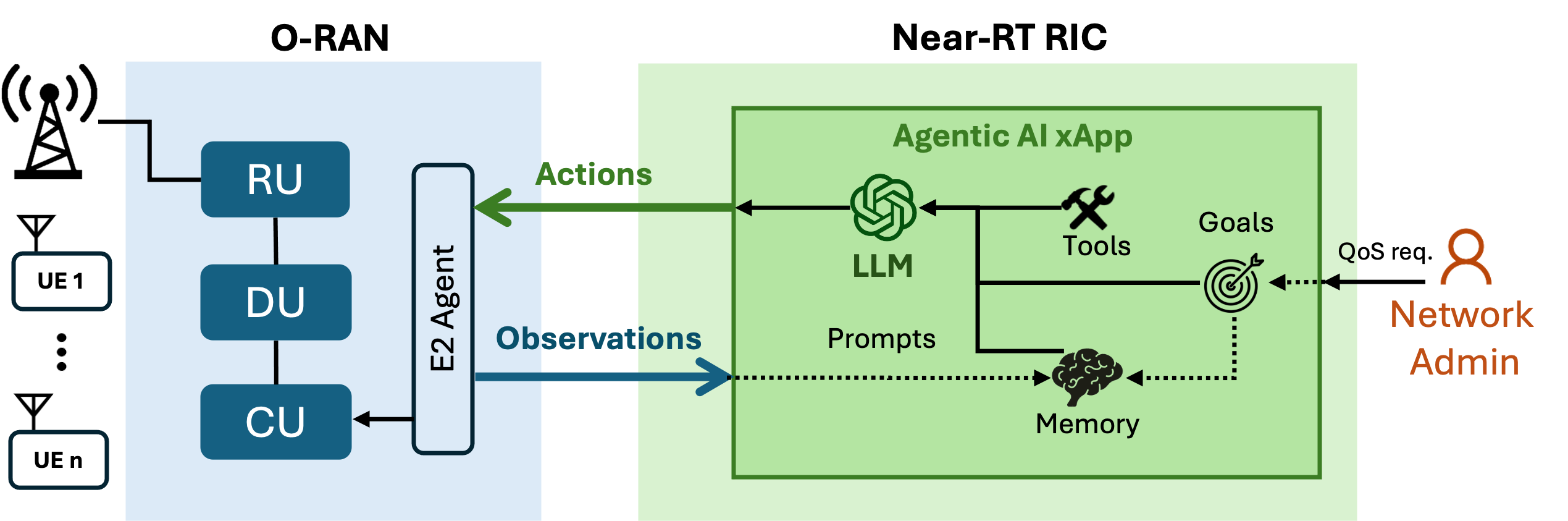}
    \caption{O-RAN system with LLM-powered xApp deployed in the near-RT RIC. \vspace{-0.0in}}
    \label{fig:oran}
\end{figure}

Fig.~\ref{fig:oran} illustrates the deployment of the LLM-powered xApp within the near-real-time RIC, where it interfaces with the O-RAN E2 control plane and operates alongside data-driven agents. The LLM observes abstracted slice-level performance summaries, historical control actions, and operator-defined intent constraints, and produces structured control recommendations that guide admission control, inter-slice prioritization, and long-term resource re-balancing.

\subsection*{Intent-to-Policy Mapping Model}

Let $\mathcal{I}_k$ denote the intent specification at reconfiguration slot $k$, expressed as a structured semantic tuple
\begin{equation}
\mathcal{I}_k = \{\mathcal{Q}_k, \mathcal{P}_k, \mathcal{C}_k\},
\end{equation}
where $\mathcal{Q}_k$ represents slice-level QoS objectives such as latency bounds or reliability targets, $\mathcal{P}_k$ encodes relative slice priorities, and $\mathcal{C}_k$ captures system-level constraints including resource budgets and policy limits. Unlike numeric optimization variables, these components may be expressed in natural language or semi-structured templates and may evolve dynamically over time.

The LLM implements a mapping
\begin{equation}
\Phi_{\text{LLM}} : (\mathcal{I}_k, \mathcal{H}_k) \rightarrow \mathcal{A}_k,
\end{equation}
where $\mathcal{H}_k$ denotes the optimization history and system feedback accumulated up to slot $k$, and $\mathcal{A}_k$ represents a candidate control action expressed as a slice-level allocation vector or policy adjustment. This mapping is realized through in-context reasoning rather than parameterized policy evaluation, allowing the LLM to generalize beyond previously observed operating points.

The generated action $\mathcal{A}_k$ is subsequently normalized and enforced through the RIC control loop, yielding concrete resource allocations $\{r_k^s\}_{s \in \mathcal{S}}$ that satisfy
\begin{equation}
\sum_{s \in \mathcal{S}} r_k^s \leq R,
\end{equation}
while respecting slice isolation and feasibility constraints.

\begin{figure}[h!]
    \centering
    \vspace{-0.0in}
    \includegraphics[width = 0.7\textwidth]{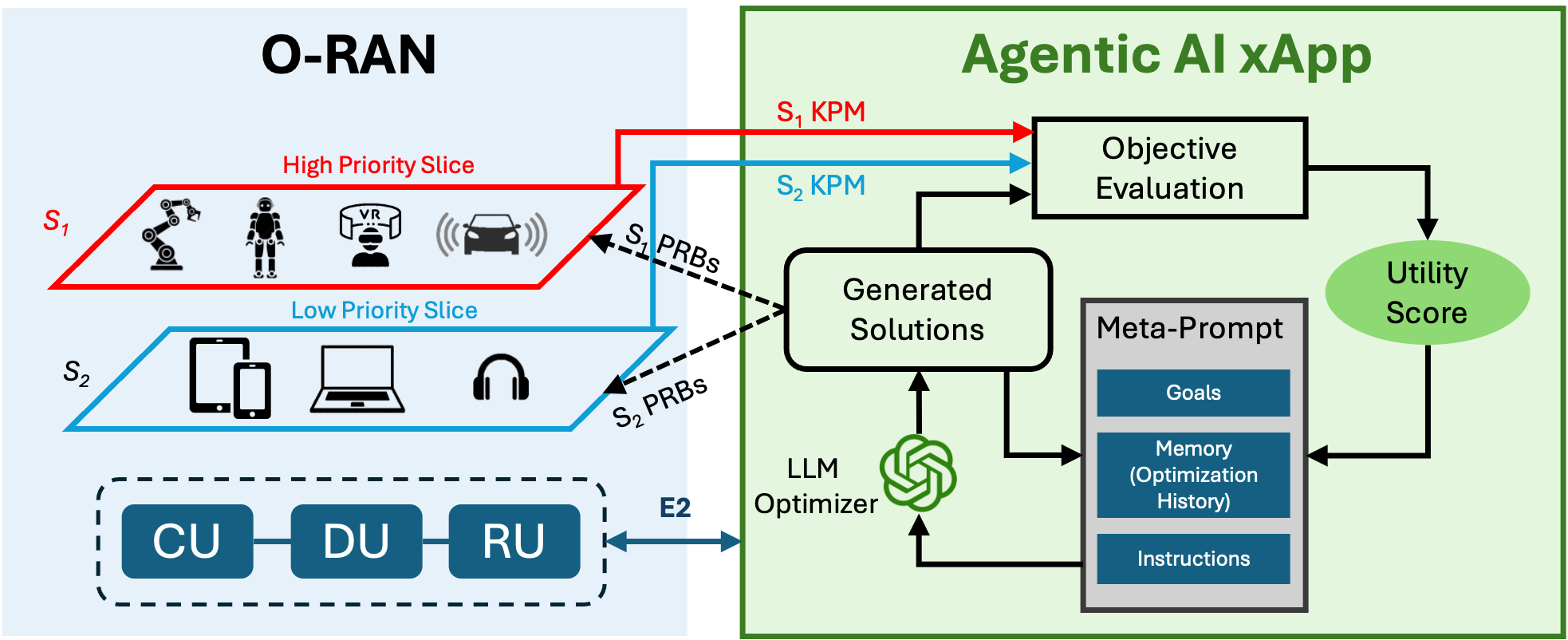}
    \caption{LLM-Driven Agentic AI-based resource provisioning framework. \vspace{-0.0in}}
    \label{fig:workflows3}
\end{figure}
Fig.~\ref{fig:workflows3} illustrates the LLM-driven agentic AI workflow for intent-aware resource provisioning within the O-RAN architecture. The framework operates by ingesting slice-level key performance measurements (KPMs) from the near real-time RIC via the E2 interface, including latency, reliability, and PRB utilization metrics for both high-priority and low-priority slices. These measurements are processed by an agentic xApp that integrates an LLM-based optimizer with a meta-prompt structure comprising optimization goals, historical memory, and operational instructions. Based on the observed system state and intent constraints, the LLM generates candidate resource allocation solutions, which are subsequently evaluated by an objective function to produce a scalar utility score. The selected solution is then translated into concrete PRB allocation commands and enforced back into the O-RAN control loop.

\begin{figure*}[h!]
  \centering
    \begin{minipage}[b]{\textwidth}
    \includegraphics[width=\textwidth]{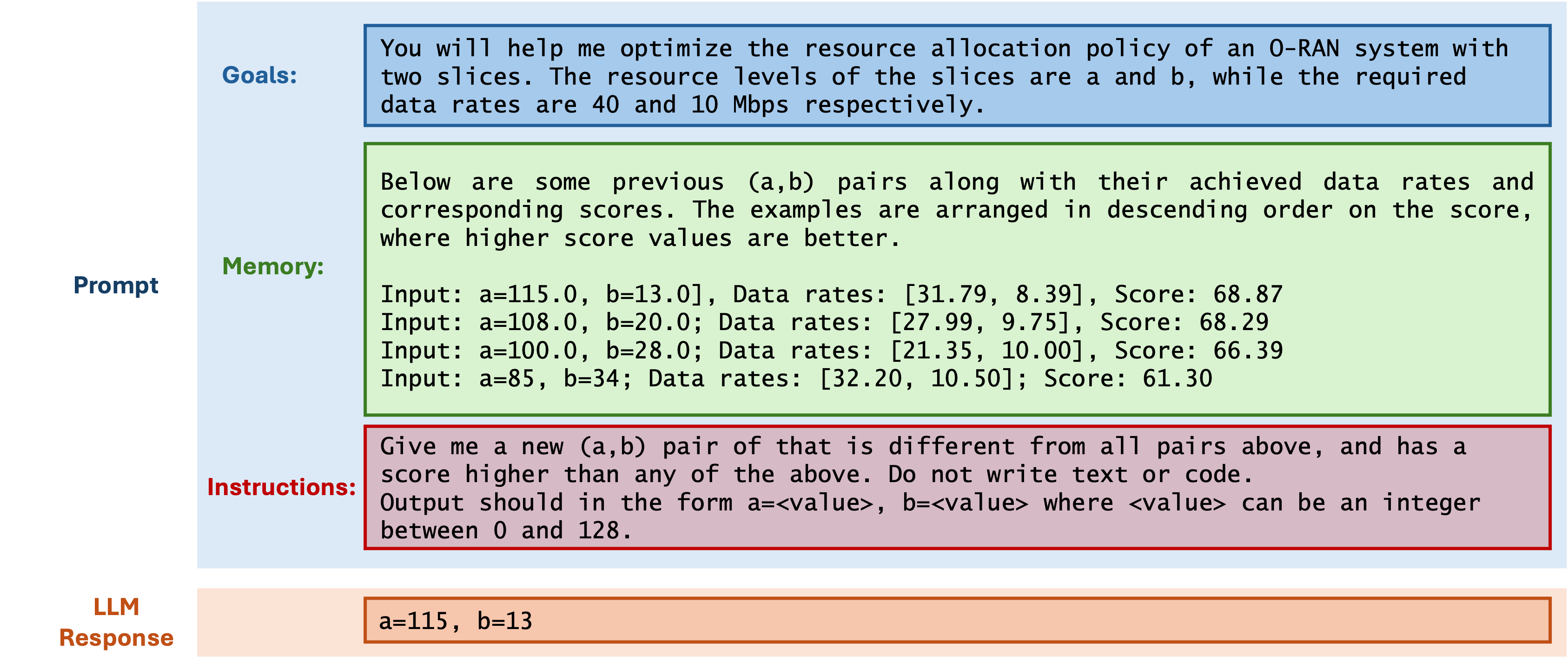}
    \caption{Examples of meta-prompt and LLM response for resource orchestration. \vspace{-0.0in}} \label{fig:prompt}
  \end{minipage}
\end{figure*}

Fig.~\ref{fig:prompt} provides concrete examples of the meta-prompt structure supplied to the LLM and the corresponding responses generated for resource orchestration. The prompts encode slice-level objectives, priority semantics, and operational constraints together with historical context, while the LLM responses produce structured allocation recommendations that are consistent with these inputs. These examples illustrate how qualitative intent and quantitative system state are jointly embedded into the decision process, bridging the semantic reasoning performed by the LLM with the utility-based evaluation and enforcement mechanisms formalized in the subsequent analysis.

\subsection*{LLM-Guided Resilience Objective}

The resilience contribution of the LLM can be formalized through its effect on long-horizon performance stability. Let $\Theta_k$ denote the system-wide reliability metric aggregated across slices at slot $k$, defined as
\begin{equation}
\Theta_k = \frac{1}{|\mathcal{S}|} \sum_{s \in \mathcal{S}} \frac{1}{|\mathcal{T}_k|} \sum_{t \in \mathcal{T}_k} \theta_s^t,
\end{equation}
where $\theta_s^t$ follows the definition in Section~5.3. The LLM-driven controller implicitly seeks to maximize the discounted resilience objective
\begin{equation}
\max_{\{\mathcal{A}_k\}} \; \mathbb{E} \left[ \sum_{k=0}^{\infty} \lambda^k \Theta_k \right],
\end{equation}
with discount factor $\lambda \in (0,1)$, without requiring an explicit model of state transitions or reward gradients.

This formulation highlights a key distinction from reinforcement learning. Rather than learning a stationary policy, the LLM continuously reinterprets the optimization objective in response to evolving context, thereby enhancing robustness under non-stationary traffic patterns, admission bursts, and slice demand shifts.

\begin{figure}[h!]
\centering

\begin{subfigure}{0.45\linewidth}
  \centering
  \includegraphics[width=\linewidth]{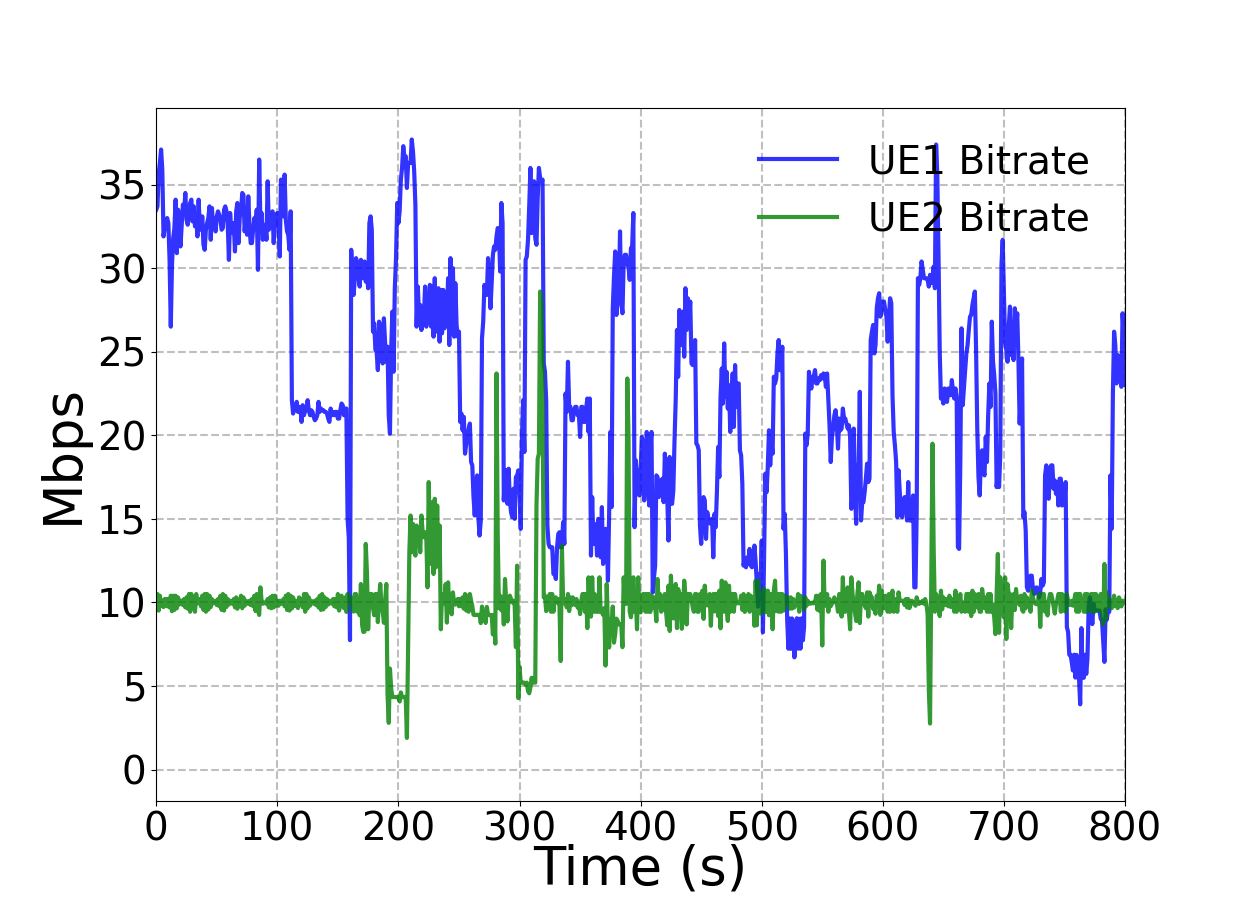}
  \caption{Random allocation.}
  \label{fig:alloc_rand}
\end{subfigure}
\hfill
\begin{subfigure}{0.45\linewidth}
  \centering
  \includegraphics[width=\linewidth]{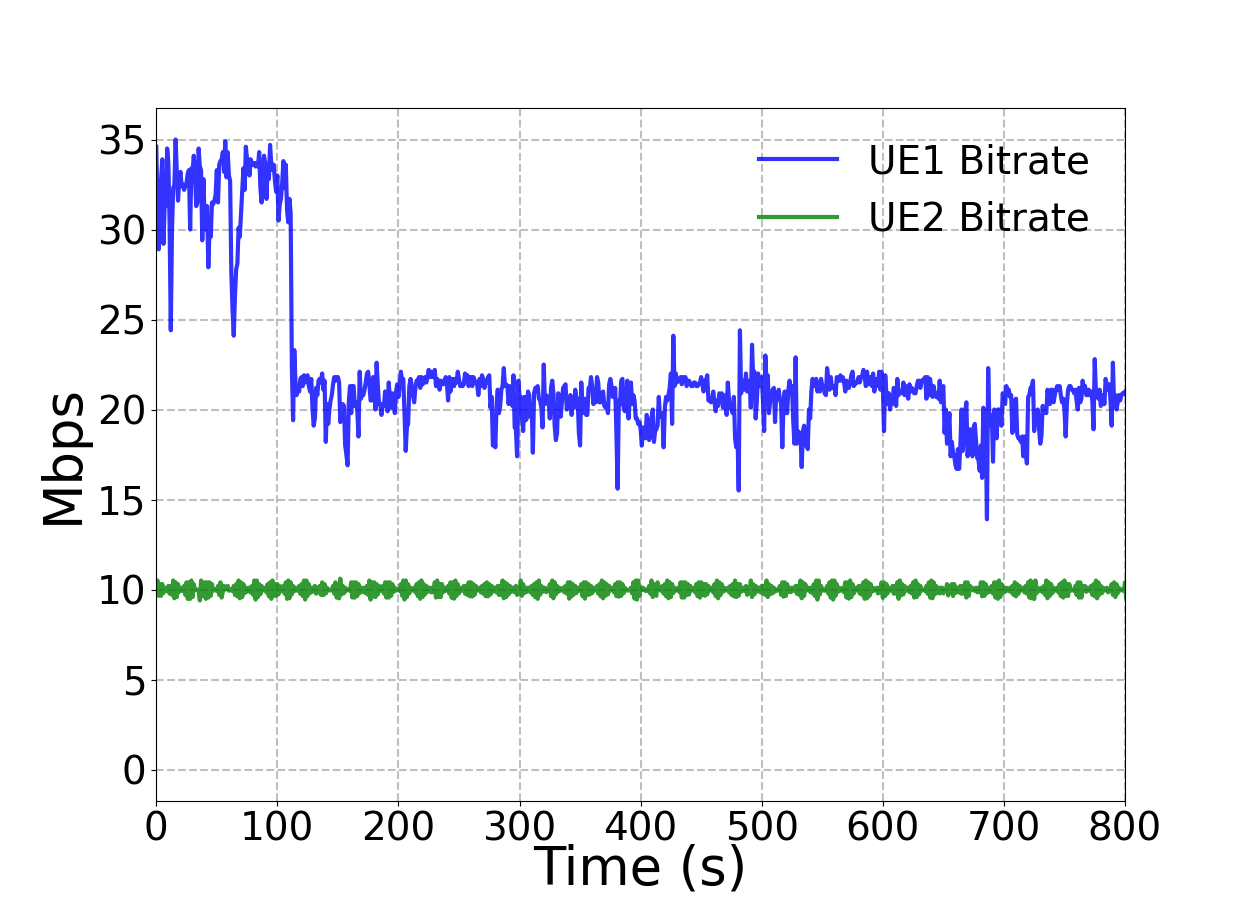}
  \caption{Equal allocation.}
  \label{fig:alloc_equal}
\end{subfigure}

\vspace{6pt}

\begin{subfigure}{0.45\linewidth}
  \centering
  \includegraphics[width=\linewidth]{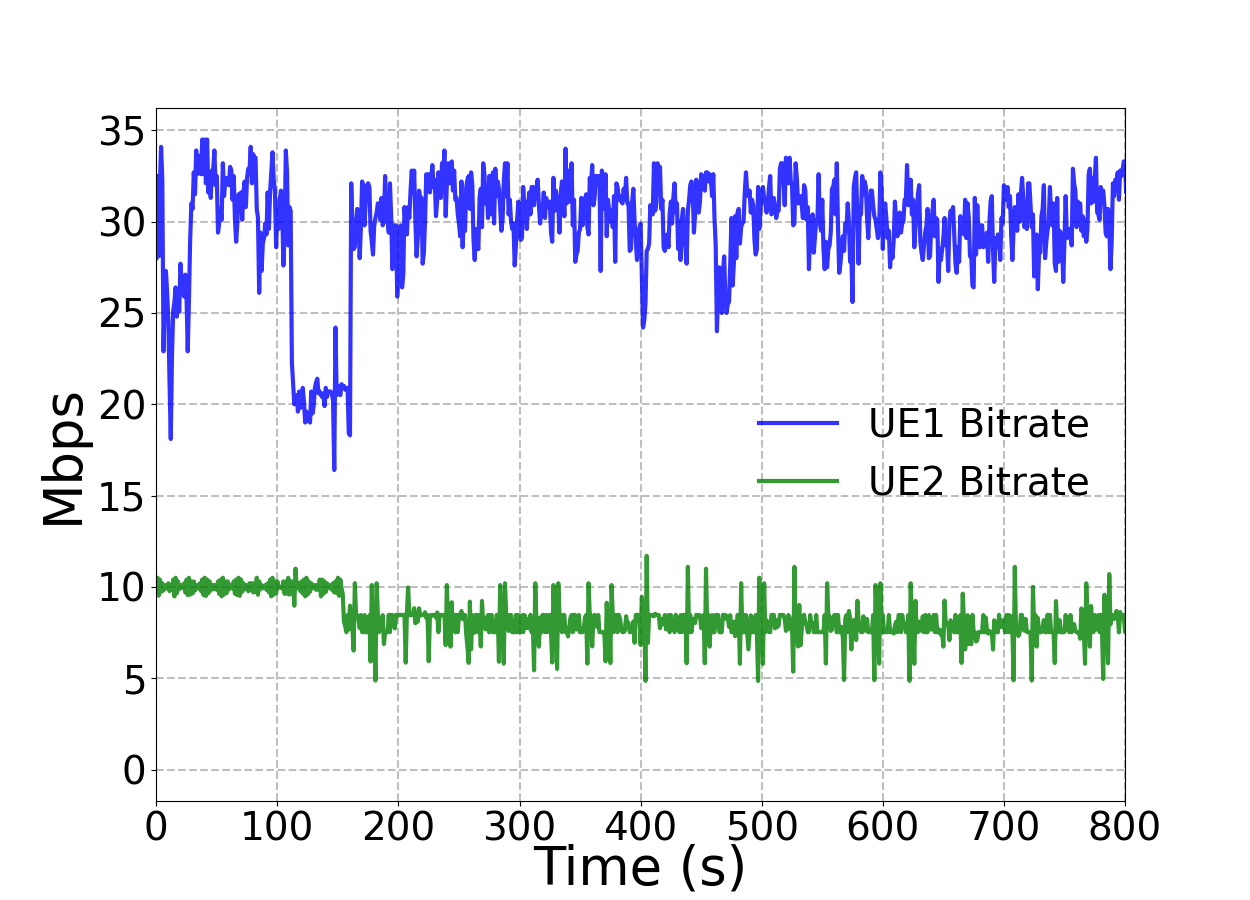}
  \caption{Proportional allocation.}
  \label{fig:alloc_prop}
\end{subfigure}
\hfill
\begin{subfigure}{0.45\linewidth}
  \centering
  \includegraphics[width=\linewidth]{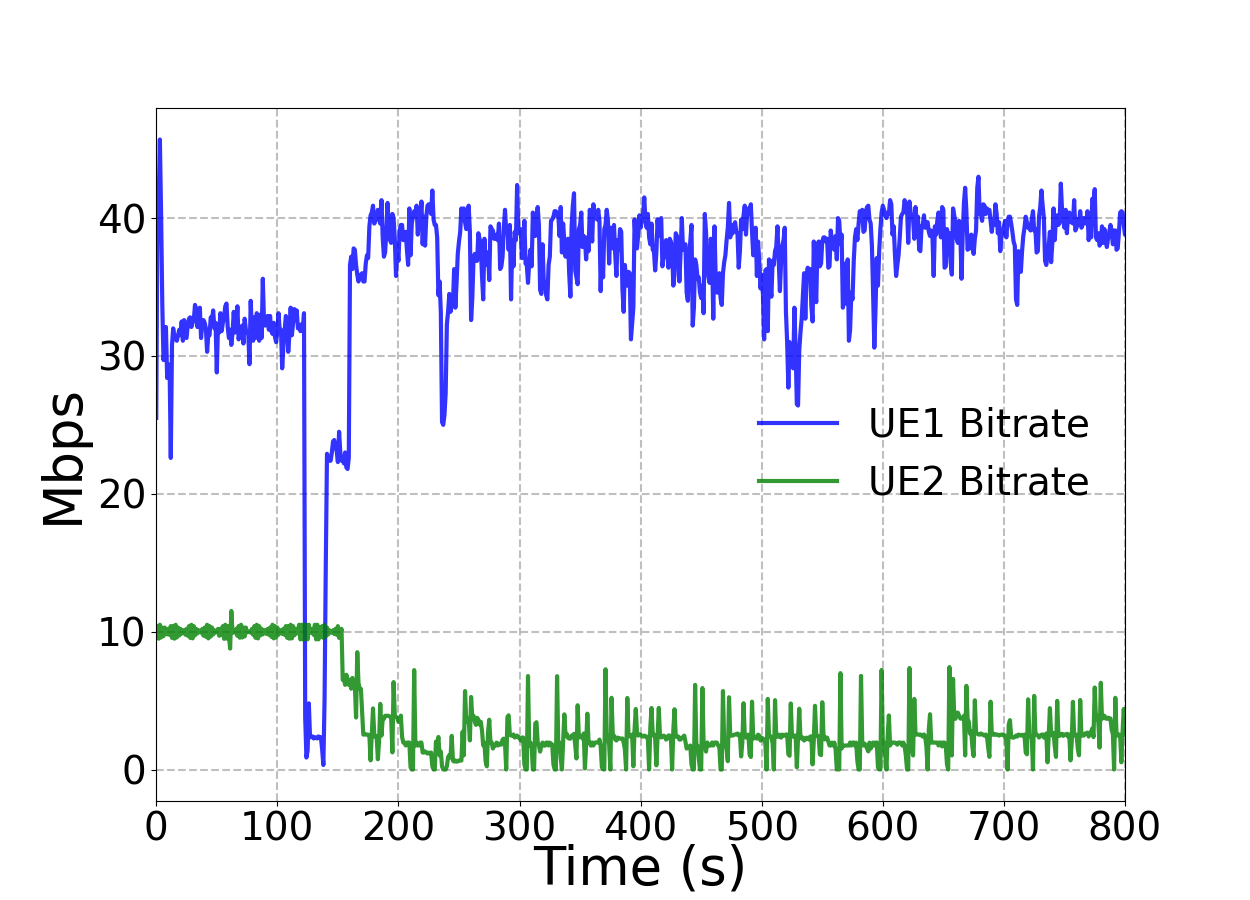}
  \caption{LLM-driven allocation.}
  \label{fig:alloc_llm}
\end{subfigure}

\caption{Comparative results of resource allocation schemes. (a) Random allocation. (b) Equal allocation. (c) Proportional allocation. (d) LLM-driven allocation.}
\label{fig:data_rate}
\end{figure}

\subsection{Performance Implications and Observed Gains}

The performance of the LLM-driven framework is first evaluated against baseline allocation schemes in Fig.~\ref{fig:data_rate}. Compared to random, equal, and proportional allocation, the LLM-driven approach consistently achieves superior performance by adapting resource distribution based on contextual reasoning rather than instantaneous demand alone. This confirms that intent-aware decision making yields tangible gains over heuristic strategies.



\begin{figure}[h!]
\centering

\begin{subfigure}{0.45\linewidth}
  \centering
  \includegraphics[width=\linewidth]{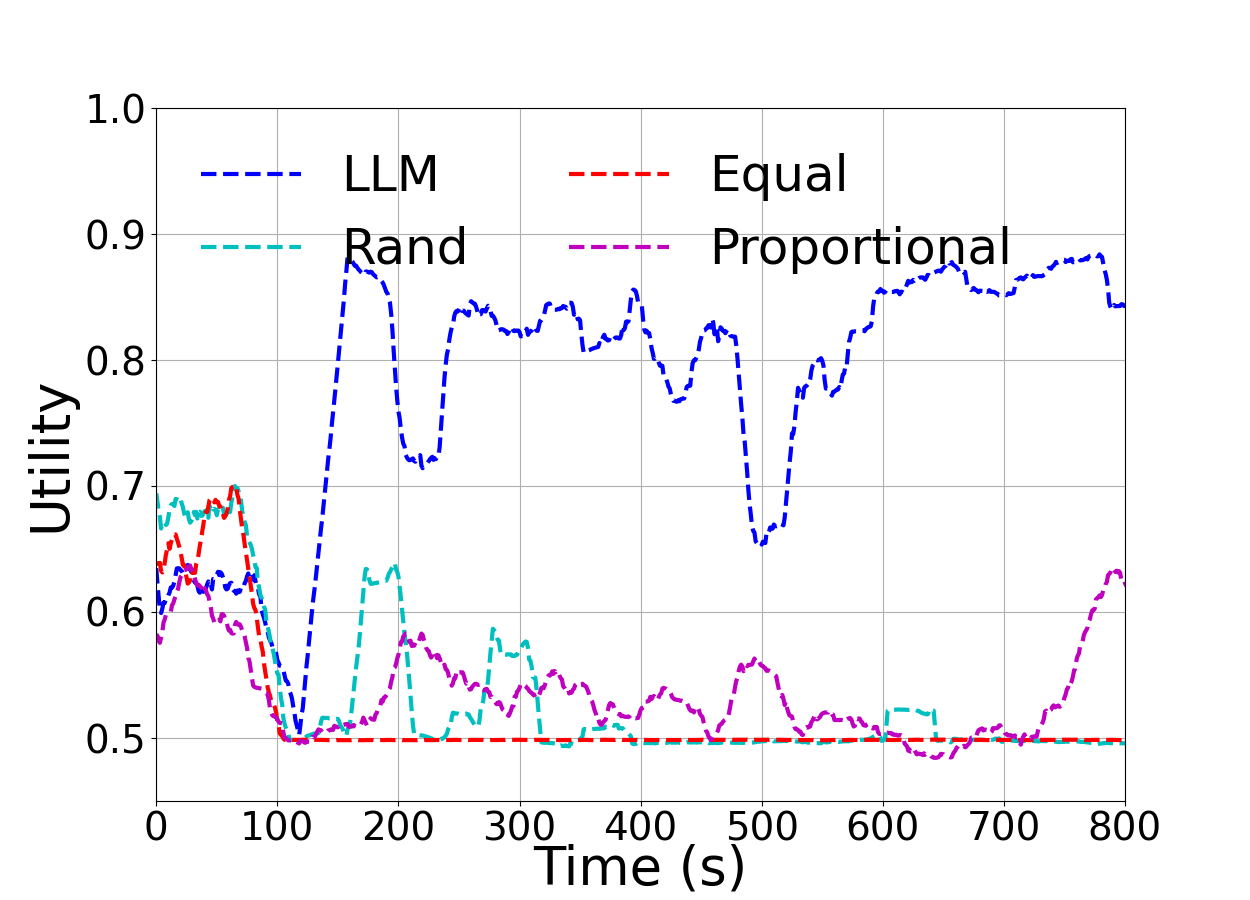}
  \caption{System utility.}
  \label{fig:system_utility}
\end{subfigure}
\hfill
\begin{subfigure}{0.45\linewidth}
  \centering
  \includegraphics[width=\linewidth]{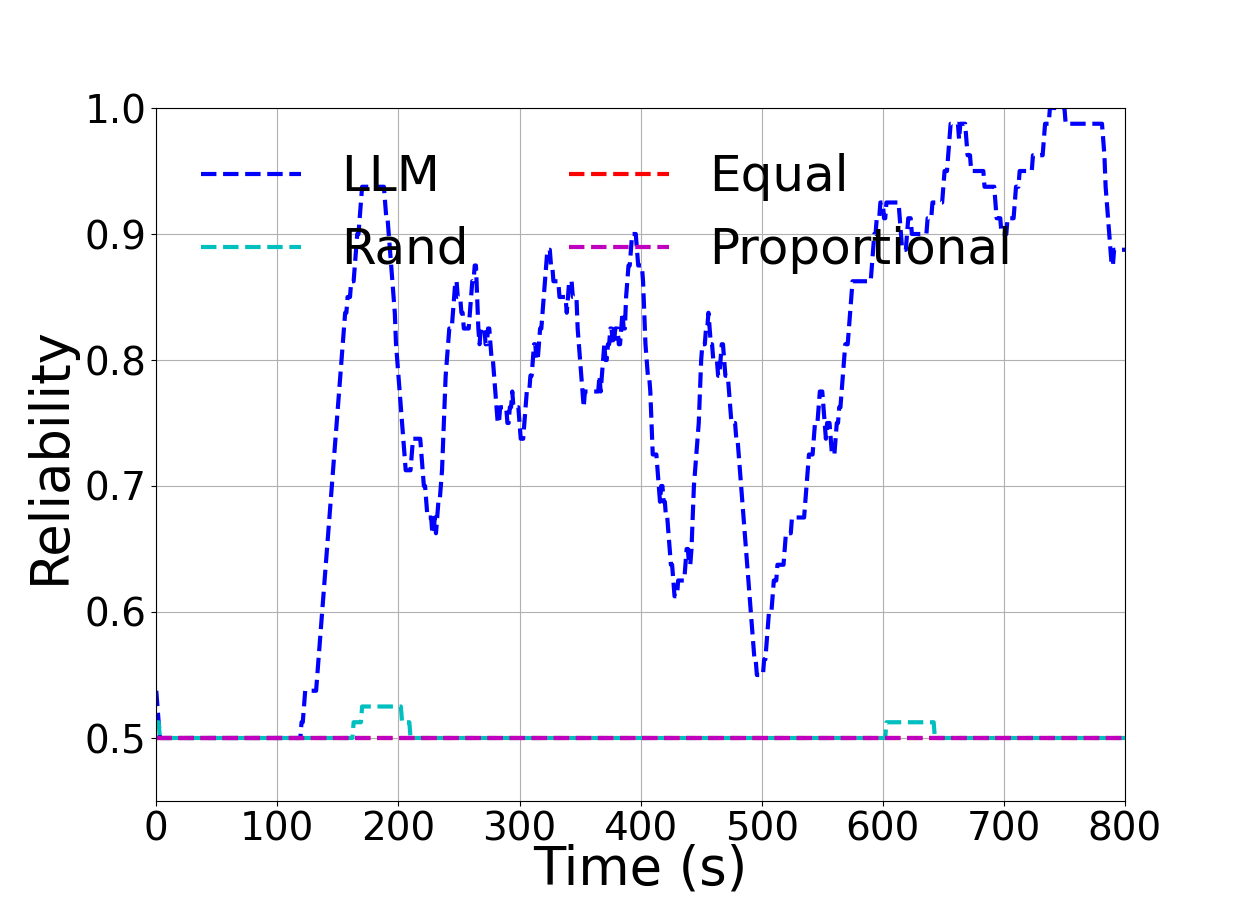}
  \caption{System reliability.}
  \label{fig:system_reliability}
\end{subfigure}

\vspace{6pt}

\begin{subfigure}{0.45\linewidth}
  \centering
  \includegraphics[width=\linewidth]{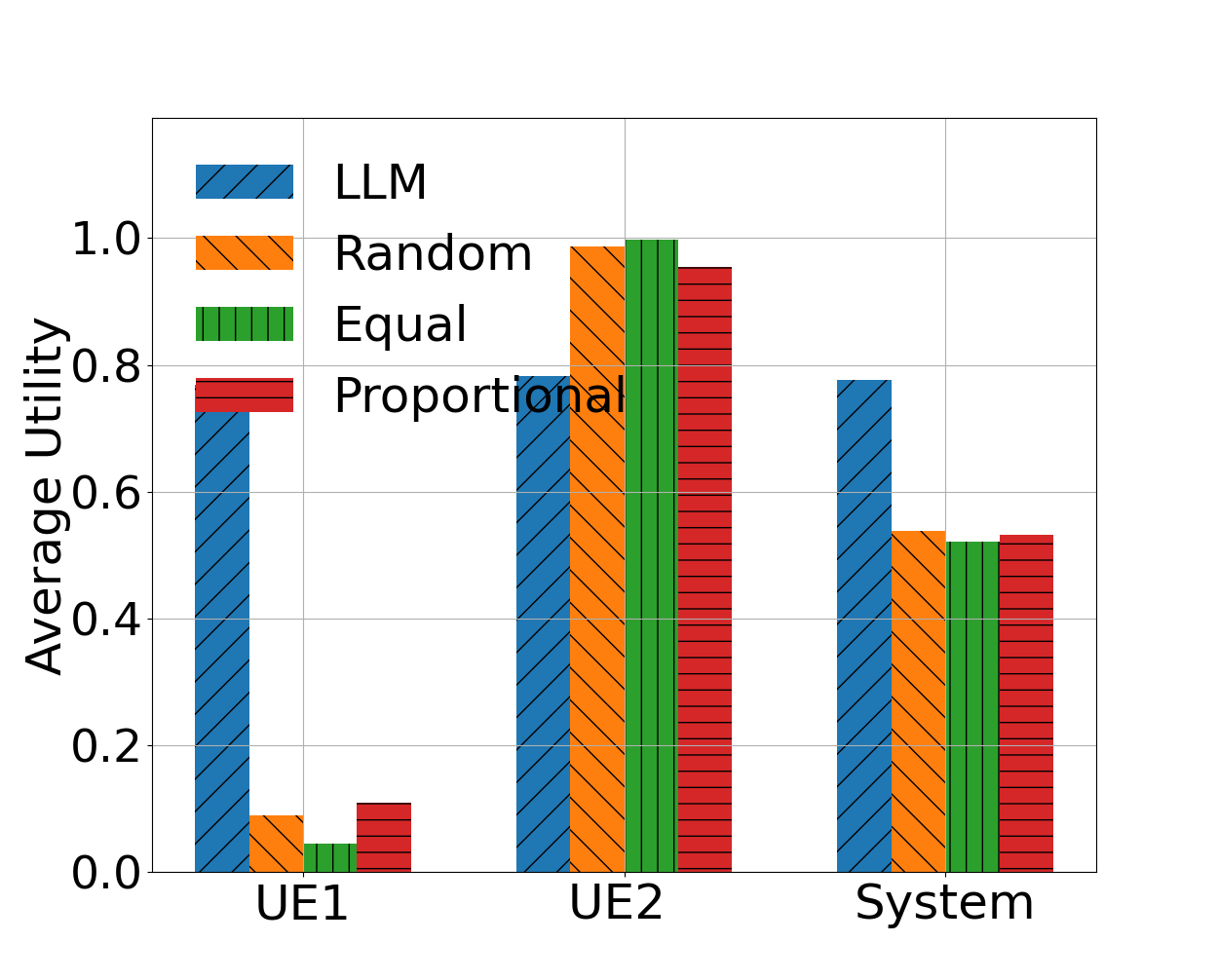}
  \caption{Average utility.}
  \label{fig:average_utility}
\end{subfigure}
\hfill
\begin{subfigure}{0.45\linewidth}
  \centering
  \includegraphics[width=\linewidth]{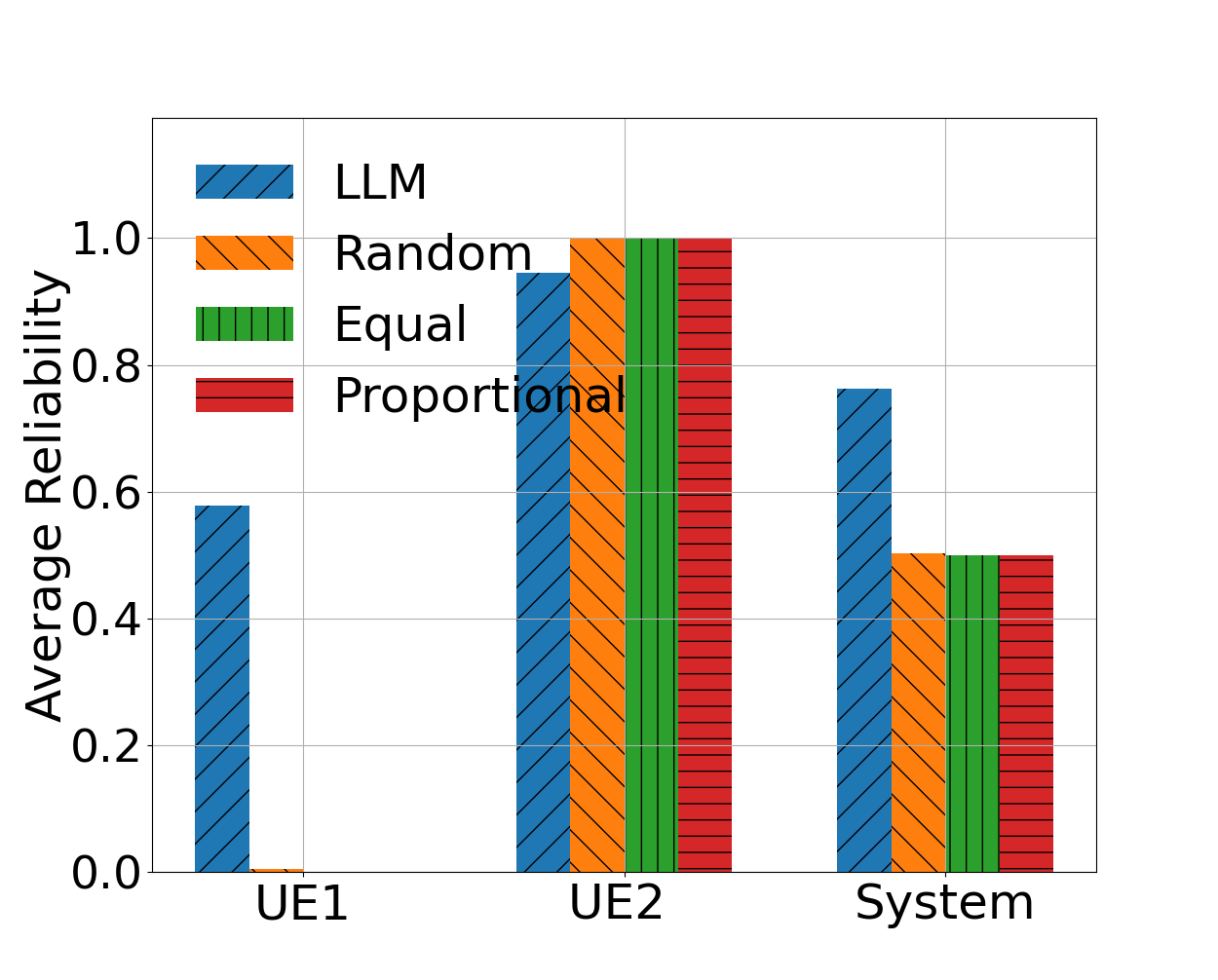}
  \caption{Average reliability.}
  \label{fig:average_reliability}
\end{subfigure}

\caption{Comparative analysis of utility and reliability. (a) System utility. (b) System reliability. (c) Average utility. (d) Average reliability.}
\label{fig:test_KPIs}
\end{figure}

Fig.~\ref{fig:test_KPIs} presents a comparative analysis of system utility and reliability. The LLM-driven approach achieves higher instantaneous and average utility while simultaneously improving reliability, indicating that performance gains are achieved without compromising resilience. The reduced variance observed in both metrics reflects improved control stability under LLM-guided operation.

\begin{figure}[h!]
\centering

\begin{subfigure}{0.45\linewidth}
  \centering
  \includegraphics[width=\linewidth]{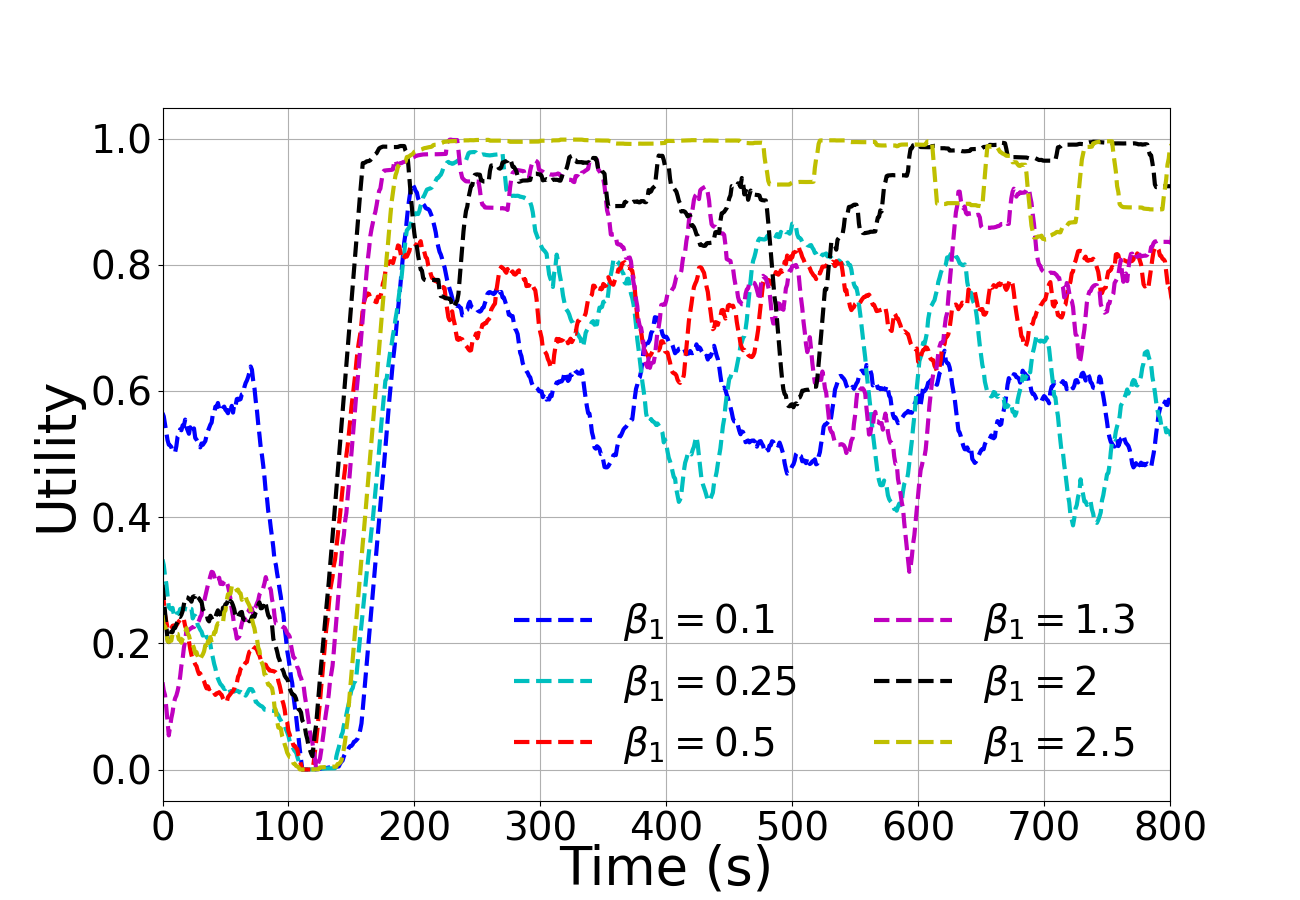}
  \caption{UE~1 utility.}
  \label{fig:ue1_utility_priority}
\end{subfigure}
\hfill
\begin{subfigure}{0.45\linewidth}
  \centering
  \includegraphics[width=\linewidth]{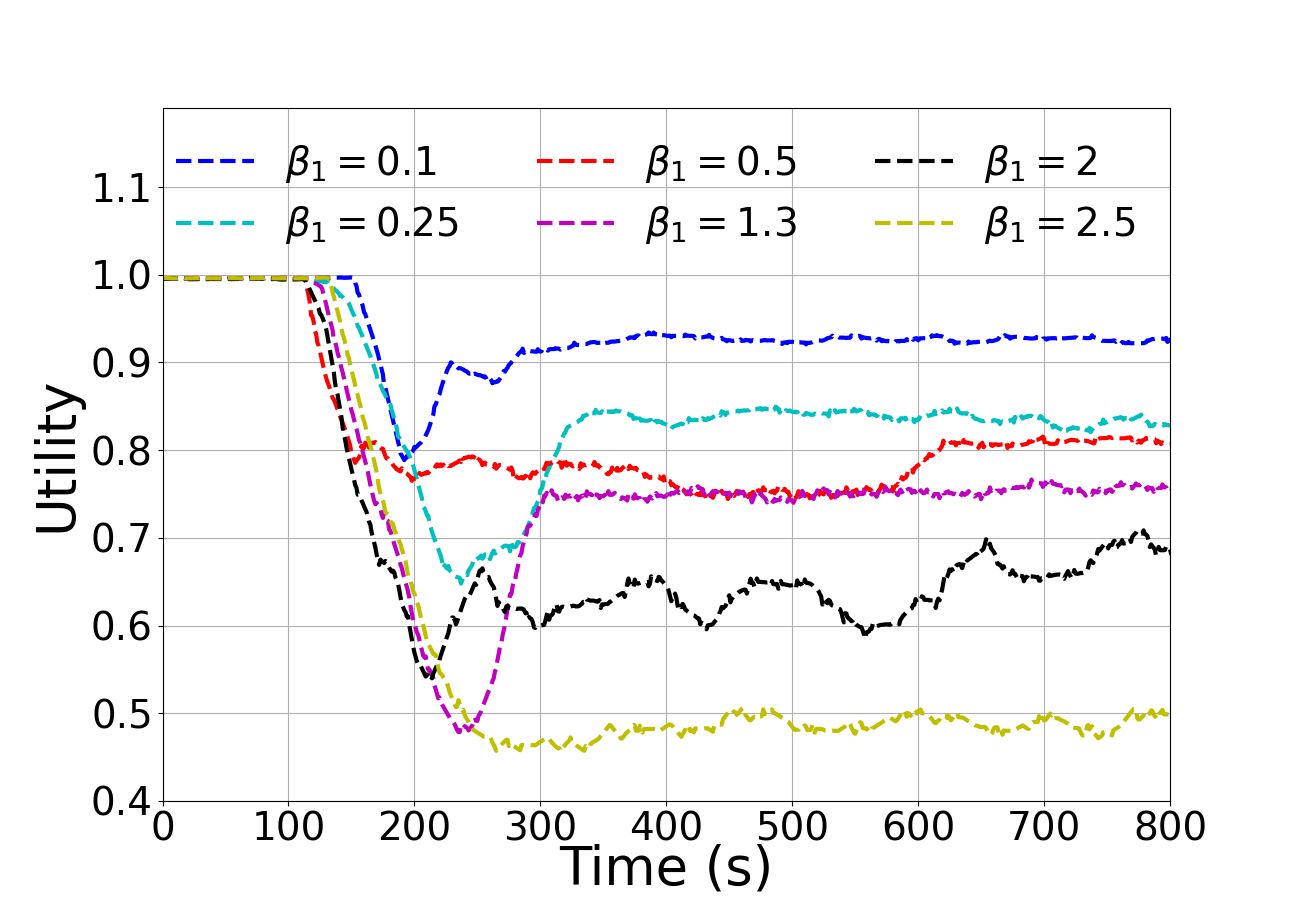}
  \caption{UE~2 utility.}
  \label{fig:ue2_utility_priority}
\end{subfigure}

\vspace{6pt}

\begin{subfigure}{0.45\linewidth}
  \centering
  \includegraphics[width=\linewidth]{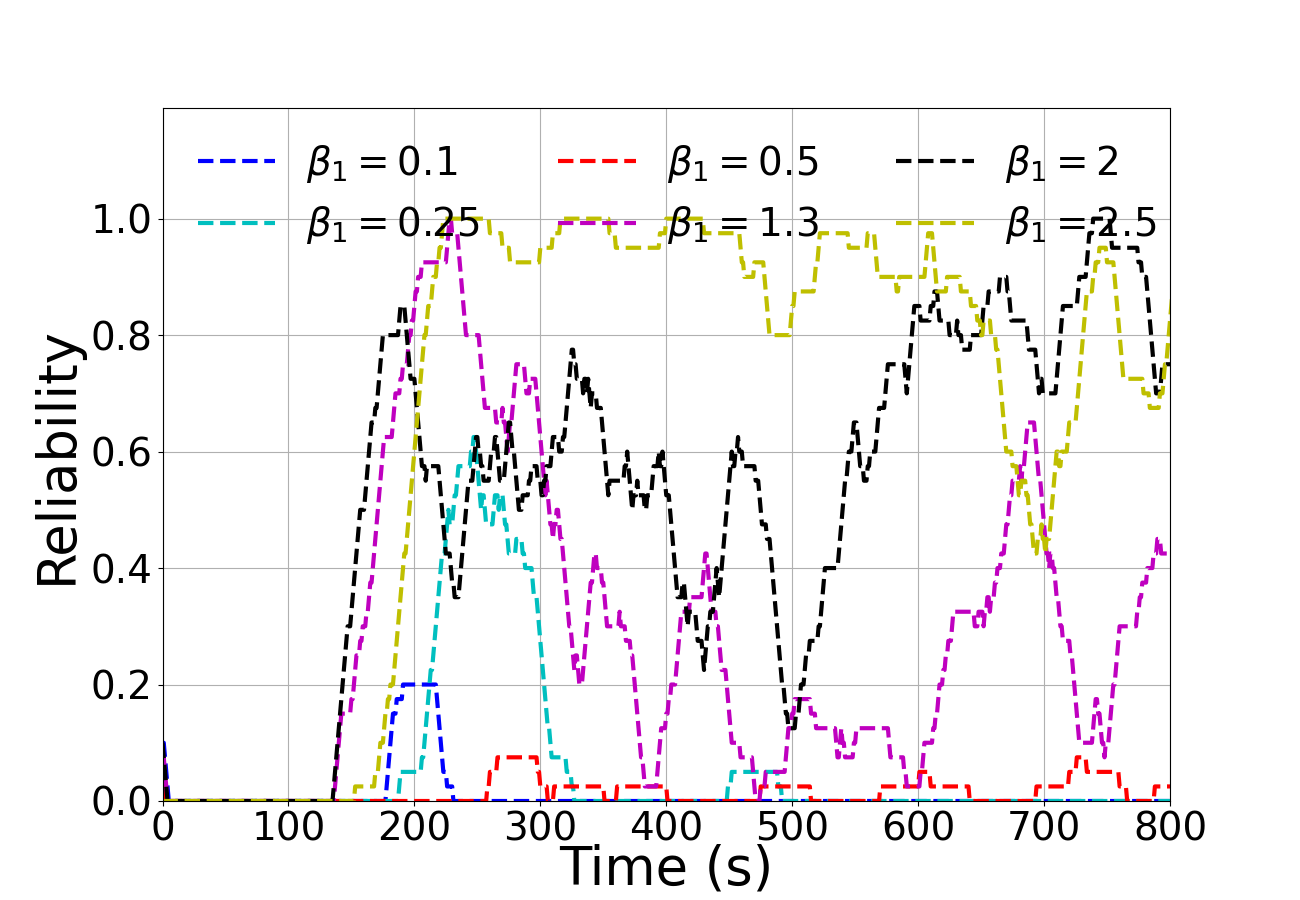}
  \caption{UE~1 reliability.}
  \label{fig:ue1_reliability_priority}
\end{subfigure}
\hfill
\begin{subfigure}{0.45\linewidth}
  \centering
  \includegraphics[width=\linewidth]{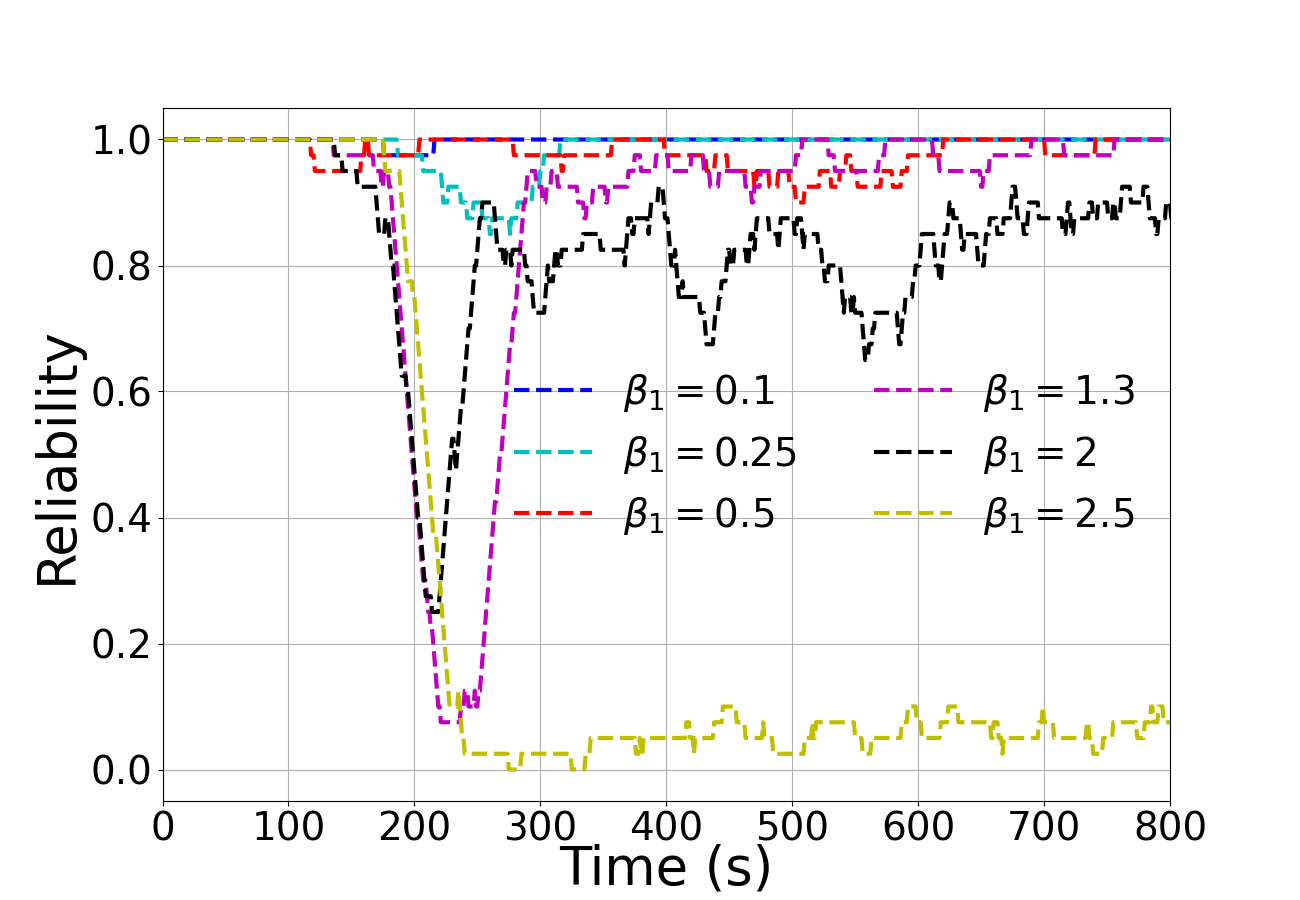}
  \caption{UE~2 reliability.}
  \label{fig:ue2_reliability_priority}
\end{subfigure}

\caption{Comparative analysis of window smoothed utility and reliability as functions of the priority factor of UE~1. (a) UE~1 utility. (b) UE~2 utility. (c) UE~1 reliability. (d) UE~2 reliability.}
\label{fig:priority-smoothed}
\end{figure}

\begin{figure}[h!]
\centering

\begin{subfigure}{0.45\linewidth}
  \centering
  \includegraphics[width=\linewidth]{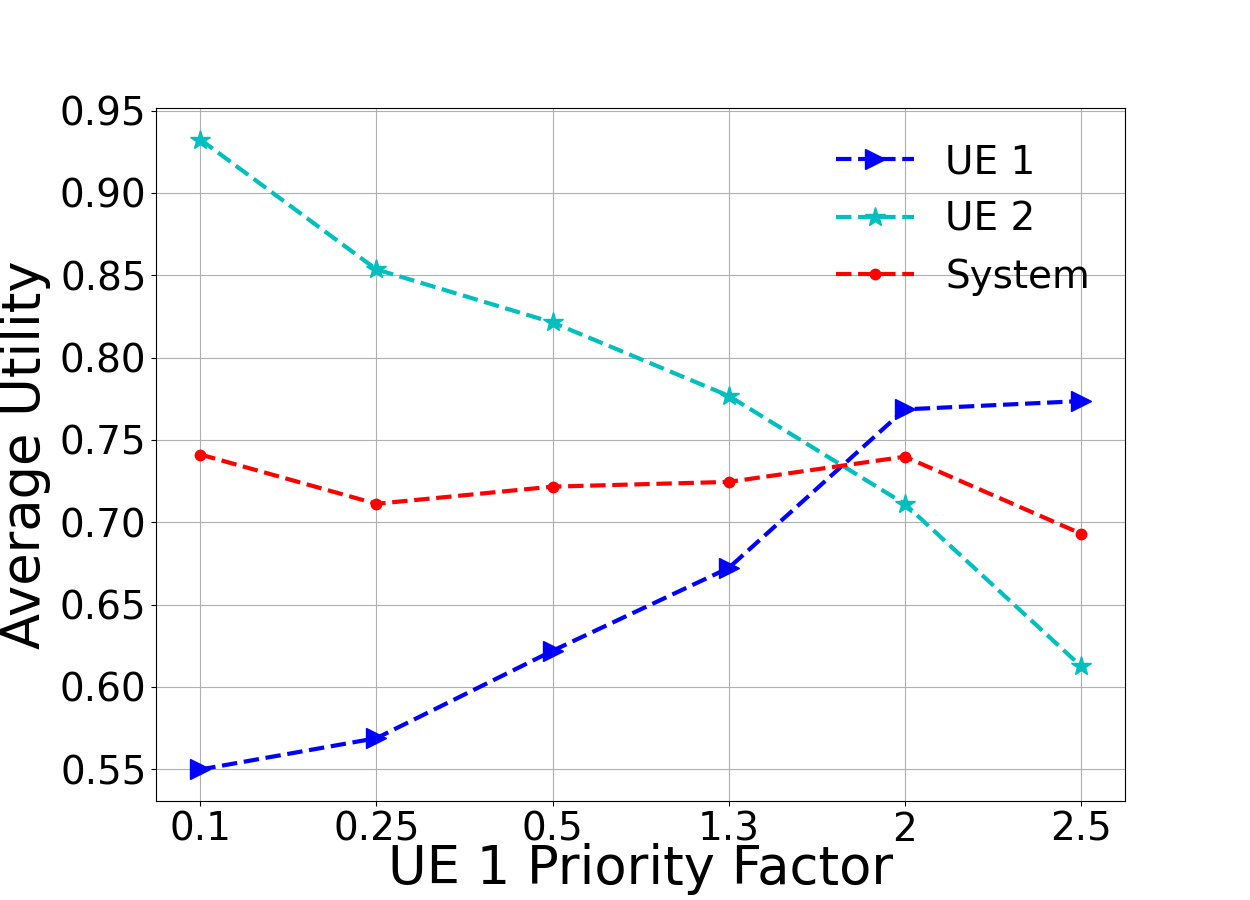}
  \caption{Time-averaged utility.}
  \label{fig:utility_priority}
\end{subfigure}
\hfill
\begin{subfigure}{0.45\linewidth}
  \centering
  \includegraphics[width=\linewidth]{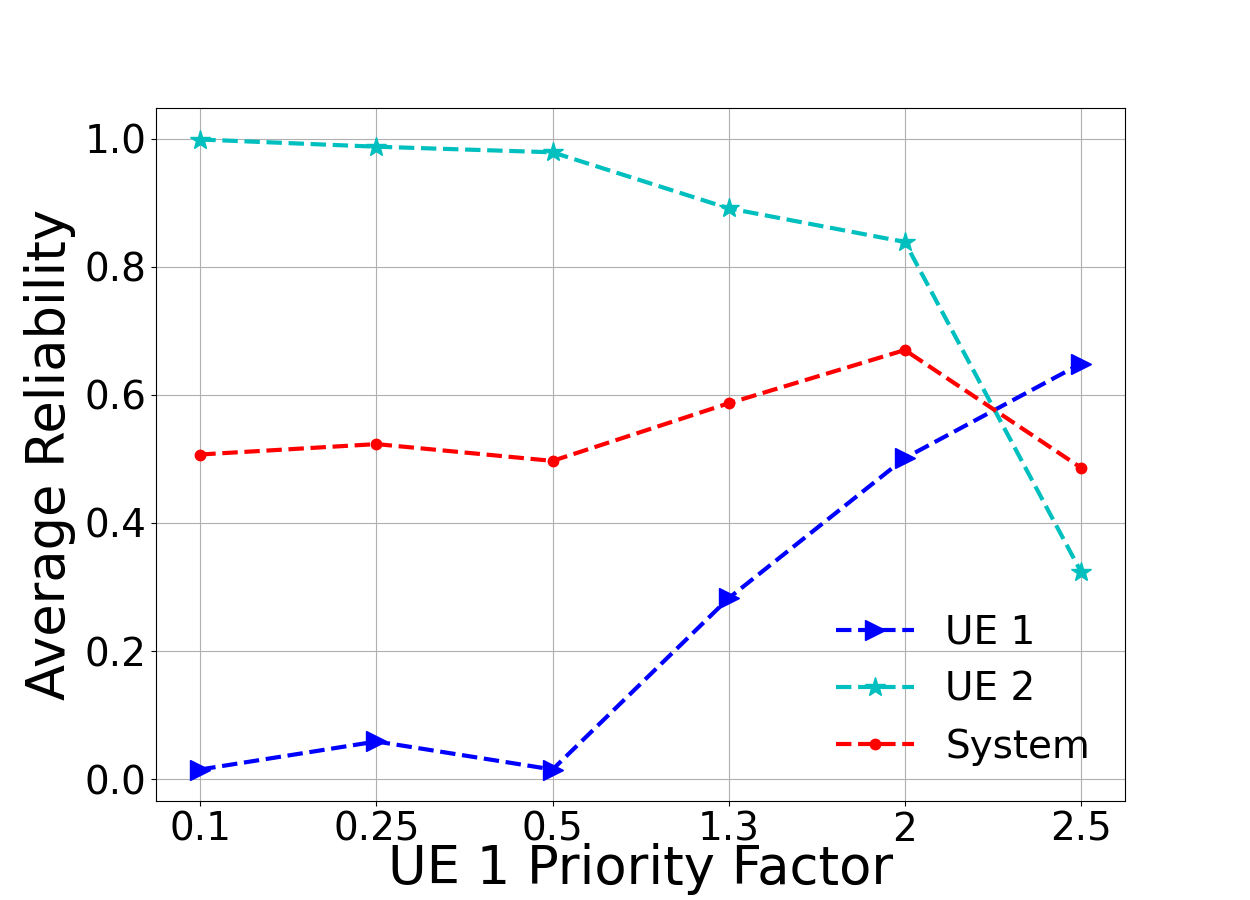}
  \caption{Time-averaged reliability.}
  \label{fig:reliability_priority}
\end{subfigure}

\caption{Comparative analysis of time-averaged utility and reliability as functions of the priority factor $\beta_1$ of UE~1. (a) Utility. (b) Reliability.}
\label{fig:priority}
\end{figure}

Priority-aware behavior is examined in Fig.~\ref{fig:priority-smoothed}, which shows window-smoothed utility and reliability for two users as the priority factor of UE~1 varies. Increasing priority yields monotonic improvements for UE~1 while inducing graceful degradation for UE~2, demonstrating controlled differentiation rather than abrupt starvation. The long-term effects of priority configuration are further quantified in Fig.~\ref{fig:priority}, where time-averaged utility and reliability vary smoothly with the priority factor $\beta_1$, indicating convergence to stable operating points.

\begin{figure}[h!]
\centering

\begin{subfigure}{0.45\linewidth}
  \centering
  \includegraphics[width=\linewidth]{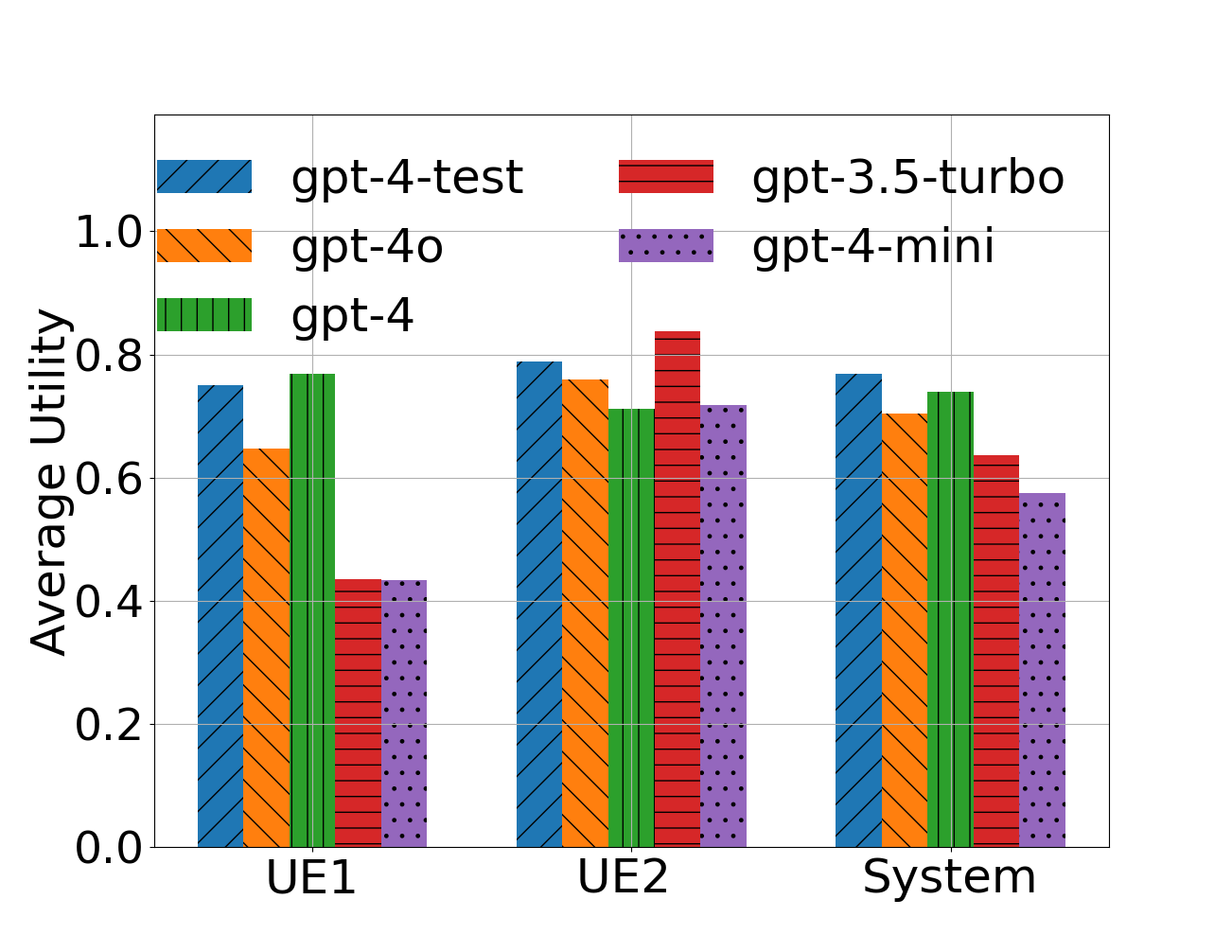}
  \caption{Time-averaged utility.}
  \label{fig:llm_utility}
\end{subfigure}
\hfill
\begin{subfigure}{0.45\linewidth}
  \centering
  \includegraphics[width=\linewidth]{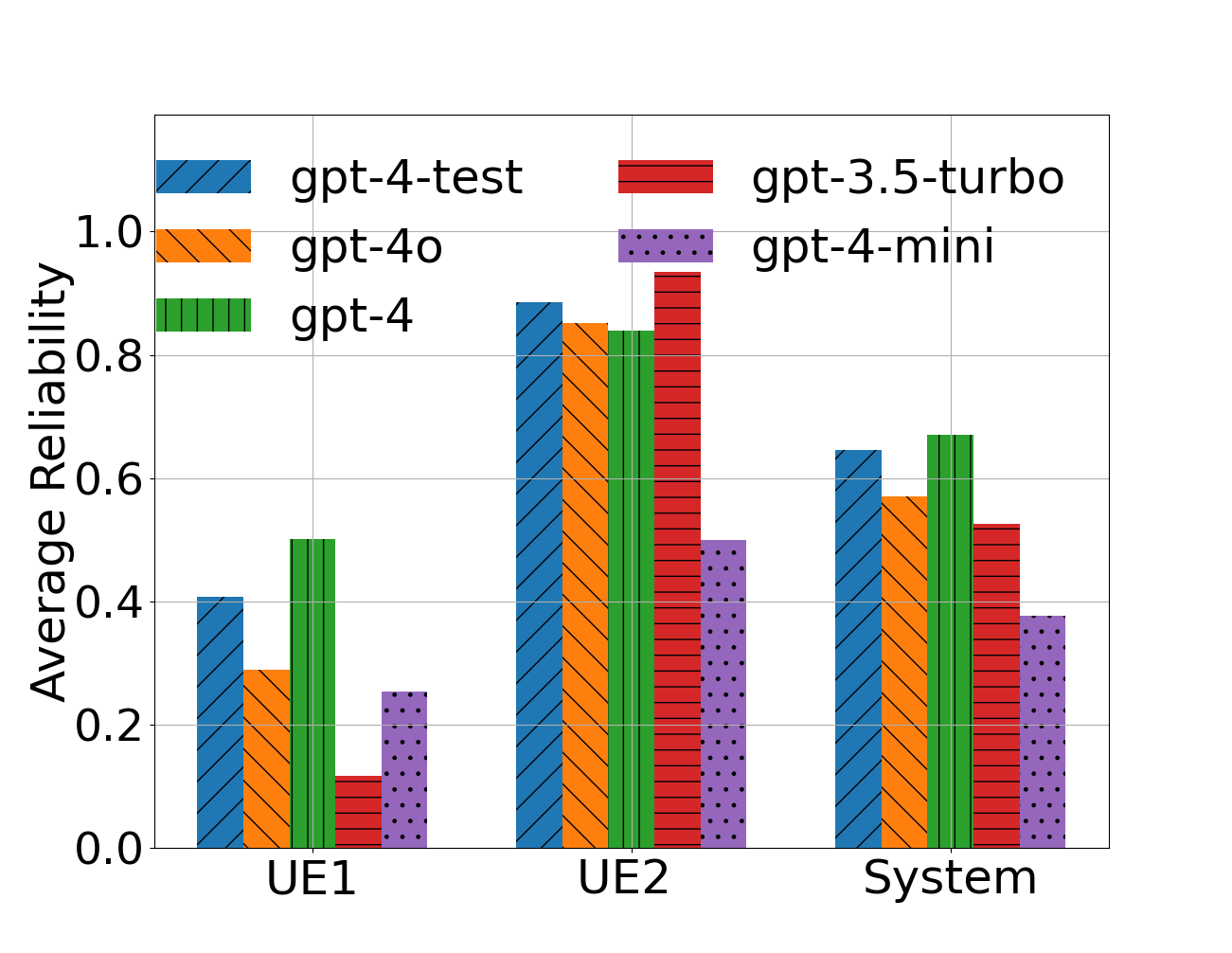}
  \caption{Time-averaged reliability.}
  \label{fig:llm_reliability}
\end{subfigure}

\caption{Comparative analysis of time-averaged utility and reliability across different LLM models. (a) Utility. (b) Reliability.}
\label{fig:llm_models}
\end{figure}

Finally, Fig.~\ref{fig:llm_models} compares different LLM models. While all evaluated models outperform heuristic baselines, more capable models achieve higher utility and improved reliability consistency, highlighting the impact of foundation model capacity on control effectiveness.

\subsection{Discussion}

This section demonstrates that LLMs can serve as effective cognitive controllers for resilient slice management by translating intent into quantitative policy guidance. Unlike optimization-based or reinforcement learning approaches that operate on fixed abstractions, LLM-driven control adapts to evolving context, priorities, and operational goals. When integrated with learning-based control layers, LLMs enable a multi-timescale resilience framework that combines semantic reasoning with fast, reactive adaptation, strengthening the robustness and interpretability of O-RAN slice management.

\chapter{Concluding Remarks}

\begin{figure*}[t]
\centering
\includegraphics[width=0.95\linewidth]{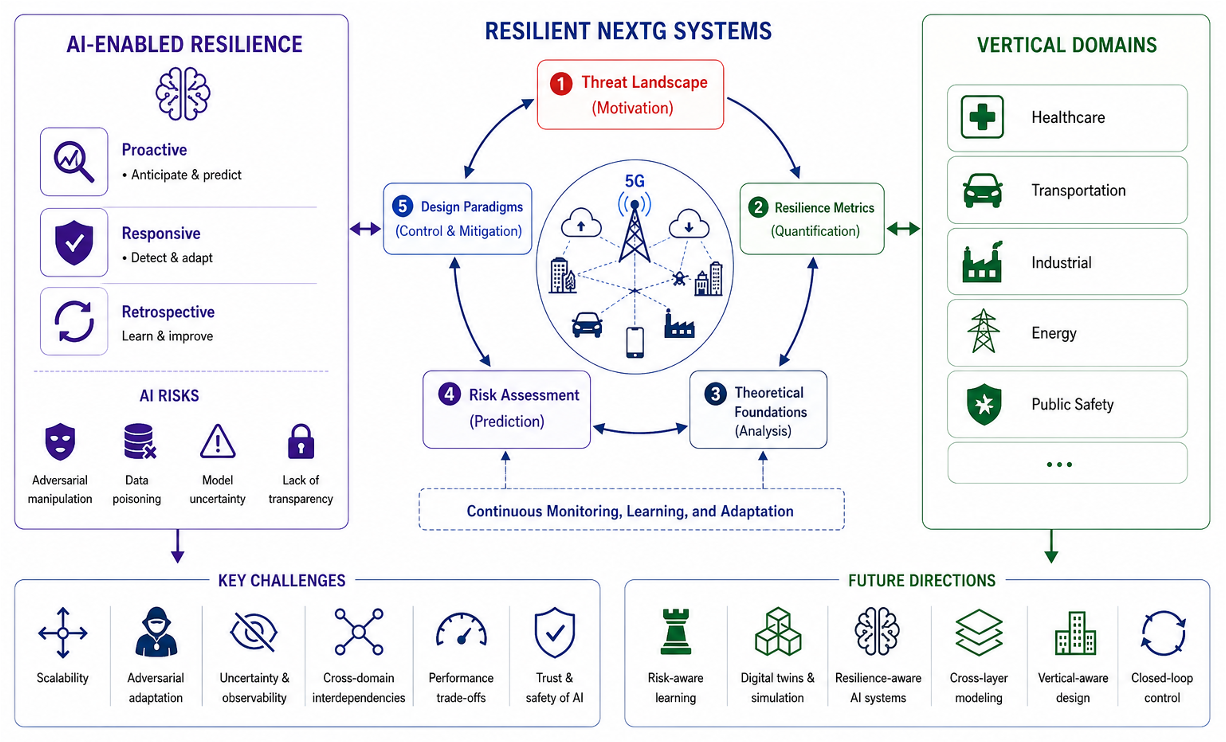}
\caption{\small 
Summary of the article, illustrating the interaction between threat landscape characterization, resilience quantification, theoretical foundations, risk assessment, and resilient system design within a continuous feedback cycle, along with key challenges and future research directions.
Summary and outlook of cyber resilience in NextG systems. 
}
\label{fig:nextg_resilience_framework}
\end{figure*}

This article has examined cyber resilience in next-generation (NextG) networks from complementary perspectives that include threat characterization, theoretical foundations, analytical methodologies, and system-level design paradigms. The chapters are organized to progressively move from problem framing to formalization and finally to actionable design insights. Figure~\ref{fig:nextg_resilience_framework} synthesizes these components into a unified system-level view, illustrating how resilience emerges from the interaction between modeling, analysis, and design under continuous learning and adaptation.

\section{Summary of the Article} 

The article begins by characterizing the evolving threat landscape of NextG networks. Unlike legacy telecommunications systems, NextG networks are deeply softwarized, virtualized, and integrated with cloud and edge computing infrastructures. This chapter identifies how architectural innovations, such as SDN, NFV, O-RAN, MEC, and network slicing, expand the attack surface and introduce new classes of cyber, physical, and AI-driven risks. By systematically mapping threats across core, transport, RAN, edge, and cloud domains, the chapter establishes that risks in NextG networks are inherently cross-domain and systemic, motivating the need for resilience-centric rather than perimeter-based security approaches. As depicted in Fig.~\ref{fig:nextg_resilience_framework}, this stage provides the foundational threat landscape that drives the need for resilience across interconnected domains.

Chapter 2 develops a rigorous conceptual foundation for resilience in NextG networks. It distinguishes resilience from robustness, fault tolerance, and reliability, emphasizing resilience as a dynamic, multi-stage capability encompassing preparation, absorption, recovery, and adaptation. The chapter introduces a lifecycle view of resilience (proactive, responsive, and retrospective) and discusses how each stage manifests itself in 5G and beyond. It also surveys metrics and evaluation approaches that go beyond static availability measures toward time-dependent performance degradation, recovery trajectories, and learning effects, providing a basis for quantitative resilience assessment. This corresponds to the resilience quantification component in Fig.~\ref{fig:nextg_resilience_framework}, which connects lifecycle stages with measurable system performance.

Chapter 3 provides the formal underpinnings necessary for principled resilience engineering. Control theory is used to model stability, disturbance rejection, and recovery under uncertainty. Dynamic and stochastic game theory captures adversarial interactions, incentive misalignment, and strategic behavior among attackers, defenders, and network stakeholders. Learning theory, particularly reinforcement learning and online adaptation—addresses nonstationarity, partial observability, and model uncertainty. Network-theoretic tools illuminate interdependencies and cascading failures. Together, these frameworks establish resilience as an emerging property that arises from feedback, strategic reasoning, and learning between interconnected subsystems. As illustrated in Fig.~\ref{fig:nextg_resilience_framework}, these theoretical foundations form the analytical core that supports both risk assessment and design.

Chapter 4 translates theory into analytical tools for assessing cyber risk. It introduces probabilistic and strategic risk assessment methods, including the use of digital twins to simulate failures, attacks, and recovery processes. The chapter emphasizes that risk in NextG networks is endogenous: adversaries adapt to defenses, and defensive actions reshape attacker incentives. It also highlights the growing role of agentic AI and large language models in automating risk analysis, threat intelligence synthesis, and scenario exploration, while cautioning that these tools themselves must be treated as part of the attack surface. In Fig.~\ref{fig:nextg_resilience_framework}, this corresponds to the risk assessment layer, which bridges theoretical analysis and system-level design under continuous monitoring and learning.

The final technical chapter focuses on concrete resilience design mechanisms. Through case studies on trust-aware resource management, risk-aware orchestration in edge–cloud systems, multi-agent reinforcement learning for slice management, and LLM-driven network control, the chapter demonstrates how resilience concepts can be operationalized. These examples illustrate how adaptive control, learning, and strategic reasoning can be embedded into real network management workflows to enable graceful degradation, rapid recovery, and long-term adaptation under uncertainty. This aligns with the design paradigms block in Fig.~\ref{fig:nextg_resilience_framework}, where mitigation and control strategies close the loop and influence future system behavior. 
Note that these components are not sequential but cyclic. Continuous monitoring, learning, and adaptation connect all stages, transforming resilience from a static property into an evolving system capability.

\section{Open Challenges and Future Directions}

As summarized in Fig.~\ref{fig:nextg_resilience_framework}, resilience in NextG systems is inherently multi-layered and cyclic. The tight coupling among threat modeling, quantification, analysis, and design introduces challenges that cannot be addressed in isolation.

Although this article presents frameworks for assessing, analyzing and designing cyber resilience in NextG networks, significant challenges remain to translate these principles into scalable, robust, and deployable systems. The preceding chapters show that resilience cannot be reduced to a single mechanism or metric. It emerges from the interaction among threat modeling, resilience quantification, theoretical analysis, risk assessment, and system design. This section synthesizes the limitations of current approaches and identifies future directions, with particular emphasis on AI-enabled resilience and vertical-specific requirements.

\subsection{Limitations of Current Approaches}

Existing resilience solutions remain fragmented across layers, domains, and methodologies. Many approaches focus on isolated components, such as the RAN, the core network, or edge-cloud orchestration, without fully accounting for the cross-domain dependencies that characterize NextG systems. However, disruptions in one layer can propagate through control-plane signaling, shared virtualized resources, network slices, and cyber-physical applications. This makes resilience fundamentally systemic rather than component-local.

A second limitation is that many methods rely on simplified assumptions about uncertainty and adversarial behavior. Classical probabilistic risk assessment and reliability models are useful for quantifying failures, but they often assume stationary distributions, known failure modes, or independent events. These assumptions become restrictive in adversarial environments where attackers observe defenses, adapt strategies, exploit misconfigurations, and combine multiple attack vectors across domains. As a result, future risk models must move beyond static failure analysis toward dynamic, strategic, and data-informed formulations.

Current resilience mechanisms are also frequently reactive and heuristic. Anomaly detection systems, orchestration policies, and recovery procedures are often tuned to known attack patterns or predefined failure scenarios. While such mechanisms are operationally useful, they may not provide formal guarantees on stability, safety, recovery time, or performance degradation under unseen attacks. This gap is especially important for mission-critical services, where delayed or incorrect responses can produce cascading consequences.

The growing use of AI introduces an additional limitation. AI-based methods can support prediction, automation, and adaptation, but they also create new forms of uncertainty and vulnerability. Learning models may drift as traffic patterns and adversarial behaviors evolve. Reinforcement learning controllers may behave unpredictably outside their training regimes. Large language models embedded in security or network management workflows may hallucinate, leak sensitive information, or be manipulated through prompts and tool interfaces. Thus, AI must be treated not merely as a tool for resilience, but also as an object of resilience analysis.

\subsection{Open Technical Challenges}

A central open challenge is scalability. NextG networks involve massive numbers of devices, dynamic slices, heterogeneous services, distributed edge resources, and cloud-native control functions. Resilience mechanisms must therefore operate at scale without introducing excessive computational overhead, signaling burden, or management complexity. The challenge is not only to design algorithms that work in principle, but to ensure that they remain feasible in operational environments where decisions must be made under latency, bandwidth, and resource constraints.

Robustness under adversarial adaptation is another fundamental challenge. In NextG systems, attackers are not passive disturbances. They learn from defensive behavior, probe system responses, and modify their strategies over time. This creates a co-evolutionary setting in which the defender’s actions reshape the attacker’s incentives and observations. Future work must develop models and algorithms that remain effective under adaptive, strategic, and partially informed adversaries. Game-theoretic learning, robust control, and risk-sensitive reinforcement learning provide promising foundations, but their integration into deployable network management systems remains an open problem.

Uncertainty and partial observability further complicate resilience engineering. Operators rarely observe the full system state, and available measurements may be noisy, delayed, incomplete, or adversarially manipulated. This limits the effectiveness of both model-based and data-driven approaches. Future frameworks must explicitly reason about uncertainty in system states, threat models, data quality, and decision outcomes. This requires combining estimation, filtering, learning, and robust decision-making in ways that can support real-time operation.

Cross-domain interdependence is also a major unresolved issue. NextG networks connect communication, computation, control, and physical infrastructures. A disruption in an edge node may affect a network slice; a slice failure may affect an industrial workflow; and a cyber incident may propagate into physical consequences. Understanding these cascading effects requires multi-layer and multi-scale models that capture interactions across network domains, application domains, and time scales. Such models are essential for designing defenses that are coordinated rather than locally optimal.

Finally, there is a persistent challenge in balancing resilience with efficiency and performance. Stronger isolation, redundancy, monitoring, and verification can improve resilience, but they also increase cost, latency, energy consumption, and operational complexity. Conversely, highly efficient systems may become brittle when exposed to unexpected shocks. Future research must therefore formulate resilience as a multi-objective design problem, where risk reduction is balanced against quality of service, resource utilization, and economic cost.

\subsection{AI-Enabled Resilience}

Artificial intelligence will play an increasingly important role in NextG resilience. AI can enhance proactive resilience by identifying latent vulnerabilities, detecting misconfigurations, analyzing supply-chain risks, and exploring large configuration spaces before deployment. It can support responsive resilience by enabling real-time anomaly detection, root-cause analysis, adaptive orchestration, and autonomous mitigation across slices, edge nodes, and cloud resources. It can also strengthen retrospective resilience by mining forensic logs, learning from past incidents, updating detection rules, and improving recovery policies.

However, the role of AI is fundamentally dual. AI systems are not only enablers of resilience; they also become part of the attack surface. This dual role is explicitly reflected in Fig.~\ref{fig:nextg_resilience_framework}, where AI-enabled resilience spans proactive, responsive, and retrospective stages while simultaneously introducing new risks that must be incorporated into the resilience loop. Learning algorithms can be manipulated through poisoned data, adversarial inputs, or distribution shifts. LLM-based agents used for network management may make incorrect inferences, over-trust incomplete evidence, or execute unsafe tool actions. These risks are especially significant when AI systems are given autonomy over operational decisions.

This motivates a shift from \emph{AI-assisted resilience} to \emph{resilience-aware AI}. In this view, AI models must be designed with explicit robustness requirements, uncertainty estimates, fallback mechanisms, and human oversight. AI should not operate as an unconstrained decision-maker, but as part of a hierarchical control architecture in which formal policies, safety constraints, and domain-specific rules bound its actions. Formal verification, game-theoretic reasoning, and risk-sensitive learning will be essential for ensuring that AI-driven autonomy improves resilience rather than amplifying systemic risk.

An important future direction is the development of AI systems that can explain their risk assessments and justify their recommended actions. Interpretability is not merely a usability concern; it is central to trust, accountability, and operational adoption. Operators must be able to understand why an AI system recommends isolating a slice, migrating a workload, throttling traffic, or changing an orchestration policy. Without such explanations, AI-enabled resilience may remain difficult to validate and govern.

\subsection{Vertical-Specific Resilience and Domain Constraints}

NextG networks increasingly support vertical domains whose requirements differ substantially from conventional consumer connectivity. Healthcare, transportation, industrial automation, energy systems, public safety, and smart cities each impose distinct resilience specifications. This means that resilience cannot be defined solely at the network level. It must be interpreted in relation to the mission, risk tolerance, safety constraints, and regulatory environment of each domain.

In healthcare, resilience is closely tied to ultra-low latency, bounded jitter, privacy, and deterministic recovery. A short communication disruption in remote surgery or patient monitoring may have consequences that are far more severe than a comparable disruption in best-effort consumer traffic. In transportation and autonomous mobility, resilience must account for interactions between communication networks, control algorithms, vehicles, infrastructure, and human safety. A network failure may not simply reduce service quality; it may alter physical behavior and require fail-safe operation.

Industrial automation and smart manufacturing introduce another set of requirements. These systems emphasize continuity, isolation, predictability, and coordination among tightly coupled workflows. A cyber incident affecting one production cell may propagate through scheduling, sensing, actuation, and supply-chain dependencies. Energy systems and smart grids present still another challenge, because disruptions may cascade across cyber and physical layers and may be subject to strict regulatory and reliability constraints.

These examples show that resilience metrics must be contextualized. Acceptable degradation, recovery time, redundancy level, and automation policy vary across verticals. A uniform resilience policy may be inefficient in one domain and unsafe in another. Future work must therefore develop vertical-aware resilience frameworks in which network slicing, orchestration, control, and recovery policies are informed by domain semantics and mission priorities.

Digital twins and simulation-based validation will be particularly important in this direction. They can provide controlled environments for testing domain-specific disruption scenarios, evaluating recovery trajectories, and comparing alternative resilience strategies before deployment. However, for digital twins to be effective, they must accurately represent not only network behavior but also the coupled application, physical, and organizational processes that define real consequences.

\subsection{Future Research Outlook}

Future research on cyber resilience in NextG networks should integrate three dimensions. The first is theoretical rigor, including control, game theory, learning, and network science, which provides the tools for modeling dynamics, adversarial behavior, uncertainty, and interdependence. The second is AI-enabled automation, which provides the computational capability to monitor, predict, and adapt at scale. The third is domain-specific grounding, which ensures that resilience objectives reflect the consequences and constraints of real applications.

The most promising direction lies in unifying these dimensions into deployable architectures. Risk-aware and game-theoretic learning can support adaptive policies under strategic attacks. Digital twins can connect models with realistic scenario testing. Resilience-aware AI can enable automation while preserving safety and accountability. Cross-layer models can reveal systemic vulnerabilities that are invisible from any single domain. Vertical-aware design can ensure that resilience is measured and optimized according to mission needs rather than generic network metrics.

Ultimately, cyber resilience in NextG networks should be viewed as an evolving systems property. It is shaped by the continuous interaction among threats, risk assessment, theoretical modeling, AI-enabled decision-making, and domain-specific design. Advancing this vision will require interdisciplinary collaboration among networking, cybersecurity, control theory, machine learning, and cyber-physical systems communities. It will also require stronger links between theory and deployment, so that resilience mechanisms can be validated not only in abstract models, but also in realistic operational environments.

\backmatter  

\renewcommand{\bibfont}{\scriptsize}
\bibliography{references,refxx}

\end{document}